
\magnification=\magstephalf

\newbox\SlashedBox 
\def\slashed#1{\setbox\SlashedBox=\hbox{#1}
\hbox to 0pt{\hbox to 1\wd\SlashedBox{\hfil/\hfil}\hss}{#1}}
\def\hboxtosizeof#1#2{\setbox\SlashedBox=\hbox{#1}
\hbox to 1\wd\SlashedBox{#2}}

\def\mathslashed#1{\setbox\SlashedBox=\hbox{$#1$}
\hbox to 0pt{\hbox to 1\wd\SlashedBox{\hfil/\hfil}\hss}#1}

\def\ifsmall{\iffalse}  
\def\titlepagefont{}  

\def\DefineTeXgraphics{%
\special{ps::[global] /TeXgraphics { } def}}  

\def\today{\ifcase\month\or January\or February\or March\or April\or May
\or June\or July\or August\or September\or October\or November\or
December\fi\space\number\day, \number\year}
\def\eatPrefix19{}
\def\Year{\expandafter\eatPrefix\the\year}
\newcount\hours \newcount\minutes
\def\monthname{\ifcase\month\or
January\or February\or March\or April\or May\or June\or July\or
August\or September\or October\or November\or December\fi}
\def\shortmonthname{\ifcase\month\or
Jan\or Feb\or Mar\or Apr\or May\or Jun\or Jul\or
Aug\or Sep\or Oct\or Nov\or Dec\fi}

\def\TimeStamp{\hours\the\time\divide\hours by60%
\minutes -\the\time\divide\minutes by60\multiply\minutes by60%
\advance\minutes by\the\time%
${\rm \shortmonthname}\cdot\if\day<10{}0\fi\the\day\cdot\the\year%
\qquad\the\hours:\if\minutes<10{}0\fi\the\minutes$}




\def\Title#1{%
\vskip 1in{\titlefont\centerline{#1}}\vskip .5in}
 
\def\Date#1{\leftline{#1}\tenrm\supereject%
\global\hsize=\hsbody\global\hoffset=\hbodyoffset%
\footline={\hss\tenrm\folio\hss}}

\newif\ifdraftmode
\newif\ifleftlabels  

\def\nolabels{\def\wrlabeL##1{}\def\eqlabeL##1{}\def\reflabeL##1{}}
\def\writelabels{\def\wrlabeL##1{\leavevmode\vadjust{\rlap{\smash%
{\line{{\escapechar=` \hfill\rlap{\sevenrm\hskip.03in\string##1}}}}}}}%
\def\eqlabeL##1{{\escapechar-1\rlap{\sevenrm\hskip.05in\string##1}}}%
\def\reflabeL##1{\noexpand\rlap{\noexpand\sevenrm[\string##1]}}}
\def\writeleftlabels{\def\wrlabeL##1{\leavevmode\vadjust{\rlap{\smash%
{\line{{\escapechar=` \hfill\rlap{\sevenrm\hskip.03in\string##1}}}}}}}%
\def\eqlabeL##1{{\escapechar-1%
\rlap{\sixrm\hskip.05in\string##1}%
\llap{\sevenrm\string##1\hskip.03in\hbox to \hsize{}}}}%
\def\reflabeL##1{\noexpand\rlap{\noexpand\sevenrm[\string##1]}}}
\nolabels

\input hyperbasics.tex

\newdimen\fullhsize
\newdimen\hstitle
\hstitle=\hsize 
\newdimen\hsbody
\hsbody=\hsize 
\newdimen\hbodyoffset
\hbodyoffset=\hoffset 
\newbox\leftpage
\def\abstract#1{#1}
\def\rotated{\special{ps: landscape}
\magnification=1000  
\baselineskip=14pt
\global\hstitle=9truein\global\hsbody=4.75truein
\global\vsize=7truein\global\voffset=-.31truein
\global\hoffset=-0.54in\global\hbodyoffset=-.54truein
\global\fullhsize=10truein
\def\DefineTeXgraphics{%
\special{ps::[global] 
/TeXgraphics {currentpoint translate 0.7 0.7 scale
              -80 0.72 mul -1000 0.72 mul translate} def}}
\let\lr=L
\def\ifsmall{\iftrue}
\def\titlepagefont{\twelvepoint}
\trueseventeenpoint
\def\almostshipout##1{\if L\lr \count1=1
      \global\setbox\leftpage=##1 \global\let\lr=R
   \else \count1=2
      \shipout\vbox{\hbox to\fullhsize{\box\leftpage\hfil##1}}
      \global\let\lr=L\fi}

\output={\ifnum\count0=1 
 \shipout\vbox{\hbox to \fullhsize{\hfill\pagebody\hfill}}\advancepageno
 \else
 \almostshipout{\leftline{\vbox{\pagebody\makefootline}}}\advancepageno 
 \fi}

\def\abstract##1{{\leftskip=1.5in\rightskip=1.5in ##1\par}} }

\def\linemessage#1{\immediate\write16{#1}}

\global\newcount\secno \global\secno=0
\global\newcount\appno \global\appno=0
\global\newcount\meqno \global\meqno=1
\global\newcount\subsecno \global\subsecno=0
\global\newcount\figno \global\figno=0

\newif\ifAnyCounterChanged
\let\terminator=\relax
\def\normalize#1{\ifx#1\terminator\let\next=\relax\else%
\if#1i\aftergroup i\else\if#1v\aftergroup v\else\if#1x\aftergroup x%
\else\if#1l\aftergroup l\else\if#1c\aftergroup c\else%
\if#1m\aftergroup m\else%
\if#1I\aftergroup I\else\if#1V\aftergroup V\else\if#1X\aftergroup X%
\else\if#1L\aftergroup L\else\if#1C\aftergroup C\else%
\if#1M\aftergroup M\else\aftergroup#1\fi\fi\fi\fi\fi\fi\fi\fi\fi\fi\fi\fi%
\let\next=\normalize\fi%
\next}
\def\makeNormal#1#2{\def\doNormalDef{\edef#1}\begingroup%
\aftergroup\doNormalDef\aftergroup{\normalize#2\terminator\aftergroup}%
\endgroup}

\def\warnIfChanged#1#2{%
\ifundef#1
\else\begingroup%
\edef\oldDefinitionOfCounter{#1}\edef\newDefinitionOfCounter{#2}%
\ifx\oldDefinitionOfCounter\newDefinitionOfCounter%
\else%
\linemessage{Warning: definition of \noexpand#1 has changed.}%
\global\AnyCounterChangedtrue\fi\endgroup\fi}

\def\Section#1{\global\advance\secno by1\relax\global\meqno=1%
\global\subsecno=0%
\bigbreak\bigskip
\centerline{\twelvepoint \bf %
\the\secno. #1}%
\par\nobreak\medskip\nobreak}
\def\tagsection#1{%
\warnIfChanged#1{\the\secno}%
\xdef#1{\the\secno}%
\ifWritingAuxFile\immediate\write\auxfile{\noexpand\xdef\noexpand#1{#1}}\fi%
}
\def\section{\Section}
\def\Subsection#1{\global\advance\subsecno by1\relax\medskip %
\leftline{\bf\the\secno.\the\subsecno\ #1}%
\par\nobreak\smallskip\nobreak}
\def\tagsubsection#1{%
\warnIfChanged#1{\the\secno.\the\subsecno}%
\xdef#1{\the\secno.\the\subsecno}%
\ifWritingAuxFile\immediate\write\auxfile{\noexpand\xdef\noexpand#1{#1}}\fi%
}

\def\subsection{\Subsection}

\def\romappno{\uppercase\expandafter{\romannumeral\appno}}
\def\makeNormalizedRomappno{%
\expandafter\makeNormal\expandafter\normalizedromappno%
\expandafter{\romannumeral\appno}%
\edef\normalizedromappno{\uppercase{\normalizedromappno}}}
\def\Appendix#1{\global\advance\appno by1\relax\global\meqno=1\global\secno=0%
\global\subsecno=0%
\bigbreak\bigskip
\centerline{\twelvepoint \bf Appendix %
\romappno. #1}%
\par\nobreak\medskip\nobreak}
\def\tagappendix#1{\makeNormalizedRomappno%
\warnIfChanged#1{\normalizedromappno}%
\xdef#1{\normalizedromappno}%
\ifWritingAuxFile\immediate\write\auxfile{\noexpand\xdef\noexpand#1{#1}}\fi%
}
\def\appendix{\Appendix}
\def\Subappendix#1{\global\advance\subsecno by1\relax\medskip %
\leftline{\bf\romappno.\the\subsecno\ #1}%
\par\nobreak\smallskip\nobreak}
\def\tagsubappendix#1{\makeNormalizedRomappno%
\warnIfChanged#1{\normalizedromappno.\the\subsecno}%
\xdef#1{\normalizedromappno.\the\subsecno}%
\ifWritingAuxFile\immediate\write\auxfile{\noexpand\xdef\noexpand#1{#1}}\fi%
}

\def\subappendix{\Subappendix}

\def\eqn#1{\makeNormalizedRomappno%
\ifnum\secno>0%
  \warnIfChanged#1{\the\secno.\the\meqno}%
  \eqno(\the\secno.\the\meqno)\xdef#1{\the\secno.\the\meqno}%
     \global\advance\meqno by1
\else\ifnum\appno>0%
  \warnIfChanged#1{\normalizedromappno.\the\meqno}%
  \eqno({\rm\romappno}.\the\meqno)%
      \xdef#1{\normalizedromappno.\the\meqno}%
     \global\advance\meqno by1
\else%
  \warnIfChanged#1{\the\meqno}%
  \eqno(\the\meqno)\xdef#1{\the\meqno}%
     \global\advance\meqno by1
\fi\fi%
\eqlabeL#1%
\ifWritingAuxFile\immediate\write\auxfile{\noexpand\xdef\noexpand#1{#1}}\fi%
}
\def\defeqn#1{\makeNormalizedRomappno%
\ifnum\secno>0%
  \warnIfChanged#1{\the\secno.\the\meqno}%
  \xdef#1{\the\secno.\the\meqno}%
     \global\advance\meqno by1
\else\ifnum\appno>0%
  \warnIfChanged#1{\normalizedromappno.\the\meqno}%
  \xdef#1{\normalizedromappno.\the\meqno}%
     \global\advance\meqno by1
\else%
  \warnIfChanged#1{\the\meqno}%
  \xdef#1{\the\meqno}%
     \global\advance\meqno by1
\fi\fi%
\eqlabeL#1%
\ifWritingAuxFile\immediate\write\auxfile{\noexpand\xdef\noexpand#1{#1}}\fi%
}
\def\anoneqn{\makeNormalizedRomappno%
\ifnum\secno>0
  \eqno(\the\secno.\the\meqno)%
     \global\advance\meqno by1
\else\ifnum\appno>0
  \eqno({\rm\normalizedromappno}.\the\meqno)%
     \global\advance\meqno by1
\else
  \eqno(\the\meqno)%
     \global\advance\meqno by1
\fi\fi%
}
\def\mfig#1#2{\ifx#20
\else\global\advance\figno by1%
\relax#1\the\figno%
\warnIfChanged#2{\the\figno}%
\xdef#2{\the\figno}%
\reflabeL#2%
\ifWritingAuxFile\immediate\write\auxfile{\noexpand\xdef\noexpand#2{#2}}\fi\fi%
}

\def\fig#1{\mfig{fig.\ ~}#1}

\catcode`@=11 

\newif\ifFiguresInText\FiguresInTexttrue
\newif\if@FigureFileCreated
\newwrite\capfile
\newwrite\figfile

\newif\ifcaption
\captiontrue
\def\captionsize{\tenrm}
\def\PlaceTextFigure#1#2#3#4{%
\vskip 0.5truein%
#3\hfil\epsfbox{#4}\hfil\break%
\ifcaption\hfil\vbox{\captionsize Figure #1. #2}\hfil\fi%
\vskip10pt}
\def\PlaceEndFigure#1#2{%
\epsfxsize=\hsize\epsfbox{#2}\vfill\centerline{Figure #1.}\eject}

\def\LoadFigure#1#2#3#4{%
\ifundef#1{\phantom{\mfig{}#1}}\else
\fi%
\ifFiguresInText
\PlaceTextFigure{#1}{#2}{#3}{#4}%
\else
\if@FigureFileCreated\else%
\immediate\openout\capfile=\jobname.caps%
\immediate\openout\figfile=\jobname.figs%
@FigureFileCreatedtrue\fi%
\immediate\write\capfile{\noexpand\item{Figure \noexpand#1.\ }{#2}\vskip10pt}%
\immediate\write\figfile{\noexpand\PlaceEndFigure\noexpand#1{\noexpand#4}}%
\fi}

\def\listfigs{\ifFiguresInText\else%
\vfill\eject\immediate\closeout\capfile
\immediate\closeout\figfile%
\centerline{{\bf Figures}}\bigskip\frenchspacing%
\catcode`@=11 
\def\captionsize{\tenrm}
\input \jobname.caps\vfill\eject\nonfrenchspacing%
\catcode`\@=\active
\catcode`@=12  
\input\jobname.figs\fi}

\font\ninerm=cmr9
\font\eightrm=cmr8
\font\sixrm=cmr6

\def\loadtrueseventeenpoint{
 \font\seventeenrm=cmr10 at 17.28truept
 \font\seventeeni=cmmi10 at 17.28truept
 \font\seventeenbf=cmbx10 at 17.28truept
 \font\seventeenit=cmti10 at 17.28truept
 \font\seventeensl=cmsl10 at 17.28truept
 \font\seventeensy=cmsy10 at 17.28truept
}
\def\loadfourteenpoint{
\font\fourteenrm=cmr10 at 14.4pt
\font\fourteeni=cmmi10 at 14.4pt
\font\fourteenit=cmti10 at 14.4pt
\font\fourteensl=cmsl10 at 14.4pt
\font\fourteensy=cmsy10 at 14.4pt
\font\fourteenbf=cmbx10 at 14.4pt
}
\def\loadtruetwelvepoint{
\font\twelverm=cmr10 at 12truept
\font\twelvei=cmmi10 at 12truept
\font\twelveit=cmti10 at 12truept
\font\twelvesl=cmsl10 at 12truept
\font\twelvesy=cmsy10 at 12truept
\font\twelvebf=cmbx10 at 12truept
}

\font\ninei=cmmi9
\font\eighti=cmmi8
\font\sixi=cmmi6
\skewchar\ninei='177 \skewchar\eighti='177 \skewchar\sixi='177

\font\ninesy=cmsy9
\font\eightsy=cmsy8
\font\sixsy=cmsy6
\skewchar\ninesy='60 \skewchar\eightsy='60 \skewchar\sixsy='60

\font\ninebf=cmbx9
\font\eightbf=cmbx8
\font\sixbf=cmbx6

\font\ninett=cmtt9
\font\eighttt=cmtt8

\hyphenchar\tentt=-1 
\hyphenchar\ninett=-1
\hyphenchar\eighttt=-1         

\font\ninesl=cmsl9
\font\eightsl=cmsl8

\font\nineit=cmti9
\font\eightit=cmti8

                      
\newskip\ttglue
\def\tenpoint{\def\rm{\fam0\tenrm}%
  \textfont0=\tenrm \scriptfont0=\sevenrm \scriptscriptfont0=\fiverm
  \textfont1=\teni \scriptfont1=\seveni \scriptscriptfont1=\fivei
  \textfont2=\tensy \scriptfont2=\sevensy \scriptscriptfont2=\fivesy
  \textfont3=\tenex \scriptfont3=\tenex \scriptscriptfont3=\tenex
  \def\it{\fam\itfam\tenit}\textfont\itfam=\tenit
  \def\sl{\fam\slfam\tensl}\textfont\slfam=\tensl
  \def\bf{\fam\bffam\tenbf}\textfont\bffam=\tenbf \scriptfont\bffam=\sevenbf
  \scriptscriptfont\bffam=\fivebf
  \normalbaselineskip=12pt
  \let\sc=\eightrm
  \let\big=\tenbig
  \setbox\strutbox=\hbox{\vrule height8.5pt depth3.5pt width\z@}%
  \normalbaselines\rm}

\def\twelvepoint{\def\rm{\fam0\twelverm}%
  \textfont0=\twelverm \scriptfont0=\ninerm \scriptscriptfont0=\sevenrm
  \textfont1=\twelvei \scriptfont1=\ninei \scriptscriptfont1=\seveni
  \textfont2=\twelvesy \scriptfont2=\ninesy \scriptscriptfont2=\sevensy
  \textfont3=\tenex \scriptfont3=\tenex \scriptscriptfont3=\tenex
  \def\it{\fam\itfam\twelveit}\textfont\itfam=\twelveit
  \def\sl{\fam\slfam\twelvesl}\textfont\slfam=\twelvesl
  \def\bf{\fam\bffam\twelvebf}\textfont\bffam=\twelvebf%
  \scriptfont\bffam=\ninebf
  \scriptscriptfont\bffam=\sevenbf
  \normalbaselineskip=12pt
  \let\sc=\eightrm
  \let\big=\tenbig
  \setbox\strutbox=\hbox{\vrule height8.5pt depth3.5pt width\z@}%
  \normalbaselines\rm}

\def\fourteenpoint{\def\rm{\fam0\fourteenrm}%
  \textfont0=\fourteenrm \scriptfont0=\tenrm \scriptscriptfont0=\sevenrm
  \textfont1=\fourteeni \scriptfont1=\teni \scriptscriptfont1=\seveni
  \textfont2=\fourteensy \scriptfont2=\tensy \scriptscriptfont2=\sevensy
  \textfont3=\tenex \scriptfont3=\tenex \scriptscriptfont3=\tenex
  \def\it{\fam\itfam\fourteenit}\textfont\itfam=\fourteenit
  \def\sl{\fam\slfam\fourteensl}\textfont\slfam=\fourteensl
  \def\bf{\fam\bffam\fourteenbf}\textfont\bffam=\fourteenbf%
  \scriptfont\bffam=\tenbf
  \scriptscriptfont\bffam=\sevenbf
  \normalbaselineskip=17pt
  \let\sc=\elevenrm
  \let\big=\tenbig                                          
  \setbox\strutbox=\hbox{\vrule height8.5pt depth3.5pt width\z@}%
  \normalbaselines\rm}

\def\seventeenpoint{\def\rm{\fam0\seventeenrm}%
  \textfont0=\seventeenrm \scriptfont0=\fourteenrm \scriptscriptfont0=\tenrm
  \textfont1=\seventeeni \scriptfont1=\fourteeni \scriptscriptfont1=\teni
  \textfont2=\seventeensy \scriptfont2=\fourteensy \scriptscriptfont2=\tensy
  \textfont3=\tenex \scriptfont3=\tenex \scriptscriptfont3=\tenex
  \def\it{\fam\itfam\seventeenit}\textfont\itfam=\seventeenit
  \def\sl{\fam\slfam\seventeensl}\textfont\slfam=\seventeensl
  \def\bf{\fam\bffam\seventeenbf}\textfont\bffam=\seventeenbf%
  \scriptfont\bffam=\fourteenbf
  \scriptscriptfont\bffam=\twelvebf
  \normalbaselineskip=21pt
  \let\sc=\fourteenrm
  \let\big=\tenbig                                          
  \setbox\strutbox=\hbox{\vrule height 12pt depth 6pt width\z@}%
  \normalbaselines\rm}

\def\ninepoint{\def\rm{\fam0\ninerm}%
  \textfont0=\ninerm \scriptfont0=\sixrm \scriptscriptfont0=\fiverm
  \textfont1=\ninei \scriptfont1=\sixi \scriptscriptfont1=\fivei
  \textfont2=\ninesy \scriptfont2=\sixsy \scriptscriptfont2=\fivesy
  \textfont3=\tenex \scriptfont3=\tenex \scriptscriptfont3=\tenex
  \def\it{\fam\itfam\nineit}\textfont\itfam=\nineit
  \def\sl{\fam\slfam\ninesl}\textfont\slfam=\ninesl
  \def\bf{\fam\bffam\ninebf}\textfont\bffam=\ninebf \scriptfont\bffam=\sixbf
  \scriptscriptfont\bffam=\fivebf
  \normalbaselineskip=11pt
  \let\sc=\sevenrm
  \let\big=\ninebig
  \setbox\strutbox=\hbox{\vrule height8pt depth3pt width\z@}%
  \normalbaselines\rm}

\def\eightpoint{\def\rm{\fam0\eightrm}%
  \textfont0=\eightrm \scriptfont0=\sixrm \scriptscriptfont0=\fiverm%
  \textfont1=\eighti \scriptfont1=\sixi \scriptscriptfont1=\fivei%
  \textfont2=\eightsy \scriptfont2=\sixsy \scriptscriptfont2=\fivesy%
  \textfont3=\tenex \scriptfont3=\tenex \scriptscriptfont3=\tenex%
  \def\it{\fam\itfam\eightit}\textfont\itfam=\eightit%
  \def\sl{\fam\slfam\eightsl}\textfont\slfam=\eightsl%
  \def\bf{\fam\bffam\eightbf}\textfont\bffam=\eightbf \scriptfont\bffam=\sixbf%
  \scriptscriptfont\bffam=\fivebf%
  \normalbaselineskip=9pt%
  \let\sc=\sixrm%
  \let\big=\eightbig%
  \setbox\strutbox=\hbox{\vrule height7pt depth2pt width\z@}%
  \normalbaselines\rm}

\def\tenbig#1{{\hbox{$\left#1\vbox to8.5pt{}\right.\n@space$}}}
\def\ninebig#1{{\hbox{$\textfont0=\tenrm\textfont2=\tensy
  \left#1\vbox to7.25pt{}\right.\n@space$}}}
\def\eightbig#1{{\hbox{$\textfont0=\ninerm\textfont2=\ninesy
  \left#1\vbox to6.5pt{}\right.\n@space$}}}

\def\footnote#1{\edef\@sf{\spacefactor\the\spacefactor}#1\@sf
      \insert\footins\bgroup\eightpoint
      \interlinepenalty100 \let\par=\endgraf
        \leftskip=\z@skip \rightskip=\z@skip
        \splittopskip=10pt plus 1pt minus 1pt \floatingpenalty=20000
        \smallskip\item{#1}\bgroup\strut\aftergroup\@foot\let\next}
\skip\footins=12pt plus 2pt minus 4pt 
\dimen\footins=30pc 

\newinsert\margin
\dimen\margin=\maxdimen
\def\titlefont{\seventeenpoint}
\loadtruetwelvepoint 
\loadtrueseventeenpoint

\def\eatOne#1{}
\def\ifundef#1{\expandafter\ifx%
\csname\expandafter\eatOne\string#1\endcsname\relax}
\def\notTrue{\iffalse}\def\isTrue{\iftrue}
\def\ifdef#1{{\ifundef#1%
\aftergroup\notTrue\else\aftergroup\isTrue\fi}}
\def\use#1{\ifundef#1\linemessage{Warning: \string#1 is undefined.}%
{\tt \string#1}\else#1\fi}



%
\catcode`"=11
\let\quote="
\catcode`"=12
\chardef\foo="22
\global\newcount\refno \global\refno=1
\newwrite\rfile
\newlinechar=`\^^J
\def\@ref#1#2{\the\refno\n@ref#1{#2}}
\def\h@ref#1#2#3{\href{#3}{\the\refno}\n@ref#1{#2}}
\def\n@ref#1#2{\xdef#1{\the\refno}%
\ifnum\refno=1\immediate\openout\rfile=\jobname.refs\fi%
\immediate\write\rfile{\noexpand\item{[\noexpand#1]\ }#2.}%
\global\advance\refno by1}
\def\nref{\n@ref} 
\def\ref{\@ref}   
\def\hrref{\h@ref}
\def\lref#1#2{\the\refno\xdef#1{\the\refno}%
\ifnum\refno=1\immediate\openout\rfile=\jobname.refs\fi%
\immediate\write\rfile{\noexpand\item{[\noexpand#1]\ }#2\semi}%
\global\advance\refno by1}
\def\cref#1{\immediate\write\rfile{#1\semi}}

\def\preref#1#2{\gdef#1{\@ref#1{#2}}}

\def\semi{;\hfil\noexpand\break}

\def\listrefs{\vfill\eject\immediate\closeout\rfile
\centerline{{\bf References}}\bigskip\frenchspacing%
\input \jobname.refs\vfill\eject\nonfrenchspacing}

\def\inputAuxIfPresent#1{\immediate\openin1=#1
\ifeof1\message{No file \auxfileName; I'll create one.
}\else\closein1\relax\input\auxfileName\fi%
}
\def\NPB{Nucl.\ Phys.\ B}
\def\PRL{Phys.\ Rev.\ Lett.\ }




\newif\ifWritingAuxFile
\newwrite\auxfile
\def\SetUpAuxFile{%
\xdef\auxfileName{\jobname.aux}%
\inputAuxIfPresent{\auxfileName}%
\WritingAuxFiletrue%
\immediate\openout\auxfile=\auxfileName}

\def\L{\left(}\def\R{\right)}

\def\LB{\left[}\def\RB{\right]}


\catcode`\@=\active
\catcode`@=12  
\catcode`\"=\active


\def\L{\left(}
\def\R{\right)}

\def\s{{1\over6}}

\def\c{\mskip 1mu\cdot\mskip 1mu }
\def\Tr{\mathop{\rm Tr}\nolimits}

\def\Re{\mathop{\rm Re}\nolimits}
\def\Im{\mathop{\rm Im}\nolimits}

\def\eps{\epsilon}

\def\d{d^{\vphantom{\dagger}}}
\def\pol{\varepsilon}

\def\dl^#1_#2{\delta^{#1}{}_{#2}}

\def\Li{\mathop{\rm Li}\nolimits}

\def\Ord{{\cal O}}

\def\A#1{{\cal A}_{#1}}

\catcode`@=11  
\def\meqalign#1{\,\vcenter{\openup1\jot\m@th
   \ialign{\strut\hfil$\displaystyle{##}$ && $\displaystyle{{}##}$\hfil
             \crcr#1\crcr}}\,}
\catcode`@=12  


\baselineskip 15pt
\overfullrule 0.5pt


\def\Tr{\mathop{\rm Tr}\nolimits}

\def\A#1{{\cal A}_{#1}}

\def\pol{\varepsilon}

\def\c{\,\cdot\,}
\def\ksl{\slashed{k}}

\def\eb{\bar{\eta}}

\def\Re{\mathop{\rm Re}}
\def\L{\left(}\def\R{\right)}

\def\spa#1.#2{\left\langle#1\,#2\right\rangle}
\def\spb#1.#2{\left[#1\,#2\right]}
\def\lor#1.#2{\left(#1\,#2\right)}
\def\sand#1.#2.#3{%
\left\langle\smash{#1}{\vphantom1}^{-}\right|{#2}%
\left|\smash{#3}{\vphantom1}^{-}\right\rangle}
\def\sandp#1.#2.#3{%
\left\langle\smash{#1}{\vphantom1}^{-}\right|{#2}%
\left|\smash{#3}{\vphantom1}^{+}\right\rangle}
\def\sandpp#1.#2.#3{%
\left\langle\smash{#1}{\vphantom1}^{+}\right|{#2}%
\left|\smash{#3}{\vphantom1}^{+}\right\rangle}
\def\sandpm#1.#2.#3{%
\left\langle\smash{#1}{\vphantom1}^{+}\right|{#2}%
\left|\smash{#3}{\vphantom1}^{-}\right\rangle}
\def\sandmp#1.#2.#3{%
\left\langle\smash{#1}{\vphantom1}^{-}\right|{#2}%
\left|\smash{#3}{\vphantom1}^{+}\right\rangle}
\catcode`@=11  
\def\meqalign#1{\,\vcenter{\openup1\jot\m@th
   \ialign{\strut\hfil$\displaystyle{##}$ && $\displaystyle{{}##}$\hfil
             \crcr#1\crcr}}\,}
\catcode`@=12  


\input epsf
\SetUpAuxFile
\loadfourteenpoint
\FiguresInTexttrue
\nopagenumbers\hsize=\hstitle\vskip1in
\overfullrule 0pt
\hfuzz 30 pt

\vbadness=10001
%
%

\newbox\charbox
\newbox\slabox
\def\s#1{{      
        \setbox\charbox=\hbox{$#1$}
        \setbox\slabox=\hbox{$/$}
        \dimen\charbox=\ht\slabox
        \advance\dimen\charbox by -\dp\slabox
        \advance\dimen\charbox by -\ht\charbox
        \advance\dimen\charbox by \dp\charbox
        \divide\dimen\charbox by 2
        \raise-\dimen\charbox\hbox to \wd\charbox{\hss/\hss}
        \llap{$#1$}
}}

\def\Tr{\mathop{\rm Tr}\nolimits}
\def\tr{\mathop{\rm tr}\nolimits}
\def\Li{\mathop{\rm Li}\nolimits}

\def\Ls{\mathop{\rm Ls}\nolimits}
\def\Ll{{\rm L}}

\def\e{\epsilon}


\def\ns{n_{\mskip-2mu s}}
\def\nf{n_{\mskip-2mu f}}
\def\tree{{\rm tree}}
\def\ib{{\; \bar\imath}}

\def\lsl{\s \ell}
\def\dlips{d^D{\rm LIPS}}
\def\Atree{A^{\rm tree}}

\def\cg{c_\Gamma}
\def\gf{{c\mskip-1mu c}}
\def\sc{{s\mskip-1mu c}}

\def\tree{{\rm tree}}

\def\ns{n_{\mskip-2mu s}}\def\nf{n_{\mskip-2mu f}}
\def\hf{{\textstyle{1\over2}}}
\def\lr{\leftrightarrow}
\def\MSbar{$\overline{\rm MS}$}

\def\Split{\mathop{\rm Split}\nolimits}

\def\spa#1.#2{\left\langle#1 \hskip .15 mm #2\right\rangle}
\def\spb#1.#2{\left[#1 \hskip .15 mm #2\right]}
\def\lor#1.#2{\left(#1 #2\right)}
\def\sand#1.#2.#3{%
  \left\langle\smash{#1}{\vphantom1}\right|{#2}%
  \left|\smash{#3}{\vphantom1}\right\rangle}
\def\sandp#1.#2.#3{%
  \left\langle\smash{#1}{\vphantom1}^{-}\right|{#2}%
  \left|\smash{#3}{\vphantom1}^{+}\right\rangle}
\def\sandpp#1.#2.#3{%
  \left\langle\smash{#1}{\vphantom1}^{+}\right|{#2}%
  \left|\smash{#3}{\vphantom1}^{+}\right\rangle}
\def\sandmm#1.#2.#3{%
  \left\langle\smash{#1}{\vphantom1}^{-}\right|{#2}%
  \left|\smash{#3}{\vphantom1}^{-}\right\rangle}
\def\sandpm#1.#2.#3{%
  \left\langle\smash{#1}{\vphantom1}^{+}\right|{#2}%
  \left|\smash{#3}{\vphantom1}^{-}\right\rangle}
\def\sandmp#1.#2.#3{%
  \left\langle\smash{#1}{\vphantom1}^{-}\right|{#2}%
  \left|\smash{#3}{\vphantom1}^{+}\right\rangle}

\def\spaa#1.#2.#3{\langle\mskip-1mu{#1} 
                  | #2 | {#3}\mskip-1mu\rangle}
\def\spbb#1.#2.#3{[\mskip-1mu{#1}
                  | #2 | {#3}\mskip-1mu]}
\def\spab#1.#2.#3{\langle\mskip-1mu{#1} 
                  | #2 | {#3}\mskip-1mu\rangle}
\def\spba#1.#2.#3{\langle\mskip-1mu{#1}^+ 
                  | #2 | {#3}^+\mskip-1mu\rangle}

\def\"#1{{\accent127 #1}}
\def\ttilde#1{{\accent126 #1}}
%

\def\eb{{\bar e}}
\def\qb{{\bar q}}
\def\Qb{{\bar Q}}
\def\Lsnew{\mathop{\rm \widetilde {Ls}}\nolimits}
\def\d#1#2{\delta_{#1 #2}}
\def\dt#1#2{\tilde\delta_{#1 #2} }
\def\del{\Delta}
\def\delt{\tilde \Delta}

\def\rtdelta{\sqrt{\Delta_3}}
\def\rtmdelta{\sqrt{-\Delta_3}}
\def\Cl{{\rm Cl}}
\def\nn{{++}}
\def\an{{+-}}
\def\bn{{+\pm}}
\def\sl{{\rm sl}}
\def\ax{{\rm ax}}
\def\vect{{\rm v}}
\def\prop#1{{\cal P}_{#1}}
\def\half{{\textstyle {1\over 2}}}

\def\q{{\vphantom{\qb}{q}}}
\def\Q{{\vphantom{\Qb}{Q}}}
\def\msq{m_t^2}
\def\flip#1{\hbox{flip}_{#1}}
\def\exch#1{\hbox{exch}_{#1}}

\def\frac#1#2{{#1\over #2}}

\def\nl{\cr &\hskip 1mm}
\def\hsa{\hskip 7mm}
\def\hsb{\hskip 3mm}
\def\tI{\tilde I \vphantom{I}}
\def\ttI{\hat I \vphantom{I}}
\def\ts{\textstyle}

\def\Tgf{T}
\def\tTgf{\widetilde T}
\def\Tsc{T}
\def\tTsc{\widetilde T}
\def\mathtag#1{}


\noindent
hep-ph/9708239 \hfill {SLAC--PUB--7529}\break
\rightline{Saclay/SPhT-T97/090}
\rightline{UCLA/97/TEP/10}

\leftlabelstrue
\vskip -0.7 in
\Title{One-Loop Amplitudes for $e^+ \, e^-$ to Four Partons}

\centerline{Zvi Bern${}^{\sharp}$}
\baselineskip12truept
\centerline{\it Department of Physics}
\centerline{\it University of California, Los Angeles}
\centerline{\it Los Angeles, CA 90024}
\centerline{\tt bern@physics.ucla.edu}

\smallskip\smallskip

\baselineskip17truept
\centerline{Lance Dixon${}^{\star}$}
\baselineskip12truept
\centerline{\it Stanford Linear Accelerator Center}
\centerline{\it Stanford University}
\centerline{\it Stanford, CA 94309}
\centerline{\tt lance@slac.stanford.edu}

\smallskip \centerline{and} \smallskip

\baselineskip17truept
\centerline{David A. Kosower}
\baselineskip12truept
\centerline{\it Service de Physique Th\'eorique${}^{\dagger}$}
\centerline{\it Centre d'Etudes de Saclay}
\centerline{\it F-91191 Gif-sur-Yvette cedex, France}
\centerline{\tt kosower@spht.saclay.cea.fr}

\vskip 0.2in\baselineskip13truept

\vskip 0.5truein
\centerline{\bf Abstract}

{\narrower We present the first explicit formul\ae\ for the complete set
of one-loop helicity amplitudes necessary for computing next-to-leading
order corrections for $e^+\,e^-$ annihilation into four jets, for $W$, $Z$
or Drell-Yan production in association with two jets at hadron colliders,
and for three-jet production in deeply inelastic scattering experiments.
We include a simpler form of the previously published amplitudes for 
$e^+\,e^-$ to four quarks.  We obtain the amplitudes using their analytic
properties to constrain their form.  Systematically eliminating spurious
poles from the amplitudes leads to relatively compact results.}

\vskip 0.3truein

\centerline{\it Submitted to Nuclear Physics B}

\vfill
\vskip 0.1in
\noindent\hrule width 3.6in\hfil\break
\noindent
${}^{\sharp}$Research supported in part by the US Department of Energy
under grant DE-FG03-91ER40662 and in part by the
Alfred P. Sloan Foundation under grant BR-3222. \hfil\break
${}^{\star}$Research supported by the US Department of
Energy under grant DE-AC03-76SF00515.\hfil\break
${}^{\dagger}$Laboratory of the
{\it Direction des Sciences de la Mati\`ere\/}
of the {\it Commissariat \`a l'Energie Atomique\/} of France.\hfil\break

\Date{}

\line{}

\baselineskip17pt
%


\preref\ThreeJetsBorn{%
J. Ellis, M.K. Gaillard and G.G. Ross, Nucl.\ Phys.\ B111:253 (1976)}

\preref\FourJetsBorn{A. Ali, et al., Phys.\ Lett.\ 82B:285 (1979); 
Nucl.\ Phys.\ B167:454 (1980)}

\preref\FiveJetsBorn{%
K. Hagiwara and D. Zeppenfeld, Nucl. Phys. B313:560 (1989)\semi 
N.K. Falck, D. Graudenz and G. Kramer, Nucl. Phys. B328:317 (1989)}

\preref\BGK{%
F.A. Berends, W.T. Giele and H. Kuijf, Nucl.\ Phys.\ B321:39
(1989)}

\preref\ThreeJetsNLOME{R.K. Ellis, D.A. Ross and A.E. Terrano,
Phys.\ Rev.\ Lett. 45:1226 (1980); Nucl.\ Phys.\ B178:421 (1981)\semi
K. Fabricius, I. Schmitt, G. Kramer and G. Schierholz, 
Phys.\ Lett.\ B97:431 (1980); Z.\ Phys.\ C11:315 (1981)}

\preref\ThreeJetsPrograms{Z. Kunszt and P. Nason, in Z Physics at
LEP1, CERN Yellow Report 89-08\semi
G. Kramer and B. Lampe, Z. Phys.\ C34:497 (1987); C42:504(E) (1989);
Fortschr. Phys. 37:161 (1989)\semi
S. Catani and M.H. Seymour, Phys.\ Lett.\ B378:287 (1996) [hep-ph/9602277]}

\preref\GieleGlover{W.T. Giele and E.W.N. Glover, 
Phys.\ Rev.\ D46:1980 (1992)}

\preref\KunsztSoper{Z. Kunszt and D.E. Soper, Phys. Rev. D46:192 (1992)} 

\preref\EventShapeAlphas{%
OPAL Collab., P.D. Acton et al., Z. Phys. C55:1 (1992)\semi
ALEPH Collab., D. Decamp et al., Phys. Lett. B284:163 (1992)\semi 
L3 Collab., O. Adriani et al., Phys. Lett. B284:471 (1992)\semi 
DELPHI Collab., P. Abreu et al., Z. Phys. C59:21 (1993)\semi
SLD Collab., K. Abe et al., Phys. Rev. D51:962 (1995)}

\preref\FourJetTests{S. Bethke, A. Ricker, P.M. Zerwas,
Z. Phys. C49:59-72 (1991);\semi
L3 Collab., B. Adeva et al., Phys. Lett. B248:227 (1990)\semi
DELPHI Collab., P. Abreu et al., Z. Phys. C59:357 (1993)\semi
OPAL Collab., R. Akers et al., Z. Phys. C65:367 (1995)}

\preref\PeskinSchroeder{%
M.E.\ Peskin and D.V.\ Schroeder, {\it An Introduction to Quantum Field Theory}
(Addison-Wesley, 1995)}

\preref\TwoLoopUnitarity{%
W.L.\ van Neerven, Nucl.\ Phys.\ B268:453 (1986)}

\preref\Adrian{A. Signer and L. Dixon, Phys. Rev. Lett. 78:811 (1997) 
[hep-ph/9609460]; preprint hep-ph/9706285, to appear in Phys. Rev. D}

\preref\FiveGluon{%
Z. Bern, L. Dixon and D.A. Kosower, Phys.\ Rev. Lett.\
70:2677 (1993) [hep-ph/9302280]}

\preref\Kunsztqqqqg{%
Z. Kunszt, A. Signer and Z. Tr\'ocs\'anyi, Phys. Lett. B336:529 (1994)
[hep-ph/9405386]}

\preref\Fermion{%
Z. Bern, L. Dixon and D.A. Kosower, Nucl.\ Phys. B437:259 (1995)
[hep-ph/9409393]}

\preref\Color{%
F.A. Berends and W.T. Giele,
Nucl.\ Phys.\ B294:700 (1987)\semi
D.A.\ Kosower, B.-H.\ Lee and V.P. Nair, Phys.\ Lett.\ 201B:85 (1988)\semi
M.\ Mangano, S. Parke and Z.\ Xu, Nucl.\ Phys.\ B298:653 (1988)\semi
Z. Bern and D.A.\ Kosower, Nucl.\ Phys.\ B362:389 (1991)}

\preref\ManganoReview{%
M. Mangano and S.J. Parke, Phys.\ Rep.\ 200:301 (1991)\semi
L. Dixon, in 
{\it QCD \& Beyond: Proceedings of TASI '95}, 
ed. D.E.\ Soper (World Scientific, 1996) [hep-ph/9601359]}

\preref\ParkeTaylor{%
S.J.\ Parke and T.R.\ Taylor, Phys.\ Rev.\ Lett.\ 56:2459 (1986)}

\preref\GG{W.T.\ Giele and E.W.N.\ Glover,
Phys.\ Rev.\ D46:1980 (1992)\semi
W.T.\ Giele, E.W.N.\ Glover and D. A. Kosower,
Nucl.\ Phys.\ B403:633 (1993)}

\preref\SpinorHelicity{%
F.A.\ Berends, R.\ Kleiss, P.\ De Causmaecker, R.\ Gastmans and T.\ T.\ Wu,
Phys.\ Lett.\ 103B:124 (1981)\semi
P.\ De Causmaeker, R.\ Gastmans,  W.\ Troost and  T.T.\ Wu,
Nucl. Phys. B206:53 (1982)\semi
R.\ Kleiss and W.J.\ Stirling, Nucl.\ Phys.\ B262:235 (1985)\semi
R.\ Gastmans and T.T.\ Wu,
{\it The Ubiquitous Photon: Helicity Method for QED and QCD}
(Clarendon Press, 1990)\semi
Z. Xu, D.-H.\ Zhang and L. Chang, Nucl.\ Phys.\ B291:392 (1987)}

\preref\StringBased{
Z. Bern and D.A.\ Kosower, \PRL 66:1669 (1991)\semi
Z. Bern and D.C.\ Dunbar,  Nucl.\ Phys.\ B379:562 (1992)}

\preref\Long{
Z. Bern and D.A.\ Kosower, \NPB 379:451 (1992)}

\preref\Cutting{L.D.\ Landau, Nucl.\ Phys.\ 13:181 (1959)\semi
 S. Mandelstam, Phys.\ Rev.\ 112:1344 (1958), 115:1741 (1959)\semi
 R.E.\ Cutkosky, J.\ Math.\ Phys.\ 1:429 (1960)}

\preref\Massive{%
Z. Bern and A.G.\ Morgan, Nucl.\ Phys.\ B467:479 (1996) [hep-ph/9511336]}

\preref\KunsztFourPoint{%
Z. Kunszt, A. Signer and Z. Tr\'ocs\'anyi, Nucl.\ Phys.\ B411:397 
(1994) [hep-ph/9305239]\semi
A. Signer, Ph.D. thesis, ETH Z\"urich (1995)\semi
S. Catani, M.H. Seymour and Z. Tr\'ocs\'anyi, 
Phys.\ Rev.\ D55:6819 (1997) [hep-ph/9610553]}

\preref\Minn{Z. Bern, L. Dixon, D.C. Dunbar and D.A. Kosower, 
in {\it Continuous advances in QCD}, ed. A.V. Smilga 
(World Scientific, 1994) [hep-ph/9405248] } 

\preref\SusyDecomp{%
Z. Bern, in {\it Proceedings of Theoretical
Advanced Study Institute in High Energy Physics (TASI 92)},
eds.\ J. Harvey and J. Polchinski (World Scientific, 1993) [hep-ph/9304249]
\semi
Z.\ Bern and A.G.\ Morgan, Phys.\ Rev.\ D49:6155 (1994) [hep-ph/9312218]\semi
A.G. Morgan, Phys.\ Lett.\ B351:249 (1995) [hep-ph/9502230]}

\preref\SusyFour{Z. Bern, L. Dixon, D.C. Dunbar and D.A. Kosower,
Nucl.\ Phys.\ B425:217 (1994) [hep-ph/9405248]}

\preref\SusyOne{Z. Bern, L. Dixon, D.C. Dunbar and D.A. Kosower,
Nucl.\ Phys.\ B435:39 (1995) [hep-ph/9409265]}

\preref\BDKconf{Z. Bern, L. Dixon and D.A. Kosower,
in {\it Proceedings of Strings 1993}, eds. M.B. Halpern, A. Sevrin
and G. Rivlis (World Scientific, 1994) [hep-th/9311026]}

\preref\AllPlus{Z. Bern, G. Chalmers, L. Dixon and D.A. Kosower,
Phys.\ Rev.\ Lett.\ 72:2134 (1994) [hep-ph/9312333]}

\preref\Mahlon{G.D.\ Mahlon, Phys.\ Rev.\ D49:2197 (1994)
[hep-ph/9311213]; Phys.\ Rev.\ D49:4438 (1994) [hep-ph/9312276]}

\preref\Siegel{W. Siegel, Phys.\ Lett.\ 84B:193 (1979)\semi
D.M.\ Capper, D.R.T.\ Jones and P. van Nieuwenhuizen, Nucl.\ Phys.\
B167:479 (1980)\semi
L.V.\ Avdeev and A.A.\ Vladimirov, Nucl.\ Phys.\ B219:262 (1983)}

\preref\CollinsBook{J.C.\ Collins, {\it Renormalization}
(Cambridge University Press, 1984)}

\preref\TreeCollinear{F.A. Berends and W.T. Giele, Nucl.\ Phys.\
B313:595 (1989)}

\preref\Factorization{%
Z. Bern and G. Chalmers, Nucl.\ Phys.\ B447:465 (1995) [hep-ph/9503236]}

\preref\Review{%
Z. Bern, L. Dixon and D.A. Kosower, Ann. Rev. Nucl. Part. Sci.
46:109 (1996) [hep-ph/9602280]}

\preref\GloverMiller{E.W.N. Glover and D.J. Miller, 
Phys.\ Lett.\ B396:257 (1997) [hep-ph/9609474]}

\preref\BijGlover{J.J. van der Bij and  E.W.N. Glover,
Nucl. Phys. B313:237 (1989)} 

\preref\ZqqggConf{ 
Z. Bern, L. Dixon and D.A. Kosower, Nucl. Phys. Proc. Suppl.
51C:243 (1996) [hep-ph/9606378]}

\preref\IntegralsLong{Z. Bern, L. Dixon and D.A.\ Kosower,
\NPB 412:751 (1994) [hep-ph/9306240]}

\preref\IntegralsShort{Z. Bern, L. Dixon and D.A.\ Kosower,
Phys.\ Lett.\ 302B:299 (1993),
 erratum {\it ibid.} 318:649 (1993) [hep-ph/9212308]}

\preref\Lewin{L.\ Lewin, {\it Dilogarithms and Associated Functions\/}
(Macdonald, 1958)}

\preref\VNV{
W. van Neerven and J.A.M. Vermaseren, Phys.\ Lett.\ 137B:241 (1984)}

\preref\FiveGluon{Z. Bern, L. Dixon and D.A.\ Kosower, Phys.\ Rev.\ Lett.\
70:2677 (1993) [hep-ph/9302280]}

\preref\ThreeMassTriangle{H.-J. Lu and C. Perez, preprint
SLAC--PUB--5809 (1992)\semi
A.I. Davydychev and J.B. Tausk, Nucl. Phys. B397:123 (1993);
Phys. Rev. D53:7381 (1996)}

\preref\UDThreeMassTriangle{
N.I.\ Ussyukina and A.I.\ Davydychev, Phys.\ Lett.\ {298B:363 (1993)}} 

\preref\Gluinos{
S. Dawson, E. Eichten and C. Quigg, Phys.\ Rev.\ D31:1581 (1985)\semi
R.M. Barnett, H.E. Haber and G.L. Kane, Nucl. Phys. B267:625
(1986)\semi
L. Clavelli, Phys.\ Rev.\ D46:2112 (1992)\semi 
J. Ellis, D. Nanopoulos, and D. Ross, Phys.\ Lett.\ B305:375 (1993)
[hep-ph/9303273]\semi
L. Clavelli, P. Coulter and K. Yuan, Phys.\ Rev.\ D47:1973 (1993)
[hep-ph/9205237]\semi
R. Mu\ttilde{n}oz-Tapia and W.J. Stirling, Phys.\ Rev.\ D49:3763 (1994)
[hep-ph/9309246]\semi
G.R. Farrar, Phys.\ Rev.\ D51:3904 (1995) [hep-ph/9407401];
preprint hep-ph/9504295;
preprint hep-ph/9508291;
preprint hep-ph/9508292; preprint hep-ph/9707467\semi
L. Clavelli, I. Terekhov, Phys.\ Rev.\ Lett.\ 77:1941 (1996) 
[hep-ph/9605463]; Phys. Lett. B385:139 (1996) [hep-ph/9603390]\semi
A. de Gouvea and H. Murayama, Phys.\ Lett.\ B400:117 (1997) 
[hep-ph/9606449]\semi
Z. Bern, A.K.\ Grant and A.G.\ Morgan, Phys. Lett. 
B387:804 (1996) [hep-ph/9606466]\semi
J.L. Hewett, T.G. Rizzo and M.A. Doncheski, preprint hep-ph/9612377}

\preref\Background{G. 't Hooft,
in Acta Universitatis Wratislavensis no.\
38, 12th Winter School of Theoretical Physics in Karpacz, {\it
Functional and Probabilistic Methods in Quantum Field Theory},
Vol. 1 (1975)\semi
B.S.\ DeWitt, in {\it Quantum gravity II}, eds. C. Isham, R.\ Penrose and
D.\ Sciama (Oxford, 1981)\semi
L.F.\ Abbott, Nucl.\ Phys.\ B185:189 (1981)\semi
L.F. Abbott, M.T. Grisaru and R.K. Schaefer,
Nucl.\ Phys.\ B229:372 (1983)}

\preref\HHKY{K. Hikasa, Mod. Phys. Lett. A5:1801 (1990)\semi
K. Hagiwara, T. Kuruma and Y. Yamada, Nucl. Phys. B358:80 (1991)}

\preref\Zqqqq{
Z. Bern, L. Dixon, D.A. Kosower and S. Weinzierl, 
Nucl. Phys. B489:3 (1997) [hep-ph/9610370]}

\preref\TwoLoopSusy{%
Z. Bern, J. Rozowsky and B. Yan, Phys.\ Lett.\ B401:273 (1997)
[hep-ph/9702424]}

\preref\SWI{%
M.T.\ Grisaru, H.N.\ Pendleton and P.\ van Nieuwenhuizen,
Phys. Rev. {D15}:996 (1977)\semi
M.T. Grisaru and H.N. Pendleton, Nucl.\ Phys.\ B124:81 (1977)\semi
S.J. Parke and T. Taylor, Phys.\ Lett.\ B157:81 (1985)\semi
Z. Kunszt, Nucl.\ Phys.\ B271:333 (1986)}

\preref\PV{%
L.M.\ Brown and R.P.\ Feynman, Phys.\ Rev.\ 85:231 (1952)\semi
L.M.\ Brown, Nuovo Cimento {21:3878} (1961)\semi
G.\ Passarino and M.\ Veltman, Nucl.\ Phys.\ {B160:151} (1979)\semi
R. Stuart, Comput. Phys. Commun. 48:367 (1988)\semi
G. Devaraj and R.G. Stuart, preprint hep-ph/9704308}

\preref\CGM{%
J.M. Campbell, E.W.N. Glover and D.J. Miller, preprint hep-ph/9612413}

\preref\CGMqggq{%
J.M. Campbell, E.W.N. Glover and D.J. Miller, preprint hep-ph/9706297}

\preref\Pittau{R. Pittau, preprint hep-ph/9607309\semi
Z. Bern, P. Gondolo and M. Perelstein, preprint hep-ph/9706538}

\preref\NT{Z. Nagy and Z. Tr\'ocs\'anyi, preprint hep-ph/9707309}

\preref\PrivateNigel{J.M. Campbell and E.W.N. Glover, private communication}

\preref\Furry{W.H. Furry, Phys. Rev. 51:125 (1937)}


\vskip -2. cm
$\null$
\section{Introduction}
\tagsection\IntroSection

The discovery of new physics at colliders relies to a large extent on our
ability to understand the known physics producing the bulk of the data.
For processes involving hadronic jets, perturbative QCD predictions are
required.  Leading-order calculations often reproduce the shapes of
distributions well but suffer from practical and conceptual problems whose
resolution requires the use of next-to-leading order (NLO) calculations.
In many processes at modern colliders, the dominant theoretical
uncertainties are due to unknown higher-order perturbative corrections.
These corrections can be enhanced by various logarithms.  For some
processes, NLO corrections are known, and programs implementing them have
already played an important role in analyzing data from a variety of
high-energy collider experiments.  Other processes have awaited the
computation of the required one-loop matrix elements.

The first type of logarithm contributing to theoretical uncertainties
is `ultraviolet' in nature.  Such logarithms are connected with the
renormalization scale $\mu$, which we are forced to introduce in order
to define the running coupling, $\alpha_s(\mu)$.  Physical quantities,
such as cross-sections or differential cross-sections, should be
independent of $\mu$.  When we compute such a quantity in perturbation
theory, however, we necessarily truncate its expansion in $\alpha_s$,
and this introduces a spurious dependence on $\mu$.  Together, these
effects can lead to anywhere from a 30\% to a factor of 2--3
normalization uncertainty in predictions of experimentally-measured
distributions.  In general, the inclusion of NLO corrections
significantly reduces the over-all sensitivity of a prediction to
variations in $\mu$.

The other type of logarithm is `infrared' in nature.  Such logarithms
are connected with the presence of soft and collinear radiation.  Jets
in a detector consist of a spray of hadrons spread over a finite
region of phase space.  Experimental measurements of jet distributions
depend on resolution parameters, such as the jet cone size and minimum
transverse energy.  In a leading-order calculation, jets are modeled
by lone partons.  As a result, these predictions either lack a
dependence on these parameters or have an incorrect dependence on
them.  In addition, the internal structure of a jet cannot be
predicted at all.

In the case of $e^+\,e^-$ annihilation into jets,
leading-order predictions for the production of up to five jets 
have been available for quite some time [\use\ThreeJetsBorn,%
\use\FourJetsBorn,\use\ThreeJetsNLOME,\use\FiveJetsBorn,\use\BGK].
The NLO matrix elements
for three-jet production and other ${\cal O}(\alpha_s)$ observables
are also known [\use\ThreeJetsNLOME], and numerical programs
implementing these corrections~[\use\ThreeJetsPrograms,\use\GieleGlover] 
have been widely used to extract a precise value of $\alpha_s$ from hadronic
event shapes at the $Z$ pole~[\use\EventShapeAlphas].

Next-to-leading order corrections for more complicated processes are
important, however, if we wish to use QCD to probe for new physics in
other standard model processes.  In $e^+\,e^-$ annihilation, for
example, four-jet production is the lowest-order process in which the
quark and gluon color charges can be measured independently.  Four-jet
production is thus sensitive to the presence of light colored fermions
such as gluinos~[\use\Gluinos].  At LEP~2 the process $e^+\,e^- \to
(\gamma^*,Z) \to 4$~jets is a background to threshold production of
$W$ pairs, when both $W$s decay hadronically. 

The calculation of $e^+\,e^-\to$~4 jets at NLO requires the tree-level
amplitudes for $e^+\,e^-\to$~5 partons~[\use\FiveJetsBorn,\use\BGK]
(at NLO two of the partons may appear inside a single jet), and the
one-loop amplitudes for $e^+\,e^-\to$~4 partons.  In a previous
paper~[\use\Zqqqq] we presented the one-loop helicity amplitudes for
electron-positron annihilation into four massless quarks, $e^+\,e^-
\to (\gamma^*,Z) \to \qb \q \Qb \Q$ ($q,Q$ may have the same or
different flavors).

In this paper, we present the $e^+\,e^- \to (\gamma^*,Z) \to
\qb \q gg$ one-loop helicity amplitudes, as well as simplified
versions of the $e^+\,e^- \to (\gamma^*,Z) \to \qb \q \Qb \Q$
amplitudes.  
We give all contributions at order $g^4$ in the strong coupling,
including those where the vector boson couples directly to a 
quark loop via a vector or axial-vector coupling.
We take all quarks to be massless, except for the top
quark, whose virtual effects we include through order $1/m_t^2$.  Thus
the list of helicity amplitudes required to construct a numerical
program for $e^+\,e^-\to$~4 jets is now complete.  Indeed, the
amplitudes presented here and in refs.~[\use\ZqqggConf,\use\Zqqqq]
have already been incorporated into the first NLO program for
$e^+\,e^- \rightarrow 4$~jets~[\use\Adrian].  Crossing symmetry and simple
coupling constant modifications allow the same amplitudes to be used
in NLO computations of the production of a vector boson ($W$, $Z$, or
Drell-Yan pair) in association with two jets at hadron colliders, and
three-jet production at $ep$ colliders. Finally, these
amplitudes will also enter the next-to-next-to-leading (NNLO) study of
three-jet production at the $Z$ pole.  Such a study (which awaits the
computation of appropriate two-loop matrix elements as well) would be
desirable in order to reduce the theoretical uncertainties in
determining $\alpha_s$ via this process.

Glover and Miller~[\use\GloverMiller] have reported on a calculation
of the helicity-summed interference term between four-quark one-loop
matrix elements and the appropriate tree-level matrix elements.
Recently, Campbell, Glover and Miller~[\use\CGMqggq] have also
calculated the analogous `squared' matrix elements
for the two-quark two-gluon final state.  In neither of these papers
did the authors provide any explicit formul\ae.  In order to compare
with the results described in refs.~[\use\GloverMiller,\use\CGMqggq], 
we considered the case of virtual photon exchange, dropped the
contributions where the photon couples directly to a quark loop, constructed
the unpolarized (helicity-summed) cross-section, and then performed an
integration over the orientation angles of the lepton pair (in the
virtual-photon rest frame).  After accounting for the different
versions of dimensional regularization used~[\use\KunsztFourPoint], we
have verified numerically that the two sets of results agree, for both
the four-quark and two-quark-two-gluon final states.%
\footnote{$^\dagger$}{We thank J.M. Campbell and E.W.N. Glover for assistance 
in comparing the results.}
Also, the squared matrix elements in
refs.~[\use\GloverMiller,\use\CGMqggq] have been incorporated into
a NLO program for four-jet fractions and shape variables by Nagy 
and Tr\'ocs\'anyi~[\use\NT], and their numerical results for the  
four-jet fractions confirm the corresponding results of ref.~[\use\Adrian].

The amplitudes we present
retain all correlations between the daughter or parent leptons of the
vector boson, and the colored partons in the process.  Such
correlations are important for computations that take into account
experimental constraints.  For example, in the production of a $W$
along with jets at a hadron collider, followed by the decay $W \to
\ell\bar{\nu}_\ell$, the longitudinal component of the $W$ momentum
cannot be observed, only that of the decay lepton, and the latter
should be isolated from the hadronic jets in order for the event to
pass detector cuts.  In the case of jet production in deep-inelastic
scattering, the orientation of the entire event with respect to the
detector is dictated by the lepton scattering angle as well as the
square of the virtual photon four-momentum.

Recent years have seen a number of technical advances in the
computation of one-loop amplitudes.  These advances have made possible
the calculation of all one-loop five-parton processes
[\use\FiveGluon,\use\Kunsztqqqqg,\use\Fermion], as well as a number of
infinite sequences of loop
amplitudes~[\use\AllPlus,\use\Mahlon,\use\SusyFour,\use\SusyOne,%
\use\TwoLoopSusy].
Many of the techniques used in the present calculation have been
reviewed in ref.~[\use\Review].  The rather complicated six-body
kinematics encountered here necessitate further techniques for
removing certain spurious singularities from the amplitudes.  The removal of
spurious singularities is essential in order to find (relatively)
compact final expressions, and also plays a role 
in improving the numerical stability of the results.

The general strategy employed in this paper is to obtain amplitudes
directly from their analytic properties instead of from Feynman
diagrams.  In particular, we use the constraints of
unitarity~[\use\Cutting,\use\SusyFour,\use\SusyOne,\use\Massive] and
factorization
[\use\ParkeTaylor,\use\ManganoReview,\use\AllPlus,\use\Factorization],
as summarized in ref.~[\use\Review].  We construct the amplitudes by
finding functions that have the correct poles and cuts in the various
channels.  The required pole and cut information is extracted from
previously obtained amplitudes (tree amplitudes or lower-point loop
amplitudes); manifest gauge invariance is therefore maintained at each
step.  This approach leads to compact expressions, especially when
compared with those obtained from a traditional diagrammatic
computation.  In a Feynman diagram approach each diagram alone is not
gauge invariant, and is often much more complicated than the sum
over all diagrams.  As a check, we performed a numerical evaluation of
the Feynman diagrams at one kinematic point, and verified that their sum
agrees with a numerical evaluation of our analytic results.

The spinor helicity method~[\use\SpinorHelicity] and color
decompositions~[\use\Color] are crucial to the success of this
approach, because they simplify the analytic structures that must be
computed.  Because of the intricate analytic structure of the
amplitudes it is rather non-trivial to remove spurious singularities
that can appear in the amplitudes.  (By a spurious singularity we mean
a kinematic pole or singularity whose residue vanishes.)  By
evaluating the cuts appropriately we can prevent the worst of the
spurious singularities from appearing.  However, some of the spurious
singularities are inherent in the amplitudes when they are expressed
in terms of logarithms and dilogarithms.  As we shall see, the spinor
helicity method is quite useful for simplifying the terms containing
spurious singularities.  

This paper is organized as follows. In section~\use\BasicToolsSection,
we briefly review spinor helicity and color decompositions and provide
formul\ae\ relating the full amplitudes to the `primitive' amplitudes in
terms of which the results are expressed.  In
section~\use\CutConstructionSection, we outline the construction of
amplitudes from their analytic properties.  Procedures for eliminating
or simplifying spurious singularities are given in
section~\use\SpuriousPoleSection.  Sample calculations are given in
section~\use\SampleSection; in particular, examples are provided for
cut constructions, rational function reconstructions, and
simplifications via numerical analysis.  The general structure of the
primitive amplitudes including regularization issues is given in
section~\use\GeneralFormSection.  The results for the primitive
amplitudes are collected in
sections~\use\MasterFunctionSection--\use\FourQuarkSection. In
section~\use\MasterFunctionSection\ the `master functions', which are
a set of functions which appear in multiple amplitudes, are given.
Section~\use\LeadingColorPrimitiveSection\ contains the amplitudes
which are leading in the number of colors and flavors.  
Subleading-in-color contributions are contained in
sections~\use\AmplitudesZqgqgSection\ and \use\AmplitudesZqqggSection.
Contributions with the vector boson coupled to a closed fermion loop
are give in section~\use\FermionLoopSection; this includes both vector
and axial-vector contributions.  Simplified versions of the four quark
amplitudes (which have been previously presented in ref.~[\use\Zqqqq])
are given in section~\use\FourQuarkSection.  Some concluding remarks
in are given in section~\use\ConclusionSection.  There are a total of
four appendices.  Appendices~\use\IntegralsAppendix\ and
\use\IntegralFunctionAppendix\ concern the evaluation of loop
integrals and their associated spurious singularities.
Appendix~\use\SpinorIdentityAppendix\ lists some spinor-product
identities that are useful for simplifying the spurious pole structure
of kinematic coefficients.  Appendix~\use\qqgAppendix\ records the
helicity amplitudes for $e^+\,e^- \to \qb q g$~[\use\GieleGlover] in
the same notation and conventions used in the paper; these amplitudes
are useful because they appear in many collinear limits of the
$e^+\,e^- \to \qb q g g$ amplitudes.

\section{Brief Review of Basic Tools} 
\tagsection\BasicToolsSection

We shall present our results in terms
of the spinor helicity method and $SU(N_c)$ color decompositions.  
The reader is referred to review articles~[\use\ManganoReview] 
and references therein for details beyond our following
brief review.

\subsection{Spinor Helicity}

We represent the gluon polarization
vectors in terms of Weyl spinors $|\,k^\pm \rangle$~[\use\SpinorHelicity],
$$
\pol^{+}_\mu (k;q) = 
     {\sandmm{q}.{\gamma_\mu}.k
      \over \sqrt2 \spa{q}.k}\, ,\hskip 1cm
\pol^{-}_\mu (k;q) = 
     {\sandpp{q}.{\gamma_\mu}.k
      \over \sqrt{2} \spb{k}.q} \, ,
\eqn\HelicityDef
$$  
where $k$ is the gluon momentum and $q$ is an arbitrary null
`reference momentum' which drops out of final gauge-invariant
amplitudes.  The plus and minus labels on the polarization vectors
refer to the gluon helicities.  Our (crossing-symmetric) convention
takes all particles to be outgoing, and labels the helicity and
particle vs.~antiparticle assignment accordingly.  For incoming
(negative energy) momenta the helicity and particle vs.~antiparticle
assignment are reversed.  It is convenient to define the following
{\it spinor strings},
$$
\eqalign{
&\spa{i}.j \equiv  \langle k_i^{-} \vert k_j^{+} \rangle\,, 
\hskip 2.0 cm 
\spb{i}.j \equiv \langle k_i^{+} \vert k_j^{-} \rangle \,, \cr
&\spab{i}.{l}.{j} \equiv 
\langle k_i^- \, |\, \ksl_l\, | \, k_j^- \rangle \,, 
\hskip 1.0 cm
\spab{i}.{(l+m)}.{j} \equiv 
\langle k_i^- \, |\, (\ksl_l+\ksl_m) \, | \, k_j^- \rangle \,, \cr
&\spbb{i}.{lm\cdots}.{j} \equiv 
\langle k_i^+ \, |\, \ksl_l\ksl_m\cdots \, | \, k_j^- \rangle \,, \cr
&\spab{i}.{lm\cdots}.{j} \equiv 
\langle k_i^- \, |\, \ksl_l\ksl_m\cdots \, | \, k_j^\pm \rangle \,, \cr
&\spbb{i}.{(l+m)(n+r)\cdots}.{j} \equiv 
\langle k_i^+ \, |\, (\ksl_l+\ksl_m)(\ksl_n+\ksl_r)\cdots
 \, | \, k_j^- \rangle \,, \cr
&\spab{i}.{(l+m)(n+r)\cdots}.{j} \equiv 
\langle k_i^- \, |\, (\ksl_l+\ksl_m)(\ksl_n+\ksl_r)\cdots 
\, | \, k_j^\pm \rangle \,, \cr
}\eqn\AngleDef
$$ 
which is the notation we shall use to quote the results. 
In the last definition we take the  $| \, k_j^\pm \rangle$ 
to mean $| \, k_j^- \rangle$ for an odd number of gamma-matrices
in the string and  $| \, k_j^+ \rangle$ when there are an even number.
All the momenta $k_i$ are massless, $k_i^2 = 0$.  
Sometimes we will also use the notation
$$
\spab{i}.{\ell_m}.{j} \equiv 
\langle k_i^- \, |\, \lsl_m \, | \, k_j^- \rangle \,, 
\anoneqn
$$
etc., where $\ell_m$ is a loop momentum.
The spinor inner products $\spa{i}.j$, $\spb{i}.j$ are antisymmetric 
and satisfy $\spa{i}.j \spb{j}.i = 2 k_i \cdot k_j \equiv s_{ij}$.
In addition to $s_{ij} \equiv (k_i+k_j)^2$ we also define
the three-particle invariants $t_{ijl} \equiv (k_i+k_j+k_l)^2$.

\subsection{Color Decomposition}
\tagsubsection\ColorDecompositionSubsection

It is convenient to decompose one-loop amplitudes in terms of
group-theoretic factors (color structures) multiplied by kinematic
functions called `partial amplitudes' [\use\Color]. 
We present results for the general gauge group $SU(N_c)$ 
($N_c=3$ for QCD), and normalize the group generators 
in the fundamental representation, $T^a$, so that
$\Tr(T^aT^b) = \delta^{ab}$.
Color decompositions are obtained by rewriting the structure
constants $f^{abc}$ as
$$
  f^{abc}\ =\ -{i\over\sqrt2} \Tr\L \LB T^a, T^b\RB T^c\R.
\eqn\structure
$$
Then one applies the $SU(N_c)$ Fierz identity
$$
   (X_1 \, T^a \, X_2)\ (Y_1 \, T^a \, Y_2)
   \ =\  (X_1 \, Y_2)\  (Y_1 \, X_2)
    - {1\over N_c} (X_1 \, X_2)\  (Y_1 \, Y_2)\ ,
\eqn\SUNFierz
$$
where $X_i,Y_i$ are strings of generator matrices $T^{a_i}$,
in order to remove contracted color indices.

Here we are interested in the amplitude
$\A{6}(1_\q,2,3,4_\qb;5_\eb,6_e)$, where legs $1,4$ are the
quark-anti-quark pair, legs $2,3$ are the gluon legs, and legs 5,6 are
the lepton pair.  We label the (outgoing) quark, anti-quark, electron
and positron lines with subscripts $q$, $\qb$, $e$, and $\eb$, while
the gluon lines do not have additional labels.  The color
decomposition of the tree-level contribution to $\A{6}$ is
$$
 \A{6}^\tree (1_\q, 2,3,4_\qb) 
  =  2 e^2 g^2 \bigl( -Q^q  + v_{L,R}^e v_{L,R}^q  \, \prop{Z}(s_{56}) \bigr) 
   \sum_{\sigma\in S_{2}}
   (T^{a_{\sigma(2)}} T^{a_{\sigma(3)}})_{i_1}^{~\ib_4}\
    \Atree_6 (1_\q,\sigma(2),\sigma(3),4_\qb)\,.
\eqn\TreeColorDecomp
$$
Here we have suppressed the $5,6$ labels of the electron pair, $e$ is
the QED coupling, $g$ the QCD coupling, $Q^\q$ is the charge of 
quark $q$ in units of $e$, and the ratio of $Z$ and photon propagators 
is given by
$$
\prop{Z}(s) = {s \over s - M_Z^2 + i \,\Gamma_Z \, M_Z}\,,
\anoneqn
$$
where $M_Z$ and $\Gamma_Z$ are the mass and width of the $Z$.  

The left- and right-handed couplings of
fermions to the $Z$ boson are
$$
\eqalign{
v_L^e & = { -1 + 2\sin^2 \theta_W \over \sin 2 \theta_W } \,, 
\hskip 2.3 cm 
v_R^e  = { 2 \sin^2 \theta_W \over  \sin 2 \theta_W } \,,  \cr 
v_L^q & = { \pm 1 - 2 Q^q\sin^2 \theta_W \over  \sin 2 \theta_W } \,,
\hskip 1.9 cm 
v_R^q = -{ 2 Q^q \sin^2 \theta_W \over \sin 2 \theta_W }  \,, \cr }
\anoneqn
$$ 
where $\theta_W$ is the Weinberg angle. 
The two signs in $v_{L,R}^q$ correspond to up $(+)$ and down $(-)$ 
type quarks.
The subscripts $L$ and $R$ refer to whether the particle to which 
the $Z$ couples is left- or right-handed. 
That is, $v_R^q$ is to be used for the configuration
where the quark (leg 1) has plus helicity and the anti-quark (leg 4) 
has minus helicity, which we denote by the shorthand $(1_\q^+, 4_\qb^-)$.
Similarly, $v_L^q$ corresponds to the configuration  $(1_\q^-, 4_\qb^+)$.
Because the electron and positron are incoming in
$e^+\,e^-$ annihilation, our outgoing-momenta notation reverses 
their helicities and particle vs.~anti-particle assignment.
Thus, $v_R^e$ corresponds to the helicity configuration $(5_\eb^-, 6_e^+)$
whereas $v_L^e$ corresponds to the configuration $(5_\eb^+, 6_e^-)$.

The one-loop color decomposition is given by 
$$
\eqalign{
 \A{6}^{1\rm -loop}(1_q, & \, 2, 3,  4_\qb) 
  =  2 e^2 \, g^4 \biggl\{ 
\bigl( -Q^q  + v_{L,R}^e v_{L,R}^q  \, \prop{Z}(s_{56}) \bigr)  \cr
& \hskip 1. cm \times 
\biggl[ N_c\, \sum_{\sigma\in S_2} 
 (T^{a_{\sigma(2)}}T^{a_{\sigma(3)}})_{i_1}^{~\ib_4}
    \ A_{6;1}(1_q,\sigma(2),\sigma(3), 4_\qb) 
   + \delta^{a_2 a_3}
   \, \delta_{i_1}^{~\ib_4}
    \,  A_{6;3}(1_q,  4_\qb; 2, 3) \biggl] \cr 
&   \hskip .5 cm 
+   \sum_{i=1}^{\nf} \Bigl( -Q^i + {1\over 2} v_{L,R}^e 
             (v_L^i+v_R^i) \prop{Z}(s_{56}) \Bigr) \cr
& \hskip 2 cm  \times
   \Bigl[ (T^{a_2}T^{a_3})_{i_1}^{~\ib_4} 
   + (T^{a_3}T^{a_2})_{i_1}^{~\ib_4} 
   - {2\over N_c}\, \delta^{a_2 a_3}
       \, \delta_{i_1}^{~\ib_4} \Bigr] 
   \, A_{6;4}^\vect(1_q,4_\qb; 2, 3)  \cr
&   \hskip .5 cm 
+ {v_{L,R}^e\over \sin 2 \theta_W}\, \prop{Z}(s_{56}) \biggr[
  \sum_{\sigma\in S_2} 
  \Bigl( (T^{a_{\sigma(2)}} T^{a_{\sigma(3)}} )_{i_1}^{~\ib_4} 
   - {1\over N_c} \delta^{a_2 a_3} 
       \, \delta_{i_1}^{~\ib_4}  \Bigr)
   \, A_{6;4}^\ax(1_q, 4_\qb; \sigma(2), \sigma(3)) \cr
& \hskip 2 cm  
   + {1\over N_c} \delta^{a_2 a_3} \, \delta_{i_1}^{~\ib_4}
      \, A_{6;5}^\ax(1_q, 4_\qb; 2, 3) \biggr]
  \biggr\} \,, \cr}
\eqn\qqggDecomp
$$
where $Q^i$ is the electric charge (in units of $e$) of the $i$th
quark and $\nf$ is the number of light quark flavors.  The partial
amplitudes $A_{6;4}^\vect$, $A_{6;4}^\ax$ and $A_{6;5}^\ax$
represent the contributions from a photon or $Z$ coupling to a
fermion loop through a vector or axial-vector coupling.
We take all quarks to be massless except the top quark.  We assume
that the top quark mass squared, $m_t^2$, is larger than the other
kinematic invariants in the process, and expand the fermion loop
contributions in $1/m_t^2$, keeping terms of order $1/m_t^2$, but
dropping those of order $1/m_t^4$.  In this approximation the top
quark contribution to $A_{6;4}^\vect$ vanishes (see
section~\use\FermionLoopSection).  On the other hand, in the axial
vector channel isodoublet cancellations for massless quarks ensure
that only the $t,b$ isodoublet contributes to $A_{6;4}^\ax$ and
$A_{6;5}^\ax$.  There are also order $1/m_t^2$ contributions to
$A_{6;1}$.

For convenience we also quote the color decomposition for the
four-quark amplitudes.  At tree level, we have
$$
\eqalign{
\A{6}^{\rm tree} (1_\q,2_\Qb,3_\Q,4_\qb) & = 
 2 e^2 g^2 \biggl[  \Bigl( - Q^\q 
+  v_{L,R}^e v_{L,R}^\q \,\prop{Z}(s_{56}) \Bigr) 
  \Atree_6(1_\q,2_\Qb,3_\Q,4_\qb) \cr
& \hskip 1.5 cm 
+  \Bigl( - Q^\Q 
+ v_{L,R}^e v_{L,R}^\Q \,\prop{Z}(s_{56})\Bigr) 
  \Atree_6(3_\Q,4_\qb,1_\q,2_\Qb) \biggr] \cr
& \hskip 1 cm \times 
\Bigl(\delta_{i_1}^{\ib_2} \,  \delta_{i_3}^{\ib_4}  \, 
-{1\over N_c} \delta_{i_1}^{\ib_4}\,  \delta_{i_3}^{\ib_2} \,\Bigr) \,,\cr}
\eqn\TreeQQQQColorDecomposition
$$
while the one-loop decomposition~[\use\Zqqqq] is
$$
\hskip -5mm\eqalign{
\A{6}^{1\rm -loop} &(1_\q ,2_\Qb,3_\Q,4_\qb)  =  \cr
&\hskip -5mm 2 e^2 g^4  \biggl[\Bigl( - Q^\q 
+  v_{L,R}^e v_{L,R}^\q \prop{Z}(s_{56}) \Bigr) 
%
\Bigl[ N_c \,  \delta_{i_1}^{\ib_2} \, \delta_{i_3}^{\ib_4}  \, 
         A_{6;1}(1_\q,2_\Qb,3_\Q,4_\qb)
   + \delta_{i_1}^{\ib_4}\,  \delta_{i_3}^{\ib_2} \,
           A_{6;2} (1_\q,2_\Qb,3_\Q,4_\qb) \Bigr] \cr
&
+ \Bigl( - Q^\Q 
+  v_{L,R}^e v_{L,R}^\Q \,\prop{Z}(s_{56}) \Bigr) 
%
\Bigl[N_c \,  \delta_{i_1}^{\ib_2} \, \delta_{i_3}^{\ib_4}  \, 
         A_{6;1}(3_\Q,4_\qb,1_\q,2_\Qb)
   + \delta_{i_1}^{\ib_4}\,  \delta_{i_3}^{\ib_2} \,
           A_{6;2} (3_\Q,4_\qb,1_\q,2_\Qb) \Bigr] \cr
&
+ { v_{L,R}^e \over \sin2\theta_W }\prop{Z}(s_{56}) \,
%
\Bigl( \delta_{i_1}^{\ib_2} \, \delta_{i_3}^{\ib_4}  
     - {1\over N_c} \delta_{i_1}^{\ib_4}\,  \delta_{i_3}^{\ib_2} \Bigr)
      A_{6;3} (1_\q,2_\Qb,3_\Q,4_\qb) \biggr] \,.\cr
}\eqn\qqqqDecomp
$$
For the case of identical quark flavors ($q=Q$) see ref.~[\use\Zqqqq].
We also include here contributions of order $1/m_t^2$ from vacuum
polarization loops to $A_{6;1}$ and $A_{6;2}$ (which are very small
at present $e^+\,e^-$ machines).
The partial amplitudes $A_{6;1}$ and $A_{6;2}$ also appear in
$W + 2$ jet production at hadron colliders; we leave the coupling 
constant and mass conversions as an exercise.

The virtual part of the next-to-leading order correction to the
parton-level cross-section is given by the sum over colors of the
interference between the tree amplitude $\A{6}^{{\rm tree}}$ and the
one-loop amplitude $\A{6}^{1\rm -loop}$.  Using the color
decompositions~(\use\TreeColorDecomp) and (\use\qqggDecomp), and the
Fierz rules~(\use\SUNFierz) the color-sum for $e^+\,e^- \to \qb q gg$
in terms of partial amplitudes is,
$$
\eqalign{
 \sum_{{\rm colors}} [\A{6}^* \A{6}]_{{\rm NLO}} 
  & =  8 e^4 \, g^6 \, (N_c^2-1) {\rm Re} \Biggl\{ 
   \bigl( -Q^q  + v_{L,R}^e v_{L,R}^q \, \prop{Z}^*(s_{56}) \bigr) 
   A_6^{{\rm tree}*}(1_q,2,3,4_\qb) \cr
& \times \biggl[   
  \bigl( -Q^q  + v_{L,R}^e v_{L,R}^q  \, \prop{Z}(s_{56}) \bigr)
  \bigl[ (N_c^2-1) A_{6;1}(1_q,2,3,4_\qb) \cr
& \hskip 3.6 cm
   - A_{6;1}(1_q,3,2,4_\qb) + A_{6;3}(1_q,4_\qb; 2,3) \bigr]\; 
   \cr
& \hskip .8 cm 
+  \sum_{i=1}^{\nf} \bigl( -Q^i + \half v_{L,R}^e 
             (v_L^i+v_R^i) \prop{Z}(s_{56}) \bigr)
   \Bigl(N_c-{4\over N_c} \Bigr) A_{6;4}^\vect(1_q,4_\qb;2,3) \cr
& \hskip .8 cm 
+         {v_{L,R}^e\over \sin 2 \theta_W}\, \prop{Z}(s_{56})
  \Bigl[ \Bigl(N_c - {2\over N_c}\Bigr) A_{6;4}^\ax(1_q,4_\qb; 2,3) \cr
& \hskip 3.6 cm
   - {2\over N_c} A_{6;4}^\ax(1_q,4_\qb; 3,2) 
   + {1\over N_c} A_{6;5}^\ax(1_q,4_\qb; 2,3) \Bigr] \biggr] \Biggr\} \; 
        + \;\{2 \leftrightarrow 3 \}
   \,. \cr}
\eqn\SquaredAmplitudeTwoQuarksTwoGluons
$$
The corresponding formula for $e^+\,e^- \to \qb q \Qb Q$ is straightforward
to obtain, but lengthier, and so we omit it here.

\subsection{Primitive Amplitudes}
\tagsubsection\PrimitiveAmplitudesSubsection

One may perform a further decomposition of the partial amplitudes in
terms of gauge invariant {\it primitive amplitudes}
[\use\Fermion,\use\Zqqqq].  Primitive amplitudes are gauge-invariant
objects from which we can build partial amplitudes.  They can be
defined as the sum of all Feynman diagrams with a fixed cyclic ordering
of the colored lines, with a definite routing of the fermion lines,
and with vertices that are given by
color-ordered Feynman rules [\use\ManganoReview].  These differ from
ordinary Feynman rules in that they have been stripped of the usual
color factors.  Here we choose {\it not} to fix the cyclic ordering of
the colorless lepton pair with respect to the colored partons.  In
this case the lepton pair plays no role in the color structure,
and so the equations expressing the partial amplitudes $A_{6;1}$
and $A_{6;3}$ as sums of primitive amplitudes are identical to those
derived in ref.~[\use\Fermion] for one-loop two-quark two-gluon
amplitudes --- see eqs.~(\use\totalpartialampl) below.

The main utility of primitive amplitudes as compared to partial
amplitudes (i.e.~the coefficients of a particular color structure) is
that their analytic structures are generally simpler because fewer
orderings of external legs appear.  They also
contain somewhat more color information, making it more
straightforward to convert the amplitudes to those for other processes
(for example, to replace gluons by photons, or quarks by gluinos).

For the $e^+\,e^-\to \qb q gg$ amplitudes, `parent diagrams' for each
gauge invariant class are depicted in \fig\PrimitiveDiagramsFigure.
By a `parent' diagram we mean a diagram from which all other diagrams
in the set can be obtained via a continuous `pinching' process, in
which two lines attached to the loop are brought together to a
four-point interaction --- if such an interaction exists --- or
further pulled out from the loop, and left as the branches of a tree
attached to the loop.  
The cyclic ordering of external legs is always preserved by pinching.
Because we do not fix the ordering of the lepton pair with respect to 
the partons, primitive amplitudes with an external gluon cyclicly 
adjacent to the vector boson ($\gamma^*,Z$) have more than one 
parent diagram.
For example, $A_6(1_q,2,3,4_\qb)$
(fig.~\use\PrimitiveDiagramsFigure{b}) has only one parent diagram,
while $A_6(1_q,2,3_\qb,4)$ (fig.~\use\PrimitiveDiagramsFigure{c}) has
two parent diagrams, and $A_6(1_q,2_\qb,3,4)$
(fig.~\use\PrimitiveDiagramsFigure{d}) has three.  Note that these
primitive amplitudes are different functions, not the same function
with a different permutation of the arguments: $A_6(1_q,2_\qb,3,4)\neq
A_6(1_q,2,3,4_\qb)$.

\vskip -.5 cm 

%
\LoadFigure\PrimitiveDiagramsFigure
{\baselineskip 13 pt
\noindent\narrower\ninerm
Parent diagrams for the various two quark and two gluon primitive
amplitudes.  Straight lines represent fermions, curly lines gluons,
and wavy lines a vector boson ($\gamma^*$ or $Z$).  
}  {\epsfxsize 5.4 truein}{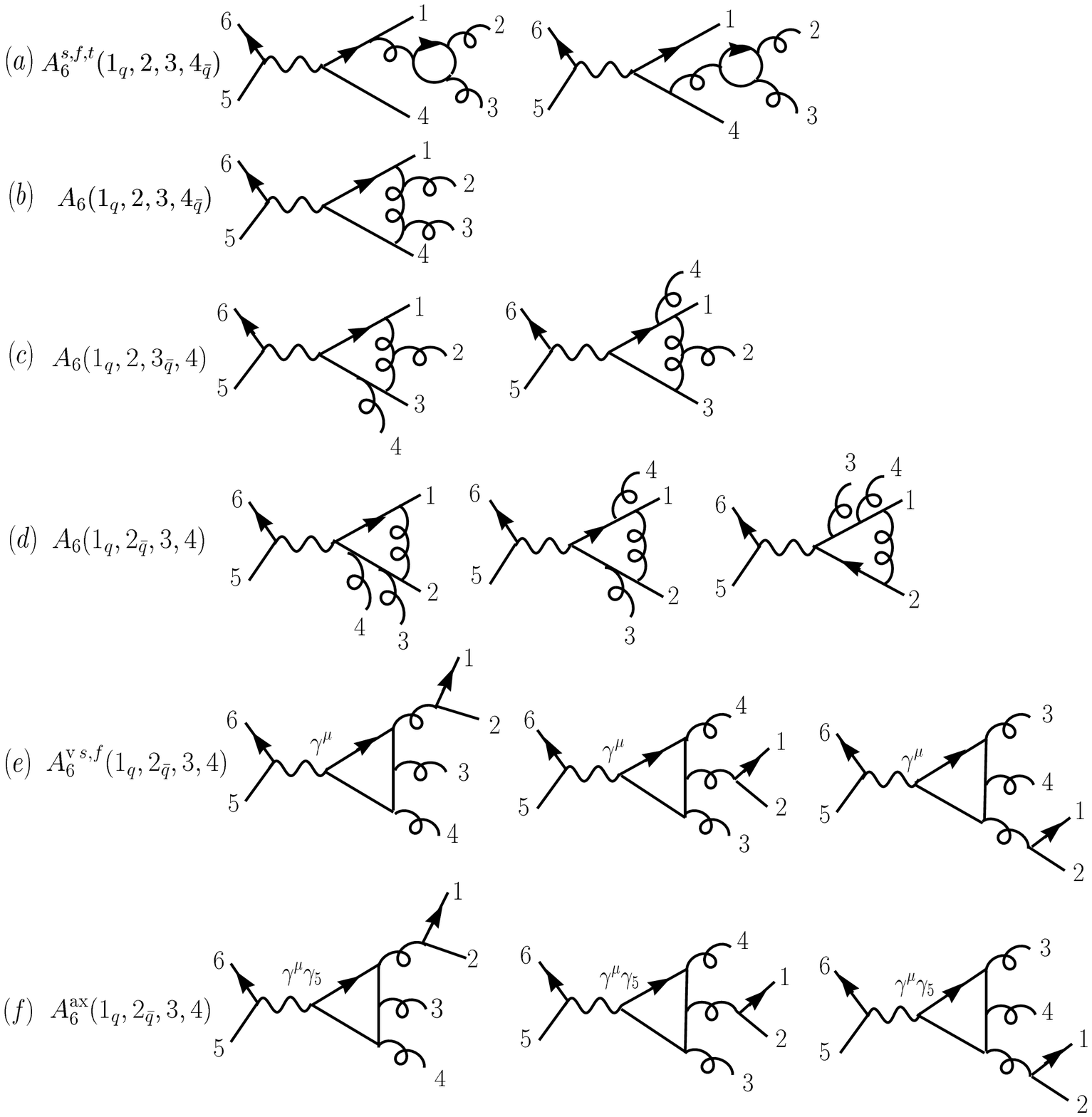}{}

It turns out to be useful, in diagrams of the type shown in
figs.~\use\PrimitiveDiagramsFigure{b}, c and d,
to replace a gluon loop contribution with two separate contributions, that of
a scalar and that of the difference of a gluon and scalar, as shown in 
\fig\ScalarGluonFigure.  This separates the gluon loop contribution into 
two gauge-invariant pieces.  As we shall discuss in
section~{\use\CutConstructionSection} this separation is advantageous
because of differing analytic properties of the two pieces.  Terms
where a scalar replaces a gluon are labeled with a superscript
`$\sc$', while those with the difference of gluon and scalar are
labeled with a superscript `$\gf$'.  (As we shall discuss in
section~\use\CuttingRulesSubsection\ the `$\gf$' terms are
`cut-constructible' meaning that they can be constructed solely from
four-dimensional cuts.)  Furthermore, terms proportional either to the
number of scalars%
\footnote{$^*$}{As in refs.~[\use\FiveGluon,\use\Fermion], each scalar 
here contains four states (to match the four states of Dirac fermions)
so that $n_s$ must be divided by two for comparisons to conventional
normalizations of scalars.}
$n_s$ (which vanishes in QCD), or fermions $n_{\!f}$, are separately
gauge invariant so we also separate these out explicitly.
(Due to Furry's theorem~[\use\Furry], or charge conjugation invariance, 
these terms appear only in $A_{6;1}$.)

\vskip -.5 cm 
%
\LoadFigure\ScalarGluonFigure
{\baselineskip 13 pt
\narrower\ninerm\noindent
The contribution from a gluon in the loop is separated
into the difference of a gluon and scalar, plus a scalar.}
{\epsfxsize 4.5 truein}{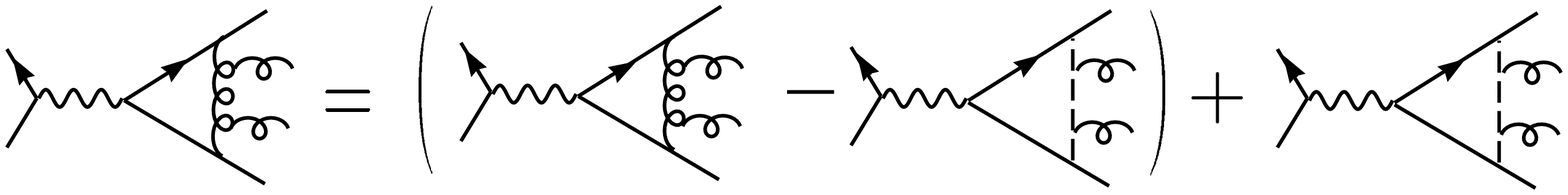}{}

Thus we take the decomposition of the partial
amplitudes in terms of the primitive amplitudes to be
$$
\eqalign{
  A_{6;1} (1_q, 2, 3, 4_\qb) &= 
     A_6 (1_q, 2, 3, 4_\qb)
  - {1\over N_c^2} A_6(1_q, 4_\qb, 3, 2)  \cr
& \hskip 1 cm
  + {n_s -n_{\! f}\over N_c} A_6^s(1_q, 2, 3, 4_\qb) 
  - {n_{\! f} \over N_c} A_6^f (1_q, 2, 3, 4_\qb) 
  + {1\over N_c} A_6^t (1_q, 2, 3, 4_\qb) \,,\cr 
 A_{6;3} (1_q, 4_\qb; 2, 3) & =  
A_6(1_q, 2, 3, 4_\qb) + A_6(1_q, 3, 2, 4_\qb) +
A_6(1_q, 2, 4_\qb, 3) + A_6(1_q, 3, 4_\qb, 2) \cr
& \hskip 1 cm 
 +  A_6(1_q, 4_\qb, 2, 3) +  A_6(1_q, 4_\qb, 3, 2)\,,  \cr
 A_{6;4}^\vect (1_q, 4_\qb; 2,3) & = -A_{6}^{\vect s} (1_q, 4_\qb, 2, 3)
         - A_{6}^{\vect f} (1_q, 4_\qb, 2, 3) \,, \cr
 A_{6;4}^\ax (1_q, 4_\qb; 2, 3) & = A_{6}^\ax (1_q, 4_\qb, 2, 3) \,, \cr
 A_{6;5}^\ax (1_q, 4_\qb; 2, 3) & = A_{6}^{\ax,\sl} (1_q, 4_\qb, 2, 3) \,.\cr
}\eqn\totalpartialampl
$$
We have decomposed the fermion loop in $A_{6;4}^\vect$ into a scalar
loop piece $A_{6}^{\vect s}$ and an additional piece $A_{6}^{\vect f}$. 
As mentioned above, we perform a further decomposition of the $A_6$ into 
$\gf$ and $\sc$ pieces, as depicted in fig.~\use\ScalarGluonFigure,
$$
\eqalign{
A_6(1_q, 2, 3, 4_\qb) & = 
         A_6^\gf(1_q, 2, 3, 4_\qb) + A_6^\sc(1_q, 2, 3, 4_\qb)\,,\cr
A_6(1_q, 2, 3_\qb, 4) & = 
         A_6^\gf(1_q, 2, 3_\qb, 4) + A_6^\sc(1_q, 2, 3_\qb, 4)\,,\cr
A_6(1_q, 2_\qb, 3, 4) & = 
         A_6^\gf(1_q, 2_\qb, 3, 4) + A_6^\sc(1_q, 2_\qb, 3, 4) \,. \cr}
\eqn\gluonminusscalardecomp
$$
Finally, $A_6^t$ gives the top quark contribution to $A_{6;1}$,
through order $1/m_t^2$.

We choose a set of helicity amplitudes from which all others may be
obtained by applying the discrete symmetries of parity and charge
conjugation.  Parity reverses all external helicities in a partial
amplitude; it is implemented by the ``complex conjugation'' operation,
which substitutes $\spa{j}.{l} \to \spb{l}.{j}$,\ \ $\spb{j}.{l} \to
\spa{l}.{j}$.  For the axial-vector fermion loop 
partial amplitudes one must multiply by an additional overall minus sign.
Charge conjugation changes the identity
of a fermion to an anti-fermion and vice-versa.  
These operations allow us to fix the helicity of the electron and the
quark to be positive, and the positron and anti-quark to be negative.
In addition, if the two gluons have the same helicity, we can fix
that common helicity to be positive.  

For the four-quark amplitudes, a similar use of charge conjugation 
and parity reduces the independent partial amplitudes to 
$A_{6;i}(1_\q^+,2_\Qb^+,3_\Q^-,4_\qb^-,5_\eb^-,6_e^+)$ and
$A_{6;i}(1_\q^+,2_\Qb^-,3_\Q^+,4_\qb^-,5_\eb^-,6_e^+)$.
The formul\ae\ for the four-quark partial amplitudes,
$A_{6;i}(1_\q^+,2_\Qb^\pm,3_\Q^\mp,4_\qb^-)$,
in terms of primitive amplitudes are~[\use\Zqqqq]
$$
\eqalign{
 A_{6;1}(1_\q^+,2_\Qb^+,3_\Q^-,4_\qb^-) 
&=  A^\nn_6(1,2,3,4)
- {2\over N_c^2} \bigl(A^\nn_6(1,2,3,4) +
                       A^\an_6(1,3,2,4) \bigr) 
+ {1\over N_c^2} A_6^{\sl}(2,3,1,4) \cr
& \hskip .5 cm 
+ {n_s - n_{\!f} \over N_c} A^{s\,,\nn}_6(1,2,3,4)
- {n_{\! f} \over N_c} A^{\!f,\,\nn}_6(1,2,3,4) 
+ {1\over N_c} A_6^{t,\,\nn}(1,2,3,4) \,,\cr\cr
A_{6;2}(1_\q^+,2_\Qb^+,3_\Q^-,4_\qb^-) 
&=  A^\an_6(1,3,2,4) 
+ {1\over N_c^2} \bigl(A^\an_6(1,3,2,4) +
                       A^\nn_6(1,2,3,4) \bigr)
- {1\over N_c^2} A_6^{\sl}(2,3,1,4) \cr
& \hskip .5 cm 
- {n_s - n_{\!f} \over N_c} A^{s,\,\nn}_6(1,2,3,4)
+ {n_{\! f} \over N_c} A^{\!f,\,\nn}_6(1,2,3,4) 
- {1\over N_c} A_6^{t,\,\nn}(1,2,3,4) \,,\cr\cr
A_{6;3}(1_\q^+,2_\Qb^+,3_\Q^-,4_\qb^-) &= A^\ax_6(1,4,2,3) \,,\cr
}\eqn\nndecomp
$$
and
$$
\eqalign{
 A_{6;1}(1_\q^+,2_\Qb^-,3_\Q^+,4_\qb^-) 
&=  A_6^\an(1,2,3,4)
- {2 \over N_c^2}\bigl( A_6^\an(1,2,3,4) 
                      + A_6^\nn(1,3,2,4) \bigr)
- {1\over N_c^2}  A_6^{\sl}(3,2,1,4) \cr
& \hskip .5 cm 
+ {n_s - n_{\!f} \over N_c} A^{s,\,\an}_6(1,2,3,4)
- {n_{\! f} \over N_c} A^{\!f,\,\an}_6(1,2,3,4) 
+ {1\over N_c} A_6^{t,\,\an}(1,2,3,4) \,,\cr\cr
A_{6;2}(1_\q^+,2_\Qb^-,3_\Q^+,4_\qb^-) 
&=  A_6^\nn(1,3,2,4)
+ {1\over N_c^2} \bigl(  A^\nn_6(1,3,2,4) 
                       + A^\an_6(1,2,3,4) \bigr)
+ {1\over N_c^2}  A_6^{\sl}(3,2,1,4) \cr
& \hskip .5 cm 
- {n_s - n_{\!f} \over N_c} A^{s,\,\an}_6(1,2,3,4)
 + {n_{\! f} \over N_c} A^{\!f,\,\an}_6(1,2,3,4) 
- {1\over N_c} A_6^{t,\,\an}(1,2,3,4) \,,\cr\cr
A_{6;3}(1_\q^+,2_\Qb^-,3_\Q^+,4_\qb^-) &= -A^\ax_6(1,4,3,2) \,.\cr
}\eqn\andecomp
$$
Although color decompositions do not depend on the helicity choices, the
sign differences in these equations appear because we have used symmetries
of the four-quark primitive amplitudes to reduce the number of independent
ones required.  In \fig\PrimitiveDiagramsZqqqqFigure\ we display the
parent diagrams associated with each of the primitive amplitudes

%
\LoadFigure\PrimitiveDiagramsZqqqqFigure
{\baselineskip 13 pt
\noindent\narrower\ninerm
Parent diagrams for the four-quark primitive
amplitudes. 
In each case the vector boson can appear on either side of a
gluon line that is attached to the same quark line.}
{\epsfxsize 4.7 truein}{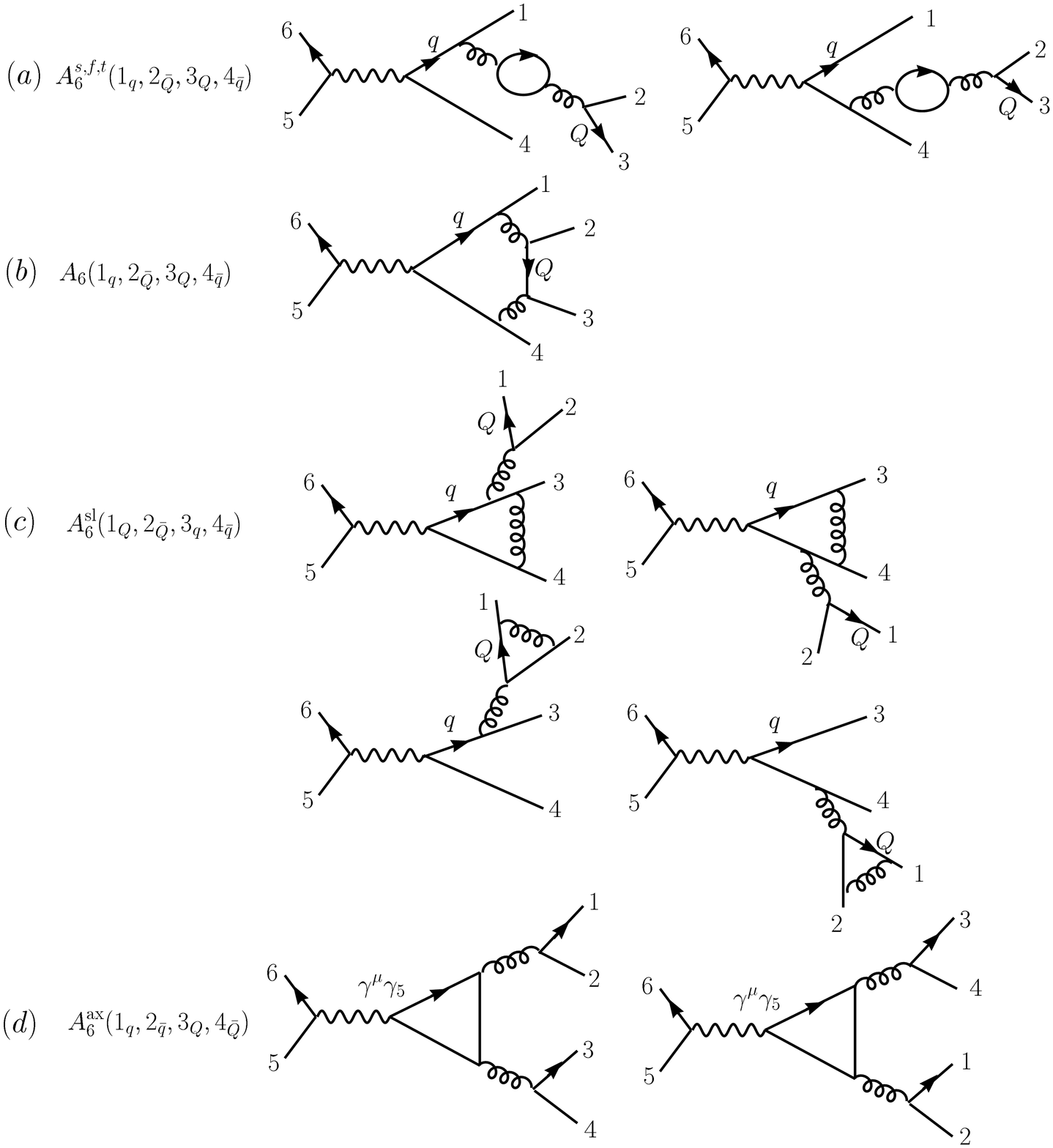}{}

\section{Analytic Construction of Amplitudes}
\tagsection\CutConstructionSection

In this section, we review the construction of one-loop amplitudes
starting from their analytic properties.  The two analytic properties we
shall use are the determination of imaginary parts by Cutkosky rules, and
factorization on particle poles.  These properties of amplitudes have, of
course, played an important role in field theory for many decades; the
recent development which we focus on here is the ability to obtain,
efficiently, complete amplitudes with no subtractions or ambiguities.
These techniques, reviewed in ref.~[\use\Review], have been applied to
obtain results for both nonsupersymmetric and supersymmetric maximally
helicity violating $n$-point
amplitudes~[\use\AllPlus,\use\SusyFour,\use\SusyOne], and more recently
for multi-loop $N=4$ supersymmetric four-point
amplitudes~[\use\TwoLoopSusy].

\subsection{Cutting Rules}
\tagsubsection\CuttingRulesSubsection

Cutkosky rules [\use\Cutting,\use\PeskinSchroeder] allow one to obtain
the imaginary parts of one-loop amplitudes directly from
products of tree amplitudes.  (By imaginary parts we mean absorptive
parts, that is discontinuities across branch cuts.) 
We apply Cutkosky rules to amplitudes instead of diagrams because
amplitudes, being gauge invariant, are simpler. In the channel with
momentum squared $K^2 \equiv (k_{m_1} + k_{m_1+1} +\cdots +
k_{m_2})^2$ the cut of an amplitude is (see \fig\CutAmplitudeFigure)
$$
\eqalign{
A^{1\rm -loop}_n (1, & 2,\ldots,n)
       \Bigr|_{ K^2 \ \rm cut}\cr
& = i \int \dlips(-\ell_1,\ell_2)
  \ A^{\rm tree}(-\ell_1,m_1,\ldots,m_2,\ell_2) \times
   A^{\rm tree}(-\ell_2,m_2+1,\ldots,m_1-1,\ell_1)\, , \cr}
\eqn\CutEquation
$$
where the integration is over $D$-dimensional Lorentz-invariant
phase space with with momenta $-\ell_1$ and $\ell_2$ for the
intermediate states.  For this channel $K^2$ 
is taken positive and all other kinematic invariants are taken
negative.

\vskip -.5 cm 
%
\LoadFigure\CutAmplitudeFigure{\baselineskip 13 pt
\noindent\narrower\ninerm
A cut amplitude, with momentum $K^2$ flowing across the cut. The lines
represent gluons, scalars or fermions.  }
{\epsfxsize 2.5 truein}{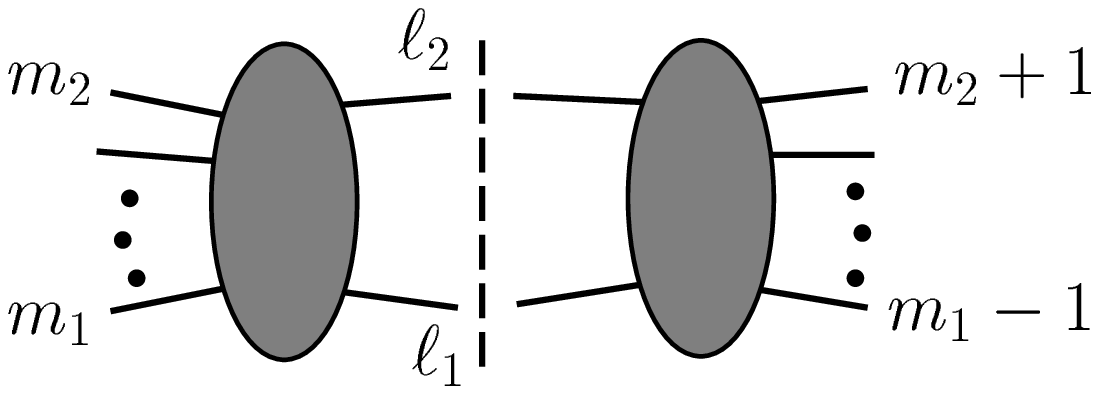}{}

The Cutkosky rules determine imaginary parts of the amplitudes.
Dispersion relations are conventionally used to reconstruct 
real parts from imaginary parts.  Here we bypass dispersion
relations; instead we replace phase-space integrals with cuts 
of unrestricted loop momentum integrals,
$$
\eqalign{
A^{1\rm -loop}_n (1, &2, \cdots n)
   \Bigr|_{K^2\ \rm cut} \cr
  & = \int {d^{D} \ell_1 \over (2\pi)^D}
  \ A^{\rm tree}(-\ell_1,m_1,\ldots,m_2,\ell_2)\, {1\over \ell_2^2}\,
   A^{\rm tree}(-\ell_2,m_2+1,\ldots,m_1-1,\ell_1)\, {1\over \ell_1^2}
   \biggr|_{K^2\ \rm cut} . \cr}
\eqn\TreeProductDef
$$
Whereas eq.~(\use\CutEquation) includes only imaginary parts,
eq.~(\use\TreeProductDef) contains both real and imaginary parts.  As
indicated, eq.~(\use\TreeProductDef) is valid only for those terms
with a $K^2$-channel branch cut; terms without such a branch cut may
not be correct. 
A useful property of this formula is that one may
continue to use on-shell conditions for the cut intermediate legs
inside the tree amplitudes without affecting the result.  Only terms
containing no cut in this channel would change. 
We are able to avoid the use of dispersion relations because we
have additional information, namely that the reconstructed analytic
functions are given by Feynman loop integrals.

A similar equation holds for every branch cut.  If one now combines 
all cuts into a single function having the correct cut in each channel, 
one obtains the full amplitude --- up to the possible addition of a
{\it rational function}, i.e.~a function having no cuts at all,
if one approximates the $(4-2\e)$-dimensional cuts by their
four-dimensional limits (see discussion below).  
The full amplitude can also be expressed as a linear combination of
various types of basic one-loop integral functions, multiplied by
rational function coefficients.  Many of the integral functions,
for example scalar box integrals (which depend on the momenta
of four external legs), have cuts in more than one channel.  
The coefficients of those integral functions as extracted from cuts in 
different channels must agree, and this provides a strong consistency 
check on the reconstructed amplitude.

In general, it is convenient to take the tree amplitudes on either
sides of the cuts to be four-dimensional.  This is natural in the
helicity formalism, which implicitly assumes that momenta are
four-dimensional.  On the other hand, we wish to regulate the
ultraviolet and infrared divergences by letting $D=4-2\eps$ in the
loop integral~(\use\TreeProductDef).  When the $(4-2\eps)$-dimensional
cut momenta are replaced by four-dimensional momenta an error may
occur.  Although the $(-2\eps)$-dimensional parts are implicitly of
$\Ord(\eps)$, if the associated integral has an ultraviolet pole in
$\eps$ an $\Ord(\eps^0)$ rational function may remain.  (As discussed
in refs.~[\use\SusyFour,\use\SusyOne], infrared poles never give rise
to such rational contributions.)  Despite this seeming error, a
complete reconstruction of an amplitude is possible even when the
$\Ord(\eps)$ parts of the cut momenta are dropped, {\it if} the
amplitude satisfies a certain power-counting criterion; we call such
amplitudes {\it cut-constructible}~[\use\SusyOne,\Review].  The
power-counting criterion is that the $n$-point integrals appearing in
the amplitude should have (for $n>2$) at most $n-2$ powers of the loop
momentum in the numerator of the integrand; two-point integrals should
have at most one power of the loop momentum.  (By an $n$-point
integral we mean an integral with $n$ propagator factors in the
denominator, as in equation~(\use\GeneralLoopIntCalI).)
Cut-constructible amplitudes are composed of a restricted set of
integral functions, and sufficient information exists from the
four-dimensional cuts to determine the coefficients of each such
function.  These integral functions automatically include the cut-free
rational functions in the amplitude~[\use\SusyOne].

The full $e^+\,e^- \to \qb q gg$ and $e^+\,e^- \to \qb q \Qb Q$ amplitudes are
not cut-constructible.  Diagrams containing a closed fermion, scalar or
gluon loop can have up to $n$ powers of the loop-momenta in the numerator
of an $n$-point integral; other diagrams typically have up to $n-1$
powers.  However, one can split the amplitudes into `scalar'
contributions, plus additional terms which are cut-constructible; 
in fact, we have already performed such a decomposition in
eqs.~(\use\totalpartialampl) and~(\use\gluonminusscalardecomp).  In the
case of a closed fermion loop, we wrote the fermion loop as the negative
of a scalar loop, plus a second term (which is the contribution of an
$N=1$ supersymmetric chiral multiplet).  As is apparent in a second-order
formalism for fermions, as reviewed in ref.~[\use\Review], the latter
contribution is cut-constructible since the leading two powers of loop 
momentum cancel.

In the case where an external quark line attaches directly to the loop,
for example the leftmost diagram in fig.~\use\ScalarGluonFigure, an
$n$-point integral has a maximum of $n-1$ powers of the loop momentum.  If
one subtracts from this diagram an identical diagram but with the gluon
replaced by a scalar (also shown in fig.~\use\ScalarGluonFigure), suitably
adjusts the scalar-fermion Yukawa coupling, and works in background-field
Feynman gauge~[\use\Background], then the leading loop-momentum terms
cancel.  Thus the difference is cut-constructible~[\use\SusyOne].

In both cases, although the scalar contributions are not
cut-constructible, they are simpler in some respects than the full
amplitudes; for example, certain cuts vanish in the scalar contribution,
but not in the full amplitude.
They also have spurious singularities of different degree from the
cut-constructible terms.  To determine the rational functions for such
amplitudes, which do not satisfy the power-counting criterion (but contain
only massless particles), one may compute to one higher power in the
dimensional regularization parameter $\eps$
[\use\TwoLoopUnitarity,\use\Massive,\use\Review].  However, a more
convenient approach here is to ignore all $\Ord(\e)$ contributions, and
instead use the amplitudes' factorization properties to reconstruct their
rational functions.

\subsection{Factorization}
\tagsubsection\FactorizationSection

The properties of tree-level QCD amplitudes as kinematic invariants
vanish have been presented in various reviews [\use\ManganoReview].
The corresponding one-loop factorization properties have also been
extensively discussed [\use\AllPlus,\use\Factorization,\use\Review], so
here we will briefly review only the salient features necessary for
obtaining the rational function parts of amplitudes.

For amplitudes with six- and higher-point kinematics, the properties of
amplitudes under factorization when a kinematic invariant vanishes are
in general sufficiently powerful to obtain the rational function parts
of amplitudes.  Although factorization is complicated by the
appearance of infrared divergences, nevertheless, as any kinematic
variable vanishes, one-loop amplitudes have a universal behavior quite
similar to that of tree-level amplitudes
[\use\SusyFour,\use\Factorization].  If one finds a function which
obeys the proper factorization equations in all channels, one has an
ansatz for the rational function part of an amplitude.  Although no
proof of the uniqueness of such a construction has been presented, for
six- and higher-point amplitudes no counterexample is known.  (For a
five-point amplitude counterexample see ref.~[\use\Minn].)
Factorization can be a particularly efficient way to obtain the
rational function terms; it avoids the need to perform loop integrals.
This complements the efficiency of the cut-construction technique for
obtaining the logarithms and dilogarithms.  (Resorting to Feynman
diagrams to obtain analytic expressions for the rational function parts
is not satisfactory because such pieces tend to have the most
complicated diagrammatic representation: they are associated with the
maximal power of loop momenta and the largest number of diagrams.)

Of particular utility are the two-particle factorization properties,
depicted in \fig\CollinearFactFigure.  As two momenta become collinear
the amplitudes behave as [\use\AllPlus,\use\SusyFour]
$$
\eqalign{
A_{n}^{1\rm -loop} \mathop{\longrightarrow}^{a \parallel b}
\sum_{\lambda=\pm}  \biggl(
  \Split^{\rm tree}_{-\lambda}(a^{\lambda_a},b^{\lambda_b})\,
&
      A_{n-1}^{1\rm -loop}(\ldots(a+b)^\lambda\ldots)
\cr
&  +\Split^{1\rm -loop}_{-\lambda}(a^{\lambda_a},b^{\lambda_b})\,
      A_{n-1}^{\rm tree}(\ldots(a+b)^\lambda\ldots) \biggr) \;,
\cr}
\eqn\Loopsplit
$$
where $k_a \rightarrow z P$ and $k_b \rightarrow (1-z) P$, with
$P=k_a+k_b$, $P^2=s_{ab} \to 0$.  
The helicity of the intermediate parton $P$ is labeled by $\lambda$.
The tree and loop splitting
amplitudes, $\Split^{\rm tree}_{-\lambda}$ and 
$\Split^{1\rm -loop}_{-\lambda}$, behave as $1/\sqrt{s_{ab}}$ in this
limit.  A complete tabulation of the splitting amplitudes appearing in 
one-loop computations in massless QCD has been given in
refs.~[\use\SusyFour,\use\Fermion,\use\Factorization].  Given the
$n-1$ point amplitude and splitting amplitudes (or `factorization
functions' in multi-particle channels), eq.~(\use\Loopsplit) provides
an extremely stringent check since one must obtain the correct limits
in all channels. A sign or labeling error, for example, will
invariably be detected in some limits.

\vskip -.5 cm 
%
\LoadFigure\CollinearFactFigure{\baselineskip 13 pt
\noindent\narrower\ninerm
A schematic representation of the behavior of one-loop amplitudes as
the momenta of two legs become collinear. }
{\epsfxsize 4.0 truein}{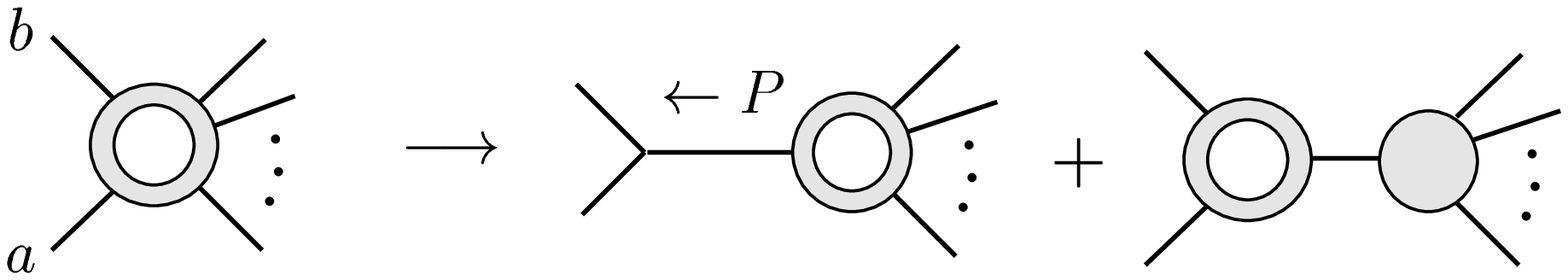}{}

The physical poles that can appear in any massless
amplitudes are square-root singularities in two-particle channels, or single
poles in multi-particle channels [\use\ManganoReview,\use\Review],
$$
{1\over \spa{i}.j} \sim {1\over \sqrt{s_{ij}}} \,, \hskip 1 cm 
{1\over \spb{i}.j} \sim {1\over \sqrt{s_{ij}}} \,, \hskip 1 cm   
{1\over t_{ijk}} \,.
\anoneqn
$$
Any other kinematic pole-type singularity that appears in individual terms
of an amplitude must be spurious; that is, the residue of the pole
must vanish for the full amplitude.  

In this paper we use factorization to construct ans\"atze for the
rational function parts of the amplitudes.  Although the construction
of such an ansatz involves a certain amount of guesswork, the
procedure can be systematized somewhat.  In general, the rational
functions contain non-removable spurious singularities (see
section~\use\SpuriousPoleSection).  In the full amplitude, this
singular behavior cancels against singular behavior in terms that have
logarithms and dilogarithms, and so it can be inferred from the
information provided by the cuts.  Therefore one can readily introduce
rational function terms that account for all, or at least most, of the
spurious singularities.  Subtracting these terms from the full
rational function ansatz leads to an ansatz for the remainder which is
(largely) free of spurious singularities.  At this stage it is
relatively straightforward to proceed channel by channel, adding terms
to the ansatz that correctly reproduce the desired singular behavior
in each channel, using the factorization information provided by
lower-point amplitudes.  The last few channels are simplest, because
by then the remaining terms have very few singularities left.  For
six-point kinematics this procedure invariably gives the correct
result, as we have verified by numerical comparisons to Feynman
diagram computations.  We determine the rational function terms 
for a simple example in section~\use\SampleRationalFunctionSubsection.

\section{Spurious Singularities.} 
\tagsection\SpuriousPoleSection

In this section we discuss the procedure for removing or at least
greatly simplifying spurious kinematic singularities appearing the
amplitudes.  This is essential in order to obtain compact results.
The presence of large numbers of such singularities leads to unwieldy
results which tend to be unsuitable for use in jet programs since they
are numerically unstable.  Although the expressions encountered in the
calculation of the cuts via eq.~(\use\TreeProductDef) are rather
compact when compared to a direct Feynman diagram calculation, the
amplitudes are sufficiently intricate that even a small number of
spurious singularities can seriously hinder attempts to obtain compact
results.

Spurious singularities fall into two categories, removable and 
non-removable:  After expressing the
amplitudes in terms of logarithms and dilogarithms there are spurious 
singularities that can be removed from the amplitudes by algebraic
simplification, and ones that cannot be removed.  
As a trivial example, consider the functions
$$
f_1 = {1-x^2 \over 1-x}\, ; \hskip 2 cm 
f_2 = {\ln x \over 1-x} \,, \hskip 1 cm
f_3 = {\ln x + 1-x \over (1-x)^2} \,,
\eqn\trivialspurious
$$
where the poles at $x=1$ have vanishing residues.  The first function
can be re-expressed as $f_1 = (1+x)$ so the spurious pole at $x=1$ is
removable.  However, the spurious poles in $f_2$ and $f_3$ are not 
removable if we wish to express the function in terms of a logarithm. 
One may, of course, define a new set of functions, of which $f_2$ 
and $f_3$ are examples, which absorb the spurious poles. 
Indeed, this is the role of the
$\Ll_i$ and $\Ls_i$ functions~[\use\FiveGluon] defined in
appendix~\use\IntegralFunctionAppendix. (Generalizations of such
functions have been presented in ref.~[\use\CGM].)  
In the $\Ll_i$ functions $x$ is a ratio of kinematic invariants;
for example, $x = s_{23}/t_{123}$ arises from the Gram determinant
$\Delta_3^{(2,5)}$ in eq.~(\use\GramThree).
In Feynman diagram
calculations both removable and non-removable singularities occur.
The removable singularities are an artifact of the integral reduction
techniques employed, but the non-removable ones are an inherent part
of the amplitudes when they are expressed in terms of logarithms and
dilogarithms.  

The complications arising from spurious denominators follow largely
from dimensional analysis and the fact that they carry 
positive dimensions.  Their appearance implies
that the numerators must have compensating powers of momenta.  In a
six-point amplitude there are five independent momenta, leading to a
substantial proliferation in the number of possible numerators.  As an
example, consider a five-point tensor integral with four powers of
loop momentum in the numerator encountered in a Feynman diagram
evaluation of $e^+\,e^- \rightarrow \qb q gg$.  If this integral were
evaluated by conventional means using a Passarino-Veltman reduction
[\use\PV], summarized in appendix~\use\IntegralReductionSubAppendix,
one would have up to four inverse powers of the pentagon Gram
determinant $\Delta_5$ defined in eq.~(\use\GramFive).  As we shall
show below, these spurious singularities can always be removed from
the amplitudes of this paper, but if they appear in intermediate
expressions, their removal is an arduous task.  The appearance of a
spurious $\Delta_5^{-4}$ denominator implies that the numerators must
contain an appropriate polynomial to cancel the poles.  Since the
terms in the numerator of a brute force calculation appear in a
seemingly haphazard pattern, one must deal with the order of $22^4
\sim 10^5$ terms to remove the spurious singularities in
$\Delta_5$. Moreover, the various spurious singularities can get
tangled together in rather intricate ways.  Our goal is to eliminate
those spurious poles which do not belong in the amplitudes and to
simplify those which are are inherently associated with the logarithms
and dilogarithms.

\subsection{Types of Spurious Singularities Appearing in the Amplitudes}
\tagsubsection\SpuriousSingularityTypeSubsection

The simplest type of spurious singularities are unphysical poles in
the kinematic variables $s_{ij}$ and $t_{ijk}$.  Even at tree level
the kinematic singularities in the amplitude can be non-trivial.  For
example, consider the tree amplitude $A_6^{\rm tree}(1_q^+, 2^+, 3^-,
4_\qb^-)$ given in eqs.~(\use\treeppmm) and (\use\treeppmmalt).  In
the first form (\use\treeppmm) the amplitude exhibits the
proper poles in the $t_{123}$ and $t_{234}$ three-particle channels,
but the behavior as $s_{23}$ or $s_{56}$ vanish is not manifest ---
the apparent full poles in these variables actually cancel down to 
square-root singularities.
The second form~(\use\treeppmmalt) of the amplitude exhibits 
the proper square-root singularities in all two-particle channels, 
at the expense of more obscure behavior in three-particle channels.

At one loop, the situation is greatly complicated by the large variety
of spurious singularities that arise from loop integrals.  
The parent diagrams for the $A_6$ primitive amplitudes in 
figs.~\use\PrimitiveDiagramsFigure\ and~\use\PrimitiveDiagramsZqqqqFigure\
require the evaluation of pentagon integrals where all internal lines are
massless, and all external legs are massless except for the leg composed 
of the lepton pair 5--6, which has invariant mass $s_{56}$.
Besides this one-mass pentagon integral, there are a number of different
types of box integrals, where either one or two external legs are
massive.  These integrals may appear directly in a diagram or a
term in a cut evaluation, or they may be generated in the reduction
of pentagon integrals, as reviewed in appendix~\use\IntegralsAppendix.
Similarly, we find triangle integrals with one, two or three external
massive legs, as well as bubble (two-point) integrals.
All these integrals have associated with them different
kinematic factors, which can appear in the denominators of
coefficients of logarithms and dilogarithms, and whose vanishings
correspond to separate spurious singularities. 
 
In appendix~\use\IntegralsAppendix\ we summarize some of the standard
integral reduction methods
[\use\PV,\use\VNV,\use\IntegralsShort,\use\IntegralsLong], and their
associated spurious singularities.   As discussed in the appendix, when
evaluating tensor $n$-point loop integrals one obtains denominators 
containing (minus) the Gram determinants
$$
\Delta_n = -\det(2 K_i \cdot K_j)\, , \hskip 2 cm i,j=1,2, \ldots, n-1,
\eqn\GramDet
$$
where the $K_i$ are external momenta or sums of external momenta.
Other kinematic denominators which appear are the determinants
$$
\det (S_{ij})\, ,  \hskip 2 cm i,j=1,2, \ldots, n,
\anoneqn
$$
where the symmetric matrix $S_{ij}$ is given by
$$
  S_{ii} = 0 \,, \hskip 2 cm 
  S_{ij} = -\hf ( K_i+K_{i+1}+\cdots+K_{j-1} )^2 \,, 
  \hskip .5 cm {\rm for}\ i<j \,.
\eqn\Sdefn
$$

In appendix~\use\IntegralsAppendix\ we collect the explicit
forms of the various determinants that can appear as poles in the
amplitudes.  Particularly important are poles in the three-mass 
triangle Gram determinant
$$
\Delta_3 \equiv
\Delta_3^{(2,4)} = 
  s_{12}^2 + s_{34}^2 + s_{56}^2 
- 2 \, s_{12}\, s_{34} - 2 \, s_{34}\, s_{56} - 2 \, s_{56}\, s_{12} \,, 
\eqn\ThreeMassDeltaThree
$$
in the spinor strings
$$
\spab1.{(2+3)}.4 \,, \hskip 2 cm 
\spab4.{(2+3)}.1 \,,
\eqn\BackToBack
$$
and in objects related to these by permutations of the external legs
$1,2,3,4$.  The spinor strings~(\use\BackToBack) vanish in the kinematic
configuration where $k_2+ k_3 = a k_1 + b k_4$ where $a$ and $b$ are
arbitrary constants.   We call this configuration `back-to-back'
because in the center-of-mass frame for particles 1 and 4, viewed
as incoming, the outgoing three-momenta $\vec{k}_2+\vec{k}_3$ 
and $\vec{k}_5+\vec{k}_6$ must be parallel to $\vec{k}_1$ and $\vec{k}_4$.
In many cases, these singularities cancel only after taking into
account the behavior of the dilogarithms and logarithms that appear, 
in analogy to the behavior of $f_2$ and $f_3$ in 
eq.~(\use\trivialspurious).
The appearance of the three-external-mass triangle Gram determinants
(\ThreeMassDeltaThree) and the back-to-back singularities
(\use\BackToBack) explains to a large extent the significant increase
in complexity of the amplitudes presented in this paper, as compared to
the massless five-parton amplitudes
[\use\FiveGluon,\use\Kunsztqqqqg,\use\Fermion].  It is essential to
simplify terms containing these singularities if we wish to obtain
(relatively) compact expressions.

\subsection{Integral Reductions}
\tagsubsection\StrategySubsection

We now describe techniques we used to help minimize the algebraic
complexity of intermediate expressions when evaluating a cut amplitude
(\use\TreeProductDef).  In particular, these techniques prevent the
appearance of the pentagon Gram determinants, which are by far the most
noxious of the unwanted determinantal denominators.  The same
techniques apply just as well to Feynman diagrams, although in 
evaluating a cut one generally begins with a much more compact 
expression, making it simpler to keep its size small.

An important aspect to the calculations performed in this paper is the
use of a helicity basis for both quark and gluon external states.
This basis simplifies general gauge theory amplitudes, as reviewed in
refs.~[\use\ManganoReview,\use\Review].  (To make effective
use of the helicity formalism at the loop level it is important to use
a compatible regularization scheme, such as the FDH scheme
[\use\Long], which at one-loop has been shown to be equivalent to a
helicity form of dimensional
reduction~[\use\Siegel,\use\KunsztFourPoint].)  The
spinor helicity method also leads to a useful procedure for evaluating
tensor integrals.  The basic observation is that certain combinations
of Lorentz invariant products, such as $s_{56}s_{23}-t_{123}t_{234}$,
which are destined to appear in the denominators of amplitudes for
reasons discussed in the previous subsection, and which cannot be
factored in terms of Lorentz invariants, {\it can} be factored in
terms of spinor strings such as $\spab1.{(2+3)}.4$, as shown in
eq.~(\use\ssmttFactor).  By multiplying and dividing by such spinor
square roots, and by further manipulating the spinor strings in the
numerator of the loop momentum integral for a cut, one can extract
inverse propagators.  These factors cancel propagators in the
denominator, leaving behind much simpler lower-point integrals to
evaluate.  This general strategy is similar to the more conventional
Passarino-Veltman reduction in that it expresses tensor integrals as
linear combinations of simpler integrals, but it differs in its
economical use of expressions which already appear in the loop
momentum integrands.  The strategy has already been applied to a
number of cases
[\SusyFour,\use\SusyOne,\use\Kunsztqqqqg,\use\Massive,\use\Pittau,%
\use\TwoLoopSusy].  The amplitudes of this paper 
have rather complicated kinematics, and so it is very important to 
exploit factorization relations such as~(\use\ssmttFactor), in order to 
appropriately arrange the spinor strings and thus maximize the 
simplifications.

When evaluating a cut (\use\TreeProductDef) one typically
encounters a pentagon loop momentum integral of the form
$$
\int {d^{4-2\eps} p \over (2 \pi)^{4-2\eps}} \,
{\spab{a}.{\ell_i}.{b} \spab{c}.{\ell_j}.{d} \cdots  \over
   \ell_1^2\, \ell_2^2\, \ell_3^2\, \ell_4^2\, \ell_5^2} \, ,
\anoneqn
$$
where $\ell_i= p - k_1 - \cdots - k_{i-1}$ is the momentum of the
$i$th loop propagator.  The kinematic configuration of this pentagon
integral is shown in \fig\PentagonKinematicsFigure. 
By multiplying and dividing by $\spab{b}.{d}.{c}$,  
we may convert the numerator factors into a single spinor string 
containing both $\lsl_i$ and $\lsl_j$, 
$$
\spab{a}.{\ell_i}.{b} \spab{c}.{\ell_j}.{d}  = 
 {1\over\spab{b}.{d}.{c}} \, \spab{a}.{\ell_i b d c \ell_j}.{b} \,.
\anoneqn
$$
We may then extract inverse
propagators by commuting $\lsl_i$ and $\lsl_j$ towards each other.  
In this way we generate terms of the form
$$
2 \, k_j \cdot \ell_m \,, \hskip 1 cm 2\, \ell_i \cdot \ell_j\,.
\anoneqn
$$
For this to be a useful rearrangement, these dot products must be
expressible as sums of inverse propagators and external kinematic
variables.  This requirement dictates some care
in deciding which pairs of strings $\spab{a}.{\ell_i}.{b}$, etc., 
to work with, and which momentum $d$ to use in the string 
$\spab{b}.{d}.{c}$ that one multiplies and divides by.
An inappropriate choice can lead to a large algebraic expression without
any reduction in powers of loop momenta in the numerators of the
integrals.

\vskip -.5 cm
%
\LoadFigure\PentagonKinematicsFigure{\baselineskip 13 pt
\noindent\narrower\ninerm
The momenta associated with the pentagon integral appearing in the 
example. }
{\epsfxsize 1.7 truein}{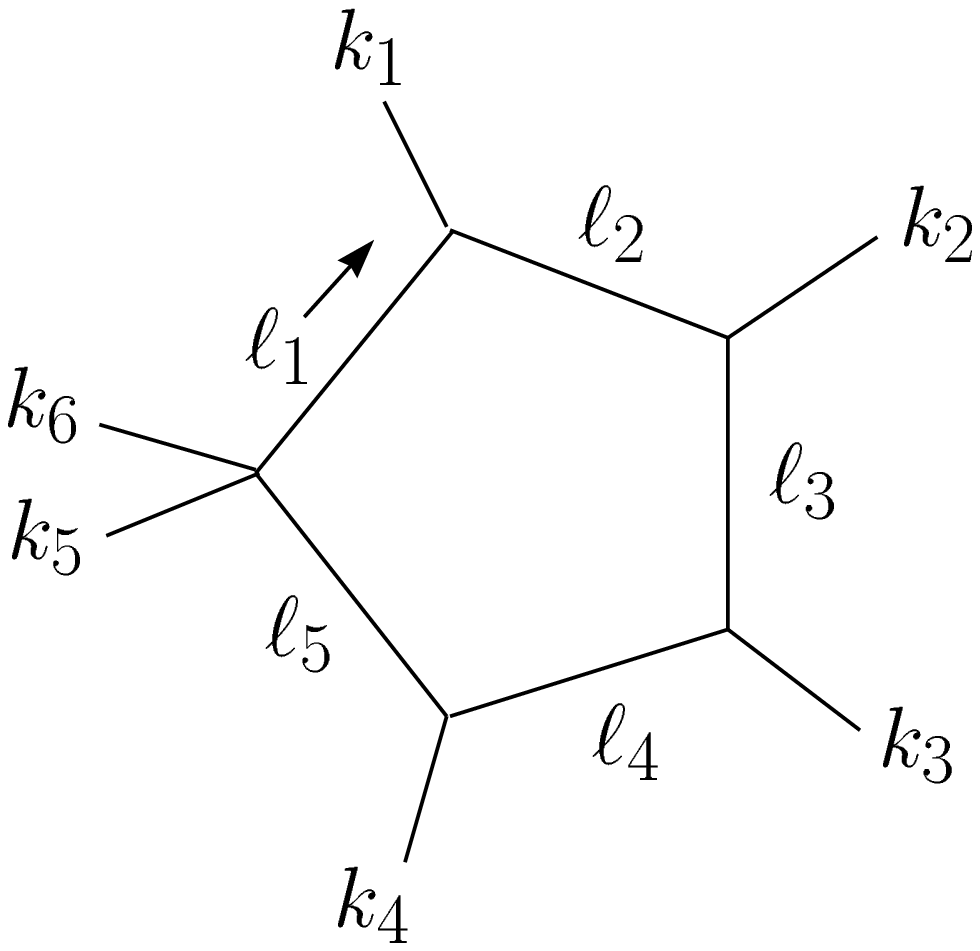}{}

For example, consider a numerator containing the product
$$
 \spab{5}.{\ell_1}.{1} \spab{4}.{\ell_5}.{6} \cdots \, ,
\anoneqn
$$
which occurs in the evaluation of the cut in the $s_{56}$ channel for
the amplitude $A_6^\sc(1_q^+,2^-,3^+,4_\qb^-,5_\eb^-,6_e^+)$.  In this
channel we may use the on-shell conditions $\ell_5^2 =0$ and $\ell_1^2
= 0$.  Before explaining a good choice for forming a spinor string we
mention first some choices which are {\it not\/} very helpful.  For
example, we might combine the spinors via
$$
 \spab{5}.{\ell_1}.{1} \spab{4}.{\ell_5}.{6} 
 = {1\over \spa1.6} \, \spab{5}.{\ell_1}.{1} \spa1.6 \spba{6}.{\ell_5}.{4}  
 = {1\over \spa1.6}\,  \spaa{5}.{\ell_1 16 \ell_5}.4 \, .
\eqn\BadChoice
$$
In this case, if we commute $\lsl_1$ or $\lsl_5$ past $\ksl_6$ we
obtain a term containing $k_6 \cdot \ell_1$ or $k_6 \cdot \ell_5$,
neither of which can be expressed in terms of inverse propagators,
since only $k_5+k_6$ appears in the loop propagators.  Other choices,
such as multiplying and dividing by $\spab1.3.4$, are better but
still introduce unwanted spurious singularities in the amplitudes, 
which must be removed at later stages in the calculation.

A much better choice is to multiply and divide by 
$\spab1.{(2+3)}.4 = - \spab1.{(5+6)}.4$, 
so that
$$
\spab{5}.{\ell_1}.{1}\spab{4}.{\ell_5}.{6} = 
- {1\over \spab1.{(2+3)}.4}
   \spab{5}.{\ell_1 1 (5+6) 4 \ell_5}.{6} \, ,
\eqn\SampleNumerator
$$
which will ensure that we commute $\lsl_1$ and $\lsl_5$ only with
neighboring momenta.  This choice is motivated by the appearance of this
kinematic singularity in the scalar pentagon integral reduction formula
(\use\IntRecursion) (with $n=5$), after expressing the determinantal
denominator (\use\PentagonSDet) in the factored form (\use\ssmttFactor).

Once a numerator term is in a form where at least two $\lsl_i$ are in 
the same inner product we can commute these terms toward each other.
In particular, for the spinor string appearing in eq.~(\use\SampleNumerator)
we have
$$
\eqalign{
\spab5.{\ell_1 1 (5+6) 4 \ell_5}.6 
& = \ell_4^2 \spab5.{\ell_1 1 5}.6
  + \ell_2^2 \spab5.{6 \ell_5 4 }.6 
  + \spab5.{1\ell_1 (5+6) \ell_5 4}.6  \cr
& = \ell_4^2 \spb6.5 \spa5.1 \spab5.{\ell_1}.1
   - \ell_2^2 \spa5.6 \spb6.4 \spab4.{\ell_5}.6 \, , \cr}
\eqn\SimplifiedSpinors
$$
where we used 
$$
\spab5.{1\ell_1 (5+6) \ell_5 4}.6 = 
\spab5.{1\ell_1 (\ell_5 - \ell_1) \ell_5 4}.6 = 0 \, ,
\anoneqn
$$
and 
$$ 
2 \ell_5 \cdot k_4 = (\ell_5 + k_4)^2 = \ell_4^2 \,, \hskip 2 cm 
 2 \ell_1 \cdot k_1 =  - \ell_2^2 \,,
\anoneqn
$$
which follow from the $s_{56}$-cut conditions $\ell_1^2 = \ell_5^2 = 0$.  
Since both terms in eq.~(\use\SimplifiedSpinors)
contain inverse propagators we
have succeeded in reducing the pentagon integral to a sum of two box
integrals.  The rather clean simplification is due to our choice of
multiplying and dividing by $\spab1.{(2 + 3)}.4$.  Of course, not all
cases are reduced as easily, but this example does illustrate the
importance of choosing appropriate factors to multiply and divide by.

Each inverse propagator appearing in a numerator cancels a
propagator, leaving a lower-point integral with one less power of loop
momentum; if there are any terms without an inverse propagator then
they are also down by one power of loop momentum.  
Thus, by combining spinor strings 
and commuting pairs of $\lsl_i$ toward each other, we can always
reduce a pentagon integral with $m>1$ powers of loop momentum to a
linear combination of pentagon and box integrals with at most $m-1$
powers of loop momentum.  When one or no powers of loop momentum are
obtained we may use the reduction formul\ae~(\use\IntRecursion) 
and (\use\SinglePowerIntRed) (with $n=5$), which are free of
pentagon Gram determinants.  In this way, we avoid encountering any
$\Delta_5$s in the calculation.

\vskip .5 cm
\hskip 2. truecm
\hbox{
\def\tend{\cr \noalign{\hrule}}

\vbox{\offinterlineskip
{
\hrule
\halign{
        &\vrule#
        &\strut\quad#\hfil\quad\vrule
        & \quad\hfil\strut # \hfil
        \cr
height15pt & {\bf Integral}  &{\bf Back-to-Back Singularities}   &\tend
height15pt &$I_5\,,\, I_4^{(3)}\,,\, \tI_4^{(3)} $
    & $\spab1.{(2+3)}.4\;, \, \spab4.{(2+3)}.1$   &\tend
height15pt &$\tI_5\,,\, I_4^{(2)} \,,\, \tI_4^{(2)} \,,\, \ttI_4^{(2)} $ 
      & $\spab4.{(1+2)}.3\;,  \, \spab3.{(1+2)}.4 $ &\tend
height15pt &$\ttI_5 \,,\, \tI_4^{(1)} \,,\, \ttI_4^{(1)} $ 
  & $\spab3.{(4+1)}.2\;, \, \spab2.{(4+1)}.3$  &\tend
height15pt &$I_4^{(4)} \,,\, \ttI_4^{(4)}$ 
  & $\spab1.{(3+4)}.2\;, \, \spab2.{(3+4)}.1$ &\tend
}
}
}
}
\vskip .2 cm
\nobreak
{\baselineskip 13 pt
\narrower\smallskip\noindent\ninerm
{\ninebf Table 1:}  The back-to-back singularities associated with 
reductions of each type of integral function. 
\smallskip}

The best quantities to multiply and divide by when forming a spinor string
are usually the back-to-back spinor products which would occur
in the Passarino-Veltman integral reductions.  In many cases, these
singularities are not removable and appear in our final expressions.
Using the results summarized in appendix~\use\IntegralsAppendix, we have
collected in table~1 the back-to-back singularities associated with
the reduction of each type of integral.  This table provides guidance
in actual calculations as to which spinor products to introduce,
although in some cases simpler alternatives are available (see below).
Since the $(\gamma^*,Z)$ is not colored, for a given color ordering 
it may appear with either cyclic ordering with respect to a 
cyclicly adjacent external gluon.
This means that when the first four legs are ordered 1234 by the color 
flow, one may still have integral functions with three possible orderings
for the six legs: 123456, 123564 and 125634.  
We denote the pentagon integrals corresponding to these three orderings
by $I_5$, $\tI_5$, and $\ttI_5$.  The box integrals are similarly
denoted by $I_4^{(i)}$, $\tI_4^{(i)}$ and $\ttI_5^{(i)}$ where the
label $(i)$ indicates that the box is obtained from the pentagon
by removing the propagator prior (in the clockwise ordering of legs)
to the $i$th external leg.  For convenience we have collected the
integrals appearing in Table~1 in \fig\PentBoxIntegralsFigure.

\vskip -.5 cm 
%
\LoadFigure\PentBoxIntegralsFigure{\baselineskip 13 pt
\noindent\narrower\ninerm
The integrals that can lead to spurious back-to-back singularities.}
{\epsfxsize 4.8 truein}{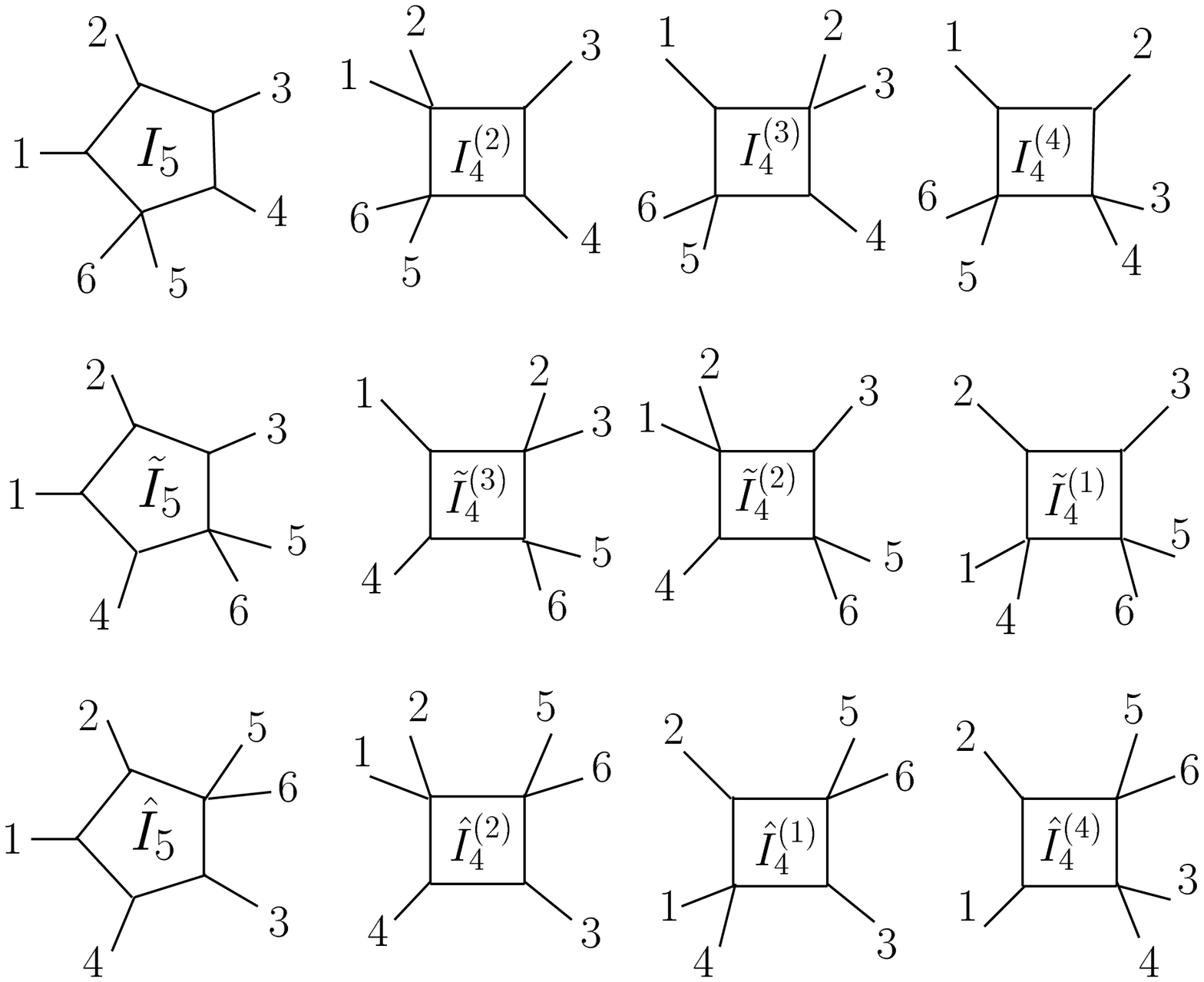}{}

Sometimes the pentagon integral $I_5$ appearing in a cut can be reduced 
to boxes without introducing the `back-to-back' factor $\spab1.{(2+3)}.4$
or its complex conjugate $\spab4.{(2+3)}.1$.  
In such cases, the required factor one should multiply and divide by 
is either $\spa2.3$, or else its complex conjugate $\spb2.3$.    
For example, the cut in the $s_{56}$
channel for the amplitude $A_6^\sc(1_q^+,2_\qb^-,3^-,4^+,5_\eb^-,6_e^+)$
contains a term with the factor
$$
 \langle 4^+ \vert \ell_3 \vert 2^+ \rangle \spab{3}.{\ell_3}.{4}
 = {1\over\spb2.3} \spbb{4}.{\ell_3 2 3 \ell_3}.{4} \, ,
\anoneqn
$$
which is easily reduced to 
$$
{1\over\spb2.3} \Bigl[ 
    (\ell_2^2-\ell_3^2) \, \spb4.3 \spab3.{\ell_3}.4
  + (\ell_4^2-\ell_3^2) \, \spb4.2 \spab2.{\ell_3}.4  
  + \ell_3^2 \, \spbb4.{23}.4 \Bigr] \, .
\anoneqn
$$

After having reduced pentagon integrals to box integrals, the next step 
is to reduce the box integrals.  
Boxes with two adjacent massive legs, depicted in columns two and four 
of fig.~\use\PentBoxIntegralsFigure, 
have the same kind of back-to-back singularities associated
with them as does the pentagon integral (see table~1);
therefore they can usually be reduced by multiplying numerator and
denominator by the string $\spab{i}.{(k+l)}.{j}$ (or its complex
conjugate), along the lines of eq.~(\use\SampleNumerator).  
Table~1 and fig.~\use\PentBoxIntegralsFigure\ 
show that for this box reduction $i$ and $j$ should be the two 
adjacent massless legs.  
In contrast, for the other two types of box integrals that appear,
boxes with only one massive leg and boxes with two diagonally opposite
massive legs, the appropriate factor to multiply and divide by turns out 
to be a single spinor product, $\spa{i}.{j}$ (or its complex conjugate), 
where now $i$ and $j$ represent the two diagonally opposite massless legs.
For a box integral with one massive leg, such as $I_4^{(5)}$,
this is suggested by the factor of 
$s_{13} = -\spa1.3 \, \spb1.3$
in the Gram determinant $\Delta_4^{(5)}$ in eq.~(\use\GramFour). 
For a box integral with diagonally opposite massive legs, such as
$I_4^{(3)}$, the corresponding factor in $\Delta_4^{(3)}$ is
$s_{14} = - \spa1.4 \, \spb1.4$.
Even though the spinor string $\spab1.{(2+3)}.4$ also appears as
a factor in $\Delta_4^{(3)}$, it is never required for the reduction
of this box integral.

The appropriateness of these factors for simplifying numerators 
is due to the fact that they terminate with spinors corresponding
to massless legs of the relevant {\it integrals}, not just the amplitude.
The key to the previous pentagon reductions was having only
massless legs of the pentagon integral interposed between two 
loop momenta in a single spinor string.  
At the level of box and lower-point integrals, fewer of the external
legs of the integral are massless, because they can instead be sums 
of massless legs of the amplitude.   
For example, in the box integral $I_4^{(2)}$ in 
fig.~\use\PentBoxIntegralsFigure, legs 3 and 4 are massless legs
of the integral, but legs 1 and 2 only appear as constituents of a 
massive leg.  If we had a string of the form $\cdots 1 \ell_1 \cdots$,
and we tried to commute $\lsl_1$ past $\ksl_1$, we would generate
$2\ell_1\cdot k_1 = \ell_1^2 - \ell_2^2$.  But $1/\ell_2^2$ is
not a propagator for $I_4^{(2)}$, hence the commutation procedure fails
to reduce the integral.
(This is the same problem encountered when the momentum $k_6$ was
introduced in the pentagon numerator~(\use\BadChoice).)

In some cases, not enough of the massless spinors terminating
the spinor strings in the numerator of an integral correspond to 
massless legs of the integral.
For example, if the string
$$
\spab{a}.{\ell_3}.{b}  \spab{3}.{\ell_3}.{c}
\anoneqn
$$
appears in the numerator of the adjacent two-mass box integral   
$I_4^{(2)}$, and none of $a,b$ and $c$ is equal to 3 or 4,
then we are seemingly blocked from using the above procedures.
One way to handle this situation is to use the Schouten identities,
$$
\eqalign{
\spa{i}.{j}\spa{k}.{l}
&= \spa{i}.{l}\spa{k}.{j} + \spa{i}.{k}\spa{j}.{l} \,, \cr
\spb{i}.{j}\spb{k}.{l}
&= \spb{i}.{l}\spb{k}.{j} + \spb{i}.{k}\spb{j}.{l} \,, 
\cr}
\eqn\SchoutenIdentity
$$
to put more `useful' momenta at the ends of strings.
In the present case, we multiply and divide by $\spb3.4$.
Then we use 
$$
\spab{a}.{\ell_3}.{b} \spb{3}.{4}
 = \spab{a}.{\ell_3}.{3} \spb{b}.{4} 
 + \spab{a}.{\ell_3}.{4} \spb{3}.{b} \,.
\anoneqn
$$
The string becomes
$$
\eqalign{
& {1\over\spb3.4} \Bigl[ 
 (\ell_3^2 - \ell_4^2) \spb{b}.{4} \spab{a}.{\ell_3}.{c}
  - \ell_3^2 \spb{b}.{4} \spab{a}.{3}.{c}
  + \spb{3}.{b} \spab{a}.{\ell_3}.{4} \spab{3}.{\ell_3}.{c} \Bigr]\,, \cr
}\anoneqn
$$
and the last term can now be reduced further, after multiplying 
and dividing by $\spab4.{(1+2)}.3$.

The trick of multiplying and dividing by spinor product factors
stops working when one has too few massless legs, which
here basically happens at the level of triangle integrals.  
Fortunately, the triangle integrals with one and two external masses 
do not generate terribly complicated expressions even when reduced
via a general (`brute force') formula, for example using Feynman
parametrization.  On the other hand,  
three-mass triangle integrals with two or three powers of the loop
momenta inserted can generate quite lengthy formul\ae.  
Indeed such terms --- coefficients of 
$I_3^{3{\rm m}}(s_{12},s_{34},s_{56})$, $\ln({-s_{12}\over-s_{56}})$, 
etc. --- account for much of the length of our final expressions.  
Part of the problem is that the three-mass triangle Gram determinant
$\Delta_3^{(2,4)}$ in eq.~(\use\GramThree) 
--- which `belongs' in the various coefficients in some form ---
has mass dimension 4, yet apparently cannot be factored at all, 
even employing spinor strings.
 
The cuts for one-loop six-point amplitudes can be divided into
$s_{ij}$ and $t_{ijk}$ cuts, according to whether the momentum flowing
across the cut is the sum of two or three external momenta.  In
general, the $t_{ijk}$ cuts are much simpler to evaluate, largely
because the three-mass triangle cannot appear in such a cut --- it has
cuts only in three $s_{ij}$ channels.  One way to handle the more
intricate $s_{ij}$ cuts is to first evaluate a `triple cut', where
three loop propagators are required to be on-shell. For example, the
$s_{12}$-$s_{34}$-$s_{56}$ triple cut, depicted in
\fig\TripleCutFigure, can be defined by the conditions
$\ell_1^2=\ell_3^2=\ell_5^2=0$.  Such triple cuts pick out those
integral functions containing cuts in all three channels, 
in particular the three-mass triangle functions.  The
utility of considering such cuts is that instead of having to evaluate
the phase-space integral of a six-point tree amplitude with a
four-point tree amplitude (as one would for an $s_{ij}$ cut), one gets
an expression where the six-point amplitude is replaced by the product
of the two four-point amplitudes that it factorizes on.  The full
$s_{56}$ cut (say) can then be written as a sum of the triple cut and
a residual term which has no $1/\ell_3^2$ propagator, and therefore no
three-mass triangle integral; that is, the three-mass triangle
complications can be confined to the triple cuts.

\vskip -.5 cm 
%
\LoadFigure\TripleCutFigure{\baselineskip 13 pt
\noindent\narrower\ninerm
The kinematics of the triple cut.  The cut lines are all on-shell.}
{\epsfxsize 2.0 truein}{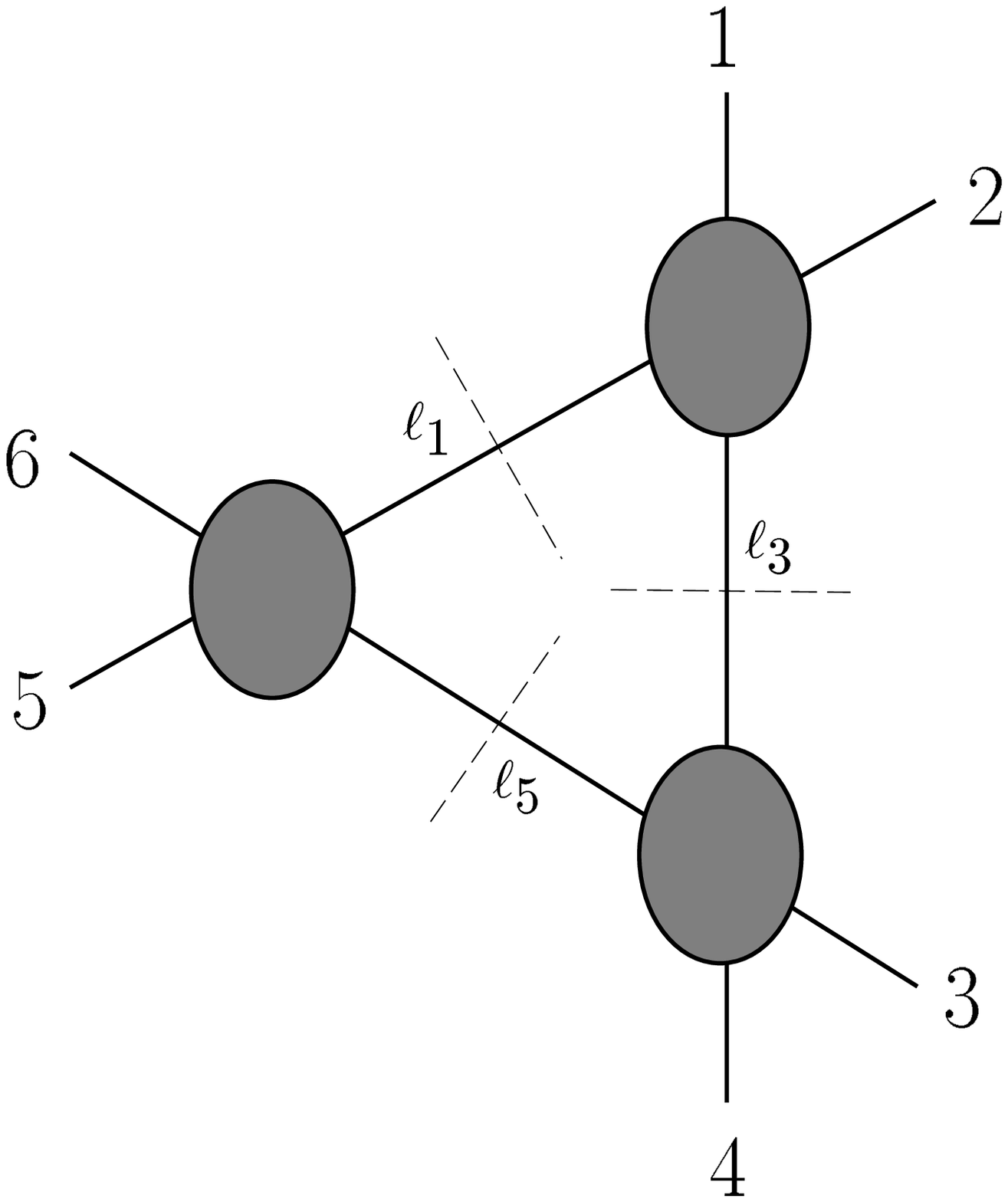}{}

\subsection{Numerical simplifications}
\tagsubsection\NumericalSimplificationSubsection

Even after employing the above spinor-product manipulations (among others) to 
help reduce the size of the expressions for cuts, one may still generate 
in intermediate steps complicated analytical expressions for the 
coefficients of $I_3^{3{\rm m}}(s_{12},s_{34},s_{56})$, $\ln(-s_{12})$, 
and so on.
It is possible to use numerical techniques to simplify such an expression,
which is some rational function of the spinor products,
{\it if} one has enough information about the analytic behavior of the
coefficient.  (The analytic information does not have to be manifest
in the complicated expression.)
The basic idea is to write down an ansatz for
the complicated expression, as a linear combination of all
possible kinematic terms that have the proper analytic behavior,
where each term is multiplied by an as-yet unknown numerical coefficient.  
The more one knows about the analytic behavior of the coefficient,
the fewer factors one has to put in the denominator of the ansatz,
and (by dimensional analysis and combinatorics) 
the fewer the possible terms. 
Then one evaluates both the 
complicated expression and the ansatz at a number of random kinematic
points, which should exceed the number of unknown coefficients.
This gives an over-determined set of linear numerical equations 
which can be solved for the unknown coefficients, which are required to be
simple rational numbers.  The solution can be checked by numerical
evaluation at further kinematic points.  An example of this procedure
is provided in section~\use\SampleNumericalSimplificationSubsection.

In practice we have been able to carry out this `numerical simplification'
procedure efficiently once the number of linearly independent
terms in the ansatz for a given coefficient is reduced to about a hundred.
To get a manageable number of terms like that,
one generally needs to know about not just the physical factorization 
limits discussed in section~\use\FactorizationSection, but also
the leading spurious singularities of the coefficient.  
However, this latter information can be typically be inferred from 
simpler cuts which have already been performed.

For example, the coefficients of both 
$I_3^{3{\rm m}}(s_{12},s_{34},s_{56})$ and $\ln(-s_{12})$ receive
contributions from the adjacent two-mass box integral $I_4^{(2)}$,
and thus they will typically have denominators containing 
$\spab3.{(1+2)}.4$.  
In the full amplitude the singularities in these terms as 
$\spab3.{(1+2)}.4 \to 0$ cancel against singularities in 
the terms containing 
$\Ls_{-1}^{2{\rm m}h}(s_{34},t_{123};s_{12},s_{56})$.
The precise way this cancellation happens is known from 
the structure of the tensor box integrals~[\use\CGM].  
In the present example we have, in the limit as 
$\spab3.{(1+2)}.4 \to 0$,
$$
\eqalign{
& \bigl[ \hbox{coefficient of }
    I_3^{3{\rm m}}(s_{12},s_{34},s_{56}) \bigr] \cr
&  \to \bigl[ \hbox{coefficient of }
   \Ls_{-1}^{2{\rm m}h}(s_{34},t_{123};s_{12},s_{56}) \bigr] 
  \times (-1) \, { \spab3.{(1+2)}.4 \spab4.{(1+2)}.3 (t_{123}-t_{124})
    s_{12} s_{56}  \over  t_{123}^2 \Delta_3 } \,, \cr\cr 
& \bigl[ \hbox{coefficient of }\ln(-s_{12}) \bigr] \cr
&  \to \bigl[ \hbox{coefficient of }
   \Ls_{-1}^{2{\rm m}h}(s_{34},t_{123};s_{12},s_{56}) \bigr] 
  \times (-1) \, { \spab3.{(1+2)}.4 \spab4.{(1+2)}.3 (t_{123}-t_{124})
    s_{12} \delta_{12}
   \over 2 s_{34} t_{123}^2 \Delta_3 } \,. \cr 
}\eqn\ithreelnlimits
$$ 
On the other hand, the coefficient of 
$\Ls_{-1}^{2{\rm m}h}(s_{34},t_{123};s_{12},s_{56})$
can be determined from the cut in the $t_{123}$ channel, which as
we have mentioned is much simpler, typically only one or two terms in
length.  The known analytic behavior of the coefficients of
$I_3^{3{\rm m}}(s_{12},s_{34},s_{56})$ and $\ln(-s_{12})$ 
as $\spab3.{(1+2)}.4 \to 0$ can also be verified
numerically by choosing kinematics close to this `back-to-back'
singularity.

Another denominator factor usually present in the coefficient of 
$\ln(-s_{12})$ is $(s_{12}-t_{123})$.  Again these terms can be inferred
from the $t_{123}$-cut; this time the relation is through the simpler
functions $\Ll_{0}({-s_{12}\over-t_{123}})$ and
$\Ll_{1}({-s_{12}\over-t_{123}})$. 

The singularity of $\ln(-s_{12})$ as 
$\Delta_3(s_{12},s_{34},s_{56}) \to 0$ is related to
that of $I_3^{3{\rm m}}(s_{12},s_{34},s_{56})$, but it cannot be related
to a $t_{ijk}$ cut, and has to be extracted from the leading loop-momentum
behavior of the $s_{12}$-$s_{34}$-$s_{56}$ triple cut.
The numerical study of the $\Delta_3 \to 0$ limit is also subtler:
In this limit the three vectors $k_1+k_2$, $k_3+k_4$ and $k_5+k_6$
all become proportional to each other; 
thus there is a simultaneous `back-to-back' 
vanishing of $\spab1.{(3+4)}.2$, $\spab3.{(1+2)}.4$ and 
$\spab5.{(1+2)}.6$.  Terms with $\Delta_3$ in the denominator often
have such vanishing factors in the numerator
(or other more complicated ones --- see eq.~(\use\covanishingfactors)), 
which mask the presence of $1/\Delta_3$.  On the bright side,
these numerator factors imply that individual kinematic coefficients 
are less singular as $\Delta_3 \to 0$ than a simple counting of
denominator $\Delta_3$s would suggest, improving their numerical 
stability near the pole.

\section{Sample Calculations}
\tagsection\SampleSection

\subsection{Evaluation of Cuts}
\tagsubsection\SampleCutSubSection

Here we describe the initial stages of evaluation of two different
cuts, in order to illustrate some of the features that are involved.

The first cut we consider is the cut in the $t_{234}$ channel
of the amplitude $A_6^\sc(1_q^+,2_\qb^-,3^-,4^+,5_\eb^-,6_e^+)$.
This amplitude has a scalar replacing the gluon in the loop, and
so the cut propagator $\ell_2$ is that of a scalar.  
(The configuration of external legs and loop momenta is shown in 
fig.~\use\PentagonKinematicsFigure.)
The emission of the scalar from the quark line is via a Yukawa coupling 
that reverses the helicity of the quark line.  
Thus the required product of tree amplitudes, to be integrated over
the two-body phase space for $\ell_2+(-\ell_5)$, is
$$
P_{234} = 
\Atree_5(1_q^+,(\ell_2)_s,(-\ell_5)_\qb^+,5_\eb^-,6_e^+)
  \times \Atree_5((-\ell_2)_s,2_\qb^-,3^-,4^+,(\ell_5)_q^-) \,. 
\anoneqn
$$  
These two five-point tree amplitudes are easily evaluated,
up to overall signs (which can always be fixed at the end of 
the calculation, using e.g. a factorization limit),
$$
\eqalign{
P_{234} &= 
\pm { \spa{\ell_2}.5^2 
    \over \spa1.{\ell_2} \spa{\ell_2}.{\ell_5} \spa5.6 }
 \times
   { \spb2.4 \spb{\ell_2}.4^2 
   \over \spb2.3 \spb3.4 \spb{\ell_2}.{\ell_5} \spb{\ell_2}.2 }
   \cr
& = \pm { \spb2.4 \over \spb2.3 \spb3.4 t_{234} \spa5.6 } 
   { \spab5.{\ell_2}.4  \spab5.{\ell_2}.1  \spab2.{\ell_2}.4 
    \over \ell_1^2 \ell_3^2 } \,. \cr    
}\eqn\cutonea
$$
In the second step we used the on-shell conditions
$\ell_2^2=\ell_5^2=0$ to replace $1/\spa1.{\ell_2}$ with
$\spb{\ell_2}.1/(2 {\ell_2\cdot k_1}) = \spb{\ell_2}.1/ \ell_1^2$,
and a similar replacement for $1/\spb{\ell_2}.2$.
Notice that the propagator $\ell_4^2$ is `missing'.  Hence no
pentagon reduction is necessary in this example; the integral
is already the box integral $I_4^{(4)}$ with two adjacent masses.
(The missing propagator can be attributed to a supersymmetry Ward
identity (SWI)~[\use\SWI]:  The limit $\ell_4^2 \to 0$ is also the 
collinear limit $\ell_5 \parallel k_4$ for the tree amplitude 
$\Atree_5((-\ell_2)_s,2_\qb^-,3^-,4^+,(\ell_5)_q^-)$, in which it
 factorizes onto
$\Atree_4((-\ell_2)_s,2_\qb^-,3^-,P_q^-)$, which vanishes by a SWI.)

Table~1 and the structure of the numerator in eq.~(\use\cutonea)
suggest that we multiply and divide this expression by $\spab1.{(3+4)}.2$, 
and then commute the pair of $\lsl_2$s toward each other in the 
following string:
$$
\eqalign{
& \spab5.{\ell_2}.1 \spab1.{(3+4)}.2 \spab2.{\ell_2}.4 \cr
& = \spab5.{\ell_2 1 (2+3+4) 2 \ell_2}.4 \cr
& =  \ell_1^2 \, \spab5.{(3+4)}.2 \spab2.{\ell_2}.4 
  + \ell_3^2 \, \spa5.1 \spbb1.{\ell_2 (2+3)}.4
  + \spa5.1 \spb2.4 
       \langle 1^+ \vert \ell_2(2+3+4)\ell_2 \vert 2^+\rangle \cr
& =  \ell_1^2 \, \spab5.{(3+4)}.2 \spab2.{\ell_2}.4 
  + \ell_3^2 \, \spa5.1 \spbb1.{\ell_2 (2+3)}.4
  + \spa5.1 \spb2.4 t_{234} \, \spab2.{\ell_2}.1 \,. \cr
}\eqn\cutoneb
$$
The first two terms in eq.~(\use\cutoneb), after multiplication by 
$\spab5.{\ell_2}.4$, are triangle integrals with two external masses
and two powers of the loop momenta in the numerator.  They can be
handled straightforwardly by Feynman parametrization.
The third term is still a quadratic box integral (i.e.~two loop momenta
in the numerator), but it can be reduced further,
using a second factor of $\spab1.{(3+4)}.2$ and a Fierz rearrangement
(since $\ell_2^2=0$),
$$
\spab1.{(3+4)}.2 \spab5.{\ell_2}.4 \spab2.{\ell_2}.1 
 = \spab5.{\ell_2}.1 \spab1.{(3+4)}.2 \spab2.{\ell_2}.4 \,, 
\eqn\cutonec
$$
and then following exactly the same steps as in eq.~(\use\cutoneb).

Now we have five terms, only one of which is a box integral, and 
that one is merely linear in the loop momenta,
$$
I_4^{(4)}[\spab2.{\ell_2}.1]\ 
=\ - { \Ls_{-1}^{{\rm 2m}h} \bigl(s_{12}, t_{234}; s_{34}, s_{56} \bigr) 
     \over \spab1.{(3+4)}.2 } 
\ +\ \hbox{(terms lacking a $t_{234}$ cut)}.
\eqn\cutoned
$$
Finally, we perform the four triangle integrals, dropping terms with
no $t_{234}$ branch cut, and assemble the pieces, thus obtaining all
terms in eqs.~(\use\VLffmpsc) and~(\use\FLffmpsc) for
$A_6^\sc(1_q^+,2_\qb^-,3^-,4^+,5_\eb^-,6_e^+)$ that have logarithms or
dilogarithms with $t_{234}$ in the argument.  (Terms representing
possible scalar one- and two-mass triangle contributions are
proportional to $(-t_{234})^{-\e}/\e^2$ and therefore can be inferred
from the known structure of the poles in
$\e$~[\use\GieleGlover,\use\KunsztSoper].  This can be used as a check
on the cut calculation, or to save labor.)

The second cut we consider is the cut in the $t_{412}$ channel
of the amplitude $A_6^\sc(1_q^+,2^+,3_\qb^-,4^-,5_\eb^-,6_e^+)$.
In this case leg 4 has to be adjacent to leg 1, as in the integral
$\tI_5$ in fig.~\use\PentBoxIntegralsFigure.
For this configuration we label the propagator momenta around the 
loop, starting just clockwise of the 5--6 lepton pair, by
$\ell_5^\prime,\ell_1,\ell_2,\ell_3,\ell_4$.
Now the required product of tree amplitudes, to be integrated over
the two-body phase space for $\ell_5^\prime+(-\ell_3)$, is
$$
\eqalign{
P_{412} & = 
 \Atree_5((\ell_5^\prime)_\q^-,(-\ell_3)_s,3_\qb^-,5_\eb^-,6_e^+)
\times \Atree_5((-\ell_5^\prime)_\qb^+,4^-,1_q^+,2^+,(\ell_3)_s) \cr
& = 
\pm { \spb{\ell_3}.6^2  
   \over \spb{\ell_5^\prime}.{\ell_3} \spb{\ell_3}.3 \spb5.6 }
  \times { \spa{\ell_3}.4^2 
    \over \spa1.2 \spa2.{\ell_3} \spa{\ell_3}.{\ell_5^\prime} } \,. \cr
}\eqn\cuttwoa
$$
Fortunately, this cut can be obtained from the first one we evaluated.
If we multiply eq.~(\use\cutonea) by $\spb3.4/\spb2.4$, and then
perform the following `flip' (pairwise permutation plus complex 
conjugation of all spinor products)
$$
\ell_1 \lr \ell_4, \quad 
\ell_2 \lr \ell_3, \quad 
\ell_5 \lr \ell_5^\prime, \quad 
1 \lr 3, \quad 5 \lr 6,  
\quad \spa{a}.{b} \lr \spb{a}.{b} \,,
\eqn\sampleflipsym
$$  
we recover eq.~(\use\cuttwoa).

This second example illustrates the principle of {\it recycling} cuts.
One can save a lot of effort by identifying different cuts that are 
actually the same up to permutations and trivial overall factors.
In many cases, it may not be possible to obtain an entire cut in 
this way, but portions of it may be recyclable.  
The `master functions' defined in section~\use\MasterFunctionSection,
which enter three different amplitudes, are the most complicated 
contributions we have been able to recycle, but there are several
other instances as well.

\subsection{Rational Function Reconstruction}
\tagsubsection\SampleRationalFunctionSubsection

In Section~\use\FactorizationSection\ we described the general
factorization properties of amplitudes, and how that information
can be used to determine the rational function parts of amplitudes.
As an example, we explicitly construct the rational function
terms proportional to $n_f$ in the leading-color helicity amplitude, 
which are given by $A_6^{s,f}(1_q^+,2^+,3^+,4_\qb^-)$ in 
eq.~(\use\Asf), and outline the construction of the scalar 
(non-cut-constructible) piece ($V^\sc$, $F^\sc$) for
the same helicity configuration, eq.~(\use\FLpppmscalar).

The first step is to account for possible spurious singularities
in the rational function terms.  These singularities 
can always be identified after all cuts have been calculated, because
they have to cancel against terms containing logarithms and dilogarithms.
For $A_6^{s,f}(1_q^+,2^+,3^+,4_\qb^-)$, the only possible cut, 
that in the $s_{23}$ channel, vanishes identically, 
because the tree-level $\qb^-q^+g^+g^+$ and $ssg^+g^+$ 
amplitudes vanish for massless quarks and scalars.
So this term is purely a rational function, and therefore can have
no spurious singularities, and we can immediately focus on the physical 
(multi-particle and collinear) singularities.

It is usually convenient to match the multi-particle behavior ---
here, the limits $t_{123},t_{234} \to 0$ 
--- before attacking the collinear poles.
The residue of a $t_{ijk}$ pole in a one-loop six-point amplitude 
receives three contributions:\footnote{$^\dagger$}{%
In the presence of infrared divergences, such as those due to virtual
gluons, the one-loop factorization is a bit more subtle, but still has a 
universal form~[\use\Factorization].}
$C_1$ from a one-loop four-point amplitude multiplying a tree-level
four-point amplitude; $C_2$ where the loop amplitude is replaced by
the corresponding tree amplitude, and the tree by the loop; and 
$C_3$ is associated with a loop correction to the
intermediate propagator, multiplied by the two tree amplitudes.  
In the case of $e^+\,e^- \to$~four partons,
two of the three contributions are easy to describe in general,
because one of the two four-point amplitudes is the relatively simple
$e^+\,e^- \to q\qb$ process.
Let us define the $C_2$ contribution to be where the $e^+\,e^- \to q\qb$
amplitude is a loop amplitude.
Then for the $n_f$-dependent term, $C_2$ and $C_3$ both vanish.
Also, for the scalar ($\sc$) part of an amplitude, 
$C_2 = {1\over2} \times \lim \Atree_6$,
and $C_3 = (-1) \times \lim \Atree_6$, where $\lim \Atree_6$
is the appropriate $t_{ijk} \to 0$ limit of the six-point tree
amplitude.

For the present $(1_q^+,2^+,3^+,4_\qb^-)$ helicity configuration,
the tree amplitude~(\use\treepppm) has no
$t_{ijk}$ poles, hence only the $C_1$ contribution survives.
Using a calculation of the $n_f$ terms in the $\qb qgg$
one-loop four-point amplitudes~[\use\KunsztFourPoint,\use\Fermion],
we have
$$
\eqalign{
{A_6^s(1_q^+,2^+,3^+,4_\qb^-) \over i \, \cg} 
&\sim 
  {\spa4.5\spb6.{P} \over s_{56}} \times {1\over t_{123}}
  \times {1\over3}{s_{12}\over s_{23}} 
    {\spab{P}.1.3 \over \spa1.2\spa2.3} 
\hskip.5cm (P \equiv 1+2+3),
\hskip.5cm \hbox{as } t_{123} \to 0, \cr
&\sim
  {-\spb1.6 \spa5.{P} \over s_{56}} \times {1\over t_{234}}
  \times {1\over3}{s_{34}\over s_{23}} 
    {\spab4.{P}.3 \over \spa{P}.2\spa2.3} 
\hskip.5cm (P \equiv 2+3+4),
\hskip.3cm \hbox{as } t_{234} \to 0. \cr
}\eqn\pppmtlimits
$$
Similarly, the limits of the scalar pieces are found to be
$$
\eqalign{
{A_6^\sc(1_q^+,2^+,3^+,4_\qb^-) \over i \, \cg} 
&\sim 
  {\spa4.5\spb6.{P} \over s_{56}} \times {1\over t_{123}}
  \times \left[ {1\over2} + {1\over3}{s_{12}\over s_{23}} \right]
    {\spab{P}.1.3 \over \spa1.2\spa2.3} 
\hskip.5cm (P \equiv 1+2+3),
\hskip.5cm \hbox{as } t_{123} \to 0, \cr
&\sim
  {-\spb1.6 \spa5.{P} \over s_{56}} \times {1\over t_{234}}
  \times \left[ {1\over2} + {1\over3}{s_{34}\over s_{23}} \right]
    {\spab4.{P}.3 \over \spa{P}.2\spa2.3} 
\hskip.5cm (P \equiv 2+3+4),
\hskip.3cm \hbox{as } t_{234} \to 0. \cr
}\eqn\pppmtlimitsc
$$

Next we need to modify the terms in eq.~(\use\pppmtlimits) to 
improve their collinear limits, while preserving their $t_{ijk}$ limits.
For example, we can improve the $k_2 \parallel k_3$ limit of the
$t_{123}$ limit in~(\use\pppmtlimits) as follows:
$$
\eqalign{
&  {1\over3}{\spa4.5 \spbb6.{(1+2+3)1}.3 s_{12} 
    \over \spa1.2\spa2.3 \, s_{23} \, s_{56} \, t_{123}}
=  {1\over3}{\spa4.5 \spbb6.{(1+2+3)1}.3 \spb1.2 
    \over \spa2.3^2 \spb2.3 \, s_{56} \, t_{123}} \cr
&=  -{1\over3}{\spa4.5 \spbb6.{(1+2+3)2}.3 \spb1.2 
    \over \spa2.3^2 \spb2.3 \, s_{56} \, t_{123}} + \cdots
=  {1\over3}{\spa4.5 \spbb6.{(1+2+3)2}.1  
    \over \spa2.3^2 \, s_{56} \, t_{123}} + \cdots\,, \cr
}\eqn\polyone
$$
where nonsingular terms in the $t_{123} \to 0$ limit are represented 
by `$\cdots$'.
A similar manipulation of the $t_{234}$ limit,
using also $s_{2P} = -s_{34}$ in that limit, gives
$$
\eqalign{
&-{1\over3} {\spb1.6 \spab5.{(2+3+4)}.2 \spab4.2.3
    \over \spa2.3 \, s_{23} \, s_{56} \, t_{234}} + \cdots  
= -{1\over3} {\spb1.6 \spaa5.{(2+3+4)2}.4 
    \over \spa2.3^2 \, s_{56} \, t_{234}} + \cdots \, . \cr 
}\eqn\polytwo
$$

Thus a first guess for $A_6^s(1_q^+,2^+,3^+,4_\qb^-)/(i \, \cg)$ would be
$$
\eqalign{
F^s_{\rm guess} = 
{1\over3}{1 \over \spa2.3^2 \, s_{56} } \left[ 
   {\spa4.5 \spbb6.{(1+2+3)2}.1 \over t_{123}}
 - {\spb1.6 \spaa5.{(2+3+4)2}.4 \over t_{234}} \right]\ . \cr 
}\eqn\polythree
$$
It might appear that the $k_2 \parallel k_3$ and $k_5 \parallel k_6$ 
collinear limits of~(\use\polythree) both need to be improved,
since the expected limits are $\sim 1/\sqrt{s_{ij}}$, not
$1/s_{ij}$.  But in fact there is a cancellation between the two
terms in eq.~(\use\polythree) in both limits.   In the $k_2 \parallel k_3$
limit, using the identity
$$
\eqalign{
s_{12} \, t_{234} - s_{24} \, t_{123} 
&= s_{12} \, s_{34} - s_{24} \, s_{13} + s_{23} \, s_{14} + \Ord(s_{23})
= \tr[2341] + \Ord(s_{23}) \cr
&= \spa2.3\spb3.4\spa4.1\spb1.2 + \spb2.3\spa3.4\spb4.1\spa1.2 
  + \Ord(s_{23}), \cr
}\anoneqn
$$
and letting $P\equiv 2+3$, we have
$$
\eqalign{
F^s_{\rm guess} &\sim 
{1\over3}{1 \over \spa2.3^2 \, s_{56} } \biggl[ 
   {\spa4.5 \spb6.1 \over t_{123} t_{234}}
   (\spa2.3\spb3.4\spa4.1\spb1.2 + \spb2.3\spa3.4\spb4.1\spa1.2) \cr
&\hskip2.5cm
 - \spa2.3  {\spa4.5 \spb6.3 \spb2.1 \over t_{123}}
 + \spb2.3  {\spb1.6 \spa5.3 \spa2.4 \over t_{234}} \biggr] \cr 
&\sim 
{1\over3}{ \sqrt{z(1-z)} \over \spa2.3^2 \, s_{56} } \left[ 
 - \spa2.3 \spa4.5 { \spab4.1.6 + \spab4.P.6 \over \spa1.{P} \spa{P}.4 }
 + \spb2.3 \spb1.6 { \spab5.4.1 + \spab5.P.1 \over \spb1.{P} \spb{P}.4 } 
 \right] \cr
&\sim 
{1\over3}{ \sqrt{z(1-z)} \over \spa2.3^2 \, s_{56} } \left[ 
   \spa2.3 { \spa4.5^2 \spb5.6 \over \spa1.{P} \spa{P}.4 }
 + \spb2.3 { \spb1.6^2 \spa5.6 \over \spb1.{P} \spb{P}.4 } 
 \right] \cr
&\sim 
{1\over3}{ \sqrt{z(1-z)} \over \spa2.3 } 
  \times { -\spa4.5^2 \over \spa1.{P} \spa{P}.4 \spa5.6 }
 \; + \; {-1\over3}{ \sqrt{z(1-z)} \spb2.3 \over \spa2.3^2 }
  \times { \spb1.6^2 \over \spb1.{P} \spb{P}.4 \spb5.6}\ ,
\hskip.5cm k_2 \parallel k_3. \cr 
}\eqn\limtwothree
$$

Not only does the leading $1/|s_{23}|$ term cancel in eq.~(\limtwothree), 
but the next terms in the expansion give precisely the desired limit,
corresponding to the $n_f$-dependent piece of the
$$
\Split^{1\rm -loop}_{\mp}(2^+,3^+) 
  \times \Atree_5(1_q^+,P^\pm,4_\qb^-,5_\eb^-,6_e^+)
\anoneqn
$$
terms in eq.~(\use\Loopsplit).   The loop splitting functions are
given in ref.~[\use\SusyFour] and the five-point $e^+\,e^- \to \qb qg$
amplitudes in appendix~\use\qqgAppendix. 
Notice that there are no $n_f$ terms in $A_{5}^{1\rm -loop}$, hence
no $\Split^{\rm tree} \times A_5^{1\rm -loop}$ terms contribute here.
Similar manipulations show that the $k_5 \parallel k_6$ limit of 
$F^s_{\rm guess}$ is also precisely correct,
$$
\eqalign{
F^s_{\rm guess} &\sim 
 {1-z\over\spb5.6} \times 
   {1\over3}{\spb2.3\spa2.4\spa3.4 \over \spa2.3^2 \spa4.{P} \spa{P}.1}
 \; + \; 
 {z\over\spa5.6} \times 
   {1\over3}{\spb1.2\spb1.3 \over \spb4.{P}\spb{P}.1\spa2.3}\ ,
\hskip.5cm k_5 \parallel k_6. \cr 
}\eqn\limfivesix
$$
Here there are only $\Split^{\rm tree} \times A_5^{1\rm -loop}$ terms,
with the relevant loop amplitudes obtainable from
ref.~[\use\Fermion].   
All singular collinear limits of $F^s_{\rm guess}$
have now been verified, so we expect the result to be correct as is.
A simple identity shows that indeed
$F^s_{\rm guess} = A_6^s(1_q^+,2^+,3^+,4_\qb^-)/(i\,\cg)$,
the result given in eq.~(\use\Asf).

A slightly more complicated example, which we only summarize
here, is the rational function terms in the scalar piece 
($V^\sc$, $F^\sc$) for the same helicity configuration,
$A_6^\sc(1_q^+,2^+,3^+,4_\qb^-)$, as given in eq.~(\use\FLpppmscalar).
By comparing the $t_{ijk}$ limits~(\use\pppmtlimits) and 
(\use\pppmtlimitsc), and similarly the collinear limits,
it is easy to see that there must be a term of the form 
$A_6^s$ in $A_6^\sc$.  
As for the remaining terms,
the first step is again to account for possible spurious singularities.
Assuming that all the cuts have previously been calculated,
the $\ln({-s_{ij}\over-t_{234}})/(s_{ij}-t_{234})^2$
terms contained in the $\Ll_1$ functions in eq.~(\use\FLpppmscalar)
are already known.  But the $\Ll_1$ functions are designed to 
cancel the spurious $(s_{ij}-t_{234})$ behavior
between logarithms and rational functions.  Hence by completing
the $\ln({-s_{ij}\over-t_{234}})/(s_{ij}-t_{234})^2$ terms into 
$\Ll_1({-s_{ij}\over-t_{234}})/t_{234}^2$ functions, we ensure
that the remaining rational function terms are free of
$1/(s_{ij}-t_{234})$ poles.  Since there are no other 
non-rational-function terms in $V^\sc$ or $F^\sc$, there are 
no other sources of spurious poles, and we next turn to the
physical singularities.

Using the $t$-channel limits~(\use\pppmtlimitsc), 
and manipulations similar to those leading to eq.~(\use\polythree),
we arrive at an ansatz for what has to be added to the $\Ll_1$ terms,
$$
F^\sc_{\rm guess} = 
{1\over2} \Biggl[ 
   { \spa4.5 \spbb6.{(2+3)1}.3 \over \spa1.2\spa2.3 \, t_{123} \, s_{56} }
 + { \spb2.3\spb1.6 \spab5.{(2+4)}.3 
     \over \spa2.3\spb3.4 \, t_{234} \, s_{56} } 
 - { {\spab5.2.3}^2 \over \spa1.2\spa2.3\spb3.4\spa5.6 \, t_{234} }
 + { {\spab4.{(2+3)}.6}^2 \over \spa1.2\spa2.3\spa3.4\spb5.6 \, t_{234} }
    \Biggr]\ . 
\eqn\Oldstart 
$$
The first two terms reproduce~(\use\pppmtlimitsc), while the last two
terms are needed to cancel off $t_{234}$ poles in the $\Ll_1$ terms.

This time, however, the $k_5 \parallel k_6$ limit,
$$
\eqalign{
F^\sc_{\rm guess} &\sim 
-{1\over2} \Biggl[ 
   { \spa4.5 \spb6.4\spa4.1\spb1.3 
     \over \spa1.2\spa2.3 \, t_{123} \, s_{56} }
 + { \spb2.3\spb1.6 \spa5.1\spb1.3 
     \over \spa2.3\spb3.4 \, t_{234} \, s_{56} } \Biggr] \cr
&\sim 
-{\sqrt{z(1-z)}\over2} {\spb1.3\over \spa2.3 \, s_{56}} \Biggl[ 
   { \spa4.{P} \spb{P}.4 \spa4.1 \over \spa1.2 \, s_{4P} }
 + { \spb2.3\spb1.{P} \spa{P}.1 
     \over \spb3.4 \, s_{1P} } \Biggr] \cr
&\sim 
{\sqrt{z(1-z)}\over2} 
  {\spb3.{P}\spa{P}.1\spb1.3 \over \spa1.2\spa2.3\spb3.4 \, s_{56}}\ , \cr
}\eqn\Oldlimfivesix 
$$
still has to be improved, for example by adding 
$$
\delta F^\sc_1 \equiv
{1\over2}{\spb3.6 \spab5.{(2+4)}.3 \over \spa1.2\spa2.3\spb3.4 \, s_{56}}
\anoneqn
$$ 
to $F^\sc_{\rm guess}$.
At this stage, the polynomial ansatz formed by the $\Ll_1$ terms,
$F^\sc_{\rm guess}$ and $\delta F^\sc_1$ has only
$1/\sqrt{s_{ij}}$ collinear singularities.   One can now systematically
add additional terms to match the known $k_1 \parallel k_2$,
$k_2 \parallel k_3$, $k_3 \parallel k_4$, and $k_5 \parallel k_6$
limits.  (One can also rewrite the answer in a form where the 
$1/\sqrt{s_{56}}$ behavior is manifest, as in eq.~(\use\FLpppmscalar).)

Other helicity amplitudes may possess more spurious singularities and/or 
physical singularities.  For these cases the determination of the 
rational function terms becomes somewhat more involved, but the
basic principles remain as illustrated above.

\subsection{Numerical Simplification}
\tagsubsection\SampleNumericalSimplificationSubsection

In order to illustrate the numerical simplification technique outlined
in section~\use\NumericalSimplificationSubsection, we consider the
particular example of the coefficient $c^{3{\rm m}}$ of the three-mass
triangle integral $I_3^{3{\rm m}}\L {s_{12}},{s_{34}},{s_{56}}\R$ in
the axial-vector fermion-loop contribution 
$A_6^\ax(1_q^+,2_\qb^-,3^+, 4^-)$, eq.~(\use\Faxffpm).
Note that the three-mass triangle contribution is contained
in the common function $C^\ax$ defined in eq.~(\use\Caxdef),
plus its image under the symmetry operation $\flip{2}$
defined in eq.~(\use\FlipTwoSym).     
Thus we determine simultaneously the corresponding contribution to 
$A_6^\ax(1_q^+,2_\qb^-,3^-, 4^+)$, eq.~(\use\Faxffmp).

The first step in the technique is to write down terms reproducing
all of the spurious and physical singularities in the various channels.
One subtlety is that a term reproducing a singularity in one channel
may be too singular, or otherwise have the wrong type of singularity,
in another channel.  In this case the term will have to be `improved'.

In the present example of $c^{3{\rm m}}$, we first write down a term
$c_1^{3{\rm m}}$ which reproduces the known $\spab3.{(1+2)}.4 \to 0$
behavior of this coefficient.  To do this we use
eq.~(\use\ithreelnlimits) and the coefficient of the hard-two-mass box
function in $C^\ax$ (which we assume has already been obtained via the
$t_{123}$ cut), to get
$$
\eqalign{
c_{1,{\rm guess}}^{3{\rm m}} & = 
{1\over2} { \spab2.{(1+3)}.4^2 \spab3.{(1+2)}.6^2
           - \spa2.3^2\spb4.6^2 \, t_{123}^2 
    \over \spa1.2\spb5.6\spab3.{(1+2)}.4^4 } \cr
& \hskip 3 cm
 \times { \spab3.{(1+2)}.4 \spab4.{(1+2)}.3 (t_{123}-t_{124})
    s_{12} s_{56}  \over  t_{123}^2 \Delta_3 } + \cdots \cr
& = 
- { \spab2.{(1+3)}.6\spb1.2\spa2.3\spb4.6\spa5.6 \, (t_{123}-t_{124})
   \, \spab4.{(1+2)}.3
  \over t_{123} \, \spab3.{(1+2)}.4^2 \Delta_3 }+ \cdots \,. \cr
}\eqn\bblimitonea
$$
In the second line of eq.~(\use\bblimitonea) we used the spinor identity 
$$
\spab2.{(1+3)}.4 \spab3.{(1+2)}.6 
= \spab3.{(1+2)}.4 \spab2.{(1+3)}.6 - \spa2.3\spb4.6\,t_{123}\,,
\eqn\bbidentone
$$ 
and dropped all but the leading terms as $\spab3.{(1+2)}.4 \to 0$.
As it stands, $c_{1,{\rm guess}}^{3{\rm m}}$ contains a pole
in $t_{123}$, but an identity similar to (\use\bbidentone) 
removes the pole while preserving the leading $\spab3.{(1+2)}.4 \to 0$
behavior, and so we take the first singular term to be
$$
c_1^{3{\rm m}} = 
  {\spb1.4\spa3.5  \, (t_{123}-t_{124})
   \, \spab4.{(1+2)}.3 \spab2.{(1+3)}.6 
   \over \spab3.{(1+2)}.4^2 \, \Delta_3 }\,.
\eqn\bblimitoneb
$$

Actually $c^{3{\rm m}}$ should contain a pole in $t_{123}$ 
(but with a different structure than that found in eq.~(\use\bblimitonea));
it has to cancel the pole in the explicit formula for the 
hard two-mass box function, eq.~(\use\Lstwomasshdef).
Thus a term reproducing that pole is given by 
$$
c_{2,{\rm guess}}^{3{\rm m}} = 
  {1\over2} {s_{12}s_{56}\over t_{123}} \times 
    { \spab2.{(1+3)}.6^2 \over \spa1.2\spb5.6 \spab3.{(1+2)}.4^2 } +\cdots \,, 
\eqn\bblimittwoa
$$
where we used identity~(\use\bbidentone) again
and dropped nonsingular terms as $t_{123} \to 0$.
We still have to remove the leading singularity as 
$\spab3.{(1+2)}.4 \to 0$ in this term, which we can do using
$$
\spba1.{2(1+3)6}.5 = -\spba1.{3(1+2+3)6}.5 + \cdots
 =  \spba1.{3(1+2)4}.5 + \cdots \,,
\anoneqn
$$
as $t_{123} \to 0$, thus obtaining
$$
c_2^{3{\rm m}} = 
  -{1\over2} { \spb1.3\spa4.5 \spab2.{(1+3)}.6 
       \over t_{123} \, \spab3.{(1+2)}.4 } 
\eqn\bblimittwob
$$
as our second inferred singular term in $c^{3{\rm m}}$.

As mentioned above, the leading singular terms as $\Delta_3\to0$
require an explicit calculation of the $s_{12}$-$s_{34}$-$s_{56}$ 
triple cut, but fortunately only the leading loop-momentum terms 
in this cut have to be retained, yielding
$$
c_3^{3{\rm m}} = 
 - 3 \, { \delta_{34} \,
    \bigl( \spab5.{2}.1 \, \delta_{12} - \spab5.{6}.1 \, \delta_{56} \bigr)
     \spab4.{(1+2)}.3  \spab2.{(1+3)}.6 \over \spab3.{(1+2)}.4 \, 
          \Delta_3^2 }\, .
\eqn\bblimitthree
$$
The remaining terms can have at most a $1/\Delta_3$ singularity.
Furthermore, there are no collinear singularities ($k_i \parallel k_j$) 
in the three-mass triangle coefficient. (This feature is general,
as long as the adjacent two-mass box function $\Ls_{-1}^{2{\rm m}h}$
is used, and not $\Lsnew^{2{\rm m}h}_{-1}$.)

At this stage we have identified enough of the singularities of
$c^{3{\rm m}}$ to write an ansatz for the remaining, less singular
terms.  The unknown coefficients in the ansatz can then be determined
by comparing it to a numerical evaluation of $c^{3{\rm m}}$ at a
number of phase-space points, equal to the number of unknown
coefficients.  The full three-mass triangle coefficient is symmetric
under both the operations $\flip{2}$ and $\flip{3}$, defined in
eqs.~(\use\FlipTwoSym) and (\use\FlipThreeSym), and the full singular
terms include also the images under $\flip{2}$ of the above terms.
(They cancel against terms containing the other box function,
$\Ls_{-1}^{2{\rm m}h}(s_{34},t_{124};s_{12},s_{56})$.)  Thus we may
write
$$
\eqalign{
c^{3{\rm m}} &= 
c_1^{3{\rm m}} + c_2^{3{\rm m}} + c_3^{3{\rm m}}  
+ {p_1 \over \Delta_3}
+ p_2 \, {\spab4.{(1+2)}.3 \over \spab3.{(1+2)}.4 \Delta_3}
  \; + \; \flip2 \ , \cr
}\eqn\axansatz
$$
where $\flip{2}$ is to be applied to all preceding terms.
In eq.~(\use\axansatz) we have explicitly identified
all denominator factors; i.e., $p_1$ and $p_2$ can only be linear
combinations of products of spinor strings.  They are further
restricted by the observation that any helicity amplitude can be
assigned a definite {\it phase weight} --- namely, an integer $n_i$
for each external leg $i$, which is equal to twice the helicity of
that leg.  For $A_6^\ax(1_q^+,2_\qb^-, 3^+, 4^-,5_\eb^-,6_e^+)$, and
hence for $C^\ax$, the phase weight is $\{ 1,-1,2,-2,-1,1\}$, 
the six entries corresponding to each of the six legs.  
The phase weight of an expression $A$ may also be calculated by making the
substitution $\spa{i}.j \to \lambda_i^{-1} \lambda_j^{-1} \spa{i}.j$,
$\spb{i}.j \to \lambda_i \lambda_j \spb{i}.j$, and reading the $n_i$
off of the resulting transformation $A \to \left( \prod_i
\lambda_i^{n_i} \right) \times A$.

Since $\Delta_3$ carries no phase weight, the phase weight of $p_1$
must match that of $C^\ax$, namely $\{ 1,-1,2,-2,-1,1\}$.
In terms of spinor string terminations, it must have the form
$$
[1, \ \langle 2, \ \bigl([3\bigr)^2, \ \bigl(\langle4\bigr)^2, \ 
  \langle 5, \ [6\ .
\eqn\poneends
$$
On the other hand, dimensional analysis implies that the mass
dimension of $p_1$ is 4.  (The mass dimension of $C^\ax$, like that of
any six-point amplitude, is $-2$; but $I_3^{3{\rm m}}$ also has
dimension $-2$, and $\Delta_3$ has dimension 4.)  Thus there are
exactly four spinor products ($\spa{i}.j,\spb{k}.l$) in $p_1$, all of
whose arguments are accounted for in eq.~(\use\poneends).  Since
$\spa{i}.{i} = \spb{i}.{i} = 0$, we see that $p_1$ has just one
independent term,
$$
p_1 = a_0 \, \spb1.3\spb3.6\spa2.4\spa4.5 \, ,
\eqn\poneansatz
$$
where $a_0$ is a constant.

Similarly, $p_2$ also has dimension 4, but its phase weight
is $\{ 1,-1,0,0,-1,1\}$, corresponding to the 
spinor string terminations
$$
[1, \ \langle 2, \ \langle 5, \ [6\ .
\eqn\ptwoends
$$
Equation~(\use\ptwoends) allows terms of the form $\spab5.i.6\spab2.j.1$,
$\spab5.i.1\spab2.j.6$, $\spa5.2\spbb1.{ij}.6$, $\spaa5.{ij}.2\spb1.6$
and $\spa5.2\spb1.6 s_{ij}$, where $i$ and $j$ are arbitrary legs.
Not all such terms are independent.  A Fierz identity lets us
eliminate all terms of the first type in favor of terms of the second
and fourth types. The symmetry $\flip{2}$ relates terms of the third
and fourth types, while combinations of these terms that are symmetrized
in $i\leftrightarrow j$ are already included in the fifth type of terms. 
Using momentum conservation and the two flip symmetries, 
we can reduce the number of independent terms in $p_2$ to
just seven,
$$
\eqalign{
p_2 &=  a_1 \, \spab5.{3}.1\spab2.{3}.6
      + a_2 \, \spab5.{3}.1\spab2.{4}.6
      + a_3 \, \spaa5.{31}.2 \spb1.6 \cr
& \hskip 1 cm                   
      + \spa5.2\spb1.6 ( a_4 \, s_{12} + a_5 \, s_{23}
                       + a_6 \, s_{24} + a_7 \, s_{25} ) 
            \;  + \; \flip{3} \, . \cr
}\eqn\ptwoansatz
$$                                   

Thus it suffices to evaluate $c^{3{\rm m}}$ numerically at eight 
phase-space points, and solve the resulting linear equations
for $a_0,a_1,\ldots,a_7$, which should be simple rational numbers.  
We find
$$
\eqalign{
p_1 &= - \spb1.3\spa4.5 \spa2.4\spb3.6 \, , \cr
p_2 &= - {3\over2} \bigl( \spab5.2.1\spab2.1.6 + \spab5.6.1 \spab2.5.6  
        - \spab5.3.1 \spab2.4.6 - \spab5.4.1 \spab2.3.6 \bigr) \,. \cr
}\eqn\ponetwo
$$
(Evaluation at additional phase-space points may be used to confirm 
this answer.)
If we combine these terms with the singular terms $c_i^{3{\rm m}}$ 
according to eq.~(\use\axansatz), we obtain the coefficient of the 
three-mass triangle given in eq.~(\use\Faxffpm), via 
the function $C^\ax$ defined in eq.~(\use\Caxdef). 

This numerical simplification procedure may be somewhat more involved
for other coefficient functions, if there are more spurious and
physical singularities to account for, but the principle is the same.
In practice we found it convenient to obtain an analytic, though
often very complicated, representation of the desired answer, before
attempting to simplify it numerically; of course, an analytic
representation is not strictly necessary.  It is also not necessary to
eliminate all the linear dependences between terms in the ansatz
before trying to solve for the unknowns.

\section{General Form of Primitive Amplitudes}
\tagsection\GeneralFormSection

The simple structure of the poles in $\e$ of the primitive 
amplitudes~[\use\GieleGlover,\use\KunsztSoper] permits us to
decompose them further into divergent ($V$) and finite ($F$) pieces,  
$$
A_6^{1\rm -loop} = \cg \Bigl[ \Atree_6 V + i \, F \Bigr]\ ,
\eqn\VFdecomp
$$
where the $A_6^{1\rm -loop}$ are any of the primitive amplitudes 
and the prefactor is
$$\eqalign{
  \cg &= {1\over (4\pi)^{2-\eps}}
 {\Gamma(1+\eps)\Gamma^2(1-\eps)\over\Gamma(1-2\eps)}\ . \cr
}\eqn\cgammadef
$$
The tree amplitudes~[\use\BGK] are denoted by $\Atree_6$.
(The $A_6^{s,f,t}$ amplitudes are rather simple so we will not bother 
with this additional decomposition for them.)

The amplitudes we present are bare ones, i.e., no ultraviolet subtraction
has been performed.  To obtain the renormalized amplitudes in 
an \MSbar -type subtraction scheme, one should subtract the quantity
$$
c_\Gamma N_c \, g^2 \left[ {1\over\e}\left( {11\over3}
  - {2\over3}{\nf\over N_c} - {1\over3}{\ns\over N_c} \right) \right]
    \A{6}^\tree\,,
\eqn\mssubtraction
$$
from the amplitudes (\use\qqggDecomp) and (\use\qqqqDecomp).

We quote the results in the four-dimensional helicity (FDH) scheme
[\use\Long,\use\Siegel], since this scheme is convenient for
performing computations in the helicity formulation.  The conversion
between the various schemes is discussed in
refs.~[\use\Long,\use\KunsztFourPoint].  (The more conventional
regularization schemes alter the number of gluon polarizations and are
therefore not natural when using a helicity basis.)  The $e^+\,e^- \to
\qb q gg$ amplitudes may be converted to the 't~Hooft-Veltman (HV)
scheme by adding the quantity
$$
- {1\over 2} \cg N_c \, g^2 \L {1 - {1\over N_c^2}} \R \A6^\tree 
\eqn\schemeshiftqqgg
$$
to the amplitude (\use\qqggDecomp) and changing the
coupling constant from the non-standard $\alpha_{\overline{DR}}$ to
the standard $\alpha_{\overline{MS}}$. 
Similarly, the $e^+\,e^- \to \qb q \Qb Q$ amplitudes may be converted by 
adding the quantity 
$$
- \cg N_c \, g^2 \L {{2\over 3} - {1\over N_c^2}} \R \A6^\tree 
\anoneqn
$$
to the amplitude (\use\qqqqDecomp) and making the same coupling
constant conversion.  The conversion of the HV scheme to the
conventional dimensional regularization (CDR) scheme is accomplished
by accounting for the fact that in the HV scheme observed gluons (at
the partonic level) are in four-dimensions but in the CDR scheme they
are in $(4-2\eps)$ dimensions.  This conversion is rather simple as it
involves only the coefficients of the poles in $\eps$ which are
proportional to the tree amplitudes.  In the final matrix elements
squared (e.g., eq.~(\use\SquaredAmplitudeTwoQuarksTwoGluons)) one simply
replaces all terms originating from the singular parts of
eqs.~(\use\VFdecomp) and (\use\mssubtraction) with their values in the
CDR scheme.  (Further details may be found in
ref.~[\use\KunsztFourPoint].)  Using this conversion recipe, our
results (integrated over lepton orientation) agree 
numerically~[\use\PrivateNigel]
with those reported in refs.~[\use\GloverMiller,\use\CGMqggq].

In order to present the amplitudes compactly, we define, in addition
to the previously-defined spinor-strings, the following combinations
of kinematic variables, related to the three-mass triangle integrals
that appear,
$$  
\eqalign{
&\d12  = s_{12} - s_{34} - s_{56}\, , \hskip 1.3 cm 
\d34  = s_{34} - s_{56} - s_{12}\, , \hskip 1.3 cm 
\d56  = s_{56} - s_{12} - s_{34}\, , \cr
& \hskip 2.8 cm 
\Delta_3 = s_{12}^2 + s_{34}^2 + s_{56}^2 
- 2 s_{12}s_{34} - 2 s_{34} s_{56} - 2 s_{56} s_{12} \,, \cr
&\dt14  = s_{14} - s_{23} - s_{56}\, , \hskip 1.3 cm 
 \dt23  = s_{23} - s_{56} - s_{14}\, , \hskip 1.3 cm 
 \dt56  = s_{56} - s_{14} - s_{23}\, , \cr
& \hskip 2.8 cm 
\delt_{3} = s_{14}^2 + s_{23}^2 + s_{56}^2 
- 2s_{14}s_{23} - 2 s_{23} s_{56} - 2 s_{56} s_{14} \,. \cr}
\eqn\deltaijdef
$$

Certain `flip' symmetries relate either various terms, or various 
amplitudes.  For later convenience, we collect the definitions
of these symmetry operations here:
$$
\flip1 : \;  1 \leftrightarrow 4 \,, \hskip .5 cm 
             2 \leftrightarrow 3 \,, \hskip .5 cm  
             5 \leftrightarrow 6 \,, \hskip .5 cm  
             \spa{a}.{b} \leftrightarrow \spb{a}.{b} \,,
\hskip .7 cm 
\eqn\FlipOneSym
$$
$$
\flip2 : \;  1 \leftrightarrow 2 \,, \hskip .5 cm 
             3 \leftrightarrow 4 \,, \hskip .5 cm  
             5 \leftrightarrow 6 \,, \hskip .5 cm  
             \spa{a}.{b} \leftrightarrow \spb{a}.{b} \,,
\hskip .7 cm 
\eqn\FlipTwoSym
$$
$$
\flip3 : \;  1 \leftrightarrow 5 \,, \hskip .5 cm 
             2 \leftrightarrow 6 \,, \hskip .5 cm  
             3 \leftrightarrow 4 \,, \hskip .5 cm  
             \spa{a}.{b} \leftrightarrow \spb{a}.{b} \,,
\hskip .7 cm 
\eqn\FlipThreeSym
$$
$$
\flip4: \;    1 \leftrightarrow 3 \,, \qquad
  5 \leftrightarrow 6 \,, \qquad
   \spa{a}.{b} \leftrightarrow \spb{a}.{b} \,,
\hskip 1.65 cm 
\eqn\FlipFourSym
$$
In general, $\spa{a}.{b} \lr \spb{a}.{b}$ denotes complex conjugation
of {\it all} spinor products, including the various strings defined in
eq.~(\use\AngleDef), $\spab{i}.{j}.{l} \lr \spab{l}.{j}.{i}$,  
$\spaa{i}.{(l+m)}.{j} \lr \spbb{i}.{(l+m)}.{j}$, and so forth.  
We also define the exchange operations
$$
\eqalign{
&\exch{16,25} : \;  1 \leftrightarrow 6 \,, \hskip .3cm 
                  2 \leftrightarrow 5 \,,  \cr 
&\exch{34} : \; 3 \leftrightarrow 4   \,.  \cr}
\eqn\ExchangeDefs
$$

\section{Master Functions}
\tagsection\MasterFunctionSection

As mentioned in section~\use\SampleCutSubSection, many of the cuts
(or portions of cuts) for different amplitudes are related to each 
other by simple permutations and overall spinor-product prefactors.
In particular, one `master function' $M_1(1,2,3,4)$ enters the
the cut-constructible part of one of the leading-color primitive amplitudes, 
as well as two of the subleading-color ones.
Two additional `master functions', $M_2(1,2,3,4)$ and $M_3(1,2,3,4)$,
enter the scalar parts of the three subleading-color amplitudes 
with opposite gluon helicities.  
We have decomposed the latter two master functions further, extracting 
$M_{2a}$ and $M_{3a}$, respectively, because these pieces appear
separately as well.

The cut-constructible master function is given by 
$$
\hskip -0.7 cm 
\eqalign{
M_{1}& (1,2,3,4) = \cr
& \hskip 0.1 cm
{\spb1.3\over \spab1.{(3+4)}.2 \spab3.{(1+2)}.4 \, \del_3}
    \Bigl(
     2\, \spa1.2 \spab5.2.6 
        (t_{123}\, \d12 + s_{56} \,\d56) 
   + 2\, \spa1.2\spa1.5 \spab4.5.6 
        \bigl( \spb1.4\, \d56 - 2\, \spbb1.{23}.4 \bigr) \cr
& \hskip 1.5 cm
   + s_{56} \spa2.4 
        \bigl(  \spa1.5\spb4.6\, \d56
         - 2\, \spab1.2.6 \spab5.3.4 \bigr) \Bigr) 
     I_3^{3{\rm m}}(s_{12},s_{34},s_{56}) \cr
& \hskip 0.1 cm
+ {2\over \spab1.{(3+4)}.2 \spab3.{(1+2)}.4 \, \del_3}  \biggl[
   {\spb1.3\over \spa5.6}
    \bigl(\spaa2.{1(3+4)}.5-\spaa2.{(3+4)(1+2)}.5 \bigr)
  \bigl( \spa1.5 t_{124} + \spaa1.{26}.5 \bigr) \cr
& \hskip 1.5 cm  - {\spb1.6 \spa2.4\over \spa3.4 \spb5.6} 
  \Bigl( (s_{14}-s_{23}) \bigl(\spab1.2.6 \, \d12
                    - \spab1.5.6 \, \d56 \bigr) 
    + \spab1.{(3+4)}.2  \bigl(\spab2.1.6 \d12
                             -\spab2.5.6\d56\bigr) \Bigr) \cr
& \hskip 1.5 cm 
    + 2\, \spab2.1.3 \spab5.{(1+2)}.6 (s_{14}-s_{23}) \biggr]
         {\ts \ln\Bigl( {-s_{12}\over -s_{56}} \Bigr)}
 \,. \cr}
\eqn\MasterOne
$$


The scalar master functions are given by
$$
\hskip -1.1 cm 
\eqalign{
M_{2a}& (1,2,3,4) = 
\nl\hsb
 {\frac{1}{2}}\frac{\spb1.2}{\spb2.3\spab4.{(1+2)}.3}
\biggl\{ 
6 \, { \spa1.2\spab3.{(1+2)}.4\spab5.{(1+2)}.6
  \bigl( \spb1.3{{\delta}_{56}} - 2\,\spbb1.{24}.3 \bigr) 
       \over \Delta_3^2 }
\nl\hsb
+ {1\over\Delta_3} \biggl[
 \frac{\spab2.1.4\spa3.5\spab5.{(1+2)}.3}{\spa5.6}
- \frac{\spa1.2\spa4.5}{\spa3.4\spa5.6}
\left(  \frac{\spb1.3\spa4.5\spab3.{(1+2)}.4{{\delta}_{12}}}
{\spab4.{(1+2)}.3}  
- \spa3.5\bigl(2\,\spb1.4{{\delta}_{12}} - \spbb1.{23}.4 \bigr)\right) 
\nl\hsa
 -\frac{\spa2.4\spa3.5{{\delta}_{12}}}{\spa3.4\spa5.6}
\bigl( \spab5.{(1+2)}.4 - \spab5.6.4 \bigr) 
+\frac{\spb4.6}{\spb5.6}
\Bigl( 2\, \bigl( \spab2.3.6 \,\d34 - \spab2.5.6 \,\d56 \bigr) 
      + \spab3.4.6 \spab2.{(1+4)}.3 \Bigr) 
\nl\hsa
- \frac{\spab2.1.3\spab5.{(3-4)}.6\, \d12}{\spab4.{(1+2)}.3}
- 4\,\spa1.2\spb3.6\, \Bigl( \spa3.5\spb1.4 
+ \frac{\spa3.4\spa5.6\spb1.3\spb4.6}{\spab4.{(1+2)}.3} \Bigr) 
 +\spb4.6\bigl(\spaa2.{46}.5 -\spaa2.{13}.5 \bigr) 
\biggr]
\nl\hsb
+ \frac{\spab5.{(1+2)}.3
      \left(\spa2.5\spa3.4 - \spa2.3\spa4.5 \right) }
      {\spa3.4\spa5.6{{\spab4.{(1+2)}.3}}}
\biggr\}
\, {\ts \ln\L \frac{-{s_{12}}}{-{s_{56}}}\R }
\nl
- {1\over2} \, \frac{1}{\spb2.3 \spab4.{(1+2)}.3\, \del_3}
\biggl\{
 - 6 \, {\spb1.2  \spab5.{(1+2)}.6 \spbb4.{3(1+2)}.4
     \bigl( \spa2.4 \d56 - 2\, \spaa2.{13}.4 \bigr) 
      \over \del_3 } 
\nl\hsb
+   \frac{\spb1.3 }{\spab4.{(1+2)}.3}  
    \bigl(\d34\, (t_{123}-t_{124})-\del_3 \bigr)
        \Bigl(\frac{\spaa5.{34}.5}{\spa5.6} + 
              \frac{\spbb6.{34}.6}{\spb5.6} \Bigr) 
\nl\hsb
   + \frac{\spab5.3.4}{\spa5.6} 
    \Bigl( 2\, \bigl( \spab5.4.1\, \d34 - \spab5.6.1 \, \d56  \bigr)
           + \spab4.{(2+3)}.1\spab5.{(1+2)}.4 \Bigr)
\nl\hsb
   - \frac{\spb4.6}{\spb5.6} 
        \Bigl( 2 \,\spab3.2.1 
        \bigl(\spb3.6 \,\d12 - 2\, \spbb3.{45}.6 \bigr)
          + \spbb1.{(2+3)4 3 (1+2)}.6
          - 3\, \spbb6.{43}.1 \, \d34 \Bigr) 
 \biggr\} \,
  {\ts \ln\L \frac{-{s_{34}}}{-{s_{56}}}\R } 
%
\,,
\cr}
\eqn\MasterTwoa
$$
$$
\hskip -1.1 cm 
\eqalign{
M_2 &(1,2,3,4)  = 
%
\nl\hsb
M_{2a}(1,2,3,4) + 
 \frac{\spab4.{(2+3)}.1{{\spab5.{(1+2)}.3}^2}}
   {\spb2.3\spa5.6{{\spab4.{(1+2)}.3}^3}}
\,\Ls_{-1}^{2{\rm m}h} 
\L {s_{34}},{t_{123}};{s_{12}},{s_{56}}\R
%
\nl
+ {\spb1.2 \over \spb2.3 \spab4.{(1+2)}.3} \,
\biggl\{ 
3\, \frac{\spa1.2\spb3.4\spab3.{(1+2)}.4\spab5.{(1+2)}.6}
           {\Delta_{3}^{2}}
 \bigl( \spab4.2.1 \, \d12 -\spab4.3.1\, \d34 \bigr) 
\nl\hsb
 - {1 \over \Delta_3}
 \biggl[ \frac{\spaa2.{1 3(1+2)4}.5\spb3.6}
   {\spab4.{(1+2)}.3}\left( {t_{124}} - {t_{123}} \right) 
 +\spa2.3 \spb4.6 \bigl( \spab 5.4.3\,{{\delta}_{34}} 
      - \spab5.6.3 \,{{\delta}_{56}} \bigr) 
\nl\hsa
 +\spab3.{(1+2)}.4\,\bigl( \spab2.4.3 \spab5.{(1+2)}.6 
             + 3\,\spab2.1.3 \spab5.4.6 \bigr) 
\nl\hsa
  -\spab2.1.4\spa3.5\,\bigl( \spbb6.{4(1+2)}.3 
   + 2\,\spbb6.{54}.3 
         + \spb3.6\,{{\delta}_{12}} \bigr) \biggr]
+ \frac{\spaa2.{3(1+2)}.5 \spb4.6}{t_{123}}
  \biggr\} \, 
    I_3^{3{\rm m}}\L {s_{12}},{s_{34}},{s_{56}}\R
\nl
%
%
 +\frac{\spb4.6\spab5.4.1}{\spb2.3\spab4.{(1+2)}.3} 
 \,\frac{{{\Ll}_1}\L \frac{-{s_{56}}}{-{t_{123}}}\R}
   {t_{123}}
%
 -\frac{\spab3.2.1 \spab5.{(1+2)}.3^2}
       {\spb2.3\spa5.6\spab4.{(1+2)}.3^2}
   \,\frac{{{\Ll}_0}\L \frac{-{t_{123}}}{-{s_{12}}}\R}{{s_{12}}}
 -\frac{ \spab5.4.1 \spab5.{(1+2)}.3{t_{123}}}
         {\spb2.3 \spa5.6{{\spab4.{(1+2)}.3}^2}}
   \,\frac{{{\Ll}_0}\L \frac{-{t_{123}}}
 {-{s_{56}}}\R}{{s_{56}}}
\nl
%
- {\frac{1}{2}}\,\frac{ \spab4.2.1\, \d12 
   - \spab4.3.1  \, \d34}
   {\spb2.3\spa3.4\spab4.{(1+2)}.3 \, \Delta_3}
\Bigl(\frac{\spa3.4{{\spb4.6}^2}}{\spb5.6}  
   + \frac{{{\spa3.5}^2}\spb3.4}{\spa5.6} \Bigr)
 -\frac{\spa3.5\spb4.6 
    \bigl( \spb1.3{{\delta}_{56}} - 2\, \spbb1.{24}.3 \bigr) }
    {\spb2.3\spab4.{(1+2)}.3 \, \Delta_3} 
 \, , 
}
\eqn\MasterTwo
$$
$$
\hskip -1.1 cm 
\eqalign{
M_{3a}&(1,2,3,4)= 
\nl\hsb
 {\frac{1}{2}} \,
  \frac{\spb2.4}{\spb2.3 s_{34} \spab1.{(3+4)}.2 } \biggl[
6\, { s_{12} s_{34} \over\Delta_{3}^2}
 ( t_{134} - t_{234} ) \spab3.{(1+2)}.4\spab5.{(1+2)}.6  
\nl\hsa
+ \frac{\spa1.2\,\d12}{\spb5.6\,\Delta_3}
 \biggl( \spab3.{(1+2)}.6\spb1.4\spb2.6 
 + \spab3.{(2+4)}.6 \spb1.2\spb4.6 
 - \spa1.3\spb2.4 \biggl( \spb1.6^2 
     + \spb2.6^2 \frac{\spab2.{(3+4)}.1}{\spab1.{(3+4)}.2} \biggr)
 \biggr)
\nl\hsa
 +\frac{\spb1.2 \,\d12}{\spa5.6\,\Delta_3}
\biggl( 
 \spab5.2.4 \bigl( \spa1.2 \spa3.5 - \spa2.3 \spa1.5 \bigr)
 - \spa1.5\spa2.5\spab3.{(1+2)}.4 
 + \frac{{{\spa1.5}^2}\spa1.3\spb2.4\spab2.{(3+4)}.1}{\spab1.{(3+4)}.2} 
\biggr) 
\nl\hsa
 - 2\, {\spa1.2\spa3.5 \over \Delta_3} \left( \spb1.6\,\spb2.4 
     + \spb1.4\spb2.6 \right) \d56
  -\Bigl( \frac{\spa3.4{{\spb4.6}^2}}{\spb5.6} 
 +  \frac{{{\spa3.5}^2}\spb3.4}{\spa5.6} \Bigr) { s_{12}\,t_{234} \over \Delta_3 } 
\nl\hsa
+ \frac{\spab3.2.6\spb4.6} {\spb5.6}
- \frac{\spa1.3\spb2.4}{\spab1.{(3+4)}.2}
 \Bigl( \frac{\spaa5.{12}.5}{\spa5.6}
      + \frac{\spbb6.{12}.6}{\spb5.6}  \Bigr) \biggr]
\,{\ts \ln\Bigl( \frac{-{s_{12}}}{-{s_{56}}}\Bigr)} 
\nl
+ {\frac{1}{2}} \, \frac{\spb2.4}{\spb2.3\spab1.{(3+4)}.2{{\Delta}_3}}
\biggl[
 3 \, { \d56 \, (t_{134} - t_{234}) \, \spab3.{(1+2)}.4\spab5.{(1+2)}.6
        \over \Delta_3 }
\nl\hsa
- \biggl( \frac{\spab3.2.6\spb4.6}{\spb5.6} 
 -\frac{\spab5.1.4\spa3.5}{\spa5.6} 
+ 2\, \frac{\spa1.5\spb2.4}{\spa5.6\spab1.{(3+4)}.2}
\Bigl( \spaa3.{(2+4)1}.5 - \spaa3.{26}.5
    \Bigr)  \biggr) \,{{\delta}_{34}}
\nl\hsa
+ \Bigl( \spab3.{(1+2)}.4 
 + 4\,\frac{\spa1.3\spb2.4{s_{56}}}{\spab1.{(3+4)}.2} \Bigr)
 \,\Bigl( \frac{\spaa5.{12}.5}{\spa5.6}
        - \frac{\spbb6.{12}.6}{\spb5.6} \Bigr)
+ 8 \, \frac{s_{12}\,\spa1.5\spb2.4\spab3.5.6}{\spab1.{(3+4)}.2}
\nl\hsa
 - 2\, \spab3.{(1-2)}.4\,\spab5.{(1+2)}.6
 + 4\, \bigl( \spbb4.{12}.6 \spa3.5 + \spaa5.{21}.3 \spb4.6 \bigr)
\biggr] {\ts \ln\Bigl( \frac{-{s_{34}}} {-{s_{56}}}\Bigr)}
%
 \,,
\cr}
\eqn\MasterThreea
$$
$$
\hskip -1.1 cm 
\eqalign{
M_3 &(1,2,3,4)  = 
%
\nl\hsb
M_{3a}(1,2,3,4) 
%
 -\frac{{{\spa1.5}^2}{{\spb2.4}^3}{t_{234}}}
  {\spb2.3\spb3.4\spa5.6{{\spab1.{(3+4)}.2}^3}}
\,\Ls_{-1}^{2{\rm m}h}\L {s_{12}},{t_{234}}; {s_{34}},{s_{56}}\R
\nl
+ { \spb2.4 \over \spb2.3 \spab1.{(3+4)}.2 \, \Delta_3 } \,
\biggl\{
  {\frac{3}{2}} \, 
 { s_{12}\,\d12\,(t_{134} - t_{234})\spab3.{(1+2)}.4\spab5.{(1+2)}.6
   \over \Delta_3 }
+ {\frac{1}{2}}\, s_{12}\,\spab3.{(1+2)}.4\spab5.{(1+2)}.6 
\nl\hsa
 -\frac{\spa1.3\spa1.5\spb2.4\spab2.{(3+4)}.1}{\spab1.{(3+4)}.2}
 \bigl( \spbb2.{1(3+4)}.6 - \spbb2.{(3+4)(1+2)}.6 \bigr)
\nl\hsa
+ \spab2.1.4 \bigl( \spab3.{(1+4)}.2\spab5.{(1+2)}.6
     + \spab5.1.2\spab3.4.6 \bigl) 
\nl\hsa
  +\spa1.5\spa2.3\biggl( \spb2.4
 \Bigl(\spbb1.{2(3+4)}.6 - \spbb1.{(3+4)(1+2)}.6 \Bigr)
 - \spb1.2 \spbb4.{(1+2)(3+4)}.6
\biggr)  \biggr\}
  I_3^{3{\rm m}}\L {s_{12}},{s_{34}},{s_{56}}\R
\nl
%
+ \frac{1}{2} \, \frac{{{\spa1.5}^2} \spb1.4 \spb2.4}
   {\spb2.3\spb3.4\spa5.6 \spab1.{(3+4)}.2 }
\biggl( - 2 \, {\spb2.4 \over \spab1.{(3+4)}.2}
   {{\Ll}_0}\Bigl( \frac{-{s_{56}}}{-{t_{234}}}\Bigr) 
+ \spb1.4
  \,\frac{{{\Ll}_1}\Bigr( \frac{-{s_{56}}}{-{t_{234}}}\Bigr)}{t_{234}}
\biggr)
\nl
 - {\frac{1}{2}}\, \frac{\spa2.3\spb2.4\spab5.{(3+4)}.2}
                        {\spb2.3\spa5.6\spab1.{(3+4)}.2}
\biggl( \biggl( \frac{\spab5.{(2+3)}.4}{{t_{234}}} 
 - 2\,\frac{\spa1.5\spb2.4}{\spab1.{(3+4)}.2} \biggr) 
 \,\frac{{{\Ll}_0}\Bigl(\frac{-{t_{234}}}{-{s_{34}}}\Bigr)} {{s_{34}}}
  + \spab5.2.4 
  \,\frac{{{\Ll}_1}\Bigl( \frac{-{s_{34}}}{-{t_{234}}}\Bigr)}{t_{234}^{2}}
\biggr) 
\nl
+ \frac{1}{2} \,
  \frac{\spb2.4}{\spb2.3\spab1.{(3+4)}.2{{\Delta}_3}}
\bigg[ -\frac{\spa3.5}{\spa3.4}\Bigl( \frac{\spa2.5}{\spa5.6}
 \bigl( \spab3.1.2\, \d12 - \spab3.4.2 \, \d34 \bigr)
  - \spb1.6 \bigl( \spa1.3 \, \d56 - 2\,\spaa1.{24}.3 \bigr) \Bigr) 
\nl\hsa 
 +\frac{\spb4.6}{\spb3.4}\Bigl( \frac{\spb1.6}{\spb5.6}
 \left( \spab1.2.4\,\d12
  - \spab1.3.4\, \d34 \right)  
  + \spa2.5 \bigl( \spb2.4 \,\d56 - 2\,\spbb2.{13}.4 \bigr) \Bigr) \bigg]
 \, . 
}
\eqn\MasterThree
$$

\section{Results for Primitive Amplitudes: $A_6(1_q, 2,3, 4_\qb)$ and 
 $A_6^{s,f,t}(1_q, 2, 3, 4_\qb)$ }
\tagsection\LeadingColorPrimitiveSection

In this section we present the independent $qgg\qb$ primitive amplitudes 
where in the parent diagrams neither of external gluons are 
attached to the external fermion lines, i.e.~the helicity configurations
$$
1_q^+,2^+,3^+,4_\qb^-,5_\eb^-,6_e^+; \qquad
1_q^+,2^+,3^-,4_\qb^-,5_\eb^-,6_e^+; \qquad
1_q^+,2^-,3^+,4_\qb^-,5_\eb^-,6_e^+. 
\eqn\leadingcolorhelconfig
$$
We suppress the lepton labels $(5_\eb^-,6_e^+)$ below.  Representative
parent diagrams are given in figs.~\use\PrimitiveDiagramsFigure{a} and
b.

\subsection{The Primitive Amplitudes: $A_6^{s,f,t}(1_q, 2, 3, 4_\qb)$ }

By far the simplest of the primitive amplitudes are those proportional
to the number of scalars $n_s$ or fermions $n_{\! f}$.  This
simplicity follows from the fact that only two Feynman diagrams
contribute, each with a triangle integral.  
(See fig.~\use\PrimitiveDiagramsFigure{a}.)  It may also be understood in
terms of the rather simple cut and factorization properties that must
be satisfied.

The results for these primitive amplitude are
$$
\eqalign{
& A_6^s(1_q^+, 2^+, 3^-, 4_\qb^-) = 
 A_6^{\! f}(1_q^+, 2^+, 3^-, 4_\qb^-) = 0 \,,\cr
& A_6^s(1_q^+, 2^-, 3^+, 4_\qb^-) = 
 A_6^{\! f}(1_q^+, 2^-, 3^+, 4_\qb^-) = 0 \,,\cr
& A_6^{\! f} (1_q^+, 2^+,3^+, 4_\qb^-) = 0 \,,\cr
& A_6^s(1_q^+, 2^+,3^+, 4_\qb^-) =  
i \, {\cg\over3} {1\over\spa2.3^2 s_{56}} \Biggl[
 - { \spa4.5 \spbb6.{(1+2)3}.1 \over t_{123} }
 + { \spb1.6 \spaa5.{(4+2)3}.4 \over t_{234} } \Biggr]
\, .\cr
}\eqn\Asf
$$
Although it is not manifest, $A_6^s(1_q^+, 2^+,3^+, 4_\qb^-)$
is antisymmetric in the exchange of $2$ and $3$, as required by
charge conjugation invariance.  
This fact accounts for the absence of $n_f$ or
$n_s$ terms in $A_{6;3}$ in eq.~(\use\totalpartialampl). 

The contribution of a virtual top quark, through bubble and triangle
graphs, is simply related to the above function $A_6^s$, 
$$
\eqalign{
& A_6^t(1_q^+, 2^+, 3^-, 4_\qb^-) = 
 A_6^t(1_q^+, 2^-, 3^+, 4_\qb^-) = 0 \,,\cr
& A_6^t(1_q^+, 2^+, 3^+, 4_\qb^-) = 
  {1\over20} {s_{23}\over m_t^2} 
   A_6^s(1_q^+, 2^+, 3^+, 4_\qb^-) \,,\cr
}\eqn\At
$$
neglecting $1/m_t^4$ corrections.

\subsection{The Helicity Configuration $q^+\, g^+ \,g^+\, \qb^-$ }

The simplest of the non-$n_{s,f}$ helicity amplitudes is the one where 
both gluons have the same helicity, $A_6(1_q^+,2^+,3^+,4_\qb^-)$.  
The tree amplitude in this case is
$$
\Atree_6 = 
-i\, {\spa4.5^2\over\spa1.2\spa2.3\spa3.4\spa5.6}
 \ .
\eqn\treepppm
$$


The contributions to the amplitude in terms of the decomposition  
(\use\VFdecomp) are 
$$
\eqalign{
V^{\gf}\ &=\  
 - {1\over\e^2} \left( 
      \left({\mu^2\over-s_{12}}\right)^\e
    + \left({\mu^2\over-s_{23}}\right)^\e
    + \left({\mu^2\over-s_{34}}\right)^\e \right) 
     - {2\over\e} \left({\mu^2\over-s_{56}}\right)^\e - 4
    \, ,
\hskip 3 cm  \cr
}\eqn\VLpppmgf  
$$
$$
\eqalign{                    
F^{\gf}\ &=\   
 {\Atree_6 \mathtag{[FPPF]} \over i} \biggl[
 - {\ts \Ls_{-1}\Bigl({-s_{12}\over-t_{123}},{-s_{23}\over-t_{123}}\Bigr)}
 - {\ts \Ls_{-1}\Bigl({-s_{23}\over-t_{234}},{-s_{34}\over-t_{234}}\Bigr)}
 - \Ls^{2{\rm m}e}_{-1}(t_{123},t_{234};s_{23},s_{56}) \biggr]\cr
&\quad 
 + 2\, { \spa4.5\spab5.2.3 \over \spa1.2\spa2.3\spa5.6 }
   { \Ll_0\Bigl({-t_{234}\over-s_{34}}\Bigr) \over s_{34} }
 + 2 { \spa4.5\spaa5.{1(2+3)}.4 \over \spa1.2\spa2.3\spa3.4\spa5.6 }
   { \Ll_0\Bigl({-s_{56}\over-t_{234}}\Bigr) \over t_{234} }
   \,, \cr
}\eqn\FLpppmgf
$$
$$
\eqalign{
V^\sc &=
{1\over2\e} \left({\mu^2\over-s_{56}}\right)^\e + {1\over2}
   \ , \cr 
F^\sc &= 
 {\Atree_6 \mathtag{[FPPF]} \over i} \Biggl[
 - {1\over2} \left({\spaa4.{32}.5 \over \spa4.5}\right)^2 
     { \Ll_1\Bigl({-s_{34}\over-t_{234}}\Bigr) \over t_{234}^2 }
 + {1\over2} \left({\spaa4.{(2+3)1}.5 \over \spa4.5}\right)^2 
     { \Ll_1\Bigl({-s_{56}\over-t_{234}}\Bigr) \over t_{234}^2 }
                              \Biggr] \cr
&\quad     
+ {1\over2} \Biggl[ 
- { \spab5.2.3 \spa5.4
     \over \spa1.2\spa2.3 t_{234} \spa5.6 } 
- { \spb2.3\spa4.5\spab5.{(2+4)}.3
     \over \spa1.2 t_{123} t_{234} \spa5.6 } \cr
&\qquad
+ {  \spab4.{(2+3)}.6^2
     \over \spa1.2\spa2.3\spa3.4 t_{234} \spb5.6 } 
+ { \spab4.2.3 \spbb6.{1(2+3)}.6
     \over \spa1.2\spa2.3 t_{123} t_{234} \spb5.6 } \Biggr]
\vphantom{+ A6s/I/cg}
+ {1\over i\,\cg} \, A_6^s(1_q^+, 2^+,3^+, 4_\qb^-)
 \,. \cr
}\eqn\FLpppmscalar  
$$

\subsection{The Helicity Configuration $q^+\, g^+\, g^-\, \qb^-$ }

\vskip5pt

The next simplest helicity configuration is $A_6(1_q^+,2^+,3^-,4_\qb^-)$.
Notice that this amplitude is symmetric under the `$\flip1$' symmetry
in eq.~(\use\FlipOneSym). 
The tree amplitude for this helicity configuration is
$$
\eqalign{
\Atree_6 &= 
i \biggl[  
   {\spa3.1 \spb1.2 \spa4.5 \spab3.{(1+2)}.6 
       \over \spa1.2 s_{23} t_{123} s_{56} }
  - {\spa3.4\spb4.2 \spb1.6 \spab5.{(3+4)}.2
       \over \spb3.4 s_{23} t_{234} s_{56} } \cr
&\qquad\quad       
  - { \spab5.{(3+4)}.2 \, \spab3.{(1+2)}.6
       \over \spa1.2 \spb3.4 s_{23} s_{56} } \biggr]
\ . \cr
}\eqn\treeppmm
$$
An alternate form for $\Atree_6$, which has manifest 
behavior in the collinear limits $k_2 \parallel k_3$
and $k_5 \parallel k_6$, at the expense of
more obscure behavior on the three-particle poles, is 
$$
\eqalign{
\Atree_6 &= 
i \biggl[  
   { \spa5.4 \spb4.2 \spb1.2 \, \spab5.{(3+4)}.2
      \over \spb2.3\spb3.4 t_{123}t_{234} \spa5.6 }
 + { \spa3.1\spb1.6 \spa3.4 \, \spab3.{(1+2)}.6
      \over \spa1.2\spa2.3 t_{123}t_{234} \spb5.6 } \cr
&\qquad\quad       
 - { \spab3.{(1+2)}.6 \, \spab5.{(3+4)}.2
      \over \spa1.2\spb3.4 t_{123}t_{234} } \biggr]
\ . \cr
}\eqn\treeppmmalt
$$


The results for the cut-constructible and scalar pieces are 
$$
\eqalign{
V^{\gf} &=  
 - {1\over\e^2} \left( 
      \left({\mu^2\over-s_{12}}\right)^\e
    + \left({\mu^2\over-s_{23}}\right)^\e
    + \left({\mu^2\over-s_{34}}\right)^\e \right) 
 - {2\over\e} \left({\mu^2\over-s_{56}}\right)^\e - 4 
\,, \cr
}\eqn\VLppmmgf  
$$
$$
\eqalign{
F^{\gf} &= 
\biggl(  { \spa1.3  \spab3.{(1+2)}.6^2
  \over \spa1.2\spa2.3\spb5.6 t_{123} \, \spab1.{(2+3)}.4 }  
+ { {\spb1.2}^3 {\spa4.5}^2 
  \over \spb2.3\spb1.3\spa5.6 t_{123} \, \spab4.{(2+3)}.1 } \biggr) 
  {\ts \Ls_{-1}\Bigl({-s_{12}\over-t_{123}},{-s_{23}\over-t_{123}}\Bigr)}
\cr
&\quad
+ \biggl({\spa1.3  \spab3.{(1+2)}.6^2
    \over \spa1.2\spa2.3\spb5.6 t_{123} \, \spab1.{(2+3)}.4 }  
+ { {\spb1.2}^2 {\spa4.5}^2 \, \spab4.{(1+3)}.2
    \over \spb2.3\spa5.6 t_{123}\,\spab4.{(2+3)}.1\,\spab4.{(1+2)}.3 } \biggr)
     \cr
& \hskip 2 cm \times
\Lsnew^{2{\rm m}h}_{-1}(s_{34},t_{123};s_{56},s_{12}) \cr
&\quad +
 {1\over2}\, {\spb1.2 \bigl(\spaa4.{(1+2)(3+4)}.5^2 
            - s_{12} s_{34} \spa4.5^2\bigr) \over
           \spa1.2 \spb3.4 \spa5.6 \spab4.{(2+3)}.1 \spab4.{(1+2)}.3} 
 \, I_3^{3{\rm m}}(s_{12},s_{34},s_{56})
\cr
&\quad 
- 2\,{ \spa1.3\spab3.{(1+2)}.6 \over \spa1.2\spb5.6\,\spab1.{(2+3)}.4} 
  \Biggl[ 
   { \spbb6.{(2+3)1}.2 \over t_{123} }
  \,  { \Ll_0\Bigl({-s_{23}\over-t_{123}}\Bigr) \over t_{123} }
 + { \spab3.4.6 \over \spa2.3 }
  \,  { \Ll_0\Bigl({-s_{56}\over-t_{123}}\Bigr) \over t_{123} } 
        \Biggr] \cr
& \hskip 3 cm \;+\; \flip1 
\,, \cr
}\eqn\FLppmmgf
$$
$$
\eqalign{
V^{\sc}\ &=\  
  {1\over2\e} \left({\mu^2\over-s_{56}}\right)^\e + {1\over2}
   \,, 
 \hskip 9. cm \cr
}\eqn\VLppmmscalar  
$$
$$
\eqalign{
F^{\sc} &= 
 - {1\over2} { \spa1.3 
     \over \spa1.2\spa2.3\spb5.6 t_{123} \, \spab1.{(2+3)}.4 } 
 \biggl[ 
    \spab3.{21(2+3)}.6^2 
  \,  { \Ll_1\Bigl({-t_{123}\over-s_{23}}\Bigr) \over s_{23}^2 } 
  +  \spab3.4.6^2 
  \,  {\ts \Ll_1\Bigl( {-s_{56}\over-t_{123}} \Bigr) }
                \biggr] 
 \cr
& \hskip 1 cm 
+ {1\over2} { {\spb6.2}^2 \over \spa1.2\spb2.3\spb3.4\spb5.6 } 
\; + \; \flip1 
\,, \hskip 3 cm \cr
}\eqn\FLppmmscalar
$$
where `$\flip1$' is to be applied to all preceding terms in the
given expression.

\subsection{The Helicity Configuration $q^+\, g^-\, g^+\, \qb^-$ }

The most complicated leading-color helicity configuration is 
$A_6(1_q^+,2^-,3^+,4_\qb^-)$.  In general, those
amplitudes where the negative and positive helicities alternate 
around the loop are the most complicated ones.
The symmetry $\flip{1}$~(\use\FlipOneSym) holds for this helicity 
configuration as well.  The tree amplitude in this case is
$$
A_6^{\rm tree} =  
i \, \biggl[
- {\spb1.3^2\spa4.5 \spab2.{(1+3)}.6 \over \spb1.2 s_{23} t_{123} s_{56}}
+ {\spa2.4^2\spb1.6 \spab5.{(2+4)}.3 \over \spa3.4 s_{23} t_{234} s_{56}}
+ {\spb1.3\spa2.4\spb1.6\spa4.5 \over \spb1.2\spa3.4 s_{23} s_{56}}
   \biggr]
 \,.
\eqn\treefmpf
$$
An alternate form with more manifest behavior in two-particle channels,
but less manifest behavior in three-particle channels, is
$$
\eqalign{
\Atree_6 &= 
 i \biggl[  
   { \spb1.3^2 \spa4.5 \spab5.{(2+4)}.3
       \over \spb1.2 \spb2.3 t_{123} t_{234} \spa5.6 }
  - { \spa2.4^2 \spb1.6 \spab2.{(1+3)}.6
       \over \spa2.3 \spa3.4 t_{123} t_{234} \spb5.6 }
  + { \spb1.3 \spa2.4 \spb1.6 \spa4.5 
       \over \spb1.2 \spa3.4 t_{123} t_{234} } \biggr]
   \ . \cr
}\eqn\treefmpfalt
$$


The results for the cut-constructible pieces are 
$$
\eqalign{
V^{\gf} &=  
 - {1\over\e^2} \left( 
      \left({\mu^2\over-s_{12}}\right)^\e
    + \left({\mu^2\over-s_{23}}\right)^\e
    + \left({\mu^2\over-s_{34}}\right)^\e \right) 
 - {2\over\e} \left({\mu^2\over-s_{56}}\right)^\e - 4
   \,, \hskip 2 cm \cr
}\eqn\VLpmpmgf  
$$
$$
\hskip -1 cm 
\eqalign{
F^{\gf} & = 
%
\biggl( {\spb1.3^3\spa4.5^2\over
    \spb1.2 \spb2.3 \spa5.6 t_{123} \spab4.{(2+3)}.1}  
+ {\spa1.2^3 \spab3.{(1+2)}.6^2 \over
    \spa2.3 \spb5.6 \spa1.3^3 t_{123} \spab1.{(2+3)}.4} \cr
& \hskip 2 cm
- {\spa1.2\spa2.3\spab1.{(2+3)}.6^2 \over
   \spb5.6 \spa1.3^3 t_{123} \spab1.{(2+3)}.4} \biggr)
 {\ts \Ls_{-1}\Bigl( {-s_{12} \over -t_{123}},
   {-s_{23}\over -t_{123}} \Bigr)} \cr
%
%
& \hskip .3 cm
+ \biggl(
 {\spb1.3^3 \spa4.5^2 \over 
    \spb1.2\spb2.3\spa5.6 t_{123} \spab4.{(2+3)}.1}
+ {\spab3.{(1+2)}.6^2\spab2.{(1+3)}.4^3 \over
   \spa2.3 \spb5.6 t_{123} \spab1.{(2+3)}.4 \spab3.{(1+2)}.4^3} \cr
& \hskip 2 cm
- {\spa2.3 \spb4.6^2 t_{123} \spab2.{(1+3)}.4 \over
    \spb5.6 \spab1.{(2+3)}.4 \spab3.{(1+2)}.4^3} \biggr)
  \Ls_{-1}^{{\rm 2m}h} \bigl(s_{34}, t_{123}; s_{12}, s_{56} \bigr) \cr
%
%
%
& \hskip .3 cm 
+ \biggl[ - 2\, {\spab2.1.3 \spab5.4.6 \spab2.{(1+3)}.4 \over
            t_{123}\spab3.{(1+2)}.4 \spab1.{(2+3)}.4} \cr
& \hskip .3 cm 
- {1\over 2}\, {t_{123}\, \d34 + 2\, s_{12} \, s_{56} \over t_{123}^2}
      \biggl( 
   {\spb1.3^3\spa4.5^2 \over \spb1.2 \spb2.3 \spa5.6 \spab4.{(2+3)}.1}
       + {\spab2.{(1+3)}.6^2\spab2.{(1+3)}.4 \over
            \spa2.3 \spb5.6 \spab3.{(1+2)}.4 \spab1.{(2+3)}.4} \biggr) \cr
& \hskip .3 cm 
+ {1\over 2} \,  {1\over \spab1.{(3+4)}.2 \spab3.{(1+2)}.4 \spab1.{(2+3)}.4}  
   \biggl( 
    {\spb3.6 \spab2.4.6 \spab2.{(1+3)}.4 \spab1.{(3+4)}.2
       \over \spb5.6} \cr
& \hskip 1.5 cm
  - {\spab2.1.6 \spab2.4.6 
       \spab4.{(1+2+3)}.4  \spab1.{(3+4)}.2
        \over \spa3.4 \spb5.6}
  - {\spab1.4.6 \spab2.3.4 \spaa5.{61}.4 \over \spa3.4} \cr
& \hskip 1.5 cm
  + \spa1.5\spab2.3.4
       \bigl( \spb3.6 t_{234} - \spbb3.{4(1+3)}.6 \bigr)
  - \spab2.1.6 \spab1.4.3 \spab5.2.4  \cr
& \hskip 1.5 cm
  - {1\over 2} \, {s_{14}\spa4.5\spb1.6 \d56 \spab1.{(2+3)}.4 \over
       \spb1.2 \spa3.4}
  + {1\over 2}\,  (s_{14} - s_{23})  \spa1.2\spb3.4 \spab5.{(2+3)}.6 \biggr)
   \biggr]
     I_3^{3{\rm m}}(s_{12},s_{34},s_{56}) \cr
%
%
& \hskip .3 cm 
- 2\, {\spab2.1.3 \spab2.{(1+3)}.6 \over
        \spb5.6\spa1.3 t_{123} }
   \biggl( {\spab3.{(1+2)}.6 \over  \spab3.{(1+2)}.4}\, 
         {\Ll_0\Bigl({-t_{123} \over - s_{12}} \Bigr) \over s_{12}}
+ {\spab1.{(2+3)}.6 \over \spab1.{(2+3)}.4} \,
         {\Ll_0 \Bigl( {-t_{123} \over -s_{23}} \Bigr) \over s_{23}} 
   \biggr)\cr
& \hskip .3 cm 
- 2\, {\spab2.4.6 \spab2.{(1+3)}.6 \spab2.{(1+3)}.4 \over
     \spa2.3 \spb5.6 \spab3.{(1+2)}.4 \spab1.{(2+3)}.4}\,
           {\Ll_0\Bigl({-t_{123} \over -s_{56}} \Bigr) \over s_{56}} 
 \; + \; M_1(1,2,3,4) \; + \; \flip1 \; 
 \, . \cr}
\eqn\FLfmpfgf
$$
The results where the gluon in the loop is replaced by a scalar are
$$
\eqalign{
V^{\sc}\ &=\  
  {1\over2\e} \left({\mu^2\over-s_{56}}\right)^\e + {1\over2}
   \ , 
 \hskip 7.7 cm \cr
}
\eqn\VLfmpfscalar  
$$
$$
\hskip -1. cm 
\eqalign{
F^{\sc} & =  
%
  {\spa1.2\spa2.3\spb1.3^2\spab1.{(2+3)}.6^2 \over
    \spb5.6\spa1.3 t_{123} \spab1.{(2+3)}.4} \, 
 \biggl[
 {\Ls_{1} \Bigl( {-s_{12} \over -t_{123}}, {-s_{23} \over -t_{123} } \Bigr) 
                  \over t_{123}^2} 
 - {1\over 2} \, 
        {\Ll_1\Bigl( {-t_{123} \over -s_{23}} \Bigr) \over s_{23}^2} \biggr]\cr
& \hskip .3 cm 
%
+ {\spa2.3 \spb4.6^2 t_{123} \spab2.{(1+3)}.4 \over
   \spb5.6\spab1.{(2+3)}.4 \spab3.{(1+2)}.4^3 } \, 
   \Ls_{-1}^{{\rm 2 m}h} \bigl(s_{34}, t_{123}; s_{12}, s_{56} \bigr) \cr
%
%
& \hskip .3 cm 
+  
{\spa1.2 \spb4.6 \over \spab1.{(2+3)}.4 \del_3} 
   \biggl[ 3 \,\spb3.4 \spa5.6  \spab2.{(3+4)}.1 
      \bigl( \spab3.{5}.6  \d56 - \spab3.{4}.6 \d34 \bigr)
      {\spab4.{(1+2)}.3 \over \spab3.{(1+2)}.4 \, \del_3}\cr
& \hskip 1.3 cm 
 + \bigl( 3 \,\spab5.6.4 \spab2.3.1 
       - \spab5.3.4 \spab2.{(3+4)}.1 \bigr)
      {\spab4.{(1+2)}.3 \over \spab3.{(1+2)}.4}  
 - \spb1.3\spa2.4\spb3.6\spa5.6 \cr
& \hskip 1.3 cm 
 + \spb1.4\spa2.3  \spab5.6.4 (t_{123} - t_{124})
      {\spab4.{(1+2)}.3 \over \spab3.{(1+2)}.4^2} \biggr]
 I_3^{3{\rm m}}(s_{12},s_{34},s_{56})
 \cr
%
%
& \hskip .3 cm 
+ { \spa2.4\spb4.6^2 \spab2.{(1+3)}.4 t_{123} \over
   \spb5.6\spab1.{(2+3)}.4 \spab3.{(1+2)}.4}  \biggl(
   - {1\over 2}  {\spa2.4 \over \spa2.3} \,
             {\Ll_1\Bigl( {-s_{56} \over -t_{123} } \Bigr) \over  t_{123}^2} 
   + {1\over \spab3.{(1+2)}.4} \,
         {\Ll_0 \Bigl( {-s_{56} \over -t_{123}} \Bigr) \over t_{123}} 
   \biggr)\cr
& \hskip .3 cm
+ {\spaa2.{13}.2 \spb4.6^2 t_{123} \over 
    \spb5.6 \spab1.{(2+3)}.4 \spab3.{(1+2)}.4^2} \,
             {\Ll_0\Bigl({-t_{123} \over -s_{12}} \Bigr) \over s_{12}} 
+ {1\over 2} \, {\spab2.1.3^2 \spab3.{(1+2)}.6^2 \over
        \spb5.6 \spa1.3 t_{123} \spab3.{(1+2)}.4} \,
            {\Ll_1\Bigl({-t_{123} \over - s_{12}} \Bigr) \over s_{12}^2} \cr
%
%
& \hskip .3 cm 
+ { \spa1.2\spb4.6 \over \spab1.{(2+3)}.4 } \biggl[ 
    3 \, \bigl( \spab5.3.4 \, \d34 - \spab5.6.4\, \d56 \bigr)\,
     { \spab2.{(3+4)}.1 \spab4.{(1+2)}.3 
    \over \spab3.{(1+2)}.4  \, \del_3^2}  \cr
& \hskip 1.3 cm 
+ {1\over 2} \,{ 1 \over
    \spa3.4\spb5.6 \spab3.{(1+2)}.4 \, \del_3}  
  \biggl( 
  - \spa2.4\, \d12 \, \bigl(\spbb6.{53}.1+\spbb6.{4(2+3)}.1 \bigr) \cr
& \hskip 1.3 cm 
  + \spab2.{(3+4)}.1 \spab4.{(1+2)}.3 \spab3.{(4-5)}.6
    - \spab2.{(3+4)}.1 \spab3.4.6 \, \d34 
         {(t_{123}-t_{124}) \over \spab3.{(1+2)}.4} \biggr) \cr
& \hskip 1.3 cm 
+ {1\over 2} \, { \spb4.6 \spab2.{(3+4)}.1 \over
        \spb5.6 \spab3.{(1+2)}.4^2} \biggr]
 {\ts \ln\Bigl({-s_{12} \over -s_{56}} \Bigr) } \cr
%
& \hskip .3 cm 
+ { \spb4.6 \over \spab1.{(2+3)}.4 \spab3.{(1+2)}.4 } \biggl[
- 3\, {\spa1.2\spb3.4 \spab2.{(3+4)}.1 \spab4.{(1+2)}.3
     \bigl( \spa3.5  \d12 - 2 \,\spaa3.{46}.5 \bigr) \over \del_3^2} \cr
& \hskip 1.3 cm 
- {1\over 2} \,  { 1 \over \spb1.2\spb5.6 \, \del_3}
    \biggl( \spab2.{(3+4)}.1 \spab4.{(1+2)}.3
          \bigl(\spbb4.{3(1+2)}.6 + \spb4.6 (\d56 - 2 s_{12}) \bigr)\cr
& \hskip 1.6 cm 
      - \d34 \, \spab2.4.3 
          \bigl(\spbb6.{53}.1+\spbb6.{4(2+3)}.1\bigr) 
- 2 \, \spb4.6 t_{123} \, \biggl( 2\, \spab2.{(3-4)}.1 \spab4.{(1+2)}.3 \cr
& \hskip 1.6 cm 
 + (t_{123}-t_{124})  \Bigl(\spb1.3\spa2.4   
    + \spb1.4 \spa2.3 {\spab4.{(1+2)}.3\over\spab3.{(1+2)}.4} \Bigr) 
    \biggr) \biggr)
- {1\over 2}\,  { \spb1.3\spab2.4.6 \over \spb1.2\spb5.6 } \biggr]
       {\ts  \ln\Bigl({-s_{34}\over  - s_{56}} \Bigr)  }\cr
%
& \hskip .3 cm 
+ {1\over 2}  \, {\spb4.6 \spab2.{(3+4)}.1 \spab4.{(1+2)}.3\,
   \bigl( \spab3.{5}.6 \,\d56 - \spab3.4.6 \, \d34\bigr) \over
    \spb1.2\spa3.4\spb5.6\spab1.{(2+3)}.4\spab3.{(1+2)}.4 \, \del_3}  \cr
& \hskip .3 cm 
-{1\over 2} \, 
    {\spab2.{4}.6 \bigl(\spbb6.{53}.1 + \spbb6.{4(2+3)}.1 \bigr)\over
    \spb1.2\spa3.4\spb5.6\spab1.{(2+3)}.4\spab3.{(1+2)}.4}
- {1\over 2} {\spb1.3^2\spab1.{(2+3)}.6^2 \over
    \spb1.2\spb2.3\spb5.6\spa1.3 t_{123} \spab1.{(2+3)}.4} 
\; + \; \flip1 
 \,. \cr}
\eqn\FLfmpfsc
$$

\section{Results for Primitive Amplitudes: $A_6(1_q, 2,3_\qb, 4)$}
\tagsection\AmplitudesZqgqgSection

In this section we present the independent $A_6(1_q, 2,3_\qb, 4)$ 
primitive amplitudes, corresponding to the helicity configurations
$$
1_q^+,2^+,3_\qb^-,4^+,5_\eb^-,6_e^+; \qquad
1_q^+,2^+,3_\qb^-,4^-,5_\eb^-,6_e^+; \qquad
\eqn\halfwayhelconfig
$$  
note that the configuration $1_q^+,2^-,3_\qb^-,4^+,5_\eb^-,6_e^+$ is
obtained from $1_q^+,2^+,3_\qb^-,4^-,5_\eb^-,6_e^+$ by the operation
`$\flip4$' defined in eq.~(\use\FlipFourSym).  For these amplitudes one
of the external gluons is cyclicly adjacent to the
vector boson, as depicted in fig.~\use\PrimitiveDiagramsFigure{c}.
These primitive amplitudes only contribute to subleading-in-color
terms.

\subsection{The Helicity Configuration $q^+\, g^+ \, \qb^-\, g^+$ }

We now give the primitive amplitudes for $A_6(1_q^+,2^+,3_\qb^-,4^+)$.
These amplitudes are relatively simple because the 
three-mass triangle does not appear.
The tree amplitude is 
$$
A_6^{\rm tree} = 
i\, {\spa1.3\spa3.5^2 \over \spa1.2 \spa2.3\spa3.4\spa4.1\spa5.6} 
\,.
\eqn\treefpfp
$$


The cut-constructible contributions are
$$
\eqalign{
V^\gf & = 
 - {1\over \e^2 } \left( \left({\mu^2 \over -s_{12}} \right)^\e
                        +\left({\mu^2 \over -s_{23}} \right)^\e \right)
 - {2\over \e} \left({\mu^2 \over - s_{56}} \right)^\e - \, 4 
\,,\cr}
\eqn\VLfpfpgf
$$
$$
\hskip -1 cm 
\eqalign{
F^{\gf} & = 
%
{\spa3.5 \bigl(\spa2.3\spa4.5 + \spa3.4\spa5.2 \bigr) 
                     \over\spa1.2\spa3.4\spa5.6\spa2.4^2}\,
 {\ts \Ls_{-1}\Bigl( {-s_{23}\over -t_{234}}, {-s_{34}\over -t_{234}} \Bigr) }
%
%
- {\spa1.3 \spa3.5^2 \over \spa1.2\spa2.3\spa3.4\spa4.1\spa5.6} \, 
 {\ts \Ls_{-1}\Bigl({-s_{12}\over -t_{123}}, 
 {-s_{23} \over -t_{123}} \Bigr)} \cr
%
& \hskip .3 cm  
+ {\spa1.3\spa3.5 \bigl( \spa1.3\spa4.5 + \spa3.4\spa5.1 \bigr) 
     \over \spa1.2\spa2.3\spa3.4\spa4.1^2\spa5.6} \,
    \Ls_{-1}^{{\rm 2m}e} \bigl(t_{123}, t_{234}; s_{23}, s_{56} \bigr)\cr
%
& \hskip .3 cm  
- {\spa1.3\spa3.5^2 \over \spa1.2\spa2.3\spa3.4\spa4.1\spa5.6}\, 
  \Ls_{-1}^{{\rm 2 m}e} \bigl(t_{124}, t_{123}; s_{12}, s_{56} \bigr)
%
- {\spa3.5^2 \over \spa2.3\spa4.1\spa5.6\spa2.4} \,
 {\ts \Ls_{-1}\Bigl({-s_{14} \over - t_{124}},
      {-s_{12} \over -t_{124}} \Bigr)} \cr
%
& \hskip .3 cm  
- 2 \, {\spa3.5\spab5.4.2 \over 
    \spa4.1\spa5.6\spa2.4} \,
         {\Ll_0\Bigl( {-t_{234} \over -s_{23} } \Bigr) \over s_{23}}
- 2\, {\spa3.5 \spab5.2.4 \over \spa1.2\spa5.6\spa2.4} \,
         {\Ll_0\Bigl({-t_{234} \over -s_{34}} \Bigr) \over s_{34}} 
+ 2 \, {\spa1.3\spa5.1\spa5.3 \spab3.{(2+4)}.1 \over
    \spa1.2\spa2.3\spa3.4\spa4.1\spa5.6 } \, 
         {\Ll_0 \Bigl( {-t_{234} \over -s_{56}} \Bigr) \over s_{56}} 
\,. \cr}
\eqn\FLfpfpgf
$$
The terms where a scalar replaces the gluon are 
$$
V^\sc = 
{1\over 2 \eps} \left( {\mu^2 \over -s_{56}}\right)^\e + {1\over 2}
\,,
\eqn\VLfpfpsc
$$
$$
\hskip -1 cm 
\eqalign{
F^\sc & = 
 {\spa3.4\spa5.2^2\spb2.4^2 \over \spa1.2 \spa5.6 \spa2.4} \, 
 {\Ls_{1} \Bigl({-s_{23} \over -t_{234}}, {-s_{34} \over -t_{234}} \Bigr) 
    \over t_{234}^2}
%
- {\spa1.3 \spa3.4 \spa5.1^2 \over \spa1.2 \spa2.3 \spa5.6 \spa4.1^3} \,
   \Ls_{-1}^{{\rm 2m} e} \bigl(t_{123}, t_{234}; s_{23}, s_{56} \bigr) \cr
%
%
& \hskip .3 cm  
- {1\over 2} {\spa1.3 \spa5.3^2 \over
    \spa1.2 \spa2.3 \spa3.4 \spa4.1 \spa5.6}
         {\ts \ln\Bigl({-t_{234} \over -s_{56}} \Bigr) }
- {\spa1.3\spa5.1^2\spb1.2\over \spa1.2 \spa5.6 \spa4.1^2}\, 
         {\Ll_0\Bigl({-t_{123} \over -s_{23}} \Bigr) \over s_{23}}
+ {\spa3.4\spa5.1^2\spb2.4 \over \spa1.2\spa5.6 \spa4.1^2}\, 
         {\Ll_0\Bigl({-t_{234} \over -s_{23}} \Bigr) \over s_{23}}\cr
& \hskip .3 cm  
- {\spa1.3\spa5.1\spa5.4\spab3.{(1+2)}.4 \over
    \spa1.2 \spa2.3 \spa5.6 \spa4.1^2}\, 
         {\Ll_0\Bigl({-t_{123} \over -s_{56}} \Bigr) \over s_{56}}
+ {\spa1.3\spb6.4\spa5.3 \over
    \spa1.2 \spa2.3 \spa4.1}\, 
         {\Ll_0\Bigl({-t_{123}\over -s_{56}} \Bigr) \over s_{56}}
+ {\spa1.3\spa5.1^2\spab3.{(2+4)}.1 \over 
    \spa1.2 \spa2.3 \spa5.6 \spa4.1^2}\, 
        { \Ll_0\Bigl({-t_{234} \over -s_{56}} \Bigr) \over s_{56}}\cr
& \hskip .3 cm  
- {\spa1.3^2\spa5.3\spb6.1 \over
     \spa1.2\spa2.3\spa3.4\spa4.1}\, 
         {\Ll_0\Bigl({-t_{234} \over -s_{56}} \Bigr) \over s_{56}}
- {1\over 2}\, {\spa2.3\spa5.4^2\spb4.2^2 \over \spa1.4\spa5.6\spa2.4}\,
        {\Ll_1 \Bigl( {-s_{23} \over -t_{234}} \Bigr) \over t_{234}^2}
- {1\over 2}\, {\spa5.2^2\spb2.4^2\spa4.3 \over\spa1.2\spa5.6\spa2.4}\,
        {\Ll_1 \Bigl({-s_{34} \over -t_{234}} \Bigr) \over t_{234}^2} \cr
& \hskip .3 cm  
- {\spa1.3\spa3.4\spa5.6\spb6.4^2 \over \spa1.2\spa2.3\spa4.1}\,
         {\Ll_1\Bigl( {-t_{123} \over -s_{56}} \Bigr) \over s_{56}^2}
- {1\over 2}\, {\spa1.3^3\spa5.6\spb6.1^2 \over 
        \spa1.2\spa2.3\spa3.4\spa4.1} \,
         {\Ll_1 \Bigl( {-t_{234} \over -s_{56}} \Bigr) \over s_{56}^2} \cr
%
%
& \hskip .3 cm  
+ {1\over 2}  
   {\spb2.4\spab1.{(2+4)}.6\spab3.{(2+4)}.6 \over
     \spa1.2 \spa4.1 \spb5.6 t_{124} t_{234}}
+ {1\over 2}  
   {\spa2.4 \spb2.4^2 \spa5.1 \spa5.3 \over
    \spa1.2\spa4.1 \spa5.6 t_{124} t_{234}}         
+ {1\over 2}  
   {\spa1.3\spa3.5^2 \over
    \spa1.2\spa2.3\spa3.4\spa4.1\spa5.6} 
\,. \cr }
\eqn\FLfpfpsc
$$

\subsection{The Helicity Configuration $q^+\,  g^+\, \qb^-\, g^-$ }

We now present the primitive amplitude $A_6(1_q^+,2^+,3_\qb^-,4^-)$.
The tree amplitude is 
$$
A_6^{\rm tree} =
i \biggl[
- {\spab3.{(1+2)}.6\spab5.{(3+4)}.1 \over
   \spa1.2\spa2.3\spb3.4\spb4.1 s_{56}}
+ {\spb1.2\spa5.3\spab4.{(1+2)}.6 \over
   \spa1.2\spb4.1 t_{124} s_{56}}
+ {\spa3.4\spb6.1\spab5.{(3+4)}.2 \over
   \spa2.3\spb3.4 t_{234} s_{56}} \biggr]
\,.
\eqn\treefpfm
$$
An alternate form is 
$$
\eqalign{
A_6^{\rm tree} & =
i \biggl[
- {\spb1.2\spb1.3\spa3.5\spab5.{(3+4)}.2 \over
   \spb3.4\spb4.1 t_{124} t_{234} \spa5.6}
- {\spa1.3\spa3.4\spb1.6\spab4.{(1+2)}.6 \over
    \spa1.2\spa2.3 t_{124} t_{234}\spb5.6} \cr
& \hskip 1.3 cm 
- {\spab4.{(1+2)}.6\spab5.{(3+4)}.2 \over
    \spa1.2\spb3.4 t_{124} t_{234}}
- {\spb1.2\spa3.4\spb1.6\spa3.5 \over
    \spa2.3\spb4.1 t_{124} t_{234}} \biggr] 
\,.
\cr}
\eqn\TreefpfmAlt
$$
The first form has manifest $t$-pole behavior while the 
second form has manifest behavior for $k_5 \parallel k_6$.

The cut-constructible contributions are 
$$
\eqalign{
V^\gf & = 
 - {1\over \eps^2} \, \left( \left({\mu^2 \over -s_{12}}\right)^\eps 
                         + \left({\mu^2 \over -s_{23}} \right)^\eps \right)
  - {2\over \eps}\, \left({\mu^2 \over -s_{56}} \right)^\eps - 4 
\,,
\cr}
\eqn\VLfpfmgf
$$
$$
\hskip -1.3 cm 
\eqalign{
F^\gf & = 
%
\frac{\spa1.3{{\spab3.{(1+2)}.6}^2}}
{\spa1.2\spa2.3\spb5.6\spab1.{(2+3)}.4\spab3.{(1+2)}.4}
 \,\Ls_{-1}^{2{\rm m}h}\bigl( {s_{34}},{t_{123}};{s_{12}},{s_{56}}\bigr)
\nl 
%
%
+\biggl( \frac{{{\spab5.{(3+4)}.2}^2}}{\spb3.4\spa5.6{t_{234}}\spab1.{(2+3)}.4}
 - \frac{{{\spa3.4}^2}{{\spb1.6}^2}}
       {\spa2.3\spb5.6{t_{234}}\spab2.{(3+4)}.1} \biggr) 
      \,\Ls_{-1}^{2{\rm m}h}\bigl( {s_{12}},{t_{234}};{s_{34}},{s_{56}}\bigr)
\nl 
%
+\biggl( \frac{{{\spa1.3}^3}{{\spb4.6}^2}t_{123}^{2}}
{\spa1.2\spa2.3\spb5.6{{\spab1.{(2+3)}.4}^3}\spab3.{(1+2)}.4}
 - \frac{\spa1.3{{\spab1.{(2+3)}.6}^2}\spab3.{(1+2)}.4}
{\spa1.2\spa2.3\spb5.6{{\spab1.{(2+3)}.4}^3}}  \biggr) 
   \,\Ls_{-1}^{2{\rm m}h}\bigl( {s_{14}},{t_{123}};{s_{23}},{s_{56}}\bigr)
\nl
- \biggl( \frac{{{\spab2.{(1+4)}.6}^2}{{\spab4.{(1+2)}.3}^2}}
{\spa1.2\spb5.6{t_{124}}{{\spab2.{(1+4)}.3}^3}}
+\frac{{{\spb1.2}^2}{{\spa3.5}^2}}{\spb1.4\spa5.6{t_{124}}\spab3.{(1+2)}.4}
-\frac{{{\spa2.4}^2}{{\spb3.6}^2}{t_{124}}}
{\spa1.2\spb5.6{{\spab2.{(1+4)}.3}^3}}\biggr) 
\nl \hsa \times
 \,\Ls_{-1}^{2{\rm m}h}\bigl( {s_{23}},{t_{124}};{s_{14}},{s_{56}}\bigr)
%
%
+\frac{\spa1.3{{\spab3.{(1+2)}.6}^2}}
 {\spa1.2\spa2.3\spb5.6\spab1.{(2+3)}.4\spab3.{(1+2)}.4}\, {\ts {{\Ls}_{-1}}
  \left(\frac{-{s_{12}}}{-{t_{123}}},
     \frac{-{s_{23}}}{-{t_{123}}}\right)} 
\nl
%
+ \frac{{{\spab5.{(3+4)}.2}^2}}
 {\spb3.4\spa5.6\spab1.{(2+3)}.4}\,
 \frac{ {\ts {{\Ls}_{-1}}
  \left(\frac{-{s_{23}}}{-{t_{234}}} 
 ,\frac{-{s_{34}}}{-{t_{234}}} \right)} } {t_{234}}
%
%
-\frac{{{\spb1.2}^2}{{\spa3.5}^2}}
{\spa5.6\spb1.4\spab3.{(1+2)}.4}\,
\frac{ \Ls_{-1}
\left(\frac{-{s_{14}}}{-{t_{124}}},\frac{-{s_{12}}}{-{t_{124}}} \right)}
{t_{124}}
\nl
+ \Tgf \mathtag{[FPFM]} \, I_3^{3{\rm m}}(s_{12}, s_{34}, s_{56})
+ \tTgf\mathtag{[FPFM]} \, I_3^{3{\rm m}}(s_{14}, s_{23}, s_{56})
%
- M_1(1,4,2,3) - \Bigl[ M_1(2,3,1,4) \bigr|_{\flip1} \Bigr] 
\nl
- 2\, \frac{\spb1.2\spa1.4\spab2.{(1+4)}.6\spab4.{(1+2)}.6}
 {\spa1.2\spb5.6{t_{124}}\spab2.{(1+4)}.3}
 \,\frac{{{\Ll}_0}\left( \frac{-{t_{124}}}{-{s_{14}}}\right)}{{s_{14}}}
+ 2\, \frac{\spb1.6\spab1.{(3+4)}.2\spab5.{(3+4)}.2}
{\spb3.4{t_{234}}\spab1.{(2+3)}.4}
\,\frac{{{\Ll}_0}\left( \frac{-{t_{234}}}{-{s_{56}}}\right)}
{{s_{56}}}
\nl 
+ 2\,\frac{\spab5.{4}.2 \spab5.{(3+4)}.2}
{\spb3.4\spa5.6\spab1.{(2+3)}.4}\,\frac{{{\Ll}_0}\left( 
 \frac{-{t_{234}}}{-{s_{23}}}\right)}{{s_{23}}}
+ 2\, \frac{\spab4.{3}.6 \spab4.{(1+2)}.6}
{\spa1.2\spb5.6\spab2.{(1+4)}.3}
\,\frac{{{\Ll}_0}\left( \frac{-{t_{124}}}{-{s_{56}}}\right)}{{s_{56}}}
\nl 
%
- 2\,\frac{{{\spab5.{(3+4)}.2}^2}}
     {\spb3.4\spa5.6{t_{234}}\spab1.{(2+3)}.4} 
   \, {\ts \ln\left( \frac{-{s_{23}}}{-{s_{56}}}\right) }
\,, \cr}
\eqn\FLfpfmgf
$$
where the three-mass-triangle coefficients are
$$
\hskip -1cm
\eqalign{
\Tgf & = 
2\, \frac{{{\spa3.4}^2}\spab5.{(3+4)}.2\spb1.6}{\spa2.3 \, t_{234}^{2}}
 +\frac{\spa1.3}{\spa2.3 \, t_{123}\, t_{234}}
\left( \frac{\spab4.{(2+3)}.1\spab5.{(3+4)}.2\spab3.{(1+2)}.6}
{\spab1.{(2+3)}.4} - \spb1.2\spa3.4\spa4.5\spb1.6 \right) 
\,, \cr}
\eqn\TLfpfmgf
$$
$$
\hskip -1cm
\eqalign{
\tTgf & = 
 \frac{1}{2}\, { 
  { t_{124} \, \dt23 + 2\,{s_{14}}{s_{56}} \over t_{124}^{2} }
 \biggl(\frac{{{\spab4.{(1+2)}.6}^2}}
           {\spa1.2\spb5.6\spab2.{(1+4)}.3} 
+ \frac{{{\spb1.2}^2}{{\spa3.5}^2}}
   {\spb1.4\spa5.6\spab3.{(1+2)}.4} \biggr)    }
\nl\hsb 
- 2\,\frac{\spaa4.{12}.4 \spab5.3.6}
{\spa1.2\, t_{124} \, \spab2.{(1+4)}.3} 
- \frac{\spab3.1.2 \spa4.5\spab3.{(1+2)}.6}
{\spa1.2\, t_{123} \, \spab3.{(1+2)}.4}
\nl\hsb
+ \frac{1}{\spab2.{(1+4)}.3\spab3.{(1+2)}.4}
\biggl[
\frac{\spaa4.{12}.3 \spab5.3.6} {\spa1.2} 
+ \frac{\spab3.1.2 \spab5.4.6}
 {\spa1.2\spab1.{(2+3)}.4}
 \bigl( \spa1.2 \left( s_{13}-s_{24} \right) + \spaa1.{3(1+4)}.2 \bigr) 
\nl\hsa
- \frac{1}{2} \,\frac{\spab4.{(1+2)}.6}{\spa1.2\spb5.6}
\bigl( \spab3.{(1+2)}.6 \, \dt23 + 2\,\spab3.1.6 {s_{56}} \bigr)
- \frac{1}{2} \, \frac{\spb1.2\spa3.5\spab5.{(3+4)}.1\, \dt23 }
       {\spb1.4\spa5.6}
\nl\hsa
+ \spb2.6 \bigl( \spab4.2.1 \spa3.5 + \spab3.4.1\spa4.5 \bigr)  
- \spa3.4\spb1.6 \spab5.{(1+3)}.2 \biggr]
 \,. \cr}
\eqn\tTLfpfmgf
$$

The contributions where a scalar replaces the gluon in the loop
are
$$
V^\sc = 
{1\over 2 \eps} \, \left( {\mu^2 \over -s_{56}} \right)^\eps
   + {1\over 2} 
\,,
\eqn\Vsfpfm
$$
$$
\hskip -1. cm 
\eqalign{
F^\sc & = 
  {\spa1.3\over \spa2.3 }\,  \Bigl[ M_2(1,2,3,4) \bigr|_{\flip4} \Bigr] 
%
+ {\spa1.4\over \spa2.4} \, \Bigl[ M_3(1,2,3,4) \bigr|_{\flip4} \Bigr] 
\nl
+ {1\over 2} \, {\spab5.1.2 \over \spb3.4 \spab1.{(2+3)}.4}
  \biggl( 
     {\spab5.{(3+4)}.2 \over \spa5.6}
        \, {\Ll_0 \Bigl( {-t_{234} \over -s_{56} }\Bigr) \over s_{56}}
    + \spbb2.{(3+4)1}.6 \,  
        {\Ll_1 \Bigl( {-s_{56} \over -t_{234}} \Bigr) \over t_{234}^2 } \biggr)
\nl
+ {1\over 2} \, {\spb2.3 \spa3.4 \spab5.{(2+3)}.4 \over 
                 \spb3.4\spa5.6\spab1.{(2+3)}.4} \biggl( 
      {\spab5.{(3+4)}.2 \over t_{234}}
        \, {\Ll_0\Bigl({-t_{234} \over -s_{23}} \Bigr) \over  s_{23}}
    + \spab5.4.2 \, 
        {\Ll_1 \Bigl( {-t_{234} \over -s_{23}}\Bigr) \over s_{23}^2} \biggr)
\nl
- {1\over 2} \,  {\spb1.2 \over \spb1.4\spab1.{(2+3)}.4 \spab2.{(1+4)}.3}
       \biggl( {\spaa5.{(3+4)1}.5 \over \spa5.6}
             - {\spbb6.{(1+2)4}.6 \over \spb5.6} \biggr) \, 
      {\ts \ln \left({-s_{23} \over -s_{56} }\right)}
\nl
+ {1\over 2}\biggl[ 
-  {\spb1.2\spa1.5\spa3.5 \over
                \spa2.3\spb3.4\spa5.6\spab1.{(2+3)}.4}
-  {\spb1.3^2\spa3.5^2 \over
                \spb1.4\spa2.3\spb3.4\spa5.6\spab2.{(1+4)}.3}
+  {\spab4.3.1\spa1.5\spa3.5 \over
                \spa2.3\spa5.6\spab1.{(2+3)}.4\spab2.{(1+4)}.3}
\nl
-  {\spa1.3\spa3.5\spa4.5 \over
                \spa1.2\spa2.3\spa5.6\spab1.{(2+3)}.4}
-  {s_{34}\spa1.5\spa4.5 \over
                \spa1.2\spa5.6\spab1.{(2+3)}.4\spab2.{(1+4)}.3}
-  {\spb1.2\spa1.5\spab5.{(3+4)}.2 \over
                \spb3.4\spa5.6 \, t_{234} \, \spab1.{(2+3)}.4 }
\nl
-   {\spa1.3\spb1.6^2 \over
                \spb1.4\spa1.2\spa2.3\spb3.4\spb5.6}
+  {\spab3.4.6\spab1.{(2+3)}.6 \over
                \spa1.2\spa2.3\spb3.4\spb5.6\spab1.{(2+3)}.4}
-  {s_{13}\spab3.4.6 \spb3.6 \over
                \spa2.3\spb3.4\spb5.6\spab1.{(2+3)}.4\spab2.{(1+4)}.3}
\nl
+ {\spa1.4\spab3.4.6\spb3.6 \over
                \spa1.2\spb5.6\spab1.{(2+3)}.4\spab2.{(1+4)}.3}
+ {\spab3.1.6\spb1.2\spb4.6 \over
                \spb1.4\spa2.3\spb3.4\spb5.6\spab1.{(2+3)}.4}
-  {\spab4.{(1+2)}.6\spb2.6\spa2.4 \over 
                \spa1.2\spb5.6 \, t_{124} \, \spab2.{(1+4)}.3  } \biggr]
\,.
\cr}
\eqn\Fsfpfm
$$


\section{Results for Primitive Amplitudes: $A_6(1_q, 2_\qb,3, 4)$}
\tagsection\AmplitudesZqqggSection

In this section we present the $A_6(1_q, 2_\qb,3, 4)$ primitive
amplitudes, corresponding to the helicity configurations
$$
1_\q^+,2_\qb^-,3^+,4^+,5_\eb^-,6_e^+; \qquad
1_\q^+,2_\qb^-,3^+,4^-,5_\eb^-,6_e^+; \qquad
1_\q^+,2_\qb^-,3^-,4^+,5_\eb^-,6_e^+; 
$$
again we shall suppress the lepton labels below.  
In the parent diagrams for these amplitudes, both external gluons are
attached to the same fermion line as the vector boson, as depicted in
fig.~\use\PrimitiveDiagramsFigure{d}.  These primitive amplitudes only
contribute to subleading-in-color terms.

\subsection{The Helicity Configuration $q^+ \,\qb^-\, g^+ \, g^+$ }

Here we present the primitive amplitude
$A_6(1_q^+,2_\qb^-,3^+,4^+)$.
The tree amplitude is 
$$
A_6^{\rm tree} =
 i\,  {\spa2.5^2 \over \spa2.3 \spa3.4 \spa4.1 \spa5.6} 
\,.
\anoneqn
$$


The cut-constructible terms are,
$$
\eqalign{
V^\gf & = 
 - {1\over \e^2} \Bigl({\mu^2 \over -s_{12}} \Bigr)^\e
 - {2\over \e} \Bigl({\mu^2 \over -s_{56}} \Bigr)^\e - 4 
\,, \cr}
\anoneqn
$$
$$
\hskip -1 cm 
\eqalign{
F^\gf & = 
 {\spa2.5 (\spa1.2\spa4.5 - \spa2.4\spa1.5)
         \over  \spa2.3 \spa3.4 \spa4.1^2 \spa5.6} \, 
   \Ls_{-1}^{{\rm 2m}e} \bigl(t_{123}, t_{234}; s_{23}, s_{56} \bigr) \cr
%
%
& \hskip .3 cm 
+ { \spa2.5 (\spa2.3\spa1.5 - \spa1.2\spa3.5)
          \over \spa2.3\spa3.4\spa4.1\spa1.3\spa5.6} \,
   {\ts \Ls_{-1}\Bigl({-s_{12} \over -t_{123}},
          {-s_{23}\over -t_{123}} \Bigr) }\cr
%
%
& \hskip .3 cm 
- {\spa2.5 (\spa2.3\spa4.5 + \spa2.4\spa3.5)
           \over\spa2.3 \spa3.4^2\spa4.1\spa5.6}\, 
     \Ls_{-1}^{{\rm 2m}e} \bigl(t_{124}, t_{123}; s_{12}, s_{56} \bigr) \cr
%
%
& \hskip .3 cm 
- {\spa5.2^2 \over \spa2.3 \spa3.4 \spa4.1 \spa5.6}\, 
 \Bigl(
  {\ts \Ls_{-1}\Bigl( {-s_{14} \over -t_{124}},
               {-s_{12}\over -t_{124}} \Bigr) }
%
%
 +  \Ls_{-1}^{{\rm 2m}e} \bigl(t_{134}, t_{124}; s_{14}, s_{56} \bigr) 
    \Bigr)\cr
%
%
& \hskip .3 cm 
- 2 \, {\spa5.2\spab5.4.3 \over 
    \spa3.4\spa4.1\spa5.6} \,
         {\Ll_0\Bigl({ -t_{234} \over -s_{23} } \Bigr) \over s_{23}}
+ 2\, {\spa2.5\spa1.5 \spab2.{(3+4)}.1 \over
    \spa2.3\spa3.4\spa4.1\spa5.6} \, 
         {\Ll_0\Bigl({-t_{234}\over -s_{56} } \Bigr) \over s_{56}} 
\,. \cr}
\anoneqn
$$

The contributions where a scalar replaces the gluon are
$$
V^\sc = 
{1\over 2 \e} \Bigl({\mu^2 \over -s_{56}} \Bigr)^\e 
       + {1\over 2} 
\,,
\anoneqn
$$
$$
\hskip -1 cm
\eqalign{
F^\sc & = 
%
- {\spa2.4^2\spa5.1^2 \over \spa2.3\spa3.4\spa4.1^3\spa5.6} \, 
 \Ls_{-1}^{{\rm 2m}e} \bigl(t_{123}, t_{234}; s_{23}, s_{56} \bigr)  \cr
%
%
& \hskip .3 cm 
- {\spa2.3\spa5.1^2 \over \spa3.4\spa4.1\spa1.3^2 \spa5.6}
 {\ts \Ls_{-1}\Bigl({-s_{12} \over -t_{123}}, {-s_{23} \over -t_{123}} \Bigr)}
%
- {\spa2.3\spa5.4^2 \over \spa3.4^3 \spa4.1 \spa5.6} \, 
   \Ls_{-1}^{{\rm 2m} e} \bigl( t_{124}, t_{123}; s_{12}, s_{56} \bigr)\cr
%
%
& \hskip .3 cm 
+ { \bigl( \spa4.1\spa5.2 - \spa2.4\spa5.1 \bigr) \spab5.4.3 \over 
     \spa3.4\spa4.1^2\spa5.6} \, 
         {\Ll_0\Bigl({-t_{234} \over -s_{23}} \Bigr) \over s_{23}}
- {1\over 2} \, {\spab5.4.3^2\spa3.2 \over
        \spa3.4\spa4.1\spa5.6 } \, 
         {\Ll_1 \Bigl({-s_{23} \over -t_{234}} \Bigr) \over  
                   t_{234}^2} \cr
& \hskip .3 cm 
- {1\over 2} \, {\spa5.2^2 \over
    \spa2.3\spa3.4\spa4.1\spa5.6} \, 
        {\ts \ln\Bigl({-t_{234} \over -s_{56}} \Bigr) } 
+ \biggl( {\spa2.4\spa5.1^2 \spab2.{(3+4)}.1 \over
        \spa2.3\spa3.4\spa4.1^2\spa5.6}\, 
- {\spa5.2 \spab2.1.6  \over 
        \spa2.3 \spa3.4 \spa4.1} \biggr) \, 
        {\Ll_0\Bigl( {-t_{234} \over -s_{56}} \Bigr) \over s_{56}} \cr
& \hskip .3 cm - {1\over 2}\, {\spa5.6 \spab2.1.6^2  \over
    \spa2.3\spa3.4\spa4.1} \, 
         {\Ll_1 \Bigl({-t_{234} \over -s_{56}} \Bigr) \over s_{56}^2} 
+  {\spa2.3 \spaa5.{(1+2)3}.5 \over 
      \spa3.4^2\spa1.3\spa5.6} \, 
       {\Ll_0\Bigl({-t_{123} \over -s_{12}} \Bigr)\over s_{12}} 
-  {\spa1.2^2 \spaa5.{(2+3)1}.5 \over
    \spa2.3\spa4.1^2\spa1.3\spa5.6} \, 
       {\Ll_0\Bigl( {-t_{123} \over -s_{23}} \Bigr) \over s_{23}} \cr
& \hskip .3 cm 
+  \biggl( {\spa2.4 \spab5.4.6 \over
               \spa3.4^2\spa4.1} 
          - { \bigl( 2\, \spa4.1\spa5.2 + \spa5.1\spa2.4 \bigr) 
                    \spab2.4.6 \over
               \spa2.3 \spa3.4 \spa4.1^2}
    + {\spa5.2\spa5.4 \spab2.{(1+3)}.4 \over
              \spa2.3\spa3.4\spa4.1\spa5.6} \biggr) \,
         {\Ll_0\Bigl({-t_{123} \over -s_{56}} \Bigr) \over s_{56}} \cr
& \hskip .3 cm 
- {\spa5.2^2 \over
    \spa2.3\spa3.4\spa4.1\spa5.6} \, 
        {\ts \ln\Bigl({-t_{123} \over -s_{56}} \Bigr) }
- {\spa5.6 \spab2.4.6^2 \over
         \spa2.3\spa3.4\spa4.1 } \,
         {\Ll_1\Bigl({-t_{123} \over -s_{56}} \Bigr)\over s_{56}^2}
- {\spa2.4\spa5.4 \spab5.{(1+2)}.4 \over 
        \spa3.4^2 \spa4.1\spa5.6 } \, 
         {\Ll_0\Bigl({-t_{124} \over -s_{12}} \Bigr) \over s_{12}} \cr
& \hskip .3 cm 
- {\spa2.4\spab5.3.6 \over 
        \spa3.4^2\spa4.1} \, 
         {\Ll_0\Bigl( {-t_{124} \over -s_{56}} \Bigr) \over s_{56}}
- {\spa2.3 \spb3.6^2 \spa5.6 \over
        \spa3.4\spa4.1} \, 
         {\Ll_1 \Bigl({-t_{124} \over -s_{56}} \Bigr) \over s_{56}^2} \cr
& \hskip .3 cm 
+ {\spa5.1\spa5.2 \over  \spa4.1 \spa1.3\spa5.6} 
   \biggl( { 1 \over \spa3.4  }
        {\ts  \ln\Bigl({-s_{12} \over -s_{56}} \Bigr) }
    - { \spa1.2 \over \spa2.3\spa4.1 }
        {\ts \ln\Bigl( {-s_{23} \over -s_{56}} \Bigr) } \biggr) \cr
%
& \hskip .3 cm 
+ {1\over 2}  \biggl[
 {\spb3.4 \spab4.{(1+3)}.6 \spab2.{(3+4)}.6 \over
   \spa3.4\spa4.1 t_{234} t_{134} \spb5.6}
+ {\spb3.4\spa5.2 \spab5.1.3 \over 
    \spa4.1 t_{234} t_{134} \spa5.6} 
+ {\spa5.2 \spab5.4.3 \over 
    \spa3.4\spa4.1 t_{234} \spa5.6}
+ {\spa2.5^2 \over \spa2.3\spa3.4\spa4.1\spa5.6} \biggr]
\,. \cr }
\anoneqn
$$

\subsection{The Helicity Configuration $q^+\, \qb^-\, g^-\, g^+$ }


Here we present the primitive amplitude $A_6(1_q^+,2_\qb^-,3^-,4^+)$.
This amplitude satisfies the symmetry `$\flip2$' defined in
eq.~(\use\FlipTwoSym). The tree amplitude is
$$
A_6^{\rm tree} =
i\, \biggl[ {\spb4.1\spa1.3\spa5.2  \spab3.{(1+4)}.6 \over
   \spa4.1 s_{34} t_{134} s_{56}}
+ {\spa2.3 \spb2.4\spb6.1 \spab5.{(2+3)}.4 \over
   \spb2.3 s_{34} t_{234} s_{56}}
- {\spab5.{(2+3)}.4 \spab3.{(1+4)}.6 \over
    \spb2.3 \spa4.1  s_{34} s_{56}} \biggr] 
 \,.
\anoneqn
$$
An alternate form with manifest $k_5 \parallel k_6$ behavior is
$$
A_6^{\rm tree} =
-i\, \biggl[
 {\spb4.1\spb2.4\spa5.2 \spab5.{(2+3)}.4 \over
   \spb2.3\spb3.4 t_{134}t_{234}\spa5.6}
+ {\spa2.3\spa1.3\spb6.1 \spab3.{(1+4)}.6 \over
   \spa3.4\spa4.1 t_{134} t_{234} \spb5.6}
+ {\spab3.{(1+4)}.6 \spab5.{(2+3)}.4 \over
   \spb2.3\spa4.1 t_{134} t_{234}} \biggr]
 \,.
\anoneqn
$$


The cut-constructible pieces are, 
$$
\hskip -1.1 cm 
V^\gf  = 
 - {1\over \eps^2} \, \left({\mu^2 \over -s_{12}}\right)^\eps
 - {2\over \eps}  \,  \left({\mu^2 \over -s_{56}} \right)^\eps - 4
\,,
\eqn\VLffmpgf
$$
$$
\hskip -1.1 cm 
\eqalign{
F^\gf & = 
\frac{{{\spab5.{(2+3)}.1}^2}}
{\spb2.3\spa5.6\spab4.{(1+2)}.3\spab4.{(2+3)}.1}\,
 \Bigl( 
\Lsnew^{2{\rm m}h}_{-1}\bigl({s_{14}},{t_{123}};{s_{23}},{s_{56}}\bigr)
 + {\ts {{\Ls}_{-1}}\left(\frac{-{s_{12}}}{-{t_{123}}} ,
       \frac{-{s_{23}}}{-{t_{123}}} \right)} \Bigr) 
\nl
+ \biggl( \frac{{{\spb2.4}^3}{{\spa1.5}^2}\,{t_{234}}}
{\spb2.3\spb3.4\spa5.6{{\spab1.{(3+4)}.2}^3}} 
-\frac{\spb2.4{{\spab1.{(2+3)}.4}^2}{{\spab5.{(3+4)}.2}^2}}
{\spb2.3\spb3.4\spa5.6\,{t_{234}}\,{{\spab1.{(3+4)}.2}^3}}
\nl\hsa
- \frac{{{\spa2.3}^2}{{\spb1.6}^2}\spab3.{(2+4)}.1}
{\spa3.4\spb5.6\,{t_{234}}\,\spab2.{(3+4)}.1\spab4.{(2+3)}.1}
\biggr) \,
\Ls_{-1}^{2{\rm m}h}\bigl({s_{12}},{t_{234}};{s_{34}},{s_{56}}\bigr)
\nl
+\biggl(
\frac{{{\spb1.3}^2}{{\spa4.5}^2}t_{123}^{2}}
{\spb2.3\spa5.6{{\spab4.{(1+2)}.3}^3}\spab4.{(2+3)}.1} 
 - \frac{\spab4.{(2+3)}.1{{\spab5.{(1+2)}.3}^2}}
 {\spb2.3\spa5.6{{\spab4.{(1+2)}.3}^3}}\biggr) 
\,\Ls_{-1}^{2{\rm m}h}\bigl({s_{34}},{t_{123}};{s_{12}},{s_{56}}\bigr)
\nl 
 -\frac{{{\spa2.3}^2}{{\spb1.6}^2}}
{\spb5.6\,{t_{234}}\,\spa2.4\spab4.{(2+3)}.1} \,
{\ts {{\Ls}_{-1}}\left(\frac{-{s_{23}}}{-{t_{234}}} 
      , \frac{-{s_{34}}}{-{t_{234}}} \right)}
%
\nl 
- {\frac{1}{2}} \, 
 { \spb1.4 
  \bigl( \spaa2.{(1+4)(2+3)}.5^2 - \spa2.5^2 s_{14}\, s_{23} \bigr)
 \over \spa1.4\spb2.3\spa5.6\spab2.{(1+4)}.3\spab2.{(3+4)}.1}
 \, I_3^{3{\rm m}}(s_{14}, s_{23}, s_{56})
\nl 
+ \biggl\{  
{\frac{1}{2}} \, { \spb2.4 {\spab5.{(2+3)}.4}^2
\bigl( 2\,{s_{34}}\, {s_{56}} + {{\delta}_{12}}\, {t_{234}} \bigr) 
  \over \spb2.3\spb3.4\spa5.6\,t_{234}^{2}\,\spab1.{(3+4)}.2 }
+ \frac{\spa2.3\spb1.6\spab3.{(2+4)}.1\spab5.{(2+3)}.4}
  {t_{234}^{2} \, \spab4.{(2+3)}.1 }
\nl\hsb
+ {\frac{1}{2}}\, \frac{\spa2.3\spb1.6\spab3.{(1+4)}.6\spab3.{(2+4)}.1}
   {\spa3.4\spb5.6\,{t_{234}}\,\spab4.{(2+3)}.1}
+ {\frac{1}{2}} \, \frac{\spb1.4\spa2.5\spb1.6\spab3.2.4}
    {\spb3.4\,{t_{234}}\,\spab4.{(2+3)}.1}
+ 2\,\frac{\spa2.3{{\spb2.4}^2}\spab5.1.6}
    {\spb2.3\,{t_{234}}\,\spab1.{(3+4)}.2}
\nl\hsb 
-\frac{\spab3.2.1 \spab5.{(2+3)}.1 \spb4.6}
     {\spb2.3\,{t_{123}}\,\spab4.{(2+3)}.1}
\nl\hsb 
+{\frac{1}{2}}\, \frac{1}{\spab1.{(3+4)}.2\spab4.{(2+3)}.1}
\biggl[ \frac{\spb1.4 \spab3.1.6 \spab5.2.4 }{\spb3.4} 
+ \frac{\spb1.4 \spab5.4.2 \spab5.{(2+3)}.4 \, \delta_{12}}
     {\spb2.3\spb3.4\spa5.6} 
\nl\hsa
- \frac{\spa2.3\spb1.6}{\spa3.4\spb5.6}
\bigl(\spab3.2.6 \spab3.{(1+4)}.2 + \spab3.1.2 \spab3.5.6 \bigr)
+ 2\,\frac{\spb1.2\spab3.5.6\spab5.{(2+3)}.4}{\spb2.3} 
\nl\hsa
  + \frac{\spb2.4\spb1.6}{\spb2.3\spb3.4}
\bigl(\spab5.2.4 (s_{56}-t_{123}) + \spab5.3.4 (s_{56}+t_{234})\bigr)
+ \spb1.4\spa3.5 \bigl( 2\,\spab3.4.6 + \spab3.2.6 \bigr) \biggr]
\nl\hsb 
+ {\frac{1}{2}}\, \frac{1}{\spab1.{(3+4)}.2
\spab4.{(1+2)}.3\spab4.{(2+3)}.1}
\biggl[  \frac{\spab4.{123}.4}{\spb2.3}
\left(\spab5.{(1-4)}.2\spb1.6 - 2\,\spb1.2 \spab5.4.6 \right)
\nl\hsa
+ \spbb1.{(3+4)2}.4 
    \bigl( \spa3.5 \spab4.1.6 + \spa4.5 \spab3.2.6\bigr) 
+ \spab5.3.6 \left(\spab3.2.1\, s_{13} + \spab3.4.1 \, s_{12} \right) 
\nl\hsa
+ \spbb1.{3(1-2)}.6 \spaa3.{(1+2)4}.5  
+ \spab5.{(3+4)}.1 \spab3.4.6 {s_{23}} 
+ \spab5.4.1 \spab3.{(1+2)}.6 \, s_{13} 
\nl\hsa
+ \bigl( 3\,\spab3.2.1 + \spab3.4.1 \bigr) \spab5.2.6\, s_{14} 
- \spab5.3.1 \spab1.{(3+4)}.6 \spab3.{(2-4)}.1
\nl\hsa
+ 2\,\spab3.4.1 \spb1.6 
   \bigl( \spa5.1 (s_{12} + t_{123}) - \spaa5.{23}.1 \bigr)
- \spab3.2.1 \spab5.4.6 \, s_{34} 
+ 4\,\spab5.{123}.1 \spab3.4.6 
\nl\hsa
 - \spa2.3 \spab4.{(2+3)}.1
 \bigl( 2\,\spab5.{(2-3)}.4\spb2.6 - \spab5.1.6\spb2.4 \bigr) \biggr]
\biggr\} \, I_3^{3{\rm m}}(s_{12}, s_{34}, s_{56})
\nl
+2\,\frac{\spa2.3\spb2.4\spab5.{(2+3)}.4\spab5.{(3+4)}.2}
{\spb2.3\spa5.6\,{t_{234}}\,\spab1.{(3+4)}.2}\,
\frac{{{\Ll}_0}\left(\frac{-{t_{234}}}{-{s_{34}}}\right)}{{s_{34}}}
-2\,\frac{\spb1.4\spb2.4\spa1.5\spab5.{(2+3)}.4}
{\spb2.3\spb3.4\spa5.6\spab1.{(3+4)}.2}\,
\frac{{{\Ll}_0}\left(\frac{-{t_{234}}}{-{s_{56}}}\right)}{{s_{56}}} 
\nl 
\hskip 3 cm + \; \flip2 
\,. \cr}
\eqn\FLffmpgf
$$
The contributions where a scalar replaces the gluon in the loop are 
$$
V^\sc = 
 {1\over 2 \eps} \left({\mu^2 \over -s_{56}} \right)^\eps
     + {1\over 2} 
\,,
\eqn\VLffmpsc
$$
$$
\hskip -1. cm 
\eqalign{
F^\sc & = 
M_2(1,2,3,4) + M_3(1,2,3,4) 
+ {1\over 2}\,  {\spa4.5\spab3.{(1+2)}.4\spab5.{(2+3)}.4 \over 
     s_{34}\spa5.6\spab1.{(3+4)}.2 \spab4.{(1+2)}.3 }
         {\ts \ln \Bigl({-s_{12} \over -s_{56}} \Bigr) }
\nl
- {1\over 2}\, { \spa3.5 \bigl(\spa2.3\spa4.5-\spa2.5\spa3.4 \bigr) \over
                  \spa1.4\spa3.4\spa5.6\spab4.{(1+2)}.3 }
+ {1\over 2} \, {\spb2.4\spa3.5 \over \spb2.3\spa5.6 \spab1.{(3+4)}.2 }
  \biggl( {\spa3.5 \over \spa3.4} + {\spab5.{(2+3)}.4 \over t_{234} } \biggr)
\; + \; \flip2 
\,,
\cr}
\eqn\FLffmpsc
$$
where $M_2$ and $M_3$ are given in eqs.~(\use\MasterTwo) 
and (\use\MasterThree).

\subsection{The Helicity Configuration $q^+\, \qb^-\, g^+\, g^-$ }


Here we present the primitive amplitude
$A_6(1_q^+,2_\qb^-,3^+,4^-)$.  This amplitude has the same flip symmetry
as the last one, as given in eq.~(\use\FlipTwoSym).
The tree amplitude is
$$
A_6^{\rm tree} =
- i \biggl[
 {\spb1.3^2\spa5.2 \spab4.{(1+3)}.6 \over
   \spb4.1 s_{34} t_{134} s_{56}}
- {\spa2.4^2\spb6.1 \spab5.{(2+4)}.3 \over
    \spa2.3 s_{34} t_{234} s_{56}}
+ {\spb1.3\spa2.4\spb6.1\spa5.2 \over
    \spa2.3 \spb4.1 s_{34} s_{56}} \biggr]
\,.
\anoneqn
$$
An alternate form with manifest $k_5 \parallel k_6$ behavior is
$$
A_6^{\rm tree} = 
-i \biggl[
 {\spb1.3^2\spa5.2 \spab5.{(2+4)}.3 \over
   \spb3.4\spb4.1\spa5.6 t_{134} t_{234} }
- {\spa2.4^2\spb6.1  \spab4.{(1+3)}.6 \over
   \spa2.3\spa3.4 \spb5.6 t_{134} t_{234}}
+ {\spb1.3\spa2.4 \spb6.1 \spa5.2 \over
   \spb4.1\spa2.3 t_{134} t_{234}} \biggr]
\,.
\anoneqn
$$

The results for the cut-constructible pieces are 
$$
\eqalign{
V^\gf  = 
 - {1\over \eps^2} \left({\mu^2 \over -s_{12}}\right)^\eps 
- {2\over \eps}  \left({\mu^2 \over -s_{56}}\right)^\eps  - 4
\,,
\cr}
\eqn\VLffpmgf
$$
$$
\hskip -1.2 cm 
\eqalign{
F^\gf & = 
\biggl( 
 \frac{{{\spa1.2}^2}{{\spb4.6}^2}\,t_{123}^{2}}
{\spa2.3\spb5.6{{\spab1.{(2+3)}.4}^3}\spab3.{(1+2)}.4} 
- \frac{{{\spab1.{(2+3)}.6}^2}
{{\spab2.{(1+3)}.4}^2}}{\spa2.3\spb5.6{{\spab1.{(2+3)}.4}^3}
\spab3.{(1+2)}.4} \biggr)
\,\Ls_{-1}^{2{\rm m}h}\bigl({s_{14}},{t_{123}};{s_{23}},{s_{56}}\bigr) 
\nl\hsb
+\biggl( \frac{{{\spab2.{(1+3)}.4}^2}{{\spab3.{(1+2)}.6}^2}}
{\spa2.3\spb5.6\spab1.{(2+3)}.4{{\spab3.{(1+2)}.4}^3}} 
- \frac{\spa2.3{{\spb4.6}^2}\,t_{123}^{2}}
{\spb5.6\spab1.{(2+3)}.4{{\spab3.{(1+2)}.4}^3}} \biggr) 
\,\Ls_{-1}^{2{\rm m}h}\bigl({s_{34}},{t_{123}};{s_{12}},{s_{56}}\bigr)
\nl\hsb
+ \biggl( 
 \frac{{{\spb2.3}^2}{{\spa1.5}^2}\,{t_{234}}\,\spab1.{(2+4)}.3}
{\spb3.4\spa5.6\spab1.{(2+3)}.4{{\spab1.{(3+4)}.2}^3}}
-\frac{{{\spab1.{(2+4)}.3}^3}{{\spab5.{(3+4)}.2}^2}}
{\spb3.4\spa5.6\,{t_{234}}\,\spab1.{(2+3)}.4{{\spab1.{(3+4)}.2}^3}}
\nl\hsa 
-\frac{{{\spa2.4}^3}{{\spb1.6}^2}}
{\spa2.3\spa3.4\spb5.6\,{t_{234}}\,\spab2.{(3+4)}.1}\biggr)\,
\Ls_{-1}^{2{\rm m}h}\bigl({s_{12}},{t_{234}};{s_{34}},{s_{56}}\bigr)
\nl\hsb
+  \bigg( 
  \frac{{{\spa1.2}^2}{{\spab3.{(1+2)}.6}^2}}
       {\spa2.3\spb5.6{{\spa1.3}^2}\spab1.{(2+3)}.4\spab3.{(1+2)}.4}
- \frac{\spa2.3{{\spab1.{(2+3)}.6}^2}}
       {\spb5.6{{\spa1.3}^2}\spab1.{(2+3)}.4\spab3.{(1+2)}.4}
   \bigg) 
\,  {\ts {{\Ls}_{-1}}\left( \frac{{-s_{12}}}{{-t_{123}}}, 
                            \frac{{-s_{23}}}{{-t_{123}}} \right) }
\nl\hsb 
+\biggl( 
  \frac{{{\spab5.{(2+3)}.4}^2}{{\spb2.3}^2}}
{\spa5.6\,{t_{234}}\,{{\spb2.4}^3}\spab1.{(2+3)}.4} 
- \frac{{{\spab5.{(3+4)}.2}^2}{{\spb3.4}^2}}
{\spa5.6\,{t_{234}}\,{{\spb2.4}^3}\spab1.{(2+3)}.4} 
\biggr)
\, {\ts {{\Ls}_{-1}}\left( \frac{-s_{23}}{-t_{234}} ,
  \frac{-s_{34}}{-t_{234}} \right) }
\nl\hsb
+ \Tgf\mathtag{[FFPM]}  \, I_3^{3{\rm m}}(s_{12}, s_{34}, s_{56})
+ \tTgf\mathtag{[FFPM]} \, I_3^{3{\rm m}}(s_{14}, s_{23}, s_{56})
 + 2 \, \frac{\spab5.1.3 \spab1.{(2+4)}.3\spab5.{(2+4)}.3}
{\spb3.4\spa5.6 \spab1.{(2+3)}.4\spab1.{(3+4)}.2}\,
\frac{{{\Ll}_0}\left(\frac{-{t_{234}}}{-{s_{56}}}\right)}{{s_{56}}}
\nl\hsb
+ 2 \, \frac{\spab4.2.3 \spab5.{(2+3)}.4\spab5.{(2+4)}.3}
{\spa5.6 \,t_{234}\, \spb2.4\spab1.{(2+3)}.4} 
\,\frac{{{\Ll}_0}\left(\frac{-{t_{234}}}{-{s_{23}}}\right)}{{s_{23}}}
+ 2 \, \frac{\spab4.2.3 \spab5.{(2+4)}.3\spab5.{(3+4)}.2}
{\spa5.6 \,{t_{234}}\, \spb2.4\spab1.{(3+4)}.2}\,\frac{{{\Ll}_0}
 \left(\frac{-{t_{234}}}{-{s_{34}}}\right)}{{s_{34}}}
\nl\hsb
+ \; M_1(1,4,3,2) \; + \; \flip2
\,. \cr}
\eqn\FLffpmgf
$$

The coefficients of the triangle integrals are 
$$
\hskip -1.1 cm 
\eqalign{
\Tgf & = 
- \frac{{{\spa2.4}^2}\spb1.6\spab5.{(2+4)}.3}{\spa2.3 \,t_{234}^{2}}
+ {\frac{1}{2}}\frac{\spb1.3\spa2.4\spa2.5 \spab5.{(2+4)}.3}
{\spa2.3\spb3.4\spa5.6\,{t_{234}}}
+{\frac{1}{2}}\frac{{{\spa2.4}^2}\spb1.6\spab4.{(1+3)}.6}
{\spa2.3\spa3.4\spb5.6\,{t_{234}}}
\nl\hsb 
- \frac{ \spa4.5\spab2.1.3\spab2.{(1+3)}.6}
{\spa2.3\,{t_{123}}\,\spab1.{(2+3)}.4}
+ {1\over2}
\frac{1}{\spab1.{(2+3)}.4\spab1.{(3+4)}.2}
\biggl[ 4\,\frac{\spab5.1.6\spab4.2.3\spab1.{(2+4)}.3}{{t_{234}}} 
\nl\hsa 
+ { \spab1.{(2+4)}.3 {\spab5.{(2+4)}.3}^2 
    ( 2\,s_{34} \, s_{56} + \d12 \, t_{234} ) \over
       \spb3.4\spa5.6 \, t_{234}^{2} }      
+ { {\spab5.1.3}^2 \bigl( \spa1.2 \d12 + \spaa1.{(2+3)4}.2 \bigr) \over
       \spa2.3\spb3.4\spa5.6 }  
\nl\hsa 
+ {\frac{{{\spa2.4}^3}{{\spb2.6}^2}\spab1.{(2+3)}.4}
    {\spa2.3\spa3.4\spb5.6}}
- \frac{\spa2.4\spa4.5\spab1.{(2+3)4(1-2)}.6}{\spa2.3\spa3.4} 
+ \frac{\spa1.2\spb3.6 \spab5.{(1-2)}.3  ({s_{23}} + {t_{134}})}
    {\spa2.3\spb3.4}
\nl\hsa  
- \frac{\spa1.2\spa4.5}{\spa2.3}
\bigl( \spb3.6 ( 2\,{s_{24}} + \d34 ) - 2\,\spbb3.{15}.6 \bigr) \biggr]
\nl\hsb
+ {1\over2}
 \frac{1}{\spab1.{(2+3)}.4 \spab1.{(3+4)}.2\spab3.{(1+2)}.4}
\biggl[ \spab1.4.6 
  \left( \spab5.1.3 \, s_{13} - \spab5.2.3 \, s_{23} \right)  
+ 5 \, \spab1.{241}.3 \spab5.4.6
\nl\hsa
+ \spab5.4.3 \spab1.{(2+3)}.6 (s_{13} - s_{24})  
+ \spab1.2.3 \Bigr( \spab5.2.6 \, s_{34} 
+ \spab5.4.6 \, s_{13} 
+ 2 \, \frac{\spaa2.{14}.3 \spab5.{(1+3)}.6}{\spa2.3}
\nl\hsa
+ 3 \, \bigl( \spab5.{431}.6 - \spab5.3.6 \, s_{24} \bigr)
- 2 \, \bigl( \spab5.1.6 ( {s_{14}} + {s_{24}}) 
            + \spab5.{143}.6 \bigr) \Bigr) \biggr] \, 
\,, \cr}
\eqn\TLffpmgf
$$
$$
\hskip -0.2 cm 
\eqalign{
\tTgf & = 
 \frac{1}{\spab1.{(2+3)}.4\spab3.{(1+4)}.2}
\biggl[ 
{1\over 2}\spab4.{(1-2)}.3\spab5.{(1+4)}.6 + \spab5.1.3\spab4.2.6
\nl\hsa  
+ \frac{\spab4.{(1+2)}.3}{\spab3.{(1+2)}.4}
\bigl( \spab5.2.4\spab3.1.6 - \spab3.1.4 \spab5.{(2-3)}.6  \bigr) 
\biggr] 
- \frac{\spb1.3\spa4.5\spab2.{(1+3)}.6}{\spab3.{(1+2)}.4 \, t_{123}} 
\,.\cr}
\eqn\tTLffpmgf
$$

The contributions where a scalar replaces the gluon in the loop are 
$$
V^\sc = 
 {1\over 2 \eps} \left({\mu^2 \over -s_{56}}\right)^\eps 
      + {1\over 2} 
\,,
\eqn\VLffpmsc
$$
$$
\hskip -1.2 cm 
\eqalign{
F^\sc &  =
{\spb2.4\spab3.{(2+4)}.1 \over \spb1.4 \spab3.{(1+2)}.4} \, 
     M_2(1,4,2,3) 
%
- {\spb2.3^2\spa2.4^2 \spab5.{(2+3)}.4^2 \over
    \spa5.6\spb2.4 t_{234}\spab1.{(2+3)}.4}
      {\Ls_1 \Bigl( {-s_{23} \over -t_{234}}, 
                 {-s_{34}\over -t_{234}} \Bigr) \over t_{234}^2}
\nl\hsb
+ {\spa2.3^2\spb4.6^2 \, t_{123}^2 \over
    \spa2.3\spb5.6\spab1.{(2+3)}.4\spab3.{(1+2)}.4^3} \,
    \Ls_{-1}^{{\rm 2m}h} \bigl( s_{34}, t_{123}; s_{12}, s_{56} \bigr)
\nl\hsb
- {\spb2.3^2\spa1.5^2 \, t_{234} \, \spab1.{(2+4)}.3 \over
    \spb3.4\spa5.6\spab1.{(2+3)}.4\spab1.{(3+4)}.2^3 }
   \Ls_{-1}^{{\rm 2m}h} \bigl(s_{12}, t_{234}; s_{34}, s_{56} \bigr)
\nl\hsb
+ {\spa2.3^2 \spab1.{(2+3)}.6^2 \over
   \spa2.3\spb5.6\spa1.3^2\spab1.{(2+3)}.4\spab3.{(1+2)}.4}
 {\ts\Ls_{-1}\Bigl( {-s_{12} \over -t_{123}},
                        {-s_{23} \over -t_{123}} \Bigr) }
%
+ \Tsc \mathtag{[FFPM]}\,  I_3^{\rm 3m}(s_{12}, s_{34}, s_{56})
+ \tTsc \mathtag{[FFPM]}\, I_3^{\rm 3m}(s_{14}, s_{23}, s_{56}) 
\nl\hsb
%
%
+ {\spb1.3 \over \spb5.6\spa1.3 \spab3.{(1+2)}.4} \biggl(
     {\spa1.2\spab3.{(1+2)}.6^2 \over \spab3.{(1+2)}.4}\, 
        {\Ll_0 \Bigl( {-t_{123} \over -s_{12} }\Bigr) \over s_{12}}  
   - {\spa2.3 \spab1.{(2+3)}.6^2 \over \spab1.{(2+3)}.4}\,
         {\Ll_0\Bigl({-t_{123} \over -s_{23}} \Big) \over s_{23}} \biggr)
\nl\hsb
+ {1\over 2}\, {\spb2.3\spa2.4^2 \over \spa5.6\spb2.4 \, t_{234}} \biggl(
      {\spb2.3\spab5.{(2+3)}.4^2\over \spab1.{(2+3)}.4} \,
       {\Ll_1 \Bigl( {-t_{234} \over -s_{23}} \Bigl) \over s_{23}^2}
    - {\spb3.4\spab5.{(3+4)}.2^2\over \spab1.{(3+4)}.2} \,
       {\Ll_1 \Bigl( {-t_{234} \over -s_{34}} \Bigr) \over s_{34}^2 } \biggr)
\nl\hsb
+ {\spa2.4\spb4.6^2\, t_{123}^2 \over
        \spb5.6\spab1.{(2+3)}.4 \spab3.{(1+2)}.4^2} \,
         {\Ll_0\Bigl({-t_{123}\over -s_{56}} \Bigr) \over s_{56}} 
- {\spb2.3^2\spa2.4\spa1.5^2 \, t_{234} \over
      \spa5.6\spab1.{(2+3)}.4 \spab1.{(3+4)}.2^2} \,
        {\Ll_0 \Bigl({-t_{234} \over -s_{34} } \Bigr) \over s_{34}}
\nl\hsb
+ {1\over 2}\, {\spb1.3\spb2.3\spa1.5^2\,t_{234}\,\spab1.{(2+4)}.3 \over
      \spb3.4\spa5.6\spab1.{(2+3)}.4 \spab1.{(3+4)}.2 } 
      \biggl(
        {\spb1.3 \over\spb2.3}\, 
           {\Ll_1\Bigl( {-s_{56}\over -t_{234} } \Bigr) \over t_{234}^2}
        - {2\over \spab1.{(3+4)}.2} \,
         {\Ll_0 \Bigl({-s_{56} \over -t_{234}} \Bigr) \over t_{234}} \biggr)
\nl\hsb
+ {1\over 2}\, {\spb1.3\spb4.6 \over
    \spab1.{(2+3)}.4\spab3.{(1+4)}.2\spab3.{(1+2)}.4}
       \Bigl( \spa1.3\spa4.5 
        + {\spab3.{(1+4)}.6 t_{123} - \spab3.{(1+2)}.6 t_{134}
           \over \spb1.4 \spb5.6}  \Bigr)
         {\ts \ln \Bigl( {-s_{23} \over -s_{56}}  \Bigr)}
\nl\hsb
%
+ \biggl\{ 
  3 \, {\spb2.3\spab2.{(3+4)}.1 \over \spab3.{(1+2)}.4 \, \delt_3^2}  
 \Bigl(\spa4.5 \bigl(\spab2.3.6\, \dt23 - \spab2.5.6 \, \dt56 \bigr)
     - \spa1.4 \spb1.6 \bigl( \spa2.5 \dt14 -2\, \spaa2.{36}.5 \bigr) \Bigr)
\nl\hsa
+ {1\over 2}\,  {1\over \spab3.{(1+2)}.4 \, \delt_3} \biggl[
     \Bigl( - {\spa2.5\spa4.5\over \spa5.6}
       + {\spb1.6 \spab2.{(1+4)}.6 \over \spb1.4 \spb5.6 } \Bigr)
          \bigl(\spb1.3 \dt23 -\spbb1.{(5+6)2}.3 \bigr)
\nl\hsa
   - 2 \, {\spb3.6\spab2.{(3+4)}.1 \over \spb1.4\spb5.6}
         \bigl(\spb1.6 \dt23 - 2 \, \spbb1.{45}.6 \bigr)
   - 4 \, \spab4.1.3 \spb1.6 \spa2.5 \biggr] \biggr\}
         {\ts\ln\Bigl( {-s_{23} \over -s_{56}} \Bigr) }
\nl\hsb
 - M_{2a}(1,2,4,3) - M_{3a}(1,2,4,3) 
\nl\hsb
- {1\over 2} \, {\spa3.5\spab4.{(1+2)}.3\spab5.{(2+4)}.3 \over
                s_{34}\spa5.6\spab1.{(3+4)}.2 \spab3.{(1+2)}.4} \,
         {\ts \ln \Bigl( {-s_{12}\over -s_{56}} \Bigr) }
%
%
+ {1\over 2} \, {\spb4.6 \spab4.{(1+2)}.3 \over
      \spb5.6 \spab3.{(1+2)}.4 \spab1.{(2+3)}.4 \, \del_3} 
\nl\hsa\times
\Bigl(
   {\spab3.{(1+2)}.6 \bigl( \spa2.4 \, \d56 
                  - 2\,\spaa2.{13}.4 \bigr) \over \spa3.4}  
  + {\spa1.2 \spb4.6 \bigl( \spb1.3\, \d56 - 2\,\spbb1.{24}.3\big) 
                        \over\spb3.4} \Bigr)
\nl \hsb
+ {1\over 2}\, {\spa1.5 \spab4.{(1+2)}.3 \over
       \spa5.6\spab1.{(3+4)}.2 \spab1.{(2+3)}.4\, \del_3 } \,
   \Bigl(\bigl(\spab5.2.1 \d12 - \spab5.6.1 \d56 \bigr) 
       {\spa1.4 \over \spa3.4} 
      - \spb2.3 \bigl( \spa2.5\d56 - 2\, \spaa2.{1(3+4)}.5 \bigr) \Bigr)
\nl\hsb
%
+ {1\over 2}\, {\spab2.{(3+4)}.1 \over \spab3.{(1+2)}.4 \, \delt_3} 
 \biggl( 
    {1\over 2} \, {\dt56\, \spa2.5 \spb1.6 \over \spb1.4\spa2.3}
  + {\dt14 \spa4.5\spa2.5\over \spa2.3 \spa5.6}
  + \spa4.5 \spb3.6 \biggr) 
\nl\hsb
%
+ {1\over 2}\, {1\over \spab3.{(1+2)}.4 \spab1.{(2+3)}.4} \biggl(  
   - {\spa1.2\spb1.3^2\spb4.6^2 \over \spb1.4 \spb3.4 \spb5.6}
   - {\spa2.5\spa4.5 \spab2.{(1+3)}.4 \over \spa2.3 \spa5.6}
   + {\spa2.4 \spa4.5 \spb4.6\over \spa3.4} \biggr)
\nl\hsb
- {1\over 2} \, {\spa2.4^2\spab5.{(3+4)}.2 \spab5.{(2+3)}.4 \over
     \spa2.3\spa3.4\spa5.6\,t_{234}\,\spb2.4\spab1.{(2+3)}.4  }
+ {1\over 2} \, {\spa1.5 \spa4.5 \spab4.{(1+2)}.3 \over
      \spa3.4\spa5.6\spab1.{(3+4)}.2 \spab1.{(2+3)}.4} \; + \; \flip2
\,.
\cr}
\eqn\FLffpmsc
$$
The three-mass triangle coefficients are 
$$
\hskip -1. cm 
\eqalign{
\Tsc & = 
 3 \, {s_{12}\, \spb4.6\spa5.6\spab4.{(1+2)}.3 \over
                \spab3.{(1+2)}.4\spab1.{(2+3)}.4 \,\del_3^2}  
     \Bigl[\spa1.2\spb5.6 
             \bigl(\spab5.6.1\, \d56  -\spab5.2.1\, \d12 \bigr) 
        + t_{124}\, 
             \bigl(\spab2.5.6\, \d56 - \spab2.1.6\, \d12 \bigr) \Bigr]
\nl\hsb
+ {\spa1.2\spb4.6 \over \spab3.{(1+2)}.4 \spab1.{(2+3)}.4\, \del_3} \,
   \biggl[ 
- \spb1.2\spa5.6 
    \Bigl( 3 \, \spab4.{(1+2)}.3 \spab2.{(1+3)}.6
         - \spa2.4\spb3.6 (t_{123} - t_{124}) \Bigr)
\nl\hsa
     + \spab4.{(1+2)}.3
        \bigl( \spab5.2.1\, \d12 - \spab5.6.1\, \d56
             + t_{123} \, \spab5.{(2+4)}.1 - t_{124}\, \spab5.{(2+3)}.1 \bigr)
\nl\hsa
+ \spa4.5 t_{123}  \bigl(\spbb1.{(3+4)(1+2)}.3 -\spbb1.{24}.3 \bigr)
      - { \spaa4.{(1+2)3}.5  \, t_{123}
         \bigl( \spbb1.{(3+4)(1+2)}.4 - \spbb1.{23}.4 \bigr)
               \over \spab3.{(1+2)}.4 }  \biggr]
\nl\hsb
  - {\spab2.1.3\spab5.4.6 \over \spab3.{(1+2)}.4\spab1.{(2+3)}.4 }
\nl\hsb
%
+ 3 \, {\spab1.5.6 \spab4.{(1+2)}.3 \over \spab1.{(2+3)}.4 \del_3^2}
     \biggl[\spab2.1.3  \bigl(\spab5.2.1\, \d12 - \spab5.6.1\, \d56 \bigr)
      + \spab1.2.3  \bigl(\spab5.6.2\, \d56 - \spab5.1.2 \,\d12 \bigr)
         {\spab2.{(3+4)}.1 \over \spab1.{(3+4)}.2 } \biggr]
\nl\hsb
+ {\spa1.5 \over \spab1.{(2+3)}.4 \, \del_3 } \biggl[
   - \spb1.3 \Bigl( \spab2.1.3 \spab4.5.6 + \spab4.2.3 \spab2.5.6
                  - \spab4.{(1+2)}.3 \spab2.{(1-5)}.6 \Bigr)
\nl\hsb
   + \spb2.3 {\spab2.{(3+4)}.1 \over \spab1.{(3+4)}.2}
  \Bigl( 3 \,\spab1.5.6 \spab4.2.3 - \spab1.2.6 \spab4.{(1+2)}.3 
   + { \spb2.3 \spa1.4 \spab1.5.6 (t_{234} - t_{134}) \over 
          \spab1.{(3+4)}.2 } \Bigr) \biggr]
\,, \cr}
\eqn\TLffpmsc
$$
$$
\hskip -1. cm 
\eqalign{
\tTsc & = 
 3\, {\spa1.4 \spb2.3\spab2.{(3+4)}.1 \over \spab3.{(1+2)}.4\, \delt_3^2}
     \biggl[\bigl( \dt14 \,\spab2.5.6 
            - s_{56} \,\spab2.3.6 \bigr) \spab5.4.1
            - {1\over2} \, \dt56 \spab2.5.6 \spab5.{6}.1 \biggr]
\nl\hsa 
- {\spab2.{(3+4)}.1 \over \spab3.{(1+2)}.4\, \delt_3}
     \biggl[  {1\over2} \, \spab4.5.6 
          \bigl( \spab5.6.3 + 2 \, \spab5.2.3 \bigr)
              - \spab5.2.3  \spab4.1.6 \biggr] 
\,.
\cr}
\eqn\tTLffpmsc
$$

\section{Contributions with the Vector Boson coupled to a Fermion Loop}
\tagsection\FermionLoopSection

We now consider the remaining contributions where the vector boson
$(\gamma^*,Z)$ couples directly to a quark loop, as depicted in
figs.~\use\PrimitiveDiagramsFigure{e} and {f}.  The contributions
proportional to the vector and axial-vector couplings of the quark to
the vector boson are separately gauge invariant and have different
symmetry properties, so we separate the two contributions.  Both
contributions are infrared and ultraviolet finite, because there is
no tree-level coupling between the vector boson and any number of gluons.
Therefore for each amplitude in this section $V=0$, and we just 
give the finite ($F$) terms in eq.~(\use\VFdecomp).

For the vector coupling case, Furry's theorem (charge conjugation)
implies that only box diagrams contribute.  The three box diagrams are
shown in fig.~\use\PrimitiveDiagramsFigure{e}.  A simple way to
understand the cancellation of triangle diagrams, depicted in
\fig\AxialCouplingFigure, is that under reversal of the fermion loop
arrow the sign of the diagrams flip so that there is a pairwise
cancellation.  (This argument also relies on the existence of only one
$SU(N_c)$-invariant combination of two gluons, that proportional to
$\delta^{ab}$.)  For the axial-vector case, Furry's theorem does not
apply, and triangle diagrams such as those in
fig.~\use\AxialCouplingFigure\ do contribute.

\vskip -.5 cm 
%
\LoadFigure\AxialCouplingFigure
{\baselineskip 13 pt \narrower\ninerm\noindent
Two examples of triangle diagrams that contribute to the axial-vector
coupling case, but cancel in the vector coupling case.}
{\epsfxsize 3.5 truein}{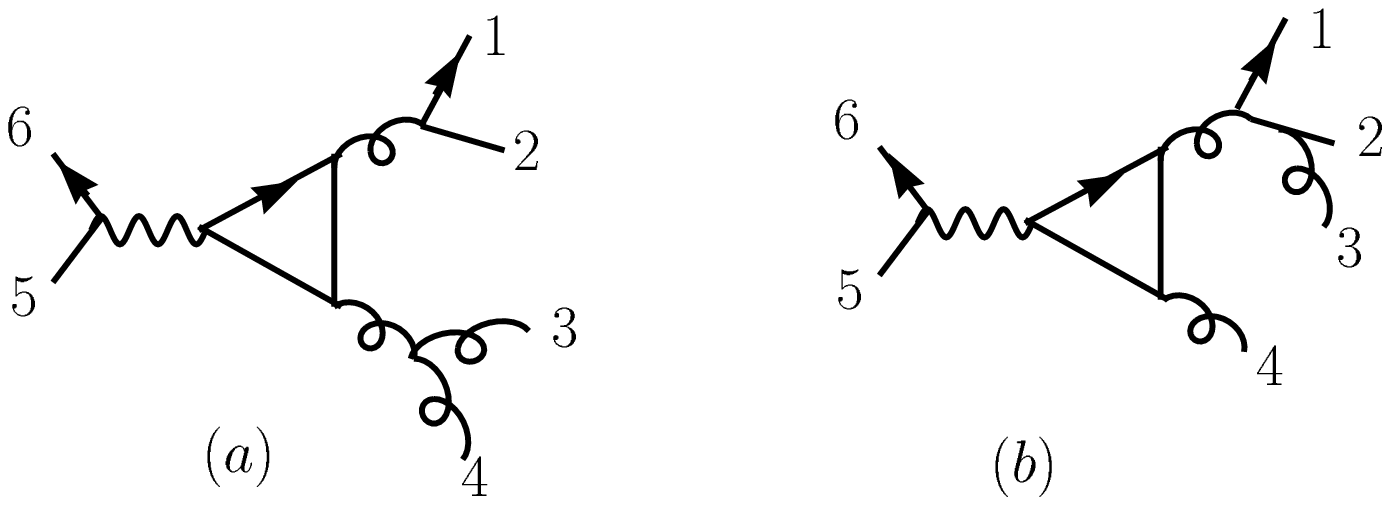}{}

For a massless isodoublet the axial-vector contribution exactly
cancels.  Since we take the $u,d,s,c,b$ quarks to be massless, but not
the $t$ quark, only the $t,b$ pair survives the isodoublet
cancellation in the loop.  As a uniform approximation we keep all
terms of order $1/m_t^2$ but drop terms of order $1/m_t^4$ and beyond.
For the vector coupling, the contribution of the top quark loop
decouples rapidly, leaving only terms of order $1/m_t^4$ which we
drop.  For the axial-vector part, terms of order $1/m_t^2$ appear
which we keep.  The $1/m_t^2$ terms are rather easily obtained
(especially when compared to a diagrammatic calculation) from an
effective Lagrangian (or operator product) analysis, where the
coefficients are fixed via the collinear limits.

\subsection{The Helicity Configuration  $q^+ \, \qb^-\, g^+\,  g^+$}

\vskip .4 cm 

We now present the results for the primitive amplitudes 
$A_6^{\vect s}(1_q^+ , 2_\qb^- , 3^+ , 4^+)$,  
$A_6^{\vect f}(1_q^+ , 2_\qb^- , 3^+ , 4^+)$,
$A_6^\ax(1_q^+ , 2_\qb^- , 3^+ , 4^+)$, and 
$A_6^{\ax,\sl}(1_q^+ , 2_\qb^- , 3^+ , 4^+)$.

The various contributions are 
%
$$
\hskip -.3 cm 
\eqalign{
F^{\vect s} & = 
- { \spa2.3\spa2.4\spa3.5\spa4.5 
        \over \spa1.2\spa5.6\spa3.4^4}
       \Ls^{{\rm 2m}e}_{-1}\bigl(t_{123}, t_{124}; 
                                       s_{12}, s_{56}\bigr) \cr
& \hskip .3 cm
- {\spa2.4\spa3.5 + \spa2.3\spa4.5 \over \spa1.2\spa5.6\spa3.4^3}
   \biggl[
      \spaa5.{31}.2  \, 
         {\Ll_0\Bigl({-t_{123}\over -s_{12}}) \over s_{12}}
    + \spaa5.{64}.2  \, 
         {\Ll_0 \Bigl({-t_{123}\over -s_{56}} \Bigr) \over s_{56}}  
    + {1\over2} \spa5.2 
          {\ts \ln\Bigl( {-t_{123}\over -t_{124} }\Bigr)}  \biggr] \cr
%
& \hskip .3 cm 
- \vphantom{dummy 1} \biggl[ { \spa1.2\spab5.3.1^2  \over \spa5.6\spa3.4^2}\,
       {  \Ll_1\Bigl({-t_{123}\over -s_{12} }\Bigr) \over s_{12}^2}
+ {\spa5.2\spab5.3.1\over \spa5.6 \spa3.4^2} \,
        {\Ll_0\Bigl({-t_{123}\over -s_{12}} \Bigr) \over s_{12}} 
      + \exch{16,25} \biggr] \cr
& \hskip .3 cm 
+ {1\over2} \, {1\over\spa3.4^2} 
  \biggl( {\spa2.5^2\over \spa1.2\spa5.6} 
        - {\spb1.6^2\over \spb1.2\spb5.6} \biggr)
+ \exch{34}
%
 \,,\cr}
\anoneqn
$$
%
$$
\hskip -1.8 cm 
\eqalign{
F^{\vect f} & = 
   - {\spa2.5^2 \over \spa1.2\spa5.6\spa3.4^2} \,
     \Ls^{{\rm 2 m} e}_{-1}\bigl( t_{123}, t_{124}; s_{12}, s_{56}\bigr)
\,,  \cr}
\hskip 4.0 cm 
\anoneqn
$$
$$
\hskip -.3 cm 
\eqalign{
F^\ax & = 
- {1\over 2}  {(\spa2.3\spa4.5+\spa2.4\spa3.5)\spa2.5 \over
               \spa1.2\spa5.6\spa3.4^3}
       \Ls^{{\rm 2m}e}_{-1}\bigl(t_{123}, t_{124}; s_{12}, s_{56}\bigr) 
%
+ \vphantom{dummy 1} \biggl[
  {\spa2.5^2\over \spa1.2 \spa5.6 \spa3.4^2} \,
         {\ts \ln\Bigl({-t_{123}\over -s_{56}} \Bigr)} \cr
& \hskip 1.5 cm 
- {\spab2.4.6 \spa2.5 \over \spa1.2 \spa3.4^2}\, \biggl(
       s_{34}\,  {\Ll_1\Bigl( {-t_{123}\over -s_{56}}\Bigr) \over s_{56}^2}
 +   {\Ll_0 \Bigl( {-t_{123} \over -s_{56} } \Bigr) \over s_{56}} \biggr) 
 - {\spab5.3.1 \spa2.5 \over \spa5.6 \spa3.4^2}
         {\Ll_0\Bigl( {-t_{123} \over -s_{12}} \Bigr) \over s_{12}} 
       - \exch{34} \biggr] \cr
& \hskip .5 cm 
+ (s_{14}+s_{34})\, {\spa2.5\spb4.6 \over \spa1.3\spa3.4}\,
         {\Ll_1\Bigl( {-t_{123} \over -s_{56} }\Bigl) \over s_{56}^2} 
+ {\spab2.3.1\spa2.5\spb3.6 \over \spa2.4 \spa3.4}\, 
         {\Ll_1\Bigl({-t_{124} \over -s_{56}} \Bigr) \over s_{56}^2} \cr
& \hskip .5 cm 
- {1\over 12\,\msq} \, {\spa2.5\over \spa3.4 s_{56}}
       \biggl( {\spb4.6 t_{134} \over \spa1.3} 
        + {\spb3.6\spab2.{(3+4)}.1\over \spa2.4} \biggr)
 \,, \cr }
\hskip 1 cm
\anoneqn
$$
and 
$$
\hskip -1.8 cm 
\eqalign{
F^{\ax,\sl} & = 
  2 \, { \spa2.5 \spb4.6 \spab2.{(1+3)}.4 \over \spa1.3 \spa2.3 s_{56} }
  \biggl[ - {1\over2} 
           { \Ll_1\Bigl({-t_{123} \over -s_{56}} \Bigr) \over s_{56} }
          + {1\over24\,\msq} \biggr] 
 + \exch{34}
 \, . \cr}
\hskip 2.0 cm 
\eqn\Faxslffpp
$$


\subsection{The Helicity Configurations  $q^+ \, \qb^-\, g^+\,  g^-$ 
and  $q^+ \, \qb^-\, g^-\,  g^+$}

It is convenient to quote the helicity amplitudes
$A_6^{\vect,\ax}(1_q^+, 2_\qb^-, 3^+, 4^-)$ and $A_6^{\vect,\ax}(1_q^+,
2_\qb^-, 3^-, 4^+)$ together because they are rather similar.  
In fact for the vector case Furry's theorem (charge conjugation) implies
that the two $A_6^\vect$ amplitudes are identical up to the 
relabelling $3\lr4$, 
$$
A_6^{\vect}(1_q^+,2_\qb^-, 3^-, 4^+) = 
A_6^{\vect}(1_q^+,2_\qb^-, 3^+, 4^-) \bigr|_{3\leftrightarrow 4} \,.
\eqn\VectorLoopSymmetry
$$
For $A_6^{\vect}(1_q^+, 2_\qb^-, 3^+, 4^-)$ we have
$$
\hskip -1. cm 
\eqalign{
F^{\vect s} & = 
%
-2 \, {\spa2.3\spb4.6 \spab3.{(1+2)}.6 \spab2.{(5+6)}.4 \, t_{123}
        \over \spa1.2 \spb5.6 \spab3.{(1+2)}.4^4} 
  \Ls_{-1}^{{\rm 2m}h} \bigl(s_{34}, t_{123}; 
                               s_{12}, s_{56} \bigr) \cr
&  \hskip .5 cm 
+ \biggl[
  2\, \spa2.3 \spb4.6 \bigl( \spb1.2\spa5.6\spa2.3\spb4.6
                   - \spab5.{(2+4)}.1 \spab3.{(1+2)}.4\bigr) 
     {\spab4.{(1+2)}.3 (t_{123} - t_{124})
      \over \spab3.{(1+2)}.4^3 \,\del_3 } \cr
%
& \hskip 1.3 cm 
-3 \, \Bigl(
     {s_{34} \,\d34 \spab2.{(3+4)}.1 \spab5.{(3+4)}.6 \over \del_3}
             - \spab2.3.1 \spab5.4.6 - \spab2.4.1 \spab5.3.6 \Bigr)
               {\spab4.{(1+2)}.3 \over \spab3.{(1+2)}.4 \,\del_3} \cr
& \hskip 1.3 cm
    -  {\spb1.4\spa3.5\spa2.3\spb4.6
              \spab4.{(1+2)}.3^2 \over \spab3.{(1+2)}.4^2 \,\del_3}
    -  {\spb1.3\spa4.5\spa2.4\spb3.6 \over \del_3} \biggr]
         I_3^{\rm 3m}(s_{12}, s_{34}, s_{56}) \cr
%
& \hskip .5 cm 
+ \biggl( 
   2 \, {\spa2.3 \spb4.6 \, t_{123} \, \spab2.{(1+4)}.6\over 
           \spa1.2 \spb5.6\spab3.{(1+2)}.4^3} 
- {\spab2.{(1+4)}.6^2 + 2 \, \spab2.3.6 \spab2.4.6 \over
          \spa1.2 \spb5.6\spab3.{(1+2)}.4^2} \biggr) \,
        {\ts \ln\Bigl({-t_{123} \over -s_{34}} \Bigr)} \cr
%
& \hskip .5 cm 
+ \vphantom{dummy 1} \biggl\{ 
2 \, {\spb1.2\spab2.3.6 \over \spb5.6 \spab3.{(1+2)}.4^2} \biggl[
   {\spab3.{(1+2)}.6 \spab2.{(1+3)}.4 \over \spab3.{(1+2)}.4} 
       {\Ll_0\Bigl( {-t_{123}\over -s_{12} }\Bigr) \over s_{12}}
  - \spab2.3.6
       \Bigl( {\Ll_0 \Bigl({-t_{123} \over -s_{12} }\Bigr) \over s_{12}}
        - {1\over 2} \, {\Ll_1\Bigl({ -t_{123} \over -s_{12} }\Bigl) 
            \over s_{12} } \Bigr) \biggr] \cr 
%
%
& \hskip 1 cm 
- \biggl[ 3 \, {\d56\, \spab2.{(3+4)}.1 \spab5.{(3+4)}.6
     \spab4.{(1+2)}.3 \over \spab3.{(1+2)}.4 \,\del_3^2}
+ 2 \, {\spab3.2.1\spb4.6 (t_{123} - t_{124})
     \bigl( \spa2.5 t_{123} + \spaa2.{16}.5 \bigr) \over
          \spab3.{(1+2)}.4^3\, \del_3}
                \cr 
& \hskip 1.5 cm  
- \Bigl( 2 \, \spab2.3.6  (t_{123}-t_{124})
       + { \d12 \bigl( \spa2.5 t_{123} + \spaa2.{16}.5 \bigr) 
            \over \spa5.6} \Bigr) \,
       {\spab5.2.1 \over \spab3.{(1+2)}.4^2 \,\del_3} \cr
& \hskip 1.5 cm 
+{1\over 2} \Bigl( {\spa1.2\spb1.6^2 \over\spb5.6} 
                 + {\spb1.2\spa2.5^2\over \spa5.6} 
                 - 2\,\spa2.5\spb1.6 \Bigr)
     {\spab4.{(1+2)}.3\over \spab3.{(1+2)}.4 \, \del_3} \biggr]
                 {\ts \ln\Bigl({-s_{12}\over -s_{34}} \Bigr)} 
\; + \; \flip3 \biggr\} \cr
%
%
& \hskip .5 cm 
+  { \spb1.6 \spab4.{(1+2)}.3 
     \bigl( \spb1.6 \d34 - 2\,\spbb1.{25}.6 \bigr) \over 
      \spb1.2 \spb5.6 \spab3.{(1+2)}.4 \, \del_3}
+ {\spab2.{(1+4)}.6^2 + \spab2.1.6\spab2.5.6 
    \over \spa1.2 \spb5.6 \spab3.{(1+2)}.4^2 }
\; + \; \flip2
 \,, \cr}
\anoneqn
$$
$$
\hskip -1. cm 
\eqalign{
F^{\vect f} & = 
-  {\spab2.{(1+3)}.6^2 \over \spa1.2\spb5.6 \spab3.{(1+2)}.4^2}
      \Ls_{-1}^{{\rm 2m}h} \bigl( s_{34}, t_{123};  
                                     s_{12},s_{56} \bigr)  \cr
%
& \hskip 2.5 cm 
- \vphantom{dummy 1} \Bigl[
{1\over 2} \, {\spb1.3\spa4.5 \spab2.{(1+3)}.6 \over
                t_{123} \, \spab3.{(1+2)}.4}
        \,  I_3^{\rm 3m}(s_{12}, s_{34}, s_{56})
\; +\; \flip3 \Bigr] 
 \;  + \; \flip2 
\,.
    \hskip 4 cm \cr}
\anoneqn
$$

The full axial-vector amplitude $A_6^\ax$ does not obey a relation as
simple as eq.~(\use\VectorLoopSymmetry) under exchange of the two 
gluons.  However, the more complicated parts of $A_6^\ax$ (the box 
diagram contributions) do obey such a relation, with an additional
minus sign due to the $\gamma_5$ insertion.
Thus it is convenient to separate the $A_6^\ax$ contributions into 
a common part $C^\ax$ obeying the relation, plus additional terms
with no special symmetry.
The common part is 
$$
\hskip -1. cm 
\eqalign{
C^\ax & =
%
 - {1\over 2} \, { \spab2.{(1+3)}.4^2\spab3.{(1+2)}.6^2
            - \spa2.3^2 \spb4.6^2 \, t_{123}^2 \over
              \spa1.2 \spb5.6 \spab3.{(1+2)}.4^4}
   \Ls_{-1}^{{\rm 2m}h} \bigl(s_{34}, t_{123}; 
                          s_{12}, s_{56}\bigr) \cr
%
& \hskip .5 cm 
+ \biggl[
- {3\over 2} \bigl( \spab5.2.1\spab2.1.6 
           + \spab5.6.1 \spab2.5.6  
      - \spab5.3.1 \spab2.4.6 - \spab5.4.1 \spab2.3.6 \bigr)
     {\spab4.{(1+2)}.3 \over \spab3.{(1+2)}.4\, \del_3} \cr
%
& \hskip 1 cm 
- 3\, { \d34 \, \bigl(\spab5.2.1 \d12 - \spab5.6.1 \d56\bigr)
        \spab4.{(1+2)}.3 \spab2.{(1+3)}.6 
          \over \spab3.{(1+2)}.4\, \del_3^2}
- {\spb1.3 \spa4.5 \spa2.4 \spb3.6 \over \del_3} \cr
& \hskip 1 cm 
+ {  \spb1.4 \spa3.5 (t_{123} - t_{124}) \spab4.{(1+2)}.3 
     \spab2.{(1+3)}.6
     \over \spab3.{(1+2)}.4^2\, \del_3} 
- {1\over 2}\, {\spb1.3 \spa4.5 \spab2.{(1+3)}.6 \over
     t_{123} \, \spab3.{(1+2)}.4} \biggr]
       \,  I_3^{\rm 3m} (s_{12}, s_{34}, s_{56}) \cr
%
& \hskip .5 cm 
+ \vphantom{dummy 1} \biggl[ \biggl(
- 6 \,{\spb1.2 \spab2.{(1+3)}.6 
      \bigl(\spa2.5 \d34 - 2 \spaa2.{16}.5 \bigr)
      \spab4.{(1+2)}.3 \over \spab3.{(1+2)}.4\, \del_3^2} 
- {\spb1.3\spb4.6 \spab2.{(1+3)}.6 \over
    \spb3.4 \spb5.6 \spab3.{(1+2)}.4^2} \cr
& \hskip 1.2 cm 
+ \spb1.4 {\spab2.{(1+3)}.6
   \bigl(3 \, \spbb4.{(1+2) 3}.6 - \spb4.6 (t_{123} - t_{124}) \bigr) 
     \spab4.{(1+2)}.3
   \over \spb3.4 \spb5.6 \spab3.{(1+2)}.4^2\, \del_3} 
- {\spb1.3 \spa2.4 \spb3.6^2 \over \spb3.4 \spb5.6\, \del_3} \biggr) 
 {\ts  \ln\Bigl( {-s_{12} \over -s_{34}} \Bigr)} \cr
& \hskip 1 cm 
+ \exch{16,25} \biggr] 
%
\; + \; {\spab2.{(1+3)}.6^2 \over \spa1.2 \spb5.6 \spab3.{(1+2)}.4^2} 
  {\ts \ln\Bigl({-s_{56} \over -s_{34}} \Bigr) } \cr
& \hskip .5 cm 
+ 
{\spa2.4 \spb3.6 \over \spab3.{(1+2)}.4 } \Bigl(
     {\spab2.4.6 \, \d34 \over \spa1.2 \spb5.6 \,\del_3}
   - {\spa2.4 \spa3.5 \d56 \over \spa1.2 \spa3.4 \, \del_3}
   - {\spb1.3 \spb4.6 \d12 \over \spb3.4 \spb5.6 \, \del_3}
   - 2 \,{\spab5.3.1 \over \del_3}  
   + {\spa2.4\spa3.5 \over \spa1.2 \spa3.4 s_{56}} \Bigr)  
 \,,  \cr}
\eqn\Caxdef
$$
where the operation $\exch{16,25}$ is to be applied to all
preceding terms within the brackets. 

For $A_6^{\ax}(1_q^+, 2_\qb^-, 3^+, 4^-)$
we have 
$$
\eqalign{
F^\ax &= 
C^\ax
%
- {\spa2.4\spab1.4.6\spab2.{(1+3)}.6 t_{123} \over
    \spa1.2 \spa1.3 \spb5.6 \spab3.{(1+2)}.4}\,
      {\Ll_1\Bigl({-s_{56} \over -t_{123}} \Bigr) \over t_{123}^2} 
- {\spab2.{(1+3)}.6 \spab3.{(1+2)}.6 \spb1.3
    \over \spb5.6  \spab3.{(1+2)}.4^2} \, 
     {\Ll_0 \Bigl({-t_{123} \over -s_{12}}\Bigr) \over s_{12}} \cr
& \hskip .5 cm 
- {\spa2.4\spab1.{(2+3)}.4 \spab2.{(1+3)}.6 \spab3.{(1+2)}.6\over
    \spa1.2 \spa1.3 \spb5.6 \spab3.{(1+2)}.4^2}\, 
      {\Ll_0 \Bigl({-t_{123}\over -s_{56}} \Bigr) \over s_{56}} \cr
%
%
& \hskip .5 cm 
+ {\spa2.4 \spa3.5 \spab4.{(1+3)}.6 \over
    \spa1.3 \spa3.4 s_{56} \spab3.{(1+2)}.4 }
- {1\over 12 \msq} \,{\spab4.{(1+3)}.6 \spa2.4 \spa4.5 \over
      \spa1.3 \spa3.4 s_{56}}
\; + \; \flip2 
\,, \cr}
\eqn\Faxffpm 
$$
where the $\flip2$ operation is to be applied to all terms including the 
$C^\ax$ terms.

Similarly, for $A_6^{\ax}(1_q^+, 2_\qb^-, 3^-, 4^+)$ we have 
$$
\eqalign{
F^\ax &= 
-C^\ax \bigr|_{3 \leftrightarrow 4}
\vphantom{Cax /. doexch34}
\vphantom{-Cax34}
%
+ {\spb1.4^2\spa4.5\spab5.{(2+3)}.1 t_{123} \over 
     \spb1.2\spb1.3\spa5.6\spab4.{(1+2)}.3} \, 
      {\Ll_1\Bigl({-s_{56} \over -t_{123}} \Bigr) \over t_{123}^2}
- {\spab5.{(2+3)}.1\spab5.{(1+2)}.3\spa2.3 \over
    \spa5.6 \spab4.{(1+2)}.3^2}\, 
      {\Ll_0\Bigl({-t_{123} \over -s_{12}}\Bigr) \over  s_{12}} \cr
& \hskip .5 cm 
+ {\spb1.4\spab4.{(2+3)}.1\spab5.{(2+3)}.1\spab5.{(1+2)}.3\over
     \spb1.2\spb1.3\spa5.6\spab4.{(1+2)}.3^2} \, 
      {\Ll_0\Bigl({-t_{123} \over -s_{56}} \Bigr) \over s_{56}} \cr
%
%
& \hskip .5 cm 
+ {\spb1.4^2\spa2.5\spb3.6 \over
    \spb1.3\spb3.4 s_{56} \spab4.{(1+2)}.3}
%
- {1\over 12\msq}  {\spb1.4^2\spa2.5\spb4.6 \over 
        \spb1.3\spb3.4 s_{56}}
\; + \; \flip2 
\,. \cr}
\eqn\Faxffmp
$$

The subleading-color axial-vector contributions do obey 
a $3\lr4$ symmetry relation,
$$
A_6^{\ax,\sl}(1_q^+,2_\qb^-, 3^-, 4^+) = 
A_6^{\ax,\sl}(1_q^+,2_\qb^-, 3^+, 4^-) \bigr|_{3\leftrightarrow 4} \,.
\eqn\SublAxialLoopSymmetry
$$
We give here $A_6^{\ax,\sl}(1_q^+,2_\qb^-, 3^+, 4^-)$:
$$
\hskip -1.8 cm 
\eqalign{
F^{\ax,\sl} & = 
  2 \, { \spa2.4 \spa4.5 \spab2.{(1+3)}.6 \over \spa1.3 \spa2.3 s_{56} }
  \biggl[ - {1\over2} 
           { \Ll_1\Bigl({-t_{123} \over -s_{56}} \Bigr) \over s_{56} }
          + {1\over24\,\msq} \biggr] 
 \; + \; \flip2
 \, . \cr}
\hskip 2.0 cm 
\eqn\Faxslffpm
$$


\section{Simplified Versions of Four Quark Amplitudes}
\tagsection\FourQuarkSection

In a previous paper we have presented the one-loop helicity amplitudes
for $e^+ \, e^- \rightarrow \qb q \Qb Q$ [\use\Zqqqq].  Here 
we present simplified versions of these amplitudes.  These
versions have fewer spurious singularities and are therefore
(slightly) better for implementing in a jet program.
We have verified that the two forms are numerically identical.
(In contrast to the $e^+ \, e^- \rightarrow \qb q gg$ case, here
there is not much to gain from decomposing the amplitude into a 
scalar contribution and a cut-constructible part.)

The primitive amplitudes $ A^{\! f}_6$ and $A^s_6$ are proportional to
tree amplitudes and are given by
$$
\eqalign{
A_6^{s,\bn}(1,2,3,4)
 &= \cg \, \Atree_6(1_\q^+,2_\Qb^\pm,3_\Q^\mp,4_\qb;5_\eb^-,6_e^+)
  \, \left[ -{1\over 3\eps} \L {\mu^2 \over -s_{23}}\R^{\eps} 
          - {8\over 9} \right] \,, \cr
A_6^{\! f,\bn}(1,2,3,4)
 &= \cg \, \Atree_6(1_\q^+,2_\Qb^\pm,3_\Q^\mp,4_\qb;5_\eb^-,6_e^+)
  \, \left[ {1\over\eps} \L {\mu^2 \over -s_{23}}\R^{\eps} 
          + 2 \right] \,, \cr}
\eqn\Afsdef
$$   
where $A_6^{\rm tree}$ is given below.

The top quark vacuum polarization contribution is
$$
A_6^{t,\bn}(1,2,3,4)
= - {2\over15} \, \cg \, {s_{23}\over m_t^2} \,
   \Atree_6(1_\q^+,2_\Qb^\pm,3_\Q^\mp,4_\qb;5_\eb^-,6_e^+) \,,
\eqn\Atdef
$$   
neglecting corrections of order $1/m_t^4$ and higher.
The remaining contributions are decomposed further into
divergent ($V$) and finite ($F$) pieces according to 
eq.~(\use\VFdecomp).

\subsection{The Helicity Configuration $q^+\Qb^+Q^-\qb^-$ }

We first give the primitive amplitude $A_6^\nn(1,2,3,4)$.
This amplitude is odd under the operation $\flip1$ defined in 
eq.~(\use\FlipOneSym).
The tree amplitude for this helicity configuration is
$$
A^{{\rm tree}, \, \nn}_6(1,2,3,4) =
  i \, \left[ {\spb1.2\spa5.4\spab3.{(1+2)}.6 \over s_{23} s_{56} t_{123}}
+  {\spa3.4\spb6.1\spab5.{(3+4)}.2 \over s_{23} s_{56} t_{234}} \right]
\, .
\anoneqn
$$
The loop amplitude is
$$
\eqalign{
V^\nn(1,2,3,4) =  
-{1\over \eps^2} \L \L{\mu^2 \over -s_{12}}\R^{\eps}
 + \L {\mu^2 \over -s_{34}}\R^{\eps} \R
 + {2\over 3\eps} \L{\mu^2 \over -s_{23}} \R^{\eps} 
 - {3\over 2} \ln\Bigl({-s_{23}\over -s_{56}} \Bigr)
 + {10\over 9} 
} \,,
\anoneqn
$$
$$
\hskip -.4 cm 
\eqalign{
F^\nn&(1,2,3,4)  = 
\biggl(
  {\spab3.{(1+2)}.6^2 \over \spa2.3 \spb5.6 t_{123} \spab1.{(2+3)}.4}
- {\spb1.2^2 \spa4.5^2 \over \spb2.3\spa5.6 t_{123} \spab4.{(2+3)}.1}
\biggr)\cr 
& \hskip 3 cm \times
\Bigl[ {\ts \Ls_{-1}\Bigl({-s_{12}\over -t_{123}}, 
                {-s_{23} \over -t_{123}} \Bigr) }
     + \Lsnew^{2{\rm m}h}_{-1} \bigl(s_{34},t_{123};
              s_{12},s_{56}\bigr) \Bigr] \cr
%
& \hskip 0.3 cm 
- 2 \, {\spab3.{(1+2)}.6 \over \spb5.6 \spab1.{(2+3)}.4 } \biggl[
    {\spab1.{(2+3)}.6 \spb1.2 \over t_{123} }
     {\Ll_0 \Bigl({-s_{23}\over -t_{123}}\Bigr) \over t_{123}}
 + {\spab3.4.6 \over \spa2.3}
     {\Ll_0 \Bigl({-s_{56} \over -t_{123}} \Bigr) \over t_{123} } 
     \biggr] \cr
& \hskip 0.3 cm 
- {1\over 2} { 1 \over \spa2.3 \spb5.6 t_{123}
       \spab1.{(2+3)}.4 } \biggl[
  \bigl( \spab3.2.1 \spab1.{(2+3)}.6 \bigr)^2
         { \Ll_1 \L {-t_{123}\over -s_{23}} \R \over s_{23}^2}
+ \spab3.4.6^2 t_{123}^2 
         { \Ll_1 \L{-s_{56} \over -t_{123}} \R \over t_{123}^2}
         \biggr] \cr 
& \hskip 2 cm 
  - \; \flip1 
\,, \cr 
}\anoneqn
$$
where $\flip1$ is to be applied to all the preceding terms in $F^\nn$.
The structure of this amplitude is already rather simple, and it
is unchanged from ref.~[\use\Zqqqq].

\subsection{The Helicity Configuration $q^+\Qb^-Q^+\qb^-$}

We now give the result for $A_6^\an(1,2,3,4)$.
This amplitude is odd under the same flip symmetry~(\use\FlipOneSym)
as $A_6^\nn$.  The tree amplitude is 
$$
A^{{\rm tree}, \, \an}_6(1,2,3,4) =
 -i \, \left[ {\spb1.3 \spa5.4 \spab2.{(1+3)}.6 \over s_{23} s_{56} t_{123}}
  +  {\spa2.4 \spb6.1 \spab5.{(2+4)}.3 \over s_{23} s_{56} t_{234}} \right]
\,.
\anoneqn
$$
Note that 
$A^{{\rm tree},\,\an}_6(1,2,3,4) = -A^{{\rm tree},\,\nn}_6(1,3,2,4)$.
The simpler form for the loop amplitude is
$$
V^\an(1,2,3,4) =  V^\nn(1,2,3,4) \,,
\hskip 9 cm
\anoneqn 
$$
$$
\hskip -.3 cm 
\eqalign{
F^\an&(1,2,3,4) = 
\biggl( -{\spb1.3^2 \spa4.5^2 \over
     \spb2.3 \spa5.6 t_{123} \spab4.{(2+3)}.1} 
+ {\spa1.2^2 \spab3.{(1+2)}.6^2
    \over \spa2.3 \spb5.6 t_{123} \spa1.3^2 \spab1.{(2+3)}.4} \biggr) 
   {\ts  \Ls_{-1}\Bigl({-s_{12}\over -t_{123}}, {-s_{23} \over -t_{123} } 
            \Bigr) }\cr
& 
+ \biggl( -{\spb1.3^2 \spa4.5^2 \over
        \spb2.3 \spa5.6 t_{123} \spab4.{(2+3)}.1 } 
+ {\spab3.{(1+2)}.6^2 \spab2.{(1+3)}.4^2 \over
   \spa2.3\spb5.6 t_{123} \spab1.{(2+3)}.4 \spab3.{(1+2)}.4^2} 
  \biggr)
 \Ls^{2 {\rm m} h}_{-1} \bigl(s_{34},t_{123};s_{12},s_{56}\bigr) \cr
%
&  
+ \biggl[ 
   {1\over2} {\d56 \spab5.{(1+2)}.6
       ( s_{56} \spab2.1.4 - t_{123} \spab2.{(1+3)}.4 )
    \over \spab1.{(2+3)}.4 \spab3.{(1+2)}.4 \Delta_3 } 
- {1\over2} { \bigl( \spab2.1.4 \spab5.{(3-4)}.6 
                 + \spab2.3.4 \spab5.{(3+4)}.6 \bigr)
              \over \spab1.{(2+3)}.4 \spab3.{(1+2)}.4 } \cr
& \hskip .7 cm              
+ {1\over2} { t_{123} \d34 + 2 s_{12} s_{56} \over t_{123}^2 } 
     \biggl( { {\spb1.3}^2 {\spa4.5}^2 
     \over \spb2.3\spa5.6 \spab4.{(2+3)}.1 }
         -  { {\spab2.{(1+3)}.6}^2 
     \over \spa2.3\spb5.6 \spab1.{(2+3)}.4 } \biggr)
- 2 \, { \spab2.1.3 \spab5.4.6 \over t_{123} \spab1.{(2+3)}.4 } \biggr]
\cr 
& \hskip 1.7 cm              
   \times I_3^{3{\rm m}}(s_{12},s_{34},s_{56})  \cr
%
%
&  
+ {\spb1.3\spa1.2 \spab3.{(1+2)}.6^2
   \over \spb5.6 t_{123} \spa1.3 \spab3.{(1+2)}.4}
       {\Ll_0\Bigl( {-t_{123}\over -s_{12}} \Bigr) \over s_{12}}
- {1\over 2} {\spb1.3^2 \spa2.3 \spab1.{(2+3)}.6^2
      \over \spb5.6 t_{123} \spab1.{(2+3)}.4 }
        {\Ll_1\Bigl( {-t_{123}\over -s_{23} } \Bigr) \over s_{23}^2 } \cr
&  
+ {\spb1.3 \spab1.{(2+3)}.6 \spab2.{(1+3)}.6
    \over \spb5.6 t_{123} \spab1.{(2+3)}.4}
        {\Ll_0\Bigl( {-t_{123}\over -s_{23}} \Bigr) \over s_{23}}
+ {\spb1.3 \spa1.2 \spab1.{(2+3)}.6 \spab3.{(1+2)}.6 
   \over \spb5.6 t_{123} \spa1.3 \spab1.{(2+3)}.4 }
        {\Ll_0 \Bigl( {-t_{123} \over -s_{23} } \Bigr) \over s_{23}} \cr
&  
- {1\over 2} {\spb6.4^2 \spa4.2^2    \, t_{123} \over
    \spa2.3 \spb5.6 \spab1.{(2+3)}.4}
       {\Ll_1\Bigl( {-s_{56}\over -t_{123}} \Bigr) \over t_{123}^2}
  - {\spb6.4^2 \spa4.2 t_{123}
      \over \spb5.6 \spab1.{(2+3)}.4 \spab3.{(1+2)}.4}
        {\Ll_0\Bigl( {-t_{123} \over -s_{56}} \Bigr) \over s_{56} } \cr
&  
- 2 \, { \spb6.4 \spa4.2 \spab2.{(1+3)}.6 \over 
     \spa2.3 \spb5.6 \spab1.{(2+3)}.4}
        {\Ll_0 \Bigl( {-t_{123}\over - s_{56}} \Bigr) \over s_{56} } \cr
&  
+ {1\over\spab3.{(1+2)}.4\Delta_3} \biggl[
    {\spa1.2 t_{123}\over\spb5.6} \biggl( {\spb1.6}^2 
    + {\spb2.6}^2 {\spab2.{(3+4)}.1 \over \spab1.{(3+4)}.2 } \biggr)
  + {\spb1.2 t_{124}\over\spa5.6} \biggl ( {\spa2.5}^2 
    + {\spa1.5}^2  {\spab2.{(3+4)}.1 \over \spab1.{(3+4)}.2 } \biggr) \cr
& \hskip 2 cm 
  + 2 s_{12} \biggl( - \spa2.5 \spb1.6 
    + \spa1.5 \spb2.6 {\spab2.{(3+4)}.1 \over \spab1.{(3+4)}.2 } \biggr)
     \biggr] {\ts \ln\Bigl({-s_{12}\over -s_{56}}\Bigr)}  \cr
& \hskip 1 cm 
 - \; \flip1 
   \,, \cr 
}
\anoneqn
$$
where $\flip1$ is to be applied to all the preceding terms in $F^\an$.

\subsection{Subleading-Color Primitive Amplitude}

The primitive amplitude $A_6^\sl(1,2,3,4)$, 
contributes only at subleading order in $N_c$.  
The ``tree amplitude'' appearing in eq.~(\use\VFdecomp) is 
$$
A_6^{\tree,\sl}(1,2,3,4) =
i\, \left[ {\spb1.3 \spa5.4 \spab2.{(1+3)}.6 \over s_{12} s_{56} t_{123}} 
- {\spa2.4 \spb6.3 \spab5.{(2+4)}.1 \over s_{12} s_{56} t_{412}} \right]
\,,
\anoneqn
$$
and satisfies 
$A^{\tree, \sl}_6(2,3,1,4) = -A^{{\rm tree}, \, \nn}_6(1,2,3,4)$.

To describe the loop amplitude, we use the operations $\flip3$ and
$\exch{16,25}$ defined in eqs.~(\use\FlipThreeSym)
and~(\use\ExchangeDefs).
The loop amplitude is given by 
$$
\eqalign{
V^\sl(1,2,3,4) & = 
\biggl[- {1\over \eps^2} \L{\mu^2 \over -s_{34}}\R^\eps 
  - {3\over 2\eps} \L {\mu^2 \over -s_{34}}\R^\eps - 4  \biggr]
\ +\ \biggl[ - {1\over \eps^2} \L{\mu^2 \over -s_{12}}\R^\eps 
  - {3\over 2 \eps} \L{\mu^2 \over -s_{12}}\R^\eps 
  - {7\over 2}\biggr] 
\,, \cr}
\anoneqn
$$
$$
\hskip -0.5 cm 
\eqalign{
F^\sl&(1,2,3,4)  = 
\biggl( {\spb1.3^2 \spa4.5^2 \over \spb1.2 \spa5.6 t_{123} \spab4.{(1+2)}.3 } 
 - {\spab3.{(1+2)}.6^2 \spab2.{(1+3)}.4^2
    \over \spa1.2 \spb5.6 t_{123}\spab3.{(1+2)}.4^3} \biggr)
     \Ls^{2 {\rm m} h}_{-1}(s_{34},t_{123};s_{12},s_{56}) \cr
%
&\hskip .3 cm
+ T\vphantom{sl} \, I_3^{3 \rm m}(s_{12}, s_{34}, s_{56}) 
%
+ \vphantom{dummy 1}
  \biggl[
  {1\over 2} {\spb6.4^2 \spa4.2^2  t_{123}
    \over \spa1.2 \spb5.6 \spab3.{(1+2)}.4}
            {\Ll_1\Bigl( {-s_{56} \over -t_{123}}\Bigr) \over  t_{123}^2 }
+ 2 \, {\spb6.4 \spa4.2 \spab2.{(1+3)}.6
   \over \spa1.2 \spb5.6 \spab3.{(1+2)}.4 }
             {\Ll_0\Bigl( {-t_{123} \over -s_{56}} \Bigr) \over s_{56} } \cr
& \hskip .7 cm 
- {\spa2.3 \spa2.4 \spb6.4^2  t_{123}
   \over \spa1.2 \spb5.6 \spab3.{(1+2)}.4^2}
             {\Ll_0\Bigl({- t_{123} \over -s_{56}} \Bigr) \over  s_{56} }
- {1\over 2} { \spa2.3 \spb6.4  \spab2.{(1+3)}.6
    \over \spa1.2 \spb5.6 \spab3.{(1+2)}.4^2 }
     \,   {\ts \ln\Bigl( {(- t_{123})(-s_{34})\over(-s_{56})^2} \Bigr)} \cr
& \hskip .7 cm 
- {3\over 4} {\spab2.{(1+3)}.6^2  
    \over \spa1.2 \spb5.6 t_{123} \spab3.{(1+2)}.4 }
      \,  {\ts \ln\Bigl( {(- t_{123})(-s_{34})\over(-s_{56})^2} \Bigr)} \cr
%
& \hskip .7 cm 
+  \biggl(
  {3\over2} { \d56 \, (t_{123}-t_{124}) \, \spab2.{(3+4)}.1 \spab5.{(3+4)}.6 
              \over \spab3.{(1+2)}.4 \Delta_3^2 } \cr
& \hskip 1.3 cm 
- {\spb1.2\spa2.3\spb4.6 \over {\spab3.{(1+2)}.4}^2 \Delta_3 }
   \Bigl( \spa2.5 (t_{123}-t_{124}) - 2 \, \spaa2.{1\,6}.5 \Bigr)
+  { \spb1.2\spa2.5 \spab2.{(3-4)}.6 \over \spab3.{(1+2)}.4 \Delta_3 } \cr
& \hskip 1.3 cm 
+ { \spb1.6 \over \spb5.6 \spab3.{(1+2)}.4 \Delta_3 }
   \Bigl( \spab2.3.6 \, t_{123} - \spab2.4.6 \, t_{124}
      + { \spa2.3\spb4.6 \, \d34 \, t_{123} \over \spab3.{(1+2)}.4 } \Bigr) 
   \biggr) {\ts \ln\Bigl({-s_{12} \over -s_{34} }\Bigl)} \cr 
%
& \hskip .7 cm 
- {1\over4} { \spb1.6 \, (t_{123}-t_{124}) 
  \bigl( \spb1.6 \, \d34 - 2 \, \spbb1.{25}.6 \bigr)
  \over \spb1.2\spb5.6 \spab3.{(1+2)}.4 \Delta_3 }
\; - \; \flip3 \biggr]  
\; + \; \exch{16,25}
\,,\cr
} \anoneqn
$$
where $\exch{16,25}$ is to be applied to all the preceding terms 
in $F^\sl$, but $\flip3$ is to be applied only to the terms 
inside the brackets ($[\ ]$) in which it appears. 
The three-mass triangle coefficient $T$ is given by
$$
\eqalign{
 T & = 
  {1\over2} { (3 s_{34} \d34 - \Delta_3 ) (t_{123}-t_{124})
               \spab2.{(3+4)}.1 \spab5.{(3+4)}.6
         \over \spab3.{(1+2)}.4 \Delta_3^2 } 
+ {1\over2} { s_{34} \, (t_{123}-t_{124}) \spa2.5\spb1.6
         \over \spab3.{(1+2)}.4 \Delta_3 }  \cr
& \hskip 1.3 cm 
- { \spb1.2\spa5.6 {\spab2.{(3+4)}.6}^2
   \over \spab3.{(1+2)}.4 \Delta_3 }
+ { \spa2.3\spb4.6 \d34 \bigl( \spab5.2.1 \, \d12 - \spab5.6.1 \, \d56 \bigr)
   \over {\spab3.{(1+2)}.4}^2 \Delta_3 }  \cr
& \hskip 1.3 cm 
+ { \spa2.3\spb4.6 \spab5.{(2+3)}.1 \over {\spab3.{(1+2)}.4}^2 } 
- 2 \, { \spb1.3\spa4.5 \spab2.{(1+3)}.4 \spab3.{(1+2)}.6
      \over t_{123}^2 \, \spab3.{(1+2)}.4 } 
  \,. \cr
}\anoneqn
$$

\subsection{Axial-Vector Quark Loop}
\tagsubsection\AxialQuarkSection

The axial-vector quark-triangle contribution $A_6^\ax(1,2,3,4)$ is 
easily obtained from the fully off-shell $Zgg$ vertex presented in
ref.~[\use\HHKY].  The infrared- and ultraviolet-finite result is 
$$
\eqalign{ 
A_6^\ax(1,2,3,4)\ &=\ 
-{ 2 i \over (4\pi)^2 } 
{ f(m_t;s_{12},s_{34},s_{56}) 
  - f(m_b;s_{12},s_{34},s_{56}) \over  s_{56} }  
\biggl( { \spb6.3\spa4.2\spa2.5 \over \spa1.2 }
      - { \spb6.1\spb1.3\spa4.5 \over \spb1.2 } \biggr) \cr
&\qquad + (1\lr3, \hskip0.2cm  2\lr4)\ , \cr
}\eqn\Aaxformula
$$
where the integral $f(m)$ is defined in eq.~(\use\fmdef).  We need
only the large mass expansion (for $m=m_t$) and the $m=0$ limit (for
$m=m_b$) of this integral; these are presented in
eq.~(\use\fmexplicit).


\section{Summary and Conclusions}
\tagsection\ConclusionSection

In this paper we presented explicit formul\ae\ for all one-loop helicity
amplitudes which enter into numerical programs for the next-to-leading order
QCD corrections to $e^+\,e^- \to (\gamma^*,Z) \to 4$ jets.  These
amplitudes have already been incorporated into the first such program
to be constructed [\use\Adrian]; the result has been a significant 
reduction in theoretical uncertainties in the four-jet cross-section 
and associated quantities.  With appropriate modifications to the
coupling constants, these same amplitudes enter into the computation
of next-to-leading order contributions to the production of a vector
boson ($W$, $Z$, or Drell-Yan pair) in association with two jets at
hadron colliders, and three-jet production at $ep$ colliders.

Following the methods reviewed in ref.~[\use\Review], we have obtained
the loop amplitudes by demanding that their functional forms satisfy
unitarity and factorization.  This approach makes use of a color
decomposition~[\use\Color] into gauge invariant primitive amplitudes
[\use\Fermion,\use\Zqqqq] which are expressed in terms of a helicity
basis~[\use\SpinorHelicity].  The color decomposition limits the
analytic functions that may appear, greatly simplifying the
reconstruction of the amplitudes from their analytic properties.  The
helicity basis results in relatively compact expressions, from which
spurious poles can be systematically removed to further simplify the
results.  A further advantage of the helicity basis is that one
retains all spin information.

As a check on the methods we have verified that the amplitudes are
numerically identical to ones obtained by a (numerical) Feynman diagram
calculation that we have performed.  This diagrammatic calculation
made use of a number of string-motivated ideas reviewed in
ref.~[\use\Review].  We have also numerically verified that the typed
form of the amplitudes appearing in this paper agree with our
Maple and Mathematica expressions for the same quantities.

We expect that the amplitudes presented in this paper should lead to
an improved knowledge of QCD predictions for a wider class of observables,
and thus of the QCD background to searches for new physics in various 
processes.


\vskip0.3in
\par\noindent
{\bf Acknowledgements}
\vskip0.1in

Z.B. wishes to thank SLAC, Saclay and the Aspen Center for Physics for
their hospitality during the period of this work; L.D. similarly wishes to
thank Rutgers University, the Aspen Center for Physics, Saclay and UCLA;
and D.A.K. wishes to thank SLAC and UCLA.  We thank Dave Dunbar, Adrian
Signer and Stefan Weinzierl for useful conversations.  This work was
supported in part by the US Department of Energy under grants
DE-FG03-91ER40662 and DE-AC03-76SF00515, by the Alfred P. Sloan Foundation
under grant BR-3222, and by the {\it Direction des Sciences de la
Mati\`ere\/} of the {\it Commissariat \`a l'Energie Atomique\/} of France.
L.D. and D.A.K. gratefully acknowledge the support of NATO Collaborative
Research Grant CRG--921322.


\appendix{One-Loop Integrals}
\tagappendix\IntegralsAppendix

\subappendix{General Properties and Reduction Formul\ae}
\tagsubappendix\IntegralReductionSubAppendix

The loop momentum integrals that appear in either a Feynman diagram 
or cut-based analysis are of the form
$$
 {\cal I}_n^{D}[\ell^{\alpha_1} \cdots \ell^{\alpha_m}] 
= \int { d^{D}\ell \over (2\pi)^{D} }
  {\ell^{\alpha_1} \cdots \ell^{\alpha_m} \over \ell^2  
    (\ell-K_1)^2 \cdots (\ell-\sum_{i=1}^{n-1} K_i )^2 } \,,
\eqn\GeneralLoopIntCalI
$$
where the $K_i$ are external momenta, or sums of external momenta.
The momentum $\ell$ flows through the propagator 
between external legs $n$ and $1$.  
For convenience we define
$$
I_n^{D}[\ell^{\alpha_1} \cdots \ell^{\alpha_m}] = 
i (-1)^{n+1} (4\pi)^{D/2} 
{\cal I}_n^{D}[\ell^{\alpha_1} \cdots \ell^{\alpha_m}] \,.
\eqn\GeneralLoopInt
$$
Integrals with powers of loop momenta $\ell^\alpha$ in the numerator 
are known as tensor integrals; integrals with no powers of loop 
momenta in the numerators are known as scalar integrals and
are denoted by $I_n^{D} \equiv I_n^{D}[1]$.  When the superscript $D$
is omitted below, $D$ is implicitly equal to $4-2\eps$.

In general, any one-loop amplitude may be expressed as a linear
combination of bubble, triangle and box (i.e.~two-, three- and
four-point) scalar integrals. This follows from the Passarino-Veltman
reduction [\use\PV] of any tensor $n$-point amplitude with $m$ powers
of loop momenta to a linear combination of $n-1$ and $n$-point
integrals with no more than $m-1$ powers of loop momenta.  The
resulting scalar integrals with $n>4$ legs can be further reduced to
scalar box integrals using an additional set of reduction 
formul\ae~[\use\VNV,\use\IntegralsShort,\use\IntegralsLong].

It is useful to review briefly the conventional reduction of integrals
to explain the origin and types of determinantal poles that can
appear.  Consider the five-point tensor integral with the kinematic
configuration depicted in fig.~\use\PentagonKinematicsFigure.
Following the Passarino-Veltman reduction technique, 
we expand the loop momentum in terms of four independent external momenta,
$$
I_5 [ \ell^{\alpha_1} \ell^{\alpha_2} \ldots \ell^{\alpha_j} ]
= \sum_{k=1}^4 p_k^{\alpha_1} A_k 
\eqn\expand
$$
where 
$$
p_1 = K_1\,, \hskip 1 cm 
p_2 = K_1+K_2\,, \hskip 1 cm 
p_3 = K_1+K_2+K_3\,, \hskip 1 cm 
p_4 = -K_5\,,
\anoneqn
$$
and we have suppressed the Lorentz indices $\alpha_2 \ldots \alpha_j$
on the right-hand side of eq.~(\use\expand) in $A_k$.  The functions
$A_i$ are found by first contracting eq.~(\use\expand) with the
momentum sums $p_{i}$, generating the four linearly independent equations,
$$
2I_5 [ \ell\c p_i \ \ell^{\alpha_2} \ldots \ell^{\alpha_j} ] =
\sum_{k=1}^4 t_{ik} A_k \, , \hskip 1 cm  (i=1,2,3,4) \, ,
\eqn\matrixeqn
$$
where $t_{ik}\equiv 2 p_{i}\c p_{k}$.  Using $2 \ell\c p_i =
-(\ell-p_i)^2 + \ell^2 + p_i^2$, and recognizing that $(\ell-p_i)^2$
and $\ell^2$ are inverse propagators of the pentagon integral, 
one can reduce the left-hand-side to a sum of two four-point integrals
plus a five-point integral with one less power of loop momentum.  
Inverting $t_{ik}$ we have
$$
\eqalign{
A_i & = \sum_{k=1}^{4} t^{-1}_{ik}\
\Bigl( I_4^{(k+1)}[\ell^{\alpha_2} \ell^{\alpha_3} \cdots \ell^{\alpha_j}]
     - I_4^{(1)}[\ell^{\alpha_2} \ell^{\alpha_3} \cdots \ell^{\alpha_j}]
  + p_k^2 I_5[\ell^{\alpha_2} \ell^{\alpha_3} 
            \cdots \ell^{\alpha_j}] \Bigr) \,,\cr}
\eqn\Asolution
$$
where $I_4^{(i)}$ denotes the box integral that is obtained from
the pentagon integral by canceling the propagator factor $1/\ell_i^2$.
The coefficients $t^{-1}_{ik}$ contain in their denominators the 
pentagon Gram determinant $\Delta_5$ given in eq.~(\use\GramFive).

The reduction of tensor boxes, triangles, and bubbles is similar except 
that one must also include the Kronecker-delta in the expansion of the 
integral function since there are less than four independent momenta.
More powerful reduction techniques which lead to square-roots of Gram 
determinants in denominators, instead of single powers, have also been 
developed~[\use\VNV]. 

The scalar pentagon integrals may be reduced to scalar box integrals
using the scalar integral recursion formula
[\use\VNV,\use\IntegralsShort,\use\IntegralsLong] (valid for $n\le 6$)
$$
 I_n = {1\over2} \Biggl[
     \sum_{i=1}^n c_i\ I_{n-1}^{(i)}
     + (n-5+2\eps)\, c_0\ I_n^{D=6-2\eps} \Biggr]\ ,
\eqn\IntRecursion
$$
where
$$
  c_i = \sum_{j=1}^n S^{-1}_{ij}\, , \hskip 2 cm 
  c_0 = \sum_{i=1}^n c_i\ =\ \sum_{i,j=1}^n S^{-1}_{ij}.
\eqn\cdef
$$
The matrix $S_{ij}$ is defined in eq.~(\use\Sdefn)
and the integral $I_n^{D=6-2\eps}$ is the scalar $n$-point integral evaluated
in $(6-2\eps)$ dimensions.  Observe that for $n=5$ the prefactor
of $I_5^{D=6-2\eps}$ is of $\Ord(\eps)$; since $I_5^{D=6-2\eps}$ is finite
as $\eps\to0$ we may drop this term.
This equation is also useful for rewriting $D=4-2\eps$ box integrals
as $D=6-2\eps$ box integrals, which turns out to be a convenient 
way to represent the amplitudes.  
Indeed, the $\Ls_{-1}$ functions defined in 
appendix~\use\IntegralFunctionAppendix\ are $D=6-2\eps$ box integrals
from which a simple overall kinematic factor has been removed.
The explicit forms of the higher-dimension box integrals 
may be obtained by solving eq.~(\use\IntRecursion) for $I_4^{D=6-2\eps}$

For the case of a single power of loop momentum we may avoid 
a Gram determinant in the denominator by making use of the 
reduction formula,
$$
  I_n[\ell^\mu]\ =\ {1\over2} \sum_{i=2}^n  p_{i-1}^\mu \Biggl[
    \sum_{j=1}^n  S^{-1}_{ij}\ I_{n-1}^{(j)}
    \ +\ (n-5+2\eps)\, c_i\ I_n^{D=6-2\eps} \Biggr]\ ,
\eqn\SinglePowerIntRed
$$
where $p_i = K_1 + K_2 + \cdots + K_i$ and the $K_j$ are the external momenta
of the integrals.  For pentagon integrals ($n=5$) the
$I_5^{D=6-2\eps}$ term may be dropped since it is of $\Ord(\eps)$.

Since the integral recursion formula (\use\IntRecursion) contains inverses
of the matrix $S$ defined in eq.~(\Sdefn), the amplitudes will contain
poles in $\det S$.  As discussed in the text, these poles do not
correspond to the propagation of physical states, and so their residues in
a full amplitude must vanish.

In summary, when reducing tensor and scalar integrals to
linear combinations of scalar box, triangle and bubble integrals, one
encounters a set of spurious determinantal poles.  Some of these
spurious poles, such as the pentagon Gram determinant 
(see section~\StrategySubsection), 
are artifacts of the reduction procedure, but others are
inherently part of the amplitude when it is expressed in terms of
logarithms and dilogarithms.

\subappendix{Determinants Appearing in Amplitudes}
\tagsubappendix\GramDetListAppendix

Here we explicitly list the determinants that appear in the
denominators of coefficients in the general integral reduction
formul\ae\ of the preceding subsection.  As discussed in
section~\use\SpuriousPoleSection, knowledge of these determinants ---
and how they can be factored --- is useful in deciding which spinor
factors to multiply and divide by in order to simplify the integral
reduction of specific cuts.

The Gram determinants are defined in eq.~(\use\GramDet).
The explicit forms of the Gram determinants associated with pentagon
and box integrals whose external legs follow the ordering 123456 is 
$$
\eqalign{
\Delta_5 & =
- s_{23}^2\, s_{34}^2
- s_{34}^2\, t_{123}^2 
- t_{123}^2\, t_{234}^2 
- s_{12}^2\, t_{234}^2 
- s_{23}^2\, s_{56}^2 
- s_{12}^2\, s_{23}^2 \cr
& \hskip .3 cm 
+ 2\, s_{34}^2\, t_{123}\, s_{23} 
+ 2\, t_{123}^2\, t_{234}\, s_{34} 
+ 2\, s_{12}^2\, t_{234}\, s_{23} 
+ 2\, s_{56}\, s_{23}^2\, s_{34} 
+ 2\, s_{56}\, s_{23}^2\, s_{12} 
+ 2\, s_{23}^2\, s_{12}\, s_{34}\cr
& \hskip .3 cm 
 -2\, t_{123}\, t_{234}\, s_{23}\, s_{34} 
- 2\, t_{234}\, s_{12}\, s_{23}\, s_{34} 
- 2\, t_{234}\, s_{34}\, t_{123}\, s_{12} 
- 2\, t_{123}\, s_{12}\, s_{23}\, s_{34} 
+ 2\, t_{234}^2\, t_{123}\, s_{12} 
\cr
& \hskip .3 cm 
- 2\, t_{123}\, t_{234}\, s_{12}\, s_{23}  
- 2\, s_{56}\, s_{23}\, s_{34}\, t_{123}
+ 2\, s_{56}\, s_{23}\, t_{123}\, t_{234}
- 2\, s_{56}\, s_{23}\, s_{12}\, t_{234}
+ 4\, s_{56}\, s_{23}\, s_{12}\, s_{34} \, ,
\cr}
\eqn\GramFive
$$
$$
\eqalign{
\Delta_4^{(1)} & = 
- 2 \, s_{23} s_{34} s_{24} \,, \cr
\Delta_4^{(2)} & = 
 2 \, s_{34} (s_{12} \, s_{56} - t_{123} \, t_{124}) \,, \cr
\Delta_4^{(3)} & = 
 2 \, s_{14} (s_{23} \, s_{56} - t_{123} \, t_{234}) \,, \cr
\Delta_4^{(4)} & = 
 2 \, s_{12} (s_{34} \, s_{56} - t_{134} \, t_{234}) \,, \cr
\Delta_4^{(5)} & =
- 2 \, s_{12}  s_{23} s_{13} \,, \cr} 
\hskip 4 cm 
\eqn\GramFour
$$
where the superscript on the box Gram determinants labels the propagator
that has been canceled in the pentagon integral to obtain the box.
Similarly, the triangle Gram determinants are labeled by a pair of
indices, for the two canceled propagators, 
$$
\eqalign{
& 
\Delta_3^{(1,2)} = s_{34}^2 \,, \hskip 1 cm 
\Delta_3^{(1,3)} = (t_{234} - s_{23})^2 \,, \hskip 1 cm 
\Delta_3^{(1,4)} = (t_{234} - s_{34})^2 \,, \hskip 1 cm 
\Delta_3^{(1,5)} = s_{23}^2 \,, \cr
&
\Delta_3^{(2,3)} = (t_{123} - s_{56})^2\,, \hskip 1 cm 
\Delta_3^{(2,4)} = 
  s_{12}^2 + s_{34}^2 + s_{56}^2 - 2\,s_{12}\,s_{34} - 2\,s_{34}\,s_{56} 
       - 2\,s_{56}\,s_{12} \,,  \cr
&
\Delta_3^{(2,5)} = (t_{123} - s_{12})^2 \,, \hskip 1 cm 
\Delta_3^{(3,4)} = (t_{234} - s_{56})^2 \,, \hskip 1 cm 
\Delta_3^{(3,5)} = (t_{123} - s_{23})^2 \,, \cr
& 
\Delta_3^{(4,5)} = s_{12}^2\,. \cr}
\eqn\GramThree
$$
The bubble Gram determinants are trivial.  
As discussed in
section~\use\SpuriousPoleSection, the pentagon Gram determinant does
not appear at all in the cut calculations (assuming the integral
reductions are performed as discussed).  However, the triangle and box
Gram determinants do appear in the final results, when expressed in
terms of logarithms and dilogarithms.

Equations~(\use\IntRecursion) and (\use\cdef) show that another source of
spurious singularities is the determinant of the matrix $S_{ij}$.  
Again labeling the pentagon and box matrices by the canceled 
propagators we have 
$$
\det S = {1\over 16} s_{12} \, s_{23} \, s_{34} 
     \bigl(s_{23}\, s_{56} - t_{123}\, t_{234} \bigr)
\eqn\PentagonSDet
$$
$$
\eqalign{
&\det S^{(1)} = {1\over 16} \, s_{23}^2 \, s_{34}^2\,, \hskip .7 cm 
\det S^{(2)} = {1\over 16} \, s_{34}^2 \, t_{123}^2\,, \hskip .7 cm 
\det S^{(3)} = {1\over 16} \bigl(s_{23} s_{56} - t_{123} t_{234}\bigr)^2\,, \cr
&\det S^{(4)} = {1\over 16} \, s_{12}^2 \, t_{234}^2\,, \hskip .6 cm 
\det S^{(5)} = {1\over 16} \, s_{12}^2 \, s_{23}^2\,. \cr}
\eqn\BoxSDet
$$

Since the primitive amplitudes contain loop integrals where the
external legs follow the orderings 123564 and 125634 instead of 123456,
an additional set of possible spurious poles are obtained
from the above set via the relabelings of external legs:
$$
1234 \rightarrow 4123\,, \hskip 2 cm 1234 \rightarrow 3412 \,.
\anoneqn
$$

Observe the appearance of the combinations 
$$
\eqalign{
& s_{12} \, s_{56} - t_{123} \, t_{124}
     = - \spab4.{(1+2)}.3 \spab3.{(1+2)}.4 \,, \cr
& s_{23}\, s_{56} - t_{123} \, t_{234} 
     = - \spab1.{(2+3)}.4 \spab4.{(2+3)}.1 \,, \cr
& s_{34} \, s_{56} - t_{134} \, t_{234}
     = - \spab1.{(3+4)}.2 \spab2.{(3+4)}.1 \,, \cr}
\eqn\ssmttFactor
$$
which we have factored into products of spinor strings.  The above
factored form is quite useful in simplifying the cuts.  As discussed
in section~\use\StrategySubsection, we insert such `back-to-back'
factors into tensor integrals by hand in order to help simplify the
expression by forming appropriate spinor strings from which inverse
propagators can be extracted.

Although the three-mass triangle Gram determinant 
$\Delta_3 \equiv \Delta_3^{(2,4)}$
cannot be factored simply like eq.~(\use\ssmttFactor),
there are many kinematic combinations that vanish whenever 
$\Delta_3$ does.   Using the fact that the three four-vectors 
$k_1+k_2$, $k_3+k_4$ and $k_5+k_6$ all become proportional in the 
limit $\Delta_3 \to 0$,
and relations like $\delta_{56} = 2\,(k_1+k_2)\cdot(k_3+k_4)$,
it is easy to verify that the expressions
$$
\eqalign{
& \spab1.{(3+4)}.2 \,, \hskip1cm
 \spa1.4 \delta_{56} - 2 \, \spaa1.{23}.4 \,, \hskip1cm
 \spab1.2.4 \, \delta_{12} - \spab1.3.4 \, \delta_{34} \,, \cr
& \spaa1.{(1+2)(3+4)}.6 - \spaa1.{(3+4)(1+2)}.6 \,, \cr
}\eqn\covanishingfactors
$$
plus their complex conjugates and a variety of permutations of them, 
all vanish in this limit.  In the amplitudes we present,
the appearance of such combinations in the numerator of coefficients, 
when $\Delta_3$ appears in the denominator, alleviates the amplitudes'
spurious singularities as $\Delta_3 \to 0$.


\appendix{Integral Functions Appearing in Amplitudes}
\tagappendix\IntegralFunctionAppendix

We collect here the integral functions appearing in the text, which
contain all logarithms and dilogarithms present in the
amplitudes. Most of the functions have already appeared in previous
papers [\use\FiveGluon,\Zqqqq], but for completeness we list them all
in this appendix.  Except for the contribution of the top quark to
vacuum polarization contributions and to the axial-vector contribution
$A_6^\ax$, all internal lines are taken to be massless.  The following
functions arise from box integrals with one external mass:
$$
\eqalign{
  \Ll_0(r) &= {\ln(r)\over 1-r}\,,\hskip 10mm
  \Ll_1(r) = {\Ll_0(r)+1\over 1-r}\,, \cr
  \Ls_{-1}(r_1,r_2) &=
      \Li_2(1-r_1) + \Li_2(1-r_2) + \ln r_1\,\ln r_2 - {\pi^2\over6}\,,\cr
  \Ls_0(r_1,r_2) &=  {1\over (1-r_1-r_2)}\, \Ls_{-1}(r_1,r_2)\,,\cr
  \Ls_1(r_1,r_2) &= {1\over (1-r_1-r_2)}\, 
  \LB \Ls_0(r_1,r_2) + \Ll_0(r_1)+\Ll_0(r_2)\RB\,,\cr
}\eqn\Lsdef
$$
where the dilogarithm is
$$
\Li_2(x) = - \int_0^x dy \, {\ln(1-y) \over y}\,.
\eqn\Lidef
$$
The function $\Ls_{-1}$ is simply related to the scalar box integral 
with one external mass, evaluated in six space-time dimensions where 
it is infrared- and ultraviolet-finite.  
The above functions have the property that they are finite as 
as their denominators vanish.  Generalizations of the $\Ls_0$ and $\Ls_1$
functions to the case of box integrals with two or more external masses 
have been presented in ref.~[\use\CGM].

The box function analogous to $\Ls_{-1}$, but for two adjacent external
masses, is
$$
\eqalign{
  \Ls^{2{\rm m}h}_{-1}(s,t;m_1^2,m_2^2) &=
    -\Li_2\left(1-{m_1^2\over t}\right)
    -\Li_2\left(1-{m_2^2\over t}\right)
    -{1\over2}\ln^2\left({-s\over-t}\right)
    +{1\over2}\ln\left({-s\over-m_1^2}\right)
              \ln\left({-s\over-m_2^2}\right) \cr
&\quad
    + \biggl[ {1\over2} (s-m_1^2-m_2^2) + {m_1^2m_2^2\over t} \biggr]
        I_3^{3{\rm m}}(s,m_1^2,m_2^2) \,,\cr}
\eqn\Lstwomasshdef
$$
where $I_3^{3{\rm m}}$ is the three-mass scalar triangle integral.
This integral vanishes in the appropriate `back-to-back' kinematic
limit.
We also employ a version of this box function 
with $I_3^{3{\rm m}}$ removed,
$$
\Lsnew^{2{\rm m}h}_{-1}(s,t;m_1^2,m_2^2) =
    -\Li_2\left(1-{m_1^2\over t}\right)
    -\Li_2\left(1-{m_2^2\over t}\right)
    -{1\over2}\ln^2\left({-s\over-t}\right)
    +{1\over2}\ln\left({-s\over-m_1^2}\right)
              \ln\left({-s\over-m_2^2}\right)\,.
\eqn\Lsnewhdef
$$
The label `$h$' refers to the fact that this integral is relatively 
hard to obtain [\use\IntegralsLong].

The box function with two non-adjacent external masses is 
$$
\eqalign{
  \Ls^{2{\rm m}e}_{-1}(s,t;m_1^2,m_3^2) &= 
    -\Li_2\left(1-{m_1^2\over s}\right)                      
    -\Li_2\left(1-{m_1^2\over t}\right)                      
    -\Li_2\left(1-{m_3^2\over s}\right)                      
    -\Li_2\left(1-{m_3^2\over t}\right) \cr                     
&\quad
    +\Li_2\left(1-{m_1^2m_3^2\over st}\right)
    -{1\over2}\ln^2\left({-s\over-t}\right) \; . \cr}
\eqn\Lstwomassedef
$$
In this case the label `$e$' refers to the fact that this integral is
relatively easy to obtain.  
This integral vanishes as $s+t-m_1^2-m_3^2 \to 0$.

The kinematic region in which an $\Ls_{-1}$ function
vanishes always turns out to be related to the spinor product (or string) 
required to reduce the corresponding box integral:
$\spa2.4$ for $I_4^{(1)}$, $\spab3.{(1+2)}.4$ for $I_4^{(2)}$,
$\spa1.4$ for $I_4^{(3)}$, $\spab1.{(3+4)}.2$ for $I_4^{(4)}$,
and $\spa1.3$ for $I_4^{(5)}$ (or their complex conjugates).
Curiously, no single helicity amplitude contains both 
$\Ls^{2{\rm m}h}_{-1}$ and $\Ls^{2{\rm m}e}_{-1}$ functions
simultaneously.  

The analytic properties of these integrals are straightforward to obtain
from the prescription of adding a small positive imaginary part to each
invariant, $s_{ij} \to s_{ij}+i\pol$.  One expands the logarithmic
ratios, $\ln(r) \equiv \ln({-s\over-s'}) = \ln(-s)-\ln(-s')$, and then 
uses
$$
  \ln(-s-i\pol)\ =\ \ln|s| - i\pi\Theta(s)\,.
\anoneqn  
$$
where $\Theta(s)$ is the step function: $\Theta(s>0) = 1$ and 
$\Theta(s<0) = 0$. 
The imaginary part of the dilogarithm $\Li_2(1-r)$ is given in terms of
the logarithmic ratio,
$$
  \Im\Li_2(1-r)\ =\ -\ln(1-r)\ \Im\ln(r)\, . 
\anoneqn  
$$
For $r>0$ the real part of $\Li_2(1-r)$ is given directly by 
eq.~(\use\Lidef).  For $r<0$ one may use~[\use\Lewin]
$$
  \Re\Li_2(1-r)\ =\  {\pi^2\over6} - \ln|r| \ln|1-r| - \Re\Li_2(r)\,, 
\anoneqn
$$ 
with $\Re\Li_2(r)$ given by eq.~(\use\Lidef).

The analytic structure of
$I_3^{3{\rm m}}$ is more complicated~[\use\ThreeMassTriangle,%
\use\UDThreeMassTriangle,\use\IntegralsLong], 
and the numerical representation we use 
depends on the kinematics.  The integral is defined by
$$
I_3^{3{\rm m}}(s_{12},s_{34},s_{56})\ =\ 
\int_0^1 d^3a_i\, \delta(1-a_1-a_2-a_3)\, 
{1 \over -s_{12}a_1a_2-s_{34}a_2a_3-s_{56}a_3a_1}\ .
\eqn\threemassintdef
$$
This integral is symmetric under any permutation of its three arguments,
and acquires a minus sign when the signs of all three arguments are  
simultaneously reversed.  Therefore we only have to consider two cases,
\par\noindent
1. The Euclidean region $s_{12},s_{34},s_{56}<0$, which is related by 
the sign flip to the pure Minkowski region ($s_{12},s_{34},s_{56}>0$) 
relevant for $e^+\,e^-$ annihilation.  Here the imaginary part vanishes.
This region has two sub-cases, depending on the sign of the 
Gram determinant $\Delta_3(s_{12},s_{34},s_{56})$ defined in
eq.~(\use\deltaijdef):
\par
1a. $\Delta_3<0$,
\par
1b. $\Delta_3>0$.
\par\noindent
2. The mixed region $s_{12},s_{56}<0$, $s_{34}>0$, for which $\Delta_3$
is always positive.  

In region 1a one may use a symmetric representation found by Lu and 
Perez~[\use\ThreeMassTriangle], which is closely related to that 
given in ref.~[\use\IntegralsLong]:
$$
\eqalign{
I_3^{3{\rm m}} &= {2\over\rtmdelta} \left[  
  \Cl_2\left(\! 2\tan^{-1}
       \left( { \rtmdelta \over \d12 } \right) \! \right)
+ \Cl_2\left(\! 2\tan^{-1}
       \left( { \rtmdelta \over \d34 } \right) \! \right)
+ \Cl_2\left(\! 2\tan^{-1}
       \left( { \rtmdelta \over \d56 } \right) \! \right)
                                       \right] , \cr
}\eqn\LuPerezForm       
$$
where the $\d ij$ are defined in eq.~(\use\deltaijdef) and
the Clausen function $\Cl_2(x)$ is defined by
$$
\Cl_2(x)\ \equiv\ \sum_{n=1}^\infty { \sin(nx) \over n^2 }
 \ =\ -\int_0^x dt\, 
   \ln\bigl( \vert 2\sin(t/2) \vert \bigr).
\eqn\Cldef
$$

In regions 1b and 2 a convenient representation is given by 
Ussyukina and Davydychev~[\use\UDThreeMassTriangle],
$$
\eqalign{
I_3^{3{\rm m}}\ &=\ 
   -{1\over \rtdelta} \Re \left[  
    2 \left( \Li_2(-\rho x) + \Li_2(-\rho y) \right)
    + \ln(\rho x) \ln(\rho y)
    + \ln\left({y\over x}\right) 
         \ln\left({1+\rho y \over 1+\rho x}\right) 
    + {\pi^2\over3} \right] \cr  
&\qquad   
 - { i\pi \Theta(s_{34}) \over \rtdelta } 
   \ln\left( { (\d12+\rtdelta)(\d56+\rtdelta) \over
               (\d12-\rtdelta)(\d56-\rtdelta) } \right)\ , \cr
}\eqn\UDForm       
$$
where
$$
 x = {s_{12}\over s_{56}}, \quad y = {s_{34}\over s_{56}},
 \quad \rho = {2s_{56} \over \d56+\rtdelta}\,. 
\eqn\xydefs
$$

Finally, in the top quark contribution to $A_6^\ax$ the
combination $f(m_t)-f(m_b)$ appears, where $f(m)$ is the integral
$$
f(m;s_{12},s_{34},s_{56})\ = 
 \int_0^1 d^3a_i\, \delta(1-a_1-a_2-a_3)\, 
{a_2a_3 \over m^2-s_{12}a_1a_2-s_{34}a_2a_3-s_{56}a_3a_1}\ .
\eqn\fmdef
$$
This integral is complicated for arbitrary mass $m$; however,
the large and small mass limits of it suffice for $m_t$ and $m_b$
respectively.
For $m=m_t$ we simply Taylor expand the integrand in $1/m$;
for $m=m_b$ we set $m_b$ to zero, and reduce $f(0)$ to a linear
combination of the massless scalar triangle integral $I_3^{3{\rm m}}$ 
given above, logarithms and rational functions.
We get
$$
\eqalign{
&f(m_t;s_{12},s_{34},s_{56}) =  
 {1 \over 24 m_t^2 } + {(2s_{34}+s_{12}+s_{56}) \over 360 m_t^4 } 
 + \cdots\,, \cr
&f(0;s_{12},s_{34},s_{56}) =  
   \left( {3 s_{34} \d34 \over \del_3^2} - {1 \over \del_3} \right)
     s_{12} s_{56} I_3^{3{\rm m}}(s_{12},s_{34},s_{56}) 
  + \left( {3 s_{56} \d56 \over \del_3^2} - {1\over2\del_3} \right)
     s_{12} \ln\left({-s_{12}\over-s_{34}}\right)\cr
&\hskip 3 cm 
  + \left( {3 s_{12} \d12 \over \del_3^2} - {1\over2\del_3} \right)
     s_{56} \ln\left({-s_{56}\over-s_{34}}\right) 
  - { \d34 \over 2\del_3 }\,. \cr  
}\eqn\fmexplicit
$$
Note that the limit where one of the invariants vanishes,
$$
f(m_t;0,t_{123},s_{56}) - f(0;0,t_{123},s_{56}) = 
- {1\over2} 
  { \Ll_1\Bigl({-t_{123} \over -s_{56}} \Bigr) \over s_{56} }
          + {1\over24\,\msq} \,,
\eqn\fmlimit
$$
appears in two amplitudes, eqs.~(\use\Faxslffpp) and (\use\Faxslffpm).


\appendix{Spinor Identities for Simplifying Spurious Poles}
\tagappendix\SpinorIdentityAppendix

In order to simplify or remove spurious poles a number of spinor
identities are of great utility. In this appendix we collect these
identities. As discussed in section~\use\SpuriousPoleSection\ some of
the spurious poles can be completely removed from the amplitudes, but
others are an inherent part of the amplitude when it is 
expressed in terms of logarithms and dilogarithms.

The most difficult spurious poles to simplify are those appearing as
coefficients of three external mass triangle integrals and associated
logarithms.  In simplifying these spurious poles, it is useful to have
identities for rewriting commonly occurring expressions to make one
behavior or another more manifest.  The following identities are
used in different `directions', depending on the situation.

Particularly important spurious singularities are where $\Delta_3$,
$\spab3.{(1+2)}.4$ and $t_{123}-s_{12}$ (and the label-permuted objects) 
vanish.
In order to obtain relatively compact expressions it is essential to 
simplify these spurious poles.

One important quantity is $(t_{123} \d34 + 2 s_{12} s_{56})$, which
appears as a factor in the coefficient of 
$I_3^{3{\rm m}}(s_{12},s_{34},s_{56})$ within the function
$\Ls^{2{\rm m}h}_{-1}(s_{34},t_{123};s_{12},s_{56})$ defined in 
eq.~(\use\Lstwomasshdef).  If we shift from using the function 
$\Ls^{2{\rm m}h}_{-1}$ in an amplitude, to $\Lsnew^{2{\rm m}h}_{-1}$ 
instead (or vice-versa), then we also have to shift the 
coefficient of the three-mass triangle by a quantity which contains this 
factor, as well as the spurious singularity factor $\spab3.{(1+2)}.4$
typically associated with the coefficient of 
$\Ls^{2{\rm m}h}_{-1}(s_{34},t_{123};s_{12},s_{56})$.
The identities
$$ 
\eqalign{
& t_{123} \, \d34 + 2 \, s_{12} \, s_{56} = 
  - t_{123} (t_{123}-t_{124}) 
  - 2 \, \spab3.{(1+2)}.4 \spab4.{(1+2)}.3\, , \cr
& t_{123} \, \d34 + 2 \,s_{12} \, s_{56} = 
   \hf \, \d34 (t_{123}-t_{124})  - \hf \, \del_3\, , \cr 
& \del_3 = - (t_{123} \, \d34 + 2\, s_{12} \, s_{56})
          - (t_{124} \, \d34 + 2\, s_{12} \, s_{56}) \cr  
& \del_3 =(\d34)^2 - 4 \, s_{12} \, s_{56}\, , \cr
& \del_3 = (t_{123}-t_{124})^2 
    + 4 \, \spab3.{(1+2)}.4 \spab4.{(1+2)}.3\, , \cr
}    
\eqn\tscombidents
$$
are useful in manipulating leading $1/\spab3.{(1+2)}.4$ singularities
in order to remove, for example, `extra' $1/t_{123}$ poles.  Also note
that
$$
 \d34 + t_{123} + t_{124} = 0 \, , \qquad 
 \d12 + t_{234} + t_{134} = 0 \, , \qquad 
 \d12 +   \d34 + 2 s_{56} = 0 \, .
\eqn\dijidents
$$

The identities
$$
\eqalign{ 
& s_{12}\, \d12 + s_{34}\, \d34 + s_{56}\, \d56 = \del_3 \, , \cr
& (\d12)^2 = 4 \, s_{34} \, s_{56} + \del_3 \, , \cr
& s_{56} \, \d56 = \hf \, \d12\, \d34 + \hf \, \del_3 \, , \cr
}     
\eqn\delthreeidents
$$
are useful in simplifying the $1/\del_3$ poles.  Also useful in this
regard are the identities,
$$
\eqalign{
& \d12 \spa3.5\spb4.6 = \spa3.4 {\spb4.6}^2 \spa6.5
                        + \spb4.3 {\spa3.5}^2 \spb5.6
    + \spab3.{(1+2)}.4 \spab5.{(3+4)}.6\, , \cr                    
& \d34 \spa2.5\spb1.6 = \spa2.1 {\spb1.6}^2 \spa6.5
                        + \spb1.2 {\spa2.5}^2 \spb5.6
    - \spab2.{(3+4)}.1 \spab5.{(3+4)}.6\, , \cr                    
& \d56 \spa3.2\spb4.1 = \spa3.4 {\spb4.1}^2 \spa1.2
                        + \spb4.3 {\spa3.2}^2 \spb2.1
    - \spab3.{(1+2)}.4 \spab2.{(3+4)}.1\, . \cr}
\eqn\spinordtidents
$$
where we show a few permutations. A consequence of the last identity is
$$
\eqalign{
 (\del_3 + 4 \, s_{12} \, s_{56}) \spa2.5\spb1.6  & = 
 \d34 \, (\spa2.1 {\spb1.6}^2 \spa6.5 + \spb1.2 {\spa2.5}^2 \spb5.6)
 - \d34 \, \spab2.{(3+4)}.1 \spab5.{(3+4)}.6\, . \cr 
}
\eqn\lastident
$$
Another related identity is (see also eq.~(\use\bbidentone))
$$
\spb6.4\spa3.2 t_{123} = 
\spab2.{(1+3)}.6 \spab3.{(1+2)}.4 + \spab3.{(1+2)}.6 \spab2.{(5+6)}.4 \,.
\anoneqn
$$



\appendix{$e^+\,e^- \to \qb q g$ Helicity Amplitudes }
\tagappendix\qqgAppendix

In this appendix we present the $e^+\,e^- \to \qb q g$ primitive
amplitudes, which appear in the collinear limits discussed in
section~\use\FactorizationSection.  These amplitudes were first
calculated in the spinor helicity formalism by Giele and
Glover~[\use\GieleGlover].  We present these amplitudes in the same
primitive amplitude format as for the $e^+\,e^- \to \qb q g g$
amplitudes discussed in section~\use\PrimitiveAmplitudesSubsection,
including a separation into cut-constructible and scalar pieces.
The parent diagrams associated with these amplitudes are depicted in
\fig\PrimitiveDiagramsZqqgFigure .  
Using parity and charge conjugation invariance, there is only one
independent helicity configuration for each of two independent
color configurations.

These amplitudes appear in the
singular collinear limits of the $e^+\,e^- \to \qb q gg$ and $e^+\,e^-
\to \qb q \Qb Q$ amplitudes, except for the $k_5 \parallel k_6$
channel where the lepton pair becomes collinear.  For the primitive
amplitudes with closed fermion loops, the amplitudes $e^+\,e^-
\rightarrow ggg$ also appear in the collinear limits.  Although we do
not present these amplitudes here, they may be obtained from
ref.~[\use\BijGlover] after using the spinor helicity representation
for the polarization vectors. One may also obtain the vector coupling
results from the $\bar q q ggg$ amplitudes of ref.~[\use\Fermion]
after summing over permutations of the quark lines, which effectively
removes their color charge.

%
\LoadFigure\PrimitiveDiagramsZqqgFigure
{\baselineskip 13 pt
\noindent\narrower\ninerm
Parent diagrams for the various two quark and one gluon primitive
amplitudes.  Straight lines represent fermions, curly lines gluons,
and wavy lines a vector boson ($\gamma^*$ or $Z$). 
}  {\epsfxsize 3.4 truein}{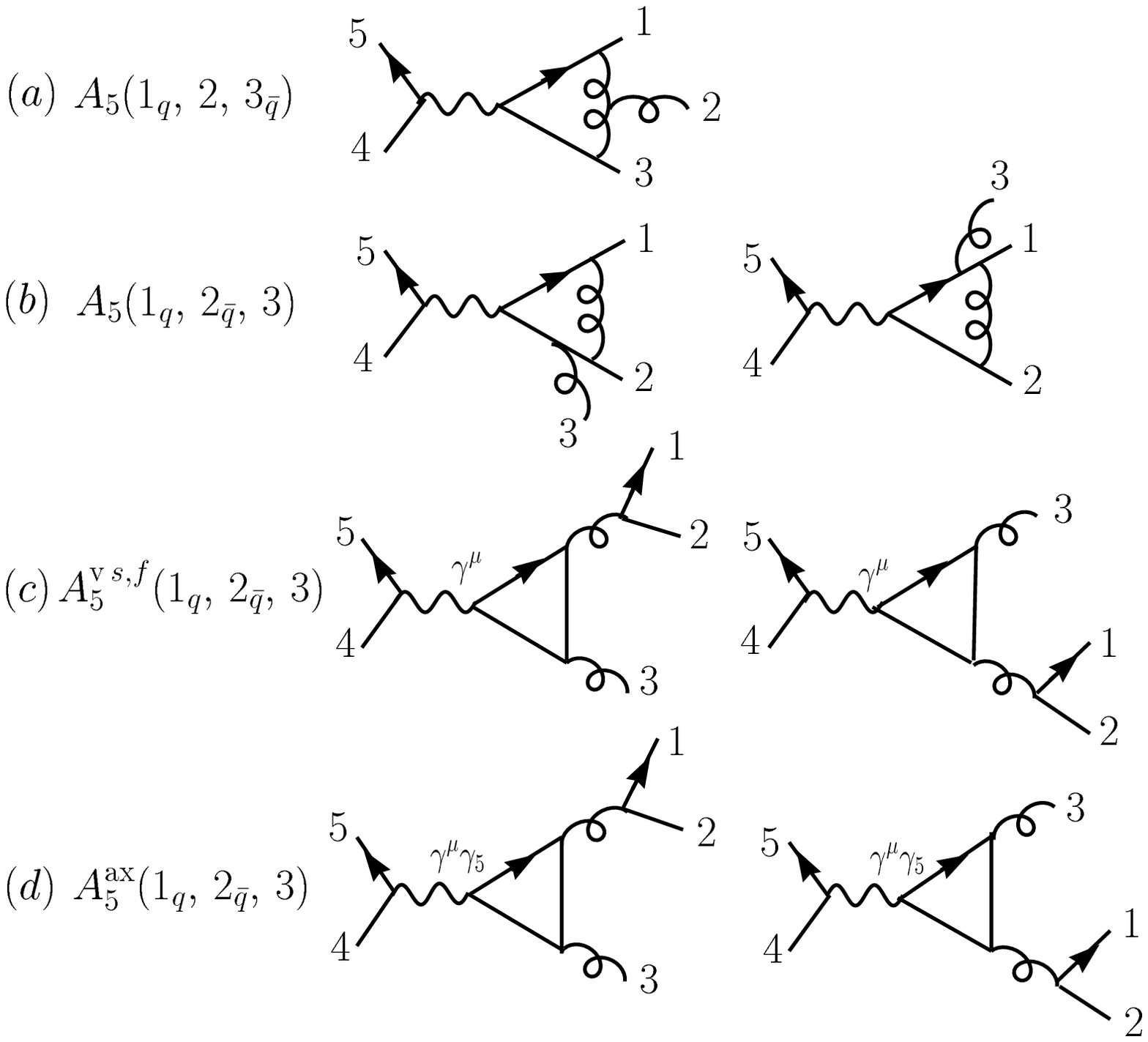}{}

First consider the primitive amplitude 
$A_5(1_q^+ , 2^+ , 3_\qb^- , 4_\eb^- , 5_e^+)$, 
which contributes at leading order in $1/N_c$.  
The tree amplitude in this case is
$$
A_5^\tree = 
 - i\, {\spa3.4^2 \over \spa1.2\spa2.3\spa4.5 } 
\,.
\eqn\GGleadtree
$$
The results for the cut-constructible and scalar pieces are 
$$
V^\gf =  
  -{1\over\e^2} \left( 
       \left( {\mu^2\over -s_{12}} \right)^\e
     + \left( {\mu^2\over -s_{23}} \right)^\e \, \right)
     - {2\over\e} \left( {\mu^2\over -s_{23}} \right)^\e  -4
\,,
\eqn\VLGGleadgf
$$
$$
F^\gf =  
   {\spa3.4^2 \over \spa1.2\spa2.3\spa4.5 }
 \biggl[
{\ts \Ls_{-1}\Bigl( {-s_{12}\over -s_{45}},{-s_{23}\over -s_{45}} \Bigr)}
 - 2 \, { \spaa3.{15}.4 \over \spa3.4 } 
      { \Ll_0\Bigl( {-s_{23}\over -s_{45}} \Bigr) \over s_{45} } \biggr]
\,,
\eqn\FLGGleadgf
$$
\vskip0.2cm
$$
V^\sc =  
  {1\over2\e} \left( {\mu^2\over -s_{23}} \right)^\e + 1 
\,,
\eqn\VLGGleadsc
$$
$$
F^\sc =  
   {\spa3.4 \spaa3.{15}.4 \over \spa1.2\spa2.3\spa4.5} \,
      { \Ll_0\Bigl( {-s_{23}\over -s_{45}} \Bigr) \over s_{45}}
  + {1\over2}\, { \spaa3.{15}.4^2 \over \spa1.2\spa2.3\spa4.5} \,
      { \Ll_1\Bigl( {-s_{23}\over -s_{45}} \Bigr) \over s_{45}^2 } 
\,.
\eqn\FLGGleadsc
$$

Similarly, for the primitive amplitude 
$A_6(1_q^+ , 2_\qb^- , 3^+ , 4_\eb^- , 5_e^+)$, 
which contributes only at subleading order in $N_c$, the tree amplitude is
$$
A_5^\tree =
 i {\spa2.4^2 \over \spa2.3\spa3.1\spa4.5}
\,.
\eqn\GGsubleadtree
$$
The cut-constructible and scalar contributions are
$$
V^\gf =  
  - {1\over\e^2} \left( {\mu^2\over -s_{12}} \right)^\e
  - {2\over\e} \left( {\mu^2\over -s_{45}} \right)^\e - 4 
\,,
\eqn\VLGGsubleadgf
$$
$$
\eqalign{
F^\gf & =  
 - {\spa2.4^2 \over \spa2.3\spa3.1\spa4.5}
 \, {\ts \Ls_{-1}\Bigl( {-s_{12}\over -s_{45}},{-s_{13}\over -s_{45}} \Bigr)}
%
  + { \spa2.4 (\spa1.2 \spa3.4 - \spa1.4 \spa2.3)
      \over \spa2.3 \spa1.3^2 \spa4.5 } \, 
   {\ts \Ls_{-1}\Bigl( {-s_{12}\over -s_{45}},{-s_{23}\over -s_{45}} \Bigr) }
\nl\hskip 1 cm 
+ 2 \, { \spb1.3\spa1.4\spa2.4 \over \spa1.3\spa4.5 } 
  \, { \Ll_0\Bigl( {-s_{23}\over -s_{45}} \Bigr) \over s_{45} } 
\,,
\cr}
\eqn\FLGGsubleadgf
$$
\vskip0.2cm
$$
V^\sc =  
  {1\over2\e} \left( {\mu^2\over -s_{45}} \right)^\e + {1\over2} 
\,,
\eqn\VLGGsubleadsc
$$
$$
\eqalign{
F^\sc  & = 
 { \spa1.4^2\spa2.3 \over \spa1.3^3\spa4.5 }\,
 {\ts \Ls_{-1}\Bigl( {-s_{12}\over -s_{45}},{-s_{23}\over -s_{45}} \Bigr)} 
- {1\over2} \,{ \spab4.1.3^2\spa2.3 \over \spa1.3\spa4.5 } \,
     { \Ll_1\Bigl( {-s_{45}\over -s_{23}} \Bigr) \over s_{23}^2 }   
+ { \spa1.4^2 \spab2.3.1 \over \spa1.3^2\spa4.5 } \,
     { \Ll_0\Bigl( {-s_{45}\over -s_{23}} \Bigr) \over s_{23} }  \cr 
& \hskip .5 cm
- { \spab2.1.3\spab4.3.5 \over \spa1.3} \,
     { \Ll_1\Bigl( {-s_{45}\over -s_{12}} \Bigr) \over s_{12}^2 }   
- { \spaa2.{13}.4 \spa1.4 \over \spa1.3^2\spa4.5} \,
     { \Ll_0\Bigl( {-s_{45}\over -s_{12}} \Bigr) \over s_{12} }  
- {1\over2}\, {\spb3.5  (\spb1.3\spb2.5+\spb2.3\spb1.5)
             \over \spb1.2\spb2.3\spa1.3\spb4.5 } 
\,.
\cr} 
\eqn\FLGGsubleadsc
$$

The contributions with a closed fermion or scalar loop are rather
simple.  By Furry's theorem, the cases with vector-like couplings
vanish, so $A_5^{{\rm v}\, s,f} (1_q, 2_\qb, 3) = 0$.  The axial
contribution $A_5^\ax(1_q^+, 2_\qb^-, 3^+)$ are also 
simple~[\use\HHKY] and are given by 
$$
F^\ax = 
 - \spb5.3\spb3.1\spa2.4  
     \biggl( {\Ll_1 \Bigl( {-s_{12} \over -s_{45}} \Bigr) \over s_{45}^2} 
       - {1\over 12 s_{45} \, \msq}  \biggr)  
\,.
\eqn\FLGGsubleadax
$$


\listrefs
\listfigs

\end